\documentclass[prd,twocolumn,preprintnumbers,showpacs,nofootinbib,superscriptaddress,notitlepage,floatfix,longbibliography]{revtex4-1}
\usepackage{hyperref}
\usepackage{graphicx}
\usepackage{amsmath}
\usepackage{amssymb}
\usepackage{slashed}
\usepackage{amsfonts}
\usepackage{xcolor}
\usepackage{multirow}
\usepackage[caption=false]{subfig}
\usepackage{array}
\usepackage{mwe}
\usepackage{dsfont}

\usepackage[utf8]{inputenc}
\usepackage[normalem]{ulem}

\newcommand{\be}{\begin{equation}}
\newcommand{\ee}{\end{equation}}
\newcommand{\bea}{\begin{eqnarray}}
\newcommand{\eea}{\end{eqnarray}}
\newcommand{\ba}{\begin{eqnarray}}
\newcommand{\ea}{\end{eqnarray}}

\begin{document}

\title{Glue in hadrons at medium resolution\\
and the QCD instanton vacuum}

\author{Wei-Yang Liu}
\email{wei-yang.liu@stonybrook.edu}
\author{Edward Shuryak}
\email{edward.shuryak@stonybrook.edu}
\author{Ismail Zahed}
\email{ismail.zahed@stonybrook.edu}
\affiliation{Center for Nuclear Theory, Department of Physics and Astronomy, Stony Brook University, Stony Brook, New York 11794–3800, USA}

\date{\today}

\begin{abstract}
We discuss a general framework for the evaluation of the gluonic form factors in light hadrons at low momentum transfer, in the QCD instanton vacuum. At  medium resolution of the order of the inverse mean instanton size, the  glue is mostly localized in single or pair of pseudoparticles, and globally constrained by the fluctuations of their topological charges. These pseudoparticles
trap light quarks, giving rise to emerging
multiflavor 't Hooft interactions.
We explicitly evaluate
the gluonic scalar, pseudoscalar, energy-momentum
tensor (EMT), and the leading C-odd and C-even three gluons hadronic form factors,
at next to leading order (NLO) in the instanton density, including molecular clusters of like and unlike instantons. We use the results for the EMT to address the contribution of the gluons in Ji$^\prime$s mass and spin sum rules, at low resolution. When evolved, our  results for the mass and spin composition of the nucleon, are shown to be in good agreement with the recently reported lattice results at higher resolution.
\end{abstract}

\maketitle

\section{Introduction}
\label{sec_intro}
We will start by explaining the title, and by recalling  the terminilogy to be used. In general, the  form factors 
are Fourier transforms of distributions of certain charges. The standard example is the electromagnetic form factors describing
the electromagnetic charge and current distributions of nucleons and nuclei. However, 
form factors can be defined for any operator, and we will focus below on gluonic probes, like $G^2_{\mu\nu}$. 

\begin{figure}
  \centering
    \includegraphics[width=8cm]{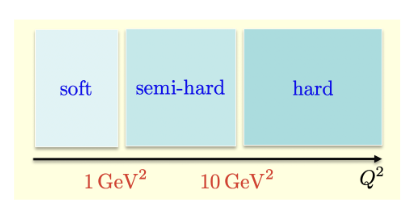}
    \caption{The soft, semi-hard and hard $Q^2$ regions characterizing different regimes of the hadronic form factors.}
    \label{fig:REGIMES}
\end{figure}

The form factors are functions of the (space-like) momentum transfer $Q$, and reflect on what the probe 'sees', if it has spatial resolution $\sim 1/Q$. 
In Fig.\ref{fig:REGIMES} we schematically identify three resolution regimes for QCD probes: hard, semi-hard and soft. We now recall the meaning  of this terminology. 

The {\bf hard} regime is given by perturbative Feynman diagrams with a number of gluon propagators $(1/Q^2)^n$. Physically it corresponds to a hadron in a very compressed form, so that the quarks are inside the perturbative Coulomb fields of each other.
In this regime,  there are no dimensional quantities involved as QCD is asymptotically free,  and the power $n$ can be obtained just by dimensional considerations. (Except for certain special cases, e.g. spin flipping amplitudes, where quark masses must also appear.)

This paper is about the 
 {\bf semi-hard} regime dominated by nonperturbative vacuum fluctuations.The latter  are assumed to be semiclassical
pseudoparticles, instantons or their pairs.  
The ensuing form factors are generically of the form  $\beta_i(Q \rho)\cdot Q^n$ with $\rho$ being a typical instanton size. At large $Q\rho\gg 1$, the form factors decrease exponentially, but they have different shapes in the $Q\rho\sim O( 1)$ region. 

In the {\bf soft}  regime, the resolution is too poor to resolve individual instantons. This regime is dominated by 'chiral' phenomena  related with the appearance of a quark condensate and a  'pion cloud'  surrounding  most hadrons. These  chiral properties 
can be described by either 
summing multi-instanton chains, or by phenomenological chiral Lagrangians. In this regime, the form factors are related with hadronic parameters,
e.g. for the scalar channel with the $2m_\pi$ threshold or  the $\sigma$ meson mass.

Different QCD operators naturally see different distributions.
However, because of the self-duality of the instantons, the scalar $O_S=G^2_{\mu\nu}$ and the pseudoscalar $O_P=G_{\mu\nu}\tilde G_{\mu\nu}$ operators see the same well-localized
 spherical ball of the shape
\be \big(G_{a\mu\nu}^{inst}\big)^2={1 \over g^2} {192 \rho^4\over (x^2+\rho^2)^4} \ee
if an instanton is alone, or a similar but more complicated  instanton is 
paired. It is important to note that
it is a spot of rather strong field, of a quite small size.

Today the mainstream of a first-principle theory of non-perturbative phenomena is lattice QCD. Therefore, it is important from the start to recall how these  three regimes are revealed on the lattice as well. Gauge lattice configurations   
are dominated by 'gluons', waves with wavelength $\sim a$, the lattice UV cutoff. Yet e.g. the {\em gradient flow} procedure (cooling) 
can be used to remove these gluons, and reveal genuine non-perturbative fields. After few coolings, the gluonic landscape looks like a (rather dense) ensemble of strongly correlated instanton-anti-instanton pairs. After further flow time or cooling, these pairs
get annihilated, leading to a (rather dilute) ensemble of individual instantons and antiinstantons,
which can withstand even 'deep cooling'. For a detailed description of this procedure see e.g. \cite{Athenodorou:2018jwu}.

For completeness, we briefly recall the history of the 'instanton vacuum'. The QCD vacuum consists 
 of semi-classical and topological instantons and anti-instantons (pseudo-particles) \cite{Belavin:1975fg}, that are described by   a gluon moduli (quenched) with additional determinantal interactions (unquenched)~\cite{Callan:1976je, Callan:1977gz}. 
In the 'instanton liquid model' (ILM) one of us  suggested that these pseudo-particles have mean size and density~\cite{Shuryak:1981ff}
\be \rho\approx \frac 13 \,fm, \,\,\,\,\,\, n_{I+A}\approx 1\, fm^{-4} \ee 
which gives the dimensionless packing fraction $\kappa=n_{I+A}\pi^2\rho^4\approx0.1$. This view of the QCD vacuum is supported by a large body of analytical and numerical results related to chiral symmetry breakings, including many aspects of the pions and the anomalous $\eta'$ mesons, for a review see e.g. \cite{Schafer:1996wv}. 

The QCD instanton vacuum  does not strictly confine at large distances, although the instanton induced central potential between heavy quarks is linear within a fm~\cite{Shuryak:2021fsu}, before turning to a constant at large distances. However, it breaks spontaneously chiral symmetry by trapping and delocalizing massless left-handed or right-handed quarks, in a narrow zero-mode-zone of about 100 MeV around the zero virtuality line. This mechanism is at the origin of mass from no mass, and plays a central role in the composition and structure of the light hadrons.

Earlier studies involving gluonic operators in the ILM, included  studies of their vacuum point-to-point correlators~\cite{Schafer:1994fd,Kacir:1996qn,Schafer:2004ke}. In particular,  the scalar $G^2_{\mu\nu}$ and pseudoscalar 
$G_{\mu\nu}\tilde G_{\mu\nu} $ operators were
found to receive large non-perturbative contributions. In contrast, the stress tensor correlator (also quadratic in the gauge field) 
was observed to be not affected. 

\begin{figure}
    \centering
    \includegraphics[scale=0.4]{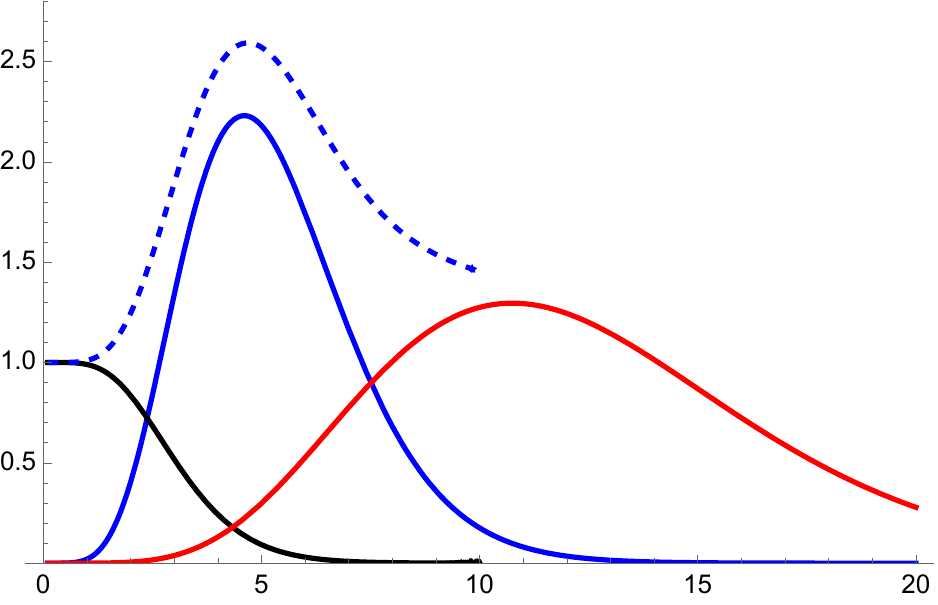}
    \caption{ Spectral function in coordinate space in units of $\rm GeV^{-1}$,
    with the perturbative contribution (black-solid line), the scalar glueball contribution (blue-solid line), the scalar meson contribution (red-solid line)
    and their sum (dashed-blue line).}
    \label{fig_G2_corr_vacX}
\end{figure}

In Fig.\ref{fig_G2_corr_vacX} we show  the ratio of the vacuum
correlator of two scalar $G^2$ operators separated by (Euclidean)
distance $x$ (in GeV$^{-1}$), normalized by the  leading perturbative 
contribution
\be \Pi_{GG}^{0}(x)={384 g^4 \over \pi^4 x^8}\ee
In the coordinate representation the``hard", "semi-hard" and "soft"
regimes appear in the opposite order (left to right) in comparison to
momentum representation in Fig.~\ref{fig:REGIMES}. 
Using estimates
from the spectral representation, we show the scalar glueball
contribution (blue-solid line), the scalar sigma contribution (red-solid line),
the perturbative contribution (black-solid line)  and their sum (blue-dashed line).
The glueball mass and coupling are $m_{0^{++}}=1.5\, \rm GeV,\lambda_{0^{++}}=17.2\, \rm GeV^3$~\cite{Schafer:1994fd}, while the sigma mass and mixing are 
 $m_\sigma=0.6\,\rm  GeV, \lambda_\sigma/\lambda_{0^{++}}\approx 0.066$~\cite{Iatrakis:2015rga}.

In Fig.\ref{fig_G2_corr_vacX} all three regimes are in display, with the
full spectral function (blue-dashed line) versus the distance in $\rm GeV^{-1}$ 
units where 5 is 1 fm. The "hard" regime is dominant at short distances (black-solid line) with a plateau, where perturbation theory holds. The rise in the spectral function for separations $x=2-5\, \rm GeV^{-1}\approx 0.4-1\, \rm fm$ 
 correspond to the ``semi-hard" regime. The  ``sigma halo" at large separation  $x>5\, \rm GeV^{-1}\approx 1 \, \rm fm $ correspond to the  ``soft" regime, familiar from e.g. nuclear forces at large distances. in fact, the transition
 between the "hard" and "semi-hard" regimes is even stronger in the ILM\cite{Schafer:1996wv}, as we will briefly review below.

Unlike chiral effects, phenomena related to the gauge fields are not only due to well-separated pseudoparticles in the dilute ILM,
but also to fields in 'incomplete tunneling' or instanton-anti-instanton
molecules. In our papers such as \cite{Shuryak:2018fjr}
(devoted to the electromagnetic pion formfactor) and the first paper of these series \cite{Shuryak:2021fsu} (potentials in quarkonia) we have found that the 'molecular'
effects can even be dominant. In this respect, we reformulated the instanton vacuum to that of a 'dense instanton ensemble', where this
contribution was included.

In this paper we will evaluate
gauge-invariant matrix elements of select gluonic operators in hadrons. Naturally, their definitions  follow from QCD factorisation, of inclusive processes (e.g.  deep inelastic scattering, DIS) or semi-inclusive processes (e.g. heavy 
quark pair production). They also emerge in the effective description of 
heavy quarkonia, and in the standard model when the electro-weak and heavy degrees of freedom  are integrated out. Physically, they describe how the glue is distributed inside
hadrons, mesons or baryons, and therefore carry important
insights into their structure.

Unfortunately, such gluonic matrix elements  are notoriously hard to  measure experimentally, owing to the confining character of QCD in the infrared. For a recent
investgation of the importance of the glue 
in semi-inclusive heavy meson production at threshold, see~\cite{Duran:2022xag,Meziani:2024cke} (and references therein).
In so far, most of the evaluations are theoretical. First principle QCD lattice simulations provide a useful framework for their evaluation, but the intricacies of renormalization with
operator mixing proves often to be quite formidable. 

While the glueball spectra in gluodynamics  are
known rather well, there are no  appropriate 'constituent gluon models'. This is not surprising, given  the  intermittent nature of the vacuum gauge fields.
Therefore,  the  explicit glue 
will not be treated as separate degrees of freedom (gluons), but as solitonic gauge fields or pseudoparticles  of semi-classical nature.  The QCD instanton vacuum captures the essentials of these pseudoparticles, 
and provide a well defined framework for the derivation of the
gluonic matrix elements, in vacuum and in hadrons. 

The matrix elements of the glue in hadrons using the QCD instanton vacuum, was first used in~\cite{Diakonov:1995qy} for few gluonic operators in forward hadronic matrix elements, and in~\cite{Kacir:1996qn} for both fermionic and gluonic  form factors.
We will
show below how the methods of \cite{Diakonov:1995qy}
can be generalized for non-forward matrix elements,  in agreement with~\cite{Kacir:1996qn}. In the process, we will analyze a large class of gluonic operators, some of which are of relevance
to select semi-inclusive processes, following factorization. Last but not least,  we will show how to include  the contributions from pairs  of like and unlike instanton  molecules. 

All matrix elements should be understood as normalized 
at the 'intermediate scale' $\mu$ fixed by the instanton size  
\be 
\mu^2\sim {1 \over \rho^2} \approx (0.6 \, \mathrm{GeV})^2 
\ee
Sometimes, this scale  appears with numerical factors, reaching
$\mu^2\sim 1\, {\rm GeV}^2 $. It should not be confused with the
smaller 'chiral' scales associated with the pion mass, or the much higher scale $\mu^2> 4\, \mathrm{GeV}^2$ 
at which perturbative evolution can be  used.



The organization of the paper is as follows: 
in section~\ref{SUMMARYX} we summarize our results 
for a number of gluonic form factors.
In section~\ref{sec_instantons} we review some aspects of the QCD instanton vacuum, and the role played by the zero modes in the spontaneous breaking of chiral symmetry. We briefly note the possible interplay of the P-vortices with the topological pseudoparticles in the
QCD vacuum. In section~\ref{sec_eff} we detail the emergence of the effective quark and gluon interactions in the QCD instanton vacuum, in the single instanton approxiation. In section~\ref{sec_hadff} we show how to use the sum ansatz in general, to evaluate the pertinent hadronic form factors at low momentum transfer. In section~\ref{sec_scalar} we detail the derivation of the 
gluonic scalar form factor at next to leading order (NLO) in the instanton density, and including instanton molecular configurations. In section~\ref{sec_pseudoscalar} we derive the pseudoscalar gluonic form factor at NLO in the instanton density, where the fluctuations in the topological charge are also included.  In section~\ref{sec_Ceven} we derive the form factor for a general C-odd and dimension-6 gluonic operator,
which appears as a leading twist-3 operator in diffractive production of heavy
pseudoscalar mesons. Its C-even and dimension-6 gluonic operator form factor is
also analysed in section~\ref{sec_Ceven}.
In section~\ref{sec_gravitationalff} we analyze the QCD gravitational form factor (GFF) in a hadronic state, in the context of the QCD instanton vacuum. The GFF is split into a traceless and tracefull part, each of which are evaluated at NLO in the instanton density. In section~\ref{sec_mass} we show that
all hadronic squared masses satisfy the scale anomaly identity in the QCD instanton vacuum. In section~\ref{sec_mass} we use Ji's mass sum rule, to detail
the various contributions of the quark and gluons in the QCD instanton vacuum at low resolution. In section~\ref{sec_sum} we extend this budget analysis to Ji's spin sum rule for the nucleon. Our conclusions are in section~\ref{sec_conclusion}. In Appendix~\ref{App:effective_Langrangian}
we outline the general structure of the emergent multi-flavor effective Lagrangian with constitutive gluons. 
In Appendix~\ref{App:tHooft} we show how the multi-flavor
interactions yield a massive eta' and a massless pion for the
case of two light flavors, and extract the scalar and pseudoscalar singlet couplings to the emerging constituent quarks. 
In Appendix~\ref{App:sing_inst} we briefly 
go over the instanton field and field strength in singular gauge, and detail
its color moduli. In Appendix~\ref{App:pairs} we detail how the color averaging 
is performed for the instanton pairs and molecules. In Appendix~\ref{App:tails} we suggest that the emergent effective vertices with LSZ reduced gluons, can
be used in Feynman graphs that include small size instantons only. In Appendix~\ref{App:grand} we briefly show how the canonical ensemble of pseudoparticles can be extended to a grand canonical ensemble, to account for the
fluctuations in their number which captures globally the scale and $U(1)$ anomalies. In Appendix~\ref{App:average} we summarize some useful identities  for averaging over the color moduli.

\begin{figure*}
    \centering
       \includegraphics[scale=0.4]{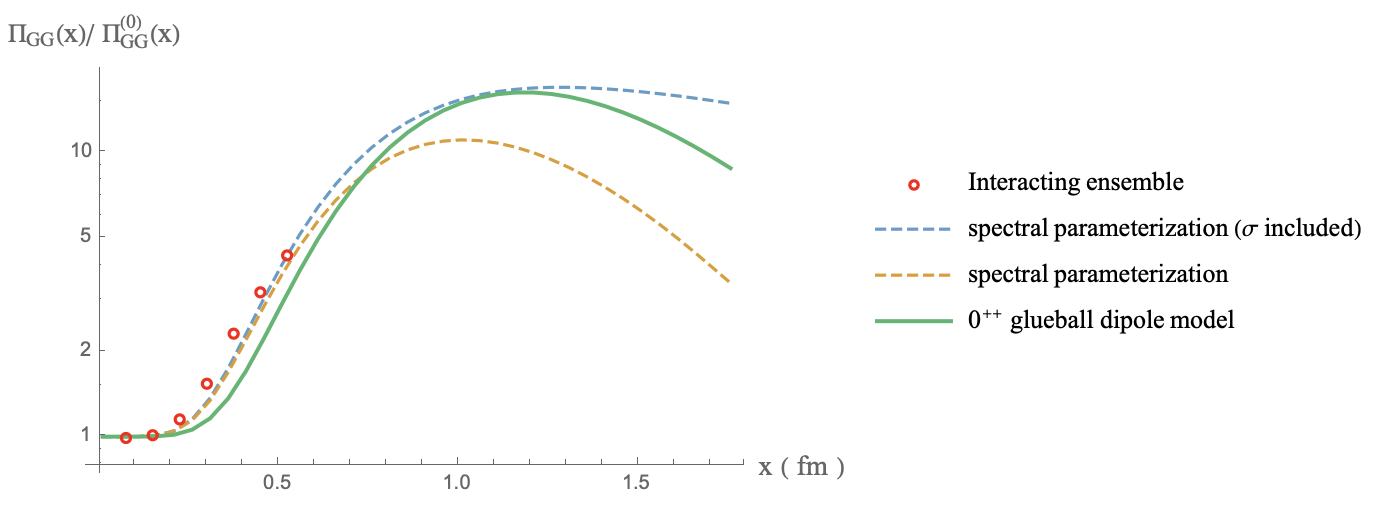}
    \caption{ Spectral function from the ILM (open red circles)~\cite{Schafer:1994fd}, 
    normalized by the perturbative contribution (\ref{})  in coordinate space. The
    comparison is to the empirical spectral parametrization  (\ref{FULLX}) with the scalar sigma (blue-dashed line) and without the scalar sigma (orange-dashed line),
     and the $0^{++}$ glueball dipole parametrization (\ref{DIPOLEX}) (green-solid line) }
    \label{fig_G2_corr_vac}
\end{figure*}

\section{Summary of the results}
\label{SUMMARYX}
The theory of semi-inclusive and exclusive QCD processes with large momentum transfer $Q^2$, is rooted in two assumptions (see~\cite{Brodsky:1973kr,Radyushkin:1977gp} for early work)
\\
\\
(i) The factorization of a soft and hard parts of the processes, appearing
 when $Q^2$ is larger than any nonperturbative scale;\\
(ii) The hard part, after it is appropriately 'factored out', can be treated using  perturbative QCD.
\\
\\
Unfortunately, there are big discrepancies between such an asymptotic theory 
and existing experimental data, which remain  
in the semi-hard regime (as  illustrated in Fig.~\ref{fig:REGIMES}). This is  particularly clear from current JLab data on the nucleon electromagnetic form factors~\cite{JeffersonLab:2008jve}, and lattice simulations~\cite{QCDSFUKQCD:2006gmg}.  The measured mesonic and baryonic form factors are well above the predictions of  the
perturbative QCD scaling laws, even when taken with maximally favorable assumptions (flat distributions and twist corrections included).  
This is not surprising since there is a drastic difference between the scales in DIS and jet physics on one hand, and exclusive processes on the other. The former are well defined above the scale $\mu^2 \sim 100\,{\rm GeV}^2$ (hard regime), while the exclusive processes are defined at much smaller scale, within $1-10\,{\rm GeV}^2$ (semi-hard regime).

In this paper we evaluate a number of gluonic form factors, e.g.
$GG, G\tilde G, fGGG, dGGG, T^{\mu\nu}_g$, in the QCD instanton vacuum at medium resolution,
within the semi-hard regime. At this resolution the gluons are described by semi-classical pseudoparticles, with the large gauge space reduced  solely to the bosonic and fermionic moduli of these pseudoparticles. Their
contribution in a given hadron, is amenable to pertinent matrix elements of the collectivized fermionic zero modes in the moduli.

The calculation 
are carried out to next order in the instanton  density in the QCD instanton vacuum,
where the contributions of like and unlike pairs of instantons are retained. 
There are many technical details associated to these calculations, most of which
can be found in the subsequent sections and Appendices. Therefore, we decided to present our results first, while leaving their derivations to the rest of the paper.

\subsection{$GG$ Vacuum form factor}
The gluonic scalar vacuum form factor, captures important aspects of the
gluonic correlation function in the QCD vacuum. Since it also shows up as
part of the glue in nucleon scalar form factor, we will start by recalling 
and then extending some of the results\cite{Kacir:1996qn,Schafer:1996wv}. The gluonic scalar vacuum form factor is 
\begin{widetext}
\bea
\label{COMP1}
\Pi_{GG}(q)=\int d^4x\,e^{-iq\cdot x}\left<G G(x)\, G G(0)\right>_{c}=(32\pi^2)^2\,\frac{{\sigma_T}(q)}V
\eea
\end{widetext}
with $\Pi_{GG}(x)$ the point-to-point correlator of the scalar source $O_S=G_{\mu\nu}^2$. (\ref{COMP1}) was discussed  in the bosonized ILM in~\cite{Kacir:1996qn}, and in the full ILM in~\cite{Schafer:1996wv}, with the latter study focusing on the transition between the semi-hard and the hard or free regime.

In Fig.~\ref{fig_G2_corr_vac} we show the $x$-space point-to-point correlator normalized to the perturbative (two gluon) version
The $x$-space spectral function (dashed line) accounts for the scalar-sigma,
plus the scalar $0^{++}$ glueball, plus the soft 2-pion cut  and the hard 2-parton cut,
\begin{widetext}
\bea
\label{FULLX}
\Pi^{(full)}_{GG}(x)=&&\lambda_{0^{++}}^2 D(m_{0^{++}},x)+\lambda_{\sigma}^2 D(m_{\sigma},x)\nonumber\\
&&+\int_{4m_\pi^2}^{\Lambda_\chi^2}ds\,\frac{3}{64\pi^2}\sqrt{1-\frac{4m_\pi^2}{s}}(s-2m_\pi^2)^2D(\sqrt{s},x)+\frac{2g^4}{\pi^2}\int_{s_0}^{\infty}ds\,s^2D(\sqrt{s},x)
\eea
\end{widetext}
The parameters used in the spectral parameterization are as  $\lambda_{0^{++}}=15.6$ \rm GeV$^{3}$, $\lambda_{\sigma}=2.6$ GeV$^{3}$, $m_{0^{++}}=1.25$ GeV, and $m_\sigma=683.1$ MeV to be consistent with the interacting ensemble calculation \cite{Schafer:1994fd}. The perturbative threshold is set to $s_0=2.4$ GeV~\cite{Schafer:1994fd},  and the chiral symmetry breaking scale range is  $\Lambda_\chi=1.1$ GeV. Note that we have re-instated the gauge coupling with a value $g^2/4\pi=0.3$, when accounting for the perturbative contributions.

To account for the dominance of the $0^{++}$ glueball at low $Q^2$,
and the free 2-parton cut at large $Q^2$, we use instead a dipole approximation
for the vaccuum form factor 
\bea
\label{GGSOFTVAC}
\sigma_T(q)\approx  \frac {\sigma_T}{(1+Q^2/m_{0^{++}}^2)^2}\\\nonumber
\eea
where $m_{0^{++}}=1.25$ GeV~\cite{Schafer:1994fd}. When translated to $x$-space 
plus the free contribution, it reads
\begin{widetext}
\bea
\label{DIPOLEX}
\Pi^{(dipole)}_{GG}(x)=\frac{384g^4}{\pi^4x^8}+\frac{128\pi^2}{b}\langle G^2\rangle\int \frac{d^4q}{(2\pi)^4}\frac{1}{(1+q^2/m^2_{0^{++}})^2}e^{-iq\cdot x}
\eea
\end{widetext}
In Fig.~\ref{fig_G2_corr_vac}  we show (\ref{DIPOLEX}) (green-solid line), which lies
between the full spectral representation with the sigma meson (dashed-blue line) and without the
sigma meson (dashed-orange line).

\begin{figure}
    \centering
    \includegraphics[scale=0.4]{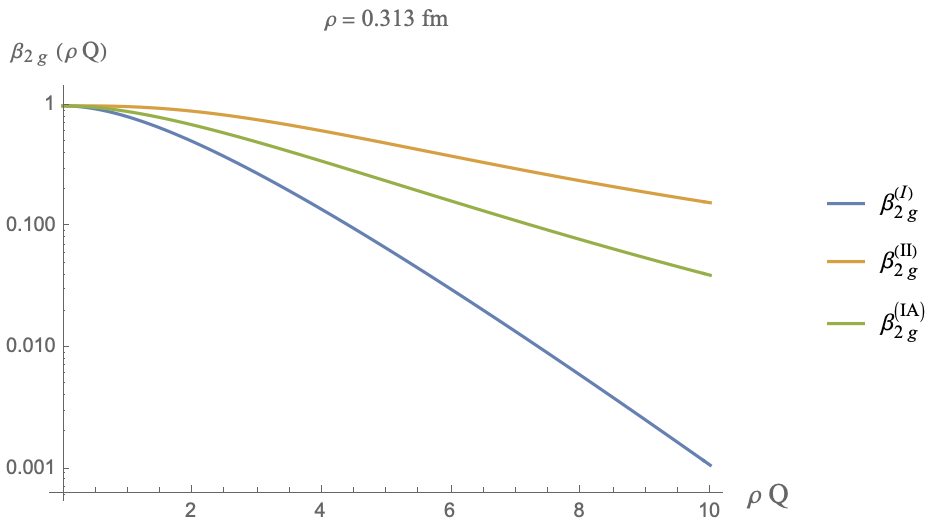}
    \caption{Induced pseudoparticle form factors $ \beta^{(I)}_{2g},\beta^{(II)}_{2g},\beta^{(IA)}_{2g}$
    versus $Q \rho$, from lower to upper 
    respectively.
    }
    \label{fig_beta_ff}
\end{figure}

\begin{figure*}
    \centering
    \includegraphics[scale=0.57]{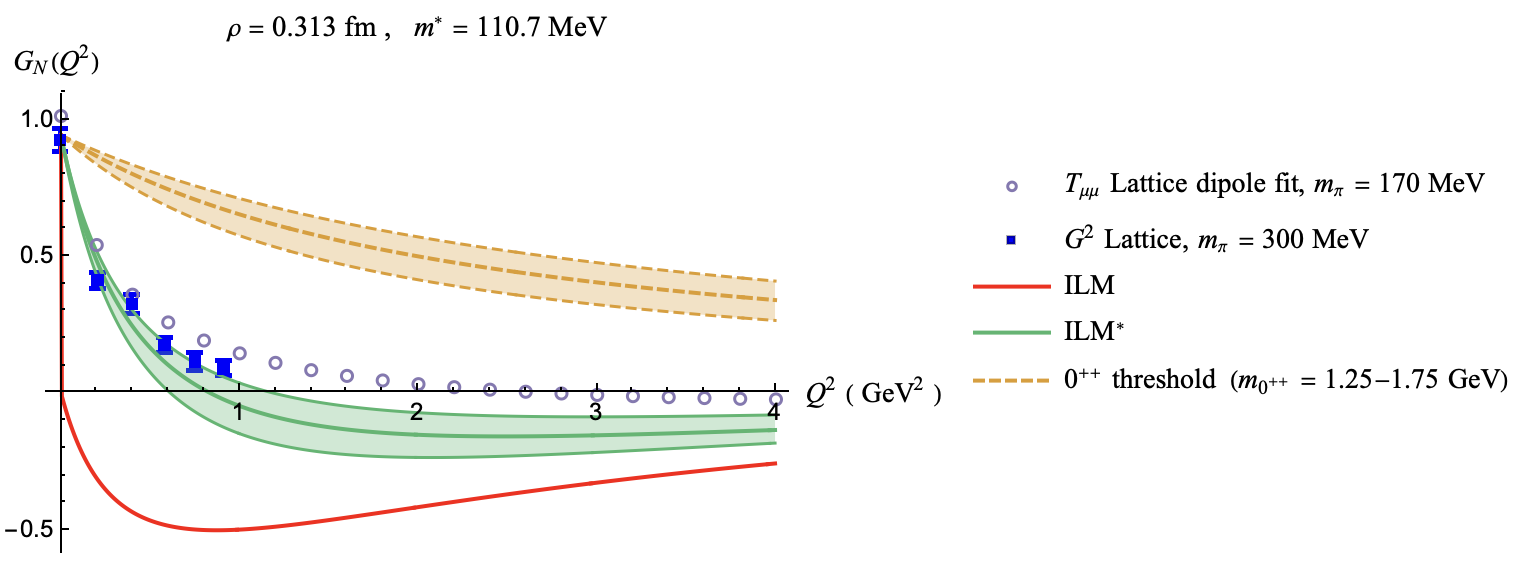}
    \caption{Nucleon gluonic scalar form factor (\ref{TMUMU}) (red-solid line) and after the substitution (\ref{SUBXX}) (green-solid line), versus the lattice results (blue data points)~\cite{Wang:2024lrm}. The lattice dipole fits to the $A$, $D$ EMT form factors from~\cite{Hackett:2023rif} are used
    to reconstruct the nucleon gluonic scalar form factor (grey-open circles), along
    with the $0^{++}$ dipole form factor (\ref{SUBXX}) (dashed-brown line) with 
    the glueball mass band $m_{0^{++}}$ ranging from  1.25 to 1.75 GeV based on the various lattice calculations on the glueball mass \cite{Schafer:1994fd,Chen:2005mg,Sun:2017ipk}}. The upper band corresponds to $m_{0^{++}}=1.75$ GeV and lower band represents $m_{0^{++}}=1.25$ GeV
    \label{fig:GG}
\end{figure*}

\subsection{ $GG$  nucleon form factor}

The nucleon gluonic scalar form factor following from the 
trace of the EMT is defined as
\begin{widetext}
\bea
\label{TMUMU}
-\frac{b}{32\pi^2}\langle P'S|g^2G^2_{\mu \nu}|PS\rangle=M_NG_N(Q^2)\bar{u}_s(P')u_s(P)
\eea
\end{widetext}
with in and out momenta momenta $P$ and $P'=P+Q$. Our main result for (\ref{TMUMU}) in the ILM is 
\begin{widetext}
\bea
\label{OFFG2X}
    &&\frac{1}{32\pi^2}\langle P'S|g^2G^2_{\mu \nu}|PS\rangle=-\left[\frac{1}{4}M^{(0)}_{\mathrm{inv}}\frac{\sigma_T}{\bar{N}}\frac{(2\pi)^4}{V}\delta^4(q)\right]\bar{u}_s(P')u_s(P)\nonumber\\
    &&-\left[\frac{1}{N_c}\left(\frac{2\kappa}{\rho^2 m^{*2}}\right)\beta^{(I)}_{2g}(\rho q)+\frac{1}{2N_c(N_c^2-1)}\left(\frac{2\kappa}{\rho^2 m^{*2}}\right)^23\rho^2m^2 T_{II}\beta^{(II)}_{2g}(\rho q)\right]\langle P'S|m\bar{\psi}\psi|PS\rangle \nonumber\\
    &&+\frac{1}{2N_c(N_c^2-1)}\left(\frac{2\kappa}{\rho^2 m^{*2}}\right)^2\frac{\rho^4m^2}{9}T_{IA}\beta^{(IA)}_{2g}(\rho q)q_\mu q_\nu\langle P'S|\bar{\psi}\left(\gamma_{(\mu}i\overleftrightarrow{\partial}_{\nu)}-\frac{1}{4}g_{\mu\nu}i\overleftrightarrow{\slashed{\partial}}\right)\psi|PS\rangle\nonumber\\
\eea
\end{widetext}
where packing fraction $\kappa$ is the dimensionless vacuum parameter defined as
\begin{equation}
    \kappa=n_{I+A}\pi^2\rho^4
\end{equation}

Here the dependence on $Q^2$ is given via induced non-local form factors, normalized to 1 and derived from certain instanton-based diagrams. Their analytic form is defined in (\ref{BETAGG}), and their specific form illustrated in Fig.\ref{fig_beta_ff}.
The quark hopping integral is defined in~(\ref{eq:quark_hopping}).

At large-$x$ or small $Q^2$, the scalar point-to-point correlator
 captures the fluctuations of the number
of pseudoparticles in the ILM, and its value at $Q^2=0$  is related to the {\bf  scale anomaly relation}, which  we will discuss in section \ref{sec_scale_anomaly}. 
It allows us to fix the normalization of the form factor. This effect can be included in (\ref{OFFG2X})  via the substitution
\bea
\label{SUBXX}
\frac{\sigma_T}{\bar{N}}\frac{(2\pi)^4}{V}\delta^4(q)
\rightarrow \frac{\sigma_T(q)}{\bar N}
\eea
which shows how the glue in the scalar vacuum form factor, exports to the nucleon scalar form factor.

In Fig.~\ref{fig:GG} we show the result for the gluonic contribution to 
the trace of the EMT in the nucleon (\ref{TMUMU}, \ref{OFFG2X}) prior to the substitution (\ref{SUBXX}) (red-solid line), and after the substitution
(green-solid line). The comparison is to the recent lattice results 
(blue squares) from~\cite{Wang:2024lrm}. The lattice results for the nucleon $A$- and $D$-form factors from \cite{Hackett:2023rif} can be used to reconstruct the gluonic contribution to the trace of the EMT in the nucleon  (gray-open circles). As expected, our results in the semi-hard region (red-solid curve)
when supplemented with the $0^{++}$-dipole glueball contribution in the
soft region (dashed-brown line) yields a good account of the reported and reconstructed lattice results in both the soft and semi-hard region (green-solid line).


\begin{figure*}
    \centering
    \includegraphics[scale=0.67]{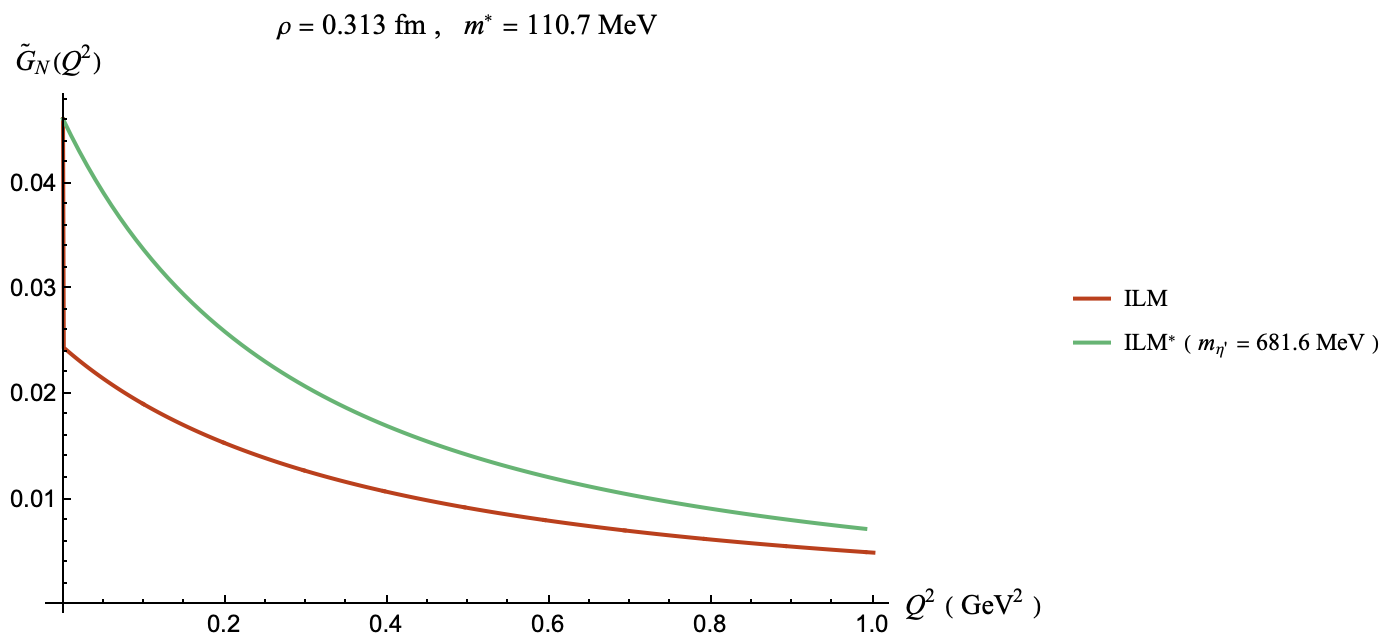}
    \caption{Nucleon gluonic pseudo-scalar form factor (\ref{HADGGTX}) in the QCD instanton vacuum (red-solid line),  and after the substitution  (\ref{SUBZZ}) (green-solid line).}
    \label{fig:GG_tilde}
\end{figure*}

\subsection{$G\tilde G$ nucleon form factor}
The gluonic pseudoscalar form factor in a hadron state 
is defined as
\bea
\label{GTGX}
&&\frac{1}{32\pi^2}\langle P'S|g^2G_{\mu\nu}\tilde G_{\mu\nu}|PS\rangle=\nonumber\\
&&M_N\tilde{G}_N(Q^2)\bar{u}_s(P')i\gamma^5u_s(P)
\eea
In the ILM 
at NLO in the instanton density, the result is
\begin{widetext}
\begin{equation}
\begin{aligned}
\label{HADGGTX}
    &\frac{1}{32\pi^2}\langle P'S|g^2G^a_{\mu \nu}\tilde{G}^a_{\mu \nu}|PS\rangle
    =\left[M_q(0)\frac{\chi_t}{\bar{N}}\frac{(2\pi)^4}{V}\delta^4(q)\right]\bar{u}_{s'}(P')i\gamma^5u_s(P)\\
    &+\left[\frac{1}{N_c}\left(\frac{2\kappa}{\rho^2 m^{*2}}\right)\tilde{\beta}^{(I)}_{2g}(\rho q)+\frac{1}{2N_c(N_c^2-1)}\left(\frac{2\kappa}{\rho^2 m^{*2}}\right)^23\rho^2m^2T_{II}\tilde{\beta}^{(II)}_{2g}(\rho q)\right]\langle P'S|m\bar{\psi}i\gamma^5\psi|PS\rangle\\
    &-\frac{1}{16N_c(N_c^2-1)}\left(\frac{2\kappa}{\rho^2 m^{*2}}\right)^2\rho^2m^2T_{IA}\tilde{\beta}^{(IA)}_{2g}(\rho q)iq_\mu\langle P'S|\bar{\psi}\gamma_\mu\gamma^5\psi|PS\rangle
\end{aligned}
\end{equation}
\end{widetext}
where the non-local form factors are given in (\ref{BETAGGTILDE}), and the hopping between pseudoparticles is defined in (\ref{eq:quark_hopping}).
The first contribution in (\ref{HADGGTX}) captures the fluctuations  of the topological charge of the pseudoparticles in the ILM in (\ref{SUS}) in the large volume limit, which we can recast as
\bea
\label{TOP1}
&&\frac{{\chi_t}(q)}V=\\
&&\int d^4x\,e^{-iq\cdot x}\left<\frac 1{32\pi^2}G\tilde G(x)\,\frac 1{32\pi^2} G \tilde G(0)\right>_{c}\nonumber
\eea
with $\chi_t(0)\approx \chi_t$ in the zero momentum limit. In Fig.~\ref{fig:GG_tilde} we  show the 
result (\ref{HADGGTX})  (red-solid line) for the gluonic pseudoscalar form factor in the nucleon state where the gluonic pseudo-scalar form factor is defined  in (\ref{GTGX}).

The expected behavior in the soft regime, near $Q=0$, now stems from the screened topological  charge fluctuations in the large volume limit, in the leading order in the pseudoparticle density.
One expects this screening to be less
singular than in the scalar case. Neglecting  the  3-pion continuum,  its range is related with the (significantly heavier) $\eta'$ meson 
\bea
\chi_t(q)\approx  \frac {\chi_t}{1+Q^2/m_{\eta'}^2}\\\nonumber
\eea
(For consistency of the setting, we use  $m_{\eta'}\approx 681.6\,{\rm MeV}$ for two flavors~\ref{MASSETAPRIME}.)

In Fig.~\ref{fig:GG}  we show the result for the topological form factor (green-solid line) following
from (\ref{HADGGTX}) after 
the substitution
\bea
\label{SUBZZ}
\frac{\chi_t}{\bar{N}}\frac{(2\pi)^4}{V}\delta^4(q)
\rightarrow \frac{\chi_t(q)}{\bar N}
\eea

\subsection{ $dGGG$ nucleon form factor}
The result for the C-odd gluonic operator $d^{abc}G^aG^bG^c$ in a hadronic state at NLO in the instanton density is given by
\begin{widetext}
\begin{equation}
\begin{aligned}
\label{eq:had_3g_2X}
    &\langle P'|g^3d^{abc}G^a_{\mu\nu}G^b_{\rho \alpha}G^{c}_{\lambda \alpha}|P\rangle=-\frac{N_c-2}{2N_c^2(N_c^2-1)}\left(\frac{2\kappa}{\rho^2 m^{*2}}\right)^2\frac{8\pi^2m}{9}\rho^2m^2 T_{II}\beta^{(II)}_{3g}(\rho q)\frac{1}{2}q^2\delta_{\rho\lambda}\langle P'|\bar{\psi}\sigma_{\mu\nu}\psi|P\rangle\\
    &+\frac{N_c-2}{2N_c^2(N_c^2-1)}\left(\frac{2\kappa}{\rho^2 m^{*2}}\right)^2\frac{8\pi^2m}{9}\rho^2m^2 T_{II}\beta^{(II)}_{3g}(\rho q)\left(\delta_{\mu\alpha}q_\beta q_\nu-\delta_{\nu\alpha}q_\beta q_\mu\right)\delta_{\rho\lambda}\langle P'|\bar{\psi}\sigma_{\alpha\beta}\psi|P\rangle\\
    &-\frac{N_c-2}{2N_c^2(N_c^2-1)}\left(\frac{2\kappa}{\rho^2 m^{*2}}\right)^2\rho^2m^2T_{IA}\frac{4\pi^2}{45}\beta^{(IA)}_{3g}(\rho q)\\
    &\times\left[\epsilon_{\beta\gamma\lambda\sigma}q_\sigma q_\nu\delta_{\mu\rho}-\epsilon_{\beta\gamma\nu\sigma}q_\sigma q_\lambda\delta_{\mu\rho}-\frac{1}{2}\epsilon_{\beta\gamma\lambda\nu}q^2\delta_{\mu\rho}+(\rho\leftrightarrow\lambda)-(\mu\leftrightarrow\nu)\right]\langle P'|\bar{\psi}\gamma_{[\beta}i\overleftrightarrow{\partial}_{\gamma]}\gamma^5\psi|P\rangle\\
    &-\frac{N_c-2}{2N_c^2(N_c^2-1)}\left(\frac{2\kappa}{\rho^2 m^{*2}}\right)^2\rho^2m^2T_{IA}\frac{4\pi^2}{15}\beta^{(IA)}_{3g}(\rho q)\delta_{\rho\lambda}\left(\epsilon_{\beta\gamma\mu\sigma}q_\sigma q_\nu-\epsilon_{\beta\gamma\nu\sigma}q_\sigma q_\mu-\frac{1}{2}\epsilon_{\beta\gamma\mu\nu}q^2\right)\langle P'|\bar{\psi}\gamma_{[\beta}i\overleftrightarrow{\partial}_{\gamma]}\gamma^5\psi|P\rangle\\
    &-\frac{N_c-2}{2N_c^2(N_c^2-1)}\left(\frac{2\kappa}{\rho^2 m^{*2}}\right)^2\rho^2m^2T_{IA}\frac{8\pi^2}{45}\beta^{(IA)}_{3g}(\rho q)\\
    &\times\left[\epsilon_{\mu\nu\rho\alpha}\left(\delta_{\beta\lambda}q_\gamma q_\alpha-\delta_{\beta\alpha}q_\gamma q_\lambda\right)-\frac{1}{2}\epsilon_{\mu\nu\rho\gamma}q^2\delta_{\beta\lambda}+(\rho\leftrightarrow\lambda)\right]\langle P'|\bar{\psi}\gamma_{[\beta}i\overleftrightarrow{\partial}_{\gamma]}\gamma^5\psi|P\rangle
\end{aligned}
\end{equation}
\end{widetext}
where the form factors normalized to unity are given in (\ref{BETACODD}.
The details of the derivation of (\ref{eq:had_3g_2X}) can be found in section~\ref{sec_Ceven}.



\subsection{ $fGGG$  nucleon form factor}
The general result for the C-even gluonic form factor $f^{abc}G^aG^bG^c$ 
in a hadronic state at NLO in the instanton density, reads
\begin{widetext}
\bea
\label{eq:had_3g_2XX}
    &&\frac{5\rho^2}{384\pi^2}\langle P'S|g^3f^{abc}G^a_{\mu\nu}G^b_{\nu\rho}G^{c}_{\rho \mu}|PS\rangle=-\left[\frac{1}{4}M^{(0)}_{\mathrm{inv}}\frac{\sigma_T}{\bar{N}}\frac{(2\pi)^4}{V}\delta^4(q)\right]\bar{u}_{s'}(P')u_s(P)\nonumber\\
    &&-\left[\frac{1}{N_c}\left(\frac{2\kappa}{\rho^2 m^{*2}}\right)-\frac{1}{2N_c(N_c^2-1)}\left(\frac{2\kappa}{\rho^2 m^{*2}}\right)^23\rho^2m^2T_{II}\right]\tilde{\beta}_{3g}(\rho q)\langle P'S|m\bar{\psi}\psi|PS\rangle\nonumber \\
\eea
\end{widetext}
where the the instanton induced form factors are given in (\ref{BETACEVEN}), with the details of the derivation given in section~\ref{sec_Ceven}. The $fGGG$ form factor is defined as
\bea
&&\langle P'S|g^3f^{abc}G^a_{\mu\nu}G^b_{\nu\rho}G^{c}_{\rho \mu}|PS\rangle=\nonumber\\
&&M^3_NA^N_{3g}(Q^2)\bar{u}_s(P')u_s(P)
\eea

\begin{figure*}
    \centering
    \includegraphics[scale=0.7]{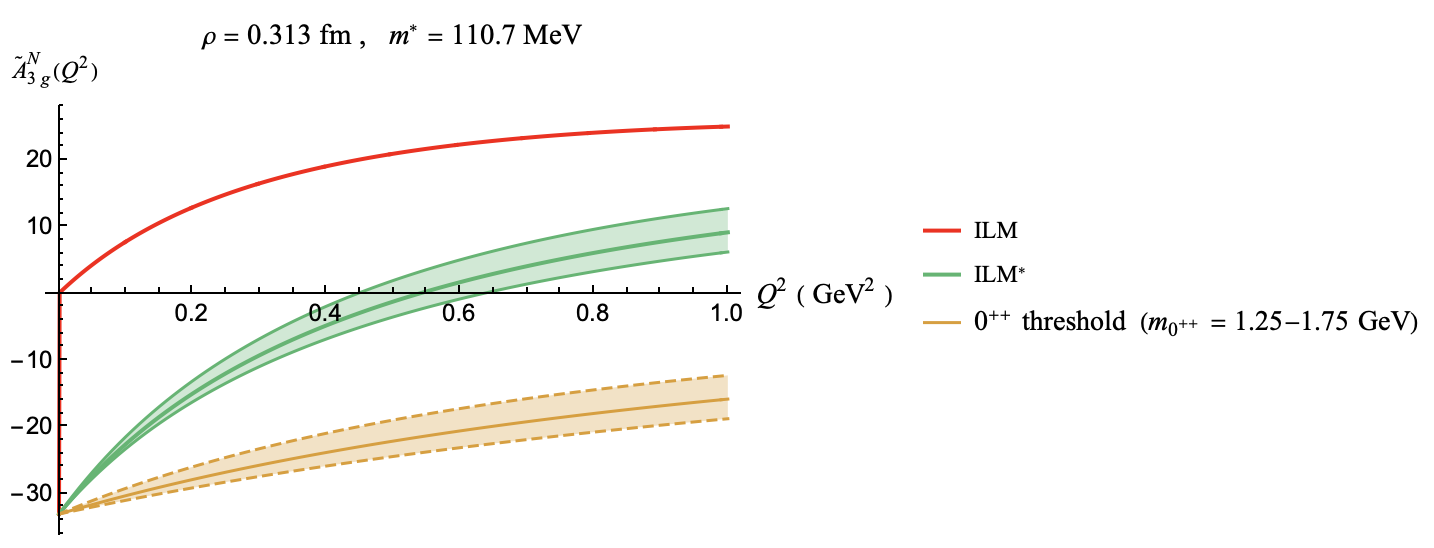}
    \caption{C-even form factor   in (\ref{eq:had_3g_2XX}) (red-solid curve), and after the substitution (\ref{sec_Ceven}) (green-solid curve), along
    with the $0^{++}$ dipole form substitution (\ref{SUBXX}) (dashed-brown line) where 
    the glueball mass band $m_{0^{++}}$ ranges from  $1.25$ to $1.75$ GeV based on the various lattice calculations on the glueball mass \cite{Schafer:1994fd,Chen:2005mg,Sun:2017ipk}. The upper band corresponds to $m_{0^{++}}=1.75$ GeV and lower band represents $m_{0^{++}}=1.25$ GeV}
    \label{fig:fGGG}
\end{figure*}

In Fig.~\ref{fig:fGGG} we show the behavior of the C-even 3-gluon form factor (\ref{eq:had_3g_2XX}) using the QCD instanton vacuum parameter (red-solid curve). The jump at $Q=0$
follows from the additional contribution stemming from the fluctuations of the 
pseudoparticles number in the large volume limit in leading order in the density. 
It is screened by higher order corrections (green-solid curve) following the
substitution (\ref{SUBXX}).

\subsection{ $T_g^{\mu\nu}$  nucleon form factor}
The hadronic form factor related with  the tensor combination of gluonic field strengths, is a very special case, since the semiclassical instanton fields have vanishing stress tensor. This implies a behavior that is very different from the previous cases, as seen already in vacuum point-to-point correlators \cite{Schafer:1994fd}.

The gluon energy momentum form factor in a hadron state only appears due to instanton-anti-instanton 'molecules'. In the pseudoparticle moduli, this form factor is amenable to quark-based observables
\begin{widetext}
\begin{equation}
\begin{aligned}
\label{eq:A_gX}
    \langle P'S|g^2\bar{T}^g_{\mu\nu}|PS\rangle=&\frac{1}{2N_c(N_c^2-1)}\left(\frac{2\kappa}{\rho^2 m^{*2}}\right)^2\rho^2m^2T_{IA}\\
    &\times\Bigg\{\frac{16\pi^2}{3}\beta^{(IA)}_{T_g,1}(\rho q)\langle P'S|\bar{\psi}\left(\gamma_{(\mu} i\overleftrightarrow{\partial}_{\nu)}-\frac{1}{4}g_{\mu\nu}i\overleftrightarrow{\slashed{\partial}}\right)\psi|PS\rangle\\
    &-\frac{4\pi^2\rho^2}{9}\beta^{(IA)}_{T_g,2}(\rho q)\left(q_\mu q_\rho g_{\nu\lambda}+q_\nu q_\rho g_{\mu\lambda}-\frac{1}{2} g_{\mu\nu}q_\rho q_\lambda\right)\langle P'S|\bar{\psi}\left(\gamma_{(\rho}i\overleftrightarrow{\partial}_{\lambda)}-\frac{1}{4}g_{\rho\lambda}i\overleftrightarrow{\slashed{\partial}}\right)\psi|PS\rangle\\
    &-4\pi^2\rho^4\beta^{(IA)}_{T_g,3}(\rho q)\left(q_\mu q_\nu-\frac{1}{4}q^2g_{\mu\nu}\right)q_\rho q_\lambda\langle P'S|\bar{\psi}\left(\gamma_{(\rho}i\overleftrightarrow{\partial}_{\lambda)}-\frac{1}{4}g_{\rho\lambda}i\overleftrightarrow{\slashed{\partial}}\right)\psi|PS\rangle\Bigg\}\\
    &+\frac{1}{2N_c(N_c^2-1)}\left(\frac{2\kappa}{\rho^2 m^{*2}}\right)^2\rho^2m^2 T_{II}\frac{8\pi^2\rho^2}{9}\beta^{(II)}_{T_g}(\rho q)\left(q_\mu q_\nu-\frac{1}{4}q^2g_{\mu\nu}\right)\langle P'S|m\bar{\psi}\psi|PS\rangle
\end{aligned}
\end{equation}
\end{widetext}
The  induced pseudoparticle form factors normalized to 1, are given in (\ref{BETAEMT}). The details regarding the derivation of (\ref{eq:A_gX}) 
in the QCD instanton vacuum,  can be found in section~\ref{sec_gravitationalff} with
the supporting  Appendices. (\ref{}) can be used to analyze the gluonic 
energy momentum form factors of the nucleon, in terms of the nucleon quark form factors. These form factors constrain the nucleon generalized parton distributions at zero skewness. They will be discussed elsewhere. Here, we will make use of 
(\ref{eq:A_gX}) to budget the mass and spin content of the nucleon in the QCD instanton vacuum as we now summarize.

\begin{figure}
\subfloat[\label{fig:mass_2000}]{%
    \includegraphics[scale=0.30]{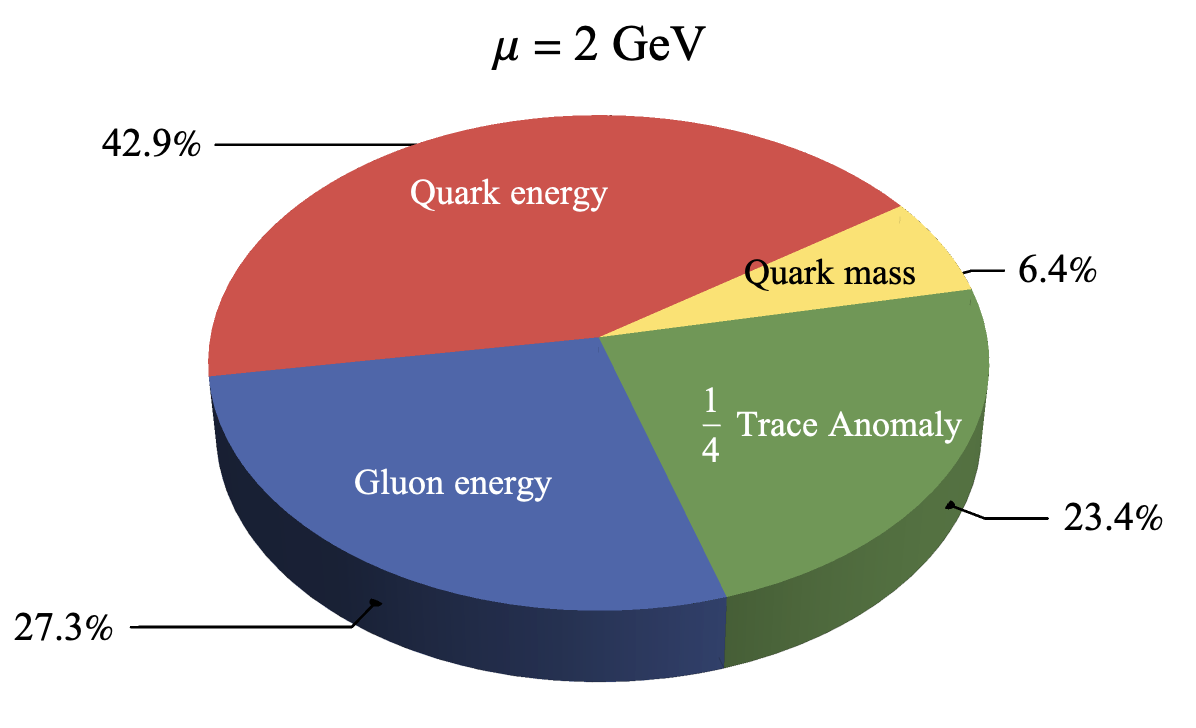}}
\hfill
\subfloat[\label{PIELAT}]{%
    \includegraphics[scale=0.30]{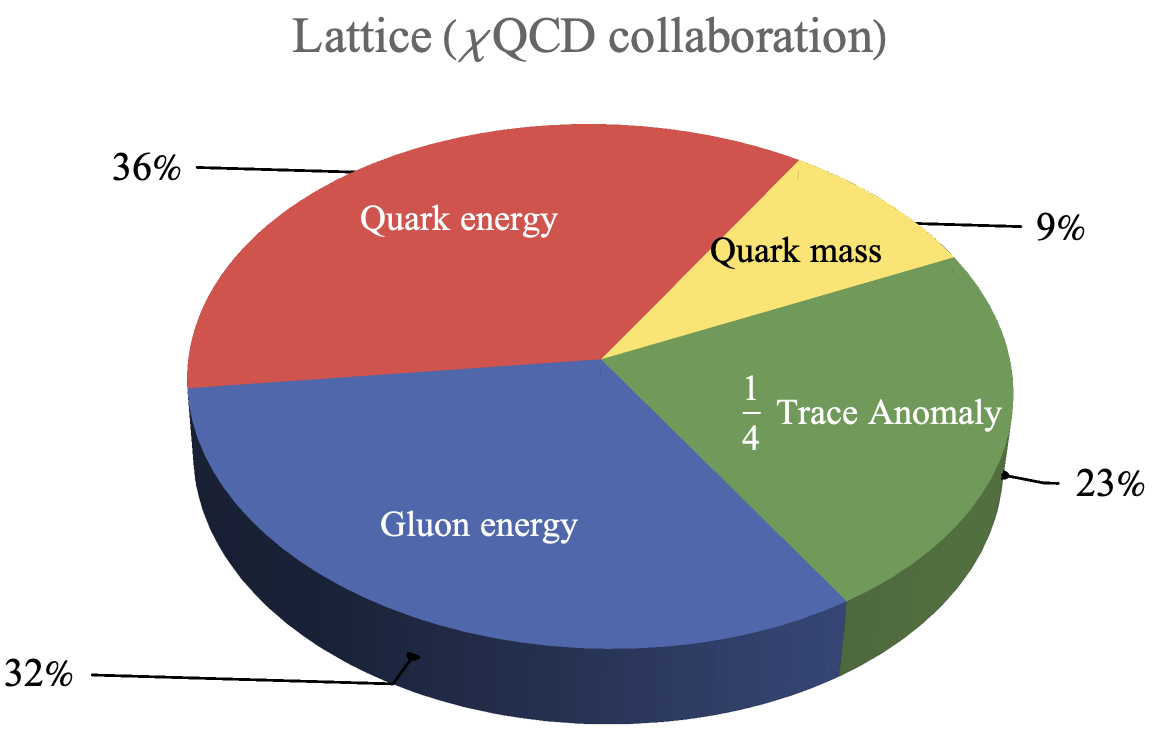}}
\caption{ 
Mass decomposition using Ji's nucleon mass sum rule, in the QCD instanton vacuum after DGLAP evolution at a resolution 
$\mu=2~\mathrm{GeV}$ (a), and the lattice results
at the same resolution from~\cite{Yang:2018nqn}
}
\label{PIEVACX}
\end{figure}

\subsection{Ji's mass sum rule}
How the nucleon mass may be assigned to the quarks and gluons is resolution dependent. The QCD instanton vacuum at low resolution, provides for a
budgeting based on semi-classics. Using Ji's mass
decomposition, whereby the nucleon mass is split to
\bea
M_N=M_Q^N+M_G^N+M_A^N+M_m^N
\eea
with the quark and gluon contributions identified in(\ref{T5X}). In particular, at NLO in the instanton density and a resolution of $\mu\approx 1/\rho\approx 600\,{\rm MeV}$, we obtain
\bea
\label{T8}
\frac{M^N_Q}{M_N}&= &\frac 34 \left(A_q(0)-\frac{\sigma_{\pi N}}{M_N}\right)\approx   69.19\%\nonumber\\
\frac{M^N_G}{M_N}&= &\frac 34 A_g(0)\approx   1.01\%\nonumber\\
\frac{M^N_A}{M_N}&=&\frac {1}4\bigg(1-\frac{\sigma_{\pi N}}{M_N}\bigg) \approx 23.40\% \nonumber\\
\frac{M^N_m}{M_M}&=&\frac{\sigma_{\pi N}}{M_N}\approx 6.39\%
\eea
with all the details given in section~\ref{sec_mass}. 
Fixing the pion-nucleon sigma term~\cite{Hoferichter:2016ocj,Alarcon:2021dlz},
shows that 69\%  of the nucleon mass
is in the valence quarks (hopping zero modes),  23.4\% in the condensate (displaced vacuum instanton field),  and 1\% in the moduli gluons. 
This budgeting evolves with the resolution. Using DGLAP evolution to $\mu\approx 2\,{\rm GeV}$ yields
a redistribution of the mass in favor of the gluons, with
\bea
\label{T9}
\frac{M^N_Q}{M_N}&\approx   42.91\%\nonumber\\
\frac{M^N_G}{M_N}&\approx   27.29\%
\eea
There is no change in the anomalous and mass contributions.
In Fig.~\ref{PIEVACX}, we show comparative pie-charts for the evolved ILM results (\ref{T8}-\ref{T9}) at the resolution of $\mu\approx 2\,{\rm GeV}$,  with the lattice results in~\cite{Yang:2018nqn} at the same resolution.

\subsection{Ji's spin sum rule}
\begin{figure}
    \subfloat[\label{}]{
    \includegraphics[scale=0.48]{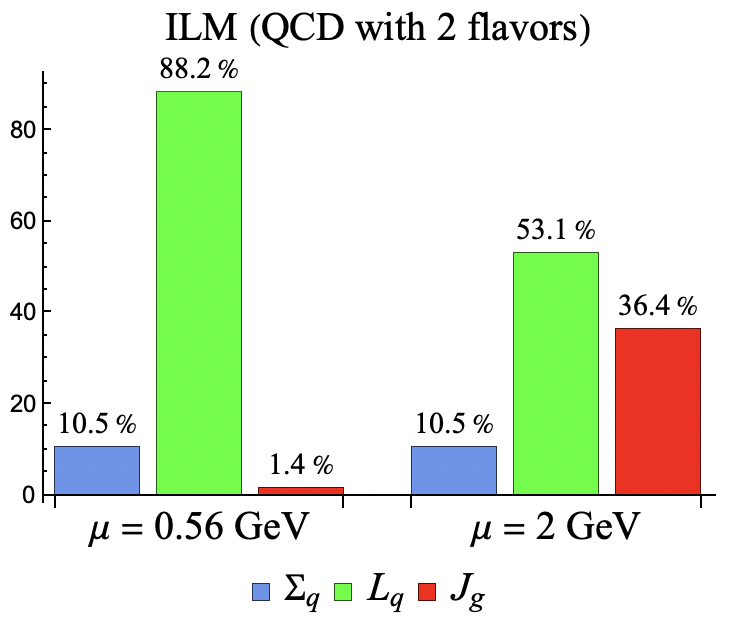}}
    \hfill
    \subfloat[\label{}]{
    \includegraphics[scale=0.48]{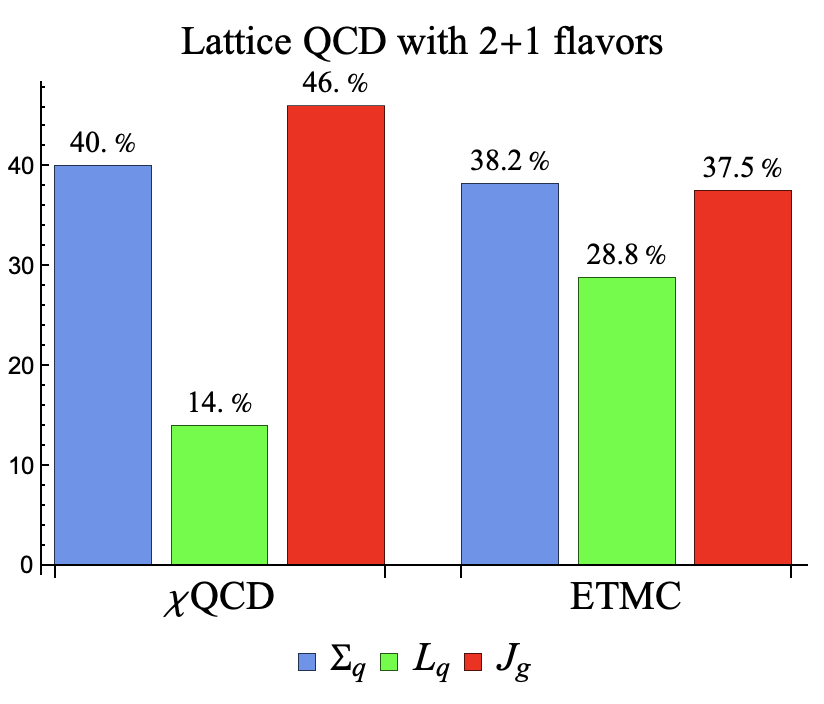}}
    \caption{a: Nucleon spin decomposition using Ji's spin sum rule, in the  QCD instanton vacuum with 2 flavors at $\mu=0.56\,{\rm GeV}$ (left) and $\mu=2\,{\rm GeV}$ (right); b: The same decomposition 
   from the $\chi$QCD lattice collaboration~\cite{Wang:2021vqy} (left) and the ETMC lattice collaboration~\cite{Alexandrou:2020sml} (right),
   at a resolution of $\mu=2\,{\rm GeV}$.}
    \label{fig:spin_sumX}
\end{figure}

Similarly to the mass decomposition, the spin decomposition can be addressed using Ji's spin sum rule
\bea
S_N=\frac 12\Sigma_q+L_q+J_g
\eea
where the quark angular momentum contribution is split into the intrinsic quark spin $\Sigma_q$ plus orbital momentum $L_q$. In the QCD instanton vacuum, the leading LO contribution to the intrinsic quark spin is mostly from the vacuum topological susceptibility. In the QCD instanton vacuum with two flavors at a resolution $\mu\approx 1/\rho\approx 600\,{\rm MeV}$, the budgeting is 
\bea
\label{S8-2}
\frac{\frac{1}{2}\Sigma_q}{S_N}&= &\Delta \tilde{q}-\frac{g^2}{8\pi^2}N_f \Delta g\approx 10.5\%\nonumber\\
\frac{L_q}{S_N}&= &A_q(0)+B_q(0)-\frac{\frac{1}{2}\Sigma_q}{S_N}\approx 88.2\%\nonumber\\
\frac{J_g}{S_N}&=&A_g(0)+B_g(0) \approx 1.4\%
\eea
with most of the derivation given in section~\ref{sec_sum}. Here we assumed $B_{q,g}(0)=0$ \cite{Mamo:2022eui}.
Again these assignments are resolution dependent,
and change with increasing $\mu$. In particular, 
for $\mu\approx 2\,{\rm GeV}$ we obtain
\bea
\label{S9-2}
\frac{\frac{1}{2}\Sigma_q}{S_N}&\approx& 10.5\%\nonumber\\
\frac{L_q}{S_N}&\approx& 53.1\%\nonumber\\
\frac{J_g}{S_N}&\approx& 36.4\%
\eea
In  Fig.~\ref{fig:spin_sumX}a we show the histograms for the spin assignments (\ref{S9-2}), in the QCD instanton vacuum with 2 flavors before evolution at $\mu=0.56\,{\rm GeV}$, and after evolution at $\mu=2\,{\rm GeV}$. They are to be compared to  the lattice results in Fig.~\ref{fig:spin_sumX}b  
from the $\chi$QCD-collboration~\cite{Wang:2021vqy} and the ETMC-collaboration~\cite{Alexandrou:2020sml}, both at the resolution of $\mu=2\,{\rm GeV}$.

\section{Semi-classical glue in the QCD  vacuum}
\label{sec_instantons}
Understanding of any quantum system starts with understanding
of its ground state, and pre-defining its excitations. Similarly, 
understanding hadrons requires an understanding of the QCD vacuum. The central aspects of the QCD vacuum are two-fold: the quantum breaking of conformal symmetry and the spontaneous breaking of chiral symmetry, both of which are tied to the topological nature of the gauge configurations 
at low resolution.

The size distribution of the pseudoparticles  is well captured semi-empirically by the original ILM \cite{Shuryak:1981ff},  confirmed then by various mean-field studies  \cite{Diakonov:1995ea,Nowak:1996aj} and statistical simulations of the ensemble \cite{Schafer:1996wv}. This distribution can be
written as
\begin{equation}
\label{dn_dist}
n(\rho) \sim  {1 \over \rho^5}\big(\rho \Lambda_{QCD} \big)^{b} \, e^{-\#\rho^2/R^2}
\end{equation}
with $b=11N_c/3-2N_f/3$ (one loop).  The small size distribution follows from the conformal nature of the
instanton moduli and perturbation theory. The large size distribution is non-perturbative, but cut-off by $R$ the mean separation of the instantons (anti-instantons) in the vacuum. 
Detailed lattice simulations using the gradient flow method \cite{Michael:1994uu,Michael:1995br} finds that the mean tunneling rate and quasiparticle size are
\begin{equation}
 \label{eqn_ILM}
n_{I+A}=\frac{\bar N}{V}\equiv \frac 1{R^4}\approx \frac 1{ {\rm fm}^{4}} \qquad\qquad\frac{\bar \rho}R \approx  \frac 13  
\end{equation}

\begin{figure}
  \centering
    \includegraphics[width=8cm]{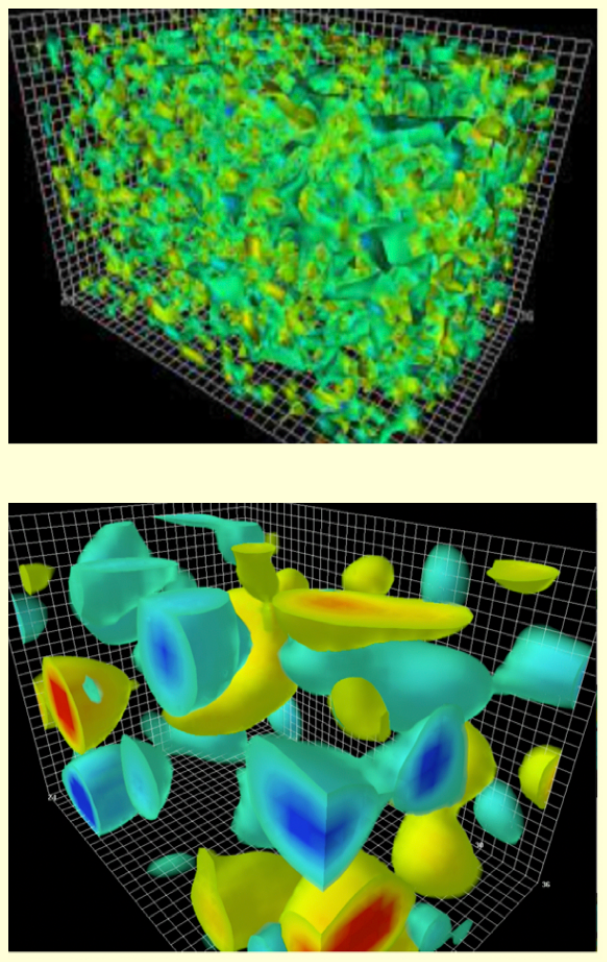}
    \caption{Visualization of the vacuum in gluodynamics, before cooling at a resolution of about $\frac 1{10}\,{\rm fm}$ (top),  and after ``deep cooling"  at a resolution of about $ \frac 13\,{\rm fm}$ (bottom)~\cite{Moran:2008xq}, where the pseudoparticles emerge.}
    \label{fig:vacuum}
\end{figure}

This distribution, as well as values of the size and density,
has been many times confirmed by lattice works, using various versions of ``deep cooling" of configurations.
In Fig.~\ref{fig:vacuum} we show e.g. results from lattice simulations
by Leinweber and his collaborators~\cite{Leinweber:1999cw},  using the gradient flow (cooling) method. At high resolution as illustrated in Fig.~\ref{fig:vacuum} (top), the vacuum is dominated by quantum or zero point motion, but as the resolution is decreased
as illustrated in Fig.~\ref{fig:vacuum} (bottom) a much smoother landscape emerges, composed
essentially of instantons and anti-instantons. These are tunneling pseudoparticles, between vacua with different Chern-Simons numbers.

Most of the hadronic correlation functions
in the QCD vacuum are little affected by the removal of the quantum gluons  by gradient flow, an indication of the central role played by these pseudoparticles~\cite{Chu:1993cn,Schafer:1996wv}. The dimensional parameters (\ref{eqn_ILM})
combine in the dimensionless parameter  packing fraction $\kappa\equiv \pi^2\bar\rho^4 n_{I+A}\approx 0.1$, a small emerging parameter that allows for a many-body analysis.

\subsection{Scale anomaly} \label{sec_scale_anomaly}
 The quantum breaking of conformal symmetry is best captured by the anomalous part of the trace of the energy-momentum tensor
\be
\label{3X}
T^\mu{}_\mu\approx -\frac{b}{32\pi^2}G^a_{\mu\nu}G^{a\mu\nu}+m\overline\psi\psi
\ee
Throughout, we will use the rescaling $gG\rightarrow G$ for operators
in the instanton or anti-instanton gauge fields.
In the QCD instanton vacuum, the gluonic operators
\bea
G^2/(32\pi^2)\rightarrow (N_++N_-)/V=N/V\nonumber\\
G\tilde G/(32\pi^2)\rightarrow (N_+-N_-)/V=\Delta N/V
\eea
 count the number-sum (first) and number-difference (second) of instantons plus anti-instantons in the 4-volume $V$.  In the canonical ensemble  with zero theta-angle,  the former is fixed by the mean instanton density  with $N_\pm/V\rightarrow \bar N/2V$
and $\Delta N=0$. As a result, the expectation value of (\ref{3X}) in the Yang-Mills 
vacuum illustrated in Fig.~\ref{fig:vacuum} is 
\be
\label{SCALE}
\left<T^\mu{}_\mu\right>\approx -b\, n_{I+A}\approx -10\,{\rm fm}^{-4}
\ee
in leading order in the packing fraction, and chiral limit.
The emerging gluon condensate $\left<G^2\right>$ which is positive, 
is at the origin of most hadronic mass in the Universe~\cite{Zahed:2021fxk}.  

\subsection{Fluctutating pseudoparticles}
In Fig.~\ref{fig:vacuum} the  tunneling quasiparticles fluctuate in numbers. Remarkably, these fluctutations are universally captured by the distributions~\cite{Diakonov:1995qy,Kacir:1996qn,Nowak:1996aj}
\be
\label{dist}
\mathbb P (N_+, N_-)\propto\bigg[e^{\frac {bN}4 }\bigg(\frac {\bar N}{N}\bigg)^{\frac {bN}4 }\bigg]\bigg[\frac 1{\big({2\pi \chi_t}\big)^{\frac 12}}e^{-\frac{\Delta N^2}{2\chi_t}}\bigg]
\ee
with mean $\bar N$, in agreement with low-energy theorems~\cite{Novikov:1981xi}.
The dominant second moments are 
\bea
\sigma_T&=&\langle(N-\bar N)^2\rangle_{\mathbb P}\nonumber\\
\chi_t&=&\langle (N_+-N_-)^2\rangle_{\mathbb P}
\eea
The variance in $N$  or vacuum compressibility, 
\be
\label{COMP}
{\sigma_T}=V\int d^4x\left<\frac 1{32\pi^2}G G(x)\,\frac 1{32\pi^2} G G(0)\right>_{\mathbb P}
\ee
vanishes in the large $N_c$ limit,
\bea
\sigma_T=\frac {4\bar N}b 
\eea

The  large volume  topological susceptibility  is
\be
\label{SUS}
{\chi_t}=V\int d^4x\left<\frac 1{32\pi^2}G\tilde G(x)\,\frac 1{32\pi^2} G\tilde G(0)\right>_{\mathbb P}
\ee
In quenched QCD, $\chi_t\rightarrow \chi_t^{(0)}$  is given by the Witten-Veneziano formula~\cite{Witten:1979vv,Veneziano:1979ec}
\be
\label{CHI}
\frac{\chi^{(0)}_t}V=\lim_{V\to\infty}\frac{\langle{\Delta N}^2(V)\rangle_{\mathbb P}}V=\frac{f_\pi^2 M_1^2}{2N_f}
\ee
with $M_1$  the quenched singlet mass,

\be
\label{CHI1}
M_1^2=m_{\eta^\prime}^2+m_\eta^2-2m_K^2
\ee
(\ref{CHI}-\ref{CHI1}) hold in the  QCD instanton vacuum~\cite{Diakonov:1995ea,Schafer:1996wv,Nowak:1996aj}.
In the unquenched QCD instanton vacuum,  (\ref{CHI}) is very sensitive to the presence of light quarks with the substitution  $M^2_1\rightarrow m_\pi^2$
(see below),
and vanishes in the chiral limit~\cite{Diakonov:1995qy,Kacir:1996qn}.

\subsection{Light quarks and zero-modes}
When a light quark crosses a tunneling configuration it develops a zero-mode that is single
handed~\cite{tHooft:1976snw}, an amazing phenomenon protected by topology and the Atiyah-Singer theorem. 
It is the delocalization of these zero modes and their interactions, that is at the origin of the spontaneous breaking of chiral symmetry, and the emergence of the light hadronic spectrum. Remarkably, this topological mechanism for mass generation leaves behind a distinct fingerprint: universal conductance-like fluctuations in the quark spectrum, predicted by random matrix theory~\cite{Verbaarschot:1993pm} and confirmed by lattice simulations~\cite{Wittig:2020jtm}.

In Fig.~\ref{fig_ZERO} we show how a light up-quark helicity in a zero-mode is  flipped when crossing an instanton (left) or anti-instanton (right). (In the zero-mode, the quark spin $\vec \sigma$  is locked to the  color $\vec \tau$, in a hedgehog like configuration with $\vec\sigma+\vec\tau=\vec 0$). 
This flipping is captured by the $^\prime$t Hooft vertex for a single quark flavor~\cite{tHooft:1976snw}. Specifically, the LSZ reduced forward scattering matrix for the
zero-mode in Fig.~\ref{fig_ZERO} is in Euclidean signature

\begin{widetext}
\bea
\label{VER1}
n_I\,
 \bigg<
\psi_R^\dagger(p)p\cdot \sigma^\dagger\bigg[\varphi^\prime(p)\hat p\cdot\sigma\epsilon U\bigg]\frac 1{m}
\bigg[\varphi^\prime(p) (\hat p\cdot\sigma\epsilon U)^\dagger\bigg]p\cdot \sigma \psi_L(p)\bigg>_{U}+{(I, L)\leftrightarrow (\bar I, R)}
\eea
with the Weyl notation subsumed $(p\cdot \sigma=p_\mu\sigma_\mu)$ where $\sigma_\mu=(-i\vec{\sigma},1)$, and
 the normalized quark zero mode

\begin{figure}
	\centering
		\includegraphics[width=14cm]{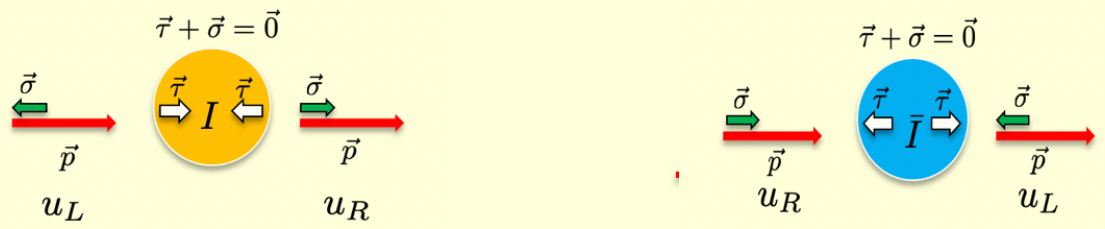}
		\caption{A quark zero mode propagating through an instanton enters left-handed and exit right-handed, to maintain a null quasi-spin $\vec\tau+\vec\sigma=\vec 0$ (left); The opposite takes place through an anti-instanton (right).}
		\label{fig_ZERO}
\end{figure}

\bea
\label{VER2}
\phi_{\alpha I}^i(p)=\varphi^\prime(p)(\hat p\cdot \sigma\epsilon U)_\alpha^i\equiv 
\bigg[\pi\rho^2\bigg(I_0K_0(z)-I_1K_1(z)\bigg)^\prime_{z=\rho p/2}\bigg]
(\hat p\cdot \sigma\epsilon U)_\alpha^i
\eea
Here the matrix  element $(\epsilon U)^i_a=\epsilon^i{}_bU_{ab}$ carries spin-$i$ and color-$a$, with $\epsilon^i{}_b$ a real antisymmetric tensor 
(hedgehog in spin-color) and $U$ an
$SU(N_c)$ valued color matrix. The averaging in (\ref{VER2}) is carried over $U$.
(\ref{VER1}) can be recast in the form  

\be
\label{VER3}
M_q(p)\bigg(\frac {N}{\bar N}\,\psi^\dagger(p) \psi(p)-\frac {\Delta N}{\bar N}\,\psi^\dagger(p) \gamma^5\psi(p)\bigg)\rightarrow
{\psi}^\dagger(p) \bigg[M_q(p)\bigg(1-\frac{\Delta N}{\bar N}\gamma^5\bigg)\bigg] \psi(p)
\ee
\end{widetext}
with the running constituent quark mass~\cite{Pobylitsa:1989uq,Kock:2020frx}  ($\tilde \kappa=\kappa/(2\pi^2 N_c)$)

\be
\label{VER4}
M_q(p)=\tilde\kappa \frac{|p\varphi^\prime (p)|^2}{m\rho^4}\rightarrow \frac{\sqrt{\tilde \kappa}}{\sqrt 2 \rho^2}\frac{|p\varphi^\prime(p)|^2}{||q\varphi^{\prime 2}||}
\ee
The singular $1/m$ effect is removed by disordering,  
with  $M_q(0)=383\pm 39$ MeV~\cite{Kock:2020frx}, which is comparable to the numerical result $M_q(0)\approx 300$ MeV~\cite{Schafer:1993ra,Schafer:1996wv}. In the QCD instanton vacuum, the running quark mass is fixed
by the same scale as the gluon condensate or $M_u(0)\approx 1/R$ since $\bar\rho\approx R/3$, with the size distribution (\ref{dn_dist}) still controlled by $R$. This ensures the renormalization group invariance of all mass scales.

The   emergent quark mass (\ref{VER4}) in the QCD instanton vacuum may remind us of the constituent quark mass 
from  the Nambu-Jona-Lasinio (NJL)model~\cite{Bernard:1987ir,Klevansky:1992qe} 
(and references therein). However, it is important to stress that the latter is a pre-QCD model, while the former
is rooted in QCD  and is now  supported by  even numerical QCD lattice visualizations  as in Fig.~\ref{fig:vacuum}.
The canonical NJL model  although useful, does not explain the vacuum gluon condensate, the running quark mass, the $\eta^\prime$ mass (unless modified)
and the universal 
spectral conductance fluctuations~\cite{Verbaarschot:1993pm}.   More importantly, the gluonic operators, their correlations and mixing with the quarks are readily described in the QCD instanton vacuum. For completeness,
we note the  non-topological  approach to the hadronic mass scale in~\cite{Roberts:2021nhw} (and references therein) where also a running quark
mass emerges by re-summing gluon rainbow diagrams.


\subsection{Center P-vortices}
The topologically active quasiparticles give rise to a linearly rising central potential till 
about 1 fm, before the potential flattens out at larger distances~\cite{Shuryak:2021fsu}. Strong lattice evidence for the disordering of large Wilson loops, points to the
center-projected vortices (center P-vortices)~\cite{Greensite:2016pfc,Biddle:2019gke,Biddle:2020eec}. Center P-vortices can  link  with large Wilson loops through $Z_{N_c}$ fluxes, leading to an emergent 
 string tension $\sigma_T$  fixed  by the  planar density of  $N_V/\sqrt{V}$ of center P-vortices. More specifically,  $\sigma_T=1/(2\pi l_s^2)$
with a string length $l_s\approx 0.2$ fm, so that $N_V/\sqrt{V}\approx 4/{\rm fm}^2$.

The center P-vortices are characterized by a number of branching points (monopoles), which are likely anchors of  topological structures or quasiparticles. Yet the latters
carry much stronger chromo-electric and -magnetic fields
$\sqrt{E}=\sqrt{B}\approx 2.5/\bar \rho\approx 1.5\,{\rm GeV}$, in comparison to $\sigma_T\bar \rho\approx 0.3\,{\rm GeV}$ carried by a center P-vortex.  This suggests
that the quantum breaking of conformal and chiral symmetry, is strongly mediated by the quasiparticles for
the low-lying hadrons in their ground state. The radial and orbitally excited states have larger sizes, hence more
susceptible to piercing by $Z_{N_c}$ fluxes threading the center P-vortices. Throughout, we will focus on the pseudoparticles for the low lying hadrons.

\section{Effective theory of instanton ensemble}
\label{sec_eff}
For a more quantitative description of the QCD vacuum at low resolution, 
we will focus on the pseudoparticles illustrated in Fig.~\ref{fig:vacuum}.
We designate by $N_+$ the number of pseudoparticles,  and by $N_-$ the number of pseudoparticles with opposite charges.  For fixed numbers $N_\pm$, the canonical partition function $Z_{N_\pm}$ is
\begin{widetext}
\begin{equation}
\begin{aligned}
\label{eq:Z_N}
   Z_{N_\pm}=\frac{1}{N_+!N_-!}\int\prod_{I=1}^{N_++N_-}d\Omega_I n_0(\rho_I)\rho_I^{N_f}e^{-S_{int}}
  \prod_{f}\mathrm{Det}(\slashed{D}+m_f)_{\mathrm{low}}
\end{aligned}
\end{equation}
\end{widetext}
where $d\Omega_I=d\rho_I d^4z_I dU_I$ is the conformal measure (size $\rho_I$, center $z_I$, and color orientation $U_I$) for each single (anti-)instanton.  The mean tunneling rate is
$$n_0(\rho)=C_{N_c}(1/\rho^5)\left(8\pi^2/g^2\right)^{2N_c} e^{-8\pi^2/g^2(\rho)}$$ 
with $C_{N_c}$ is the number dependent on color number $N_c$ and the gauge interaction between instantons and anti-instantons is $S_{int}$.

\subsection{Emergent $^\prime$t Hooft vertices}
The fermion determinant receives contribution from the high momentum modes
as well as the low momentum modes. The contribution of the higher modes are localized on the pseudoparticles. They renormalize the mean-density rate,  with an additional factor of $\rho^{N_f}$. The low momentum modes in the form of quasi-zero modes, are delocalized among the pseudoparticles. The chief result of the delocalization is the emerging constituent mass (\ref{VER3}-\ref{VER4}), and 
$^\prime$t Hooft vertices~\cite{Diakonov:1995ea, Diakonov:1995qy,Schafer:1996wv,doi:10.1142/1681}
\begin{widetext}
\begin{align}
\label{eq:tHooft}
\Theta_{I}=&\prod_f\left[\frac{m_f}{4\pi^2\rho^2}+i\psi^\dagger_f(x)U_I\frac{1}{2}\left(1+\frac{1}{4}\tau^a\bar{\eta}^a_{\mu\nu}\sigma^{\mu\nu}\right)U_I^\dagger\frac{1-\gamma^5}{2}\psi_f(x)\right]e^{-\frac{2\pi^2}{g}\rho^2R^{ab}(U_I)\bar{\eta}^b_{\mu\nu}G^a_{\mu\nu}}\nonumber\\
\Theta_{A}=&\prod_f\left[\frac{m_f}{4\pi^2\rho^2}+i\psi^\dagger_f(x)U_A\frac{1}{2}\left(1+\frac{1}{4}\tau^a\eta^a_{\mu\nu}\sigma^{\mu\nu}\right)U_A^\dagger\frac{1+\gamma^5}{2}\psi_f(x)\right]e^{-\frac{2\pi^2}{g}\rho^2R^{ab}(U_A)\eta^b_{\mu\nu}G^a_{\mu\nu}} 
\end{align}
\end{widetext}
to lowest order in the current quark masses $m_f$.
The gluonic field strength $G^a_{\mu\nu}$ follows from the LSZ reduction of pseudoparticle field strength, and is sourced by the color-magnetic moment~\cite{Vainshtein:1981wh,Kochelev:1996pv}. 
The rigid color rotation $R^{ab}(U)$ is defined as
$$
    R^{ab}(U)=\frac{1}{2}\mathrm{Tr}(\tau^aU\tau^bU^\dagger)
$$

With this in mind, the ensuing  canonical partition function $Z_{N_\pm}$ in \eqref{eq:Z_N}, reads
\begin{widetext}
\bea
\label{eq:Z_N2}
   Z_{N_\pm}=Z^{(g)}_{N_\pm}&&\int \prod_{I=1}^{N_++N_-}d^4z_IdU_I\frac{(4\pi^2\rho^3)^{N_f}}{V}\nonumber\\
   &&\times \int \mathcal{D}\psi \mathcal{D}\psi^\dagger DA_\mu\left(\prod_{I=1}^{N_+}\Theta_I\prod_{A=1}^{N_-}\Theta_A\right) \exp\left[-\int d^4x \left(-\psi^\dagger i\slashed{\partial}\psi+\frac{1}{4}(G^a_{\mu\nu})^2\right)\right]
\eea
with
\begin{equation}
   Z^{(g)}_{N_\pm}=\frac{1}{N_+!N_-!}\left(\int d\rho n_+(\rho)V\right)^{N_+}\left(\int d\rho n_-(\rho)V\right)^{N_-} e^{-\bar{S}_{int}}
\end{equation}
\end{widetext}
Here $n_\pm(\rho)$ is the effective instanton size distribution, including the pseudoparticle binary interaction $\bar{S}_{int}$ which can be estimated by Feynman variational principle \cite{Diakonov:1995qy,Diakonov:1983hh}.

\subsection{Single instanton approximation}
Since most of the gluonic matrix elements will be assessed in hadronic states, we can ignore
 $\bar{S}_{int}$, and each emerging vertex $\Theta_{I,A}$ in (\ref{eq:Z_N2}) can be randomly averaged over the single pseudoparticle moduli with mean size fixed,
\begin{equation}
    \theta_\pm=\int d^4z_{I,A} dU_{I,A}\Theta_{I,A}
\end{equation}
The explicit form of the vertices in the single instanton approximation (SIA) can be found in Appendix \ref{App:effective_Langrangian}.
In the large volume limit with fixed pseudoparticle
density, the emergent vertices $\theta_\pm$
exponentiate, giving
\begin{equation}
\label{ZNEFF}
   Z_{N_\pm}=Z^{(g)}_{N_\pm}
   \int \mathcal{D}\psi \mathcal{D}\psi^\dagger \mathcal{D}A_\mu\exp\left(-S_{\mathrm{eff}}\right)
\end{equation}
where the effective action in Euclidean space reads 
\begin{widetext}
\begin{equation}
\begin{aligned}
\label{eq:effective_action}
    S_{\mathrm{eff}}(N_+,N_-)=&\int d^4x\left[-\psi^\dagger (i\slashed{\partial}-m^*_f)\psi+\frac{1}{4}(G^a_{\mu\nu})^2\right]
    -G(1+\delta)\theta_+-G(1-\delta)\theta_-
\end{aligned}
\end{equation}
\end{widetext}
The emergent parameters $G$ and $\delta$ are fixed by the saddle point approximation. The effective coupling $G$ 
\bea
    G=\frac{N}{2V}
    \frac{(4\pi^2\rho^3)^{N_f}}{\prod_f(\rho m^*_f)}
\eea
is tied to the mean instanton size $\rho$, density $N/V$,  and   determinantal mass $m^*_f$~\cite{Schafer:1995pz,Faccioli:2001ug,Shuryak:2021fsu}
\bea
\label{eq:m_det}
    m_f^*=m_f-\frac{2\pi^2\rho^2}{N_c}\langle\bar{\psi}_f\psi_f\rangle
\eea
It follows from the mean-field approximation to the effective action in the SIA summarized in~Appendix \ref{App:effective_Langrangian}.

We note that the determinantal mass $m_f^*$ does not run with momentum, and is much smaller than the running constituent  quark mass $M_q(0)$ used in~\cite{Liu:2023yuj,Liu:2023fpj} (and references therein). The latter resums all pseudoparticle contributions  (close and far) to the quark propagator in leading order in the packing fraction. 

The determinantal mass  follows from the SIA, by retaining only the closest  pseudoparticle in the inverted quark propagator, for a given zero mode (see \cite{Faccioli:2001ug} and Eq. ($72$--$74$) in \cite{Shuryak:2021fsu}). It is appropriate for the description of
the hopping of fermions at distances $|x-y|\leq R\approx 1\,{\rm fm}$, e.g. in the local clustering of the zero modes in the effective $^\prime$t Hooft vertices, and pairing of pseudoparticles in molecules. The larger constituent quark mass $M_q(0)$ describes long range 
propagation of the emerging quarks for $|x-y|\gg R\approx 1\,{\rm fm}$, and is more appropriate 
in the description of long range hadronic correlators.

The screened topological charge $\delta$ is fixed to
\begin{equation}
\delta=\frac{\Delta N}{N}\sum_f\frac{m_f^*}{m_f}
\end{equation}
For a canonical ensemble of pseudoparticles, the instanton number sum $N$ and difference $\Delta N$ are fixed to  $N=Vn_{I+A}$ and $\Delta N=0$, respectively.  In a  grand canonical ensemble, the instanton number sum and difference are allowed to fluctuate.

Using (\ref{ZNEFF}), the  mean  values of the instanton determinantal vertices $\langle\theta_\pm\rangle$, are
\begin{equation}
\langle\theta_\pm\rangle=\prod_{f}\left(\frac{m^*_f}{4\pi^2\rho^2}\right)V
\end{equation}
The effective instanton vertices are composed of the $2N_f$-quark 't-Hooft interaction and the emission of multiple gluons. The effective action can be decomposed into the fermionic instanton-induced interactions ('t-Hooft Lagrangian), with or without  multi-gluon tail-emission. The corresponding effective Lagrangian 
is given in  Appendix~\ref{App:effective_Langrangian}. Note that  multi-gluon tail-emissions are further suppressed by the small instanton size $\rho$.



The vacuum parameters are fixed to $\rho=0.313~\mathrm{fm}$ and $n_{I+A}=1~ \mathrm{fm}^{-4}$.  The table for the parameters used in ILM is shown in Table \ref{tab:ILM}. 

\begin{table}[]
   \centering
   \begin{tabular}{ccc}
   \hline
     &  Covariant & Light front\cite{Liu:2023fpj} \\
   \hline
   $G$  & 610.3~$\mathrm{GeV}^{-2}$ & 567.8~$\mathrm{GeV}^{-2}$   \\
   $m$ & $12.2$ MeV & $16.17$ MeV\\
   $m^*$ & $110.70$ MeV & $114.76$ MeV \\
   $M_q(0)$ & $395.17$ MeV & $398.17$ MeV\\
   $\langle\bar{\psi}_f\psi_f\rangle$ & -($208.39$ MeV)$^3$ & -($208.58$ MeV)$^3$\\
   \hline
   \end{tabular}
    \caption{Emergent parameters}
    \label{tab:ILM}
\end{table}

The determinantal mass is to be compared to the heavier constituent 
quark mass $M_q(0)\approx 395\,{\rm MeV}$. Both masses are close those used in~\cite{Faccioli:2001ug,Shuryak:2021fsu,Liu:2023yuj,Liu:2023fpj}. The quark condensate $\langle \bar{\psi}\psi\rangle$ is also close to the one given in \cite{Ioffe:2002ee}. The values of the current quark mass and quark condensate can also be compared to the FLAG lattice calculation by renormalization group evolution. At $\mu=2$ GeV, the current mass is $m\simeq6.9$ MeV and the quark condensate is $\langle\bar{\psi}_f\psi_f\rangle\simeq-(251.7 \mathrm{MeV})^3$ comparable to the result in FLAG lattice $N_f=2+1+1$ calculation \cite{FlavourLatticeAveragingGroup:2019iem}. For completeness, we refer to~\cite{Diakonov:1995qy, Schafer:1996wv}, for more details regarding the phenomenology of the QCD instanton vacuum. 

\subsection{Pseudoparticles form factors}
The emergent vertices (\ref{eq:tHooft}) can be generalized to include further finite size effects of the pseudoparticles. More specifically, each  quark field in the interaction vertices $\Theta_{I,A}$ get dressed
\begin{align}
&\psi(k)\rightarrow\sqrt{\mathcal{F}(\rho k)}~\psi(k)
\end{align}
with 
\begin{equation}
    \sqrt{\mathcal{F}(k)}=z\frac{d}{dz}[I_0(z)K_0(z)-I_1(z)K_1(z)]\bigg|_{z=\frac{k}{2}}
\end{equation}
which is essentially the profiling of the instanton by the quark zero mode.

Also each emitted gluon gets dressed by an induced  non-local form factor. 
For that we recall that the BPST instanton in singular gauge is given by
\begin{equation}
\label{INSINGULAR}
A_\mu^a(x)=\frac{1}{g}\frac{2\bar\eta^{a}_{\mu\nu}x_\nu\rho^2}{x^2(x^2+\rho^2)}
\end{equation}
which is seen to satisfy both fixed-point and covariant gauge.  In  momentum space  it reads
\begin{equation}
\label{INSINGULARQ}
gA_\mu^a(q)=i4\pi^2\frac{\bar\eta^a_{\mu\nu}q_\nu}{q^2}\mathcal{F}_g(\rho q)
\end{equation}
with the gluonic form factor induced by the finite instanton size~\cite{Qian:2015wyq,Diakonov:2002fq}
\begin{equation}
    \mathcal{F}_g(q)=\frac{4}{q^2}-2K_2(q)
\end{equation}
As a result,  each instanton in the interaction vertices $\Theta_{I,A}$, is regulated by
\begin{align}
\frac{2\pi^2}{g}\bar{\eta}^a_{\mu\nu}\rightarrow \frac{2\pi^2}{g}\bar{\eta}^a_{\mu\nu}\mathcal{F}_g(\rho q)
\end{align}
which follows from (\ref{INSINGULARQ}) by LSZ reduction.

The use of the gluonic vertices $\Theta_{I,A}$ is justified in momentum space
diagrams, when the exchange "tail" gluons carry energies below the sphaleron mass
(the top of the tunneling barrier)
\begin{equation}
M_S=\int d^3x \frac 18{G^2_{\mu\nu}(0,\vec x)}
=\frac{3\pi}{4\alpha_s\rho}
\end{equation}
Using the above vacuum parameters, we have $8\pi^2/g^2(\rho)=10$--$15$~\cite{Schafer:1996wv}. This fixes the sphaleron mass $M_S\sim 2.5$ GeV, for $\alpha_s(1/\rho)\sim 0.42$--$0.7$.

\section{Hadronic Form Factors}
\label{sec_hadff}
The hadronic form factors are characterized by several regimes:\\
1/ a soft energy regime with $Q^2<1\,{\rm GeV}^2$ where meson
exchanges induced by the emerging vacuum interactions are dominant;\\
2/ a semi-hard regime with $1< Q^2<10\,{\rm GeV}^2$ where
scaling is still largely violated, where nonperturbative vacuum fields are still
important~\cite{Shuryak:2020ktq};\\
3/ a hard regime $Q^2\gg 10\,{\rm GeV}^2$ where pertubative scaling ultimately takes place.\\ Here we will focus on the soft energy regime, where the emergent multiflavor interactions in the QCD vacuum at low resolution are dominant, and manifest in the form of effective meson exchanges.

With this in mind, and to evaluate the pertinent hadronic form factors with gluonic operators in the QCD instanton vacuum, we will trade matrix elements of the gluonic operators for effective fermionic operators. This can be done in two ways, either by averaging the gluonic operators in the presence of the zero modes for forward matrix elements~\cite{Weiss:2021kpt}, or by using the semi-classical bosonization for forward and off-forward matrix elements~\cite{Kacir:1996qn}. This trading was shown to be exact in two-dimensional QCD in the large  $N_c$ limit in~\cite{Ji:2020bby} (see Eqs. 91-92). Here  we will show how to generalize~\cite{Weiss:2021kpt}, by showing how it can be extended to off-forward matrix elements, and also at NLO in the mean instanton density to account for like and unlike instanton molecules. The latters
play a dominant role in  the gluonic contribution to the energy momentum tensor,  and in general most gluonic operators on the light front.

\subsection{Sum ansatz}
Let ${\cal O}[A]$ be a generic gluonic operator, sourced by a multi-pseudoparticle gluon field  given by the sum ansatz
\begin{equation}
\label{eq:gluon_field}
    A(x)=\sum_{I=1}^{N_++N_-} A_I(x)
\end{equation} 
The ensuing gluonic operator ${\cal O}[A]$ is seen to split
into a sum of multi-instanton contributions
\begin{equation}
\label{eq:gluo_op}
    \mathcal{O}[A]=\sum_{I}\mathcal{O}[A_I]+\sum_{I\neq J}\mathcal{O}[A_I,A_J]+\cdots
\end{equation}
of increasing complexity. In the QCD instanton vacuum, the gauge fields and their quark zero modes,  are reduced to a quantum moduli. The vacuum averaging  over the quantum moduli in the absence of source, is the  effective Lagrangian  given in Appendix~\ref{App:effective_Langrangian}. In the presence of the split form of the gluonic  source (\ref{eq:gluo_op}), the averaging over the quantum moduli
trade the gluonic source for multi-flavored fermionic vertices.

\subsection{Vacuum averages}
The vacuum averages of local gluonic operators using (\ref{eq:Z_N2}) for  fixed-$N_\pm$ configurations (canonical ensemble), follow from
\begin{widetext}
\bea
\label{eq:op_N}
\langle\mathcal{O}[A]\rangle_{N_\pm}=&&\frac{Z^{(g)}_{N_\pm}}{Z_{N_\pm}}\int \prod_{I=1}^{N_++N_-}d^4z_IdU_I\frac{(4\pi^2\rho^3)^{N_f}}{V}\nonumber\\
&&\times\int \mathcal{D}\psi \mathcal{D}\psi^\dagger DA_\mu\mathcal{O}[A]\left(\prod_{I=1}^{N_+}\Theta_I\prod_{A=1}^{N_-}\Theta_A\right) e^{-\int d^4x \left(-\psi^\dagger i\slashed{\partial}\psi+\frac{1}{4}(G^a_{\mu\nu})^2\right)}
\eea
\end{widetext}
The evaluation of the gluonic operators from the instanton vacuum is two-fold. When the operator probes the gluonic content inside the hadronic state with a momentum transfer, the contribution can be calculated by replacing the gluonic field in the operator by the semi-classical background of the instantons \cite{Diakonov:1995qy,Weiss:2021kpt}. The averages over the operators and the quark-instantonic vertices convert the gluonic operator into the corresponding effective quark operator. Thus, in the instanton ensemble, the gluonic operators are mapped into some effective fermionic operators associated with the multi-instanton configuration in the instanton vacuum.

Using (\ref{eq:gluo_op}), the vacuum expectation value of $\mathcal{O}[A]$ in the instanton ensemble can be organized 
in diluteness contributions using the instanton density $n_{I+A}$.
\begin{widetext}
\begin{equation}
\label{eq:inst_contribution}
\begin{aligned}
    \langle \mathcal{O}[A]\rangle_{N_\pm}=&\sum_{n=1}^\infty\frac{1}{n!}\left[\sum_{k=0}^n\binom{N}{k}N_+^{n-k}N_-^k\frac{\langle\mathcal{O}_{++\cdots-}\rangle_{\mathrm{eff}}}{\langle\theta_+\rangle^{n-k}\langle\theta_-\rangle^k}\right]\\[5pt]
    =&N_+\frac{\langle\mathcal{O}_+\rangle_{\mathrm{eff}}}{\langle\theta_+\rangle}+N_-\frac{\langle\mathcal{O}_-\rangle_{\mathrm{eff}}}{\langle\theta_-\rangle}+\frac{N^2_+}{2}\frac{\langle\mathcal{O}_{++}\rangle_{\mathrm{eff}}}{\langle\theta_+\rangle^2}+N_+N_-\frac{\langle\mathcal{O}_{+-}\rangle_{\mathrm{eff}}}{\langle\theta_+\rangle\langle\theta_-\rangle}+\frac{N^2_-}{2}\frac{\langle\mathcal{O}_{--}\rangle_{\mathrm{eff}}}{\langle\theta_-\rangle^2}+\cdots
\end{aligned}
\end{equation}
where the effective fermionic operator $\mathcal{O}_{++\cdots-}$ is obtained by simultaneously connecting $\mathcal{O}[A]$ to the $n$ instantons by sharing the classical fields: 
\begin{equation}
\label{eq:op_avg}
    \mathcal{O}_{++\cdots-}=\int d^4z_{I_1}dU_{I_1}\cdots d^4z_{I_n}dU_{I_n}\mathcal{O}[A_{I_1},A_{I_2},\cdots,A_{I_n}]\Theta_{I_1}\cdots\Theta_{I_n}
\end{equation}
\end{widetext}

Now the canonical ensemble average effectively reduces to the path integral of the effective field theory. The calculations become the vacuum expectation values of a bunch of effective multi-instanton fermionic operators over the effective Lagrangian $\mathcal{L}_{\mathrm{eff}}$. This is the consequences of the diluteness of the instanton vacuum. The calculations now can be done order by order in the framework of the instanton density expansion. As the same idea of the diluteness, the correlation between the instantons becomes irrelevant. Therefore, the fermion and gluon exchanges among the instanton vertices $\Theta_I$ will be neglected.
The extension to a grand canonical ensemble of pseudoparticles with varying $N_\pm$, will follow by inspection (see below).

\subsection{Form factors}
The preceding arguments for the vacuum averages, can be extended to hadronic matrix elements,
provided that the resulting effective vertices are localized within the hadronic size. This
is true for most light hadrons, since the instanton size is comparable to even to the pion, the smallest of all light hadrons. With this in mind, the transition matrix element of the gluon operator $\mathcal{O}[A]$ in a hadron  state, will be given by an ensemble average similar to (\ref{eq:inst_contribution}) with in-out
on-shell hadronic states,
\begin{widetext}
\bea
\label{eq:o_had_exp}
    \langle P'| \mathcal{O}[A]|P\rangle_{N_\pm}=&&\sum_{n=1}^\infty\frac{1}{n!}\left[\sum_{k=0}^n\binom{n}{k}\left(\frac{N_+}{\langle\theta_+\rangle}\right)^{n-k}\left(\frac{N_-}{\langle\theta_-\rangle}\right)^k\langle P'|\mathcal{O}_{++\cdots-}|P\rangle_{\mathrm{eff}}\right]\nonumber\\
    =&&\frac{N_+}{\langle\theta_+\rangle}\langle P'|\mathcal{O}_+|P\rangle_{\mathrm{eff}}+\frac{N_-}{\langle\theta_-\rangle}\langle P'|\mathcal{O}_-|P\rangle_{\mathrm{eff}}+\frac{1}{2}\frac{N^2_+}{\langle\theta_+\rangle^2}\langle P'|\mathcal{O}_{++}|P\rangle_{\mathrm{eff}}\nonumber\\
    &&+\frac{N_+}{\langle\theta_+\rangle}\frac{N_-}{\langle\theta_-\rangle}\langle P'|\mathcal{O}_{+-}|P\rangle_{\mathrm{eff}}+\frac{1}{2}\frac{N^2_-}{\langle\theta_-\rangle^2}\langle P'|\mathcal{O}_{--}|P\rangle_{\mathrm{eff}}+\cdots
\eea
\end{widetext}
The form factors following from (\ref{eq:o_had_exp}) can be expanded systematically, in terms of the instanton density, which is commensurate with a  book-keeping in $1/N_c$. 
Translational symmetry  relates the hadronic matrix element of $\mathcal{O}[A]$ to the momentum transfer  between the hadronic states, 
\bea
     \langle P'|\mathcal{O}[A]|P\rangle = \frac{1}{V}\int d^4x\langle P|\mathcal{O}[A(x)]|P\rangle e^{-iq\cdot x}\nonumber\\
\eea
The recoiling hadron momentum is defined as $P'=P+q$, and the forward limit follows from $q\rightarrow0$.
(\ref{eq:o_had_exp}) generalizes the arguments in~\cite{Weiss:2021kpt} to off-forward and multi-instanton
contributions. 

Graphically, the color-averaging in (\ref{eq:op_avg}) connects $\mathcal{O}[A]$ to $n$ instantons through the  classical field backgrounds. Each matrix element in (\ref{eq:o_had_exp}) is evaluated by the effective Lagrangian $\mathcal{L}_{\mathrm{eff}}$, with only the connected diagrams retained. The hadronic matrix element effectively reduces to the path integral of the effective field theory, thanks to the diluteness of the pseudoparticles in the vacuum state.  The calculations can be carried order by order  in the instanton density expansion.

More specifically, each of the external fermion (antifermion) lines in the diagram contributes a pair of $UU^\dagger$ in the color group integral. Each of the $UU^\dagger$ pair gives a $1/N_c$ factor in the large $N_c$ limit. Therefore, the $1/N_c$ counting of each diagram is $1/N_c^{N_f}$ where $N_f$ is the external unattached fermion number (the number of the external (anti)-fermion line unattached to the operator).
Note  that in some cases the leading $1/N_c$ contribution gives the disconnected diagrams in the matrix element \cite{Diakonov:1995qy}. In this case, the fluctuation in the instanton numbers comes into play,  and the ensemble formulation has to be generalized to the grand canonical ensemble.

\section{Gluonic Scalar Form factors}
\label{sec_scalar}
 We will start
our analysis, by illustrating how this averaging carries 
for the simplest 2-gluon scalar operator
\begin{equation}
\label{eq:O2gs}
    \mathcal{O}_{2g}[A]=G^2_{\mu \nu}(x)
\end{equation}

\subsection{One-instanton contribution}
In  leading order in the density expansion, the effective fermionic operator for $\mathcal{O}_{2g}$ is obtained by averaging the leading expansion of $\mathcal{O}_{2g}$ with one instanton vertex
\begin{equation}
\label{eq:O2g}
    \mathcal{O}_{2g\pm}(x)=\int d^4z_{I}dU_{I}\mathcal{O}_{2g}[A_I]\Theta_{I}
\end{equation}
with $\mathcal{O}_{2g}[A_I]=G^2_{\mu \nu}[A_I]$. 
The classical field of the single instanton plays an important role in the expectation value of the gluonic operators, due to its  strong and localized nature. 

\begin{figure*}
\subfloat[\label{fig:inst_2g-1}]{%
    \includegraphics[scale=0.75]{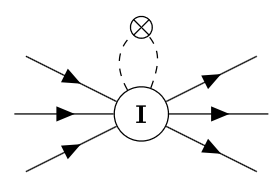}
}\hfill
\subfloat[\label{fig:inst_op_2}]{%
    \includegraphics[width=0.8\textwidth]{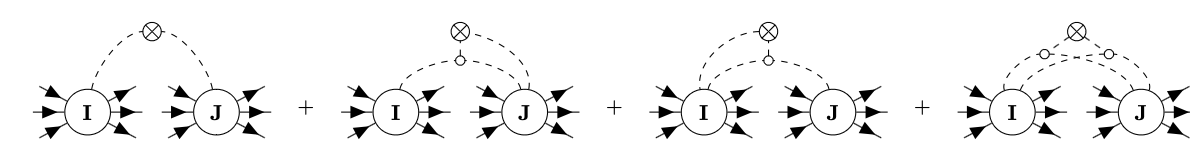}
}
\caption{ 
a: The diagrams of $\mathcal{O}[I]$ in the multi-instanton expansion of the two gluon operator $\mathcal{O}_{2g}$ and $\tilde{\mathcal{O}}_{2g}$. Each dashed line connected to the instanton represents the classical background field;
b: The diagrams of $\mathcal{O}[A_I,A_J]$ in the multi-instanton expansion of the two-gluon operator $\mathcal{O}_{2g}$ and $\tilde{\mathcal{O}}_{2g}$. Each line connected to the instantons $I$ and $J$ represents the gluon fields in the operator. Each of the ringdots in the diagrams represent the insertion of the non-Abelian cross term $G_{\mu\nu}[A_I,A_J]$ in \eqref{eq:cross_term}.
}
\end{figure*}

The calculations are illustrated by the diagrammatic contribution in~Fig.\ref{fig:inst_2g-1}, where each of the dash lines denotes the replacement of the instanton classical fields into the gluon field in the operator. In the Feynman diagram, each of the external fermion flavor $N_f$, contributes a pair of $UU^\dagger$ in the color group integral, giving a $1/N_c$ factor. The $1/N_c$ counting of each diagram is $1/N_c^{N_f}$. Hence, we only consider the one flavor contribution at each order of the instanton density expansion.

In leading order in the instanton density, the $1/N_c$ expansion of the result in (\ref{eq:O2g}) gives
\begin{widetext}
\bea
\label{eq:O_2g}
    &&\frac{1}{V}\int d^4x\mathcal{O}_{2g\pm}(x)e^{-iq \cdot x}=\nonumber\\
    &&\frac{1}{V}\int d^4xF^{(I,A)}_{2g}(x)e^{-i\rho q \cdot x}\left[\left(\frac{m^*}{4\pi^2\rho^2}\right)^{N_f}(2\pi)^4\delta^4(q)-\frac{1}{N_c}\left(\frac{m^*}{4\pi^2\rho^2}\right)^{N_f-1}\int d^4z\bar{\psi}(z)\frac{1\mp\gamma^5}{2}\psi(z)e^{-iq \cdot z}\right]\nonumber\\
\eea
\end{widetext}
where the profile function for the single (anti)-instanton $I$ ($A$) is defined as
\begin{equation}
    F^{(I,A)}_{2g}(x)=\frac{192}{(x^2+1)^4}
\end{equation}

The leading-order contribution of the $1/N_c$ expansion comes from the diagrams where all of the $N_f$ flavors looped up. These diagrams do not have any contribution in the canonical ensemble as they are disconnected to the hadronic source. Similar to the calculation of the $G^2_{\mu\nu}$ operator in \cite{Diakonov:1995qy}, we have to consider the fluctuations in the instanton vacuum. However, this part will only contribute to the forward matrix element. In the off-forward matrix element, the nontrivial contribution has to be connected to the hadronic source and therefore is down by $1/N_c$.

\subsection{Two-instanton contribution}
At next-to-leading order of the instanton density expansion, the effective fermionic operator for $\mathcal{O}_{2g}$ can be obtained by averaging the leading expansion of $\mathcal{O}_{2g}$ with a two-instanton vertex
\bea
    &&\mathcal{O}_{2g\pm\pm}(x)=\nonumber\\
    &&\int d^4z_{I} d^4z_{J} dU_{I}dU_{J}\mathcal{O}_{2g}[A_I,A_J]\Theta_{I}\Theta_{J}
\eea
with
\begin{widetext}
\begin{equation}
\begin{aligned}
\label{eq:O2g_2}
    \mathcal{O}_{2g}[A_I,A_J]=&2G^{a}_{\mu \nu}[I]G^{a}_{\mu \nu}[J]+2G^{a}_{\mu \nu}[J]G^{a}_{\mu \nu}[I,J]+2G^{a}_{\mu \nu}[I]G^{a}_{\mu \nu}[I,J]+G^{a}_{\mu \nu}[I,J]G^{a}_{\mu \nu}[I,J]
\end{aligned}
\end{equation}
\end{widetext}
Here $G_{\mu\nu}[I]$ is short for $G_{\mu\nu}[A_I]$, the single instanton field strength.

In Fig. \ref{fig:inst_op_2}, we also show the diagrams that contribute to the gluonic scalar operator $\mathcal{O}_{2g}$  at the second order in the instanton density. Each diagram corresponds to each terms of the terms given in  (\ref{eq:O2g_2}). Each of the dash lines denotes the replacement of the instanton classical fields by the gluon field in the operator. In the case of a two-instanton cluster, at the leading order in the $1/N_c$ contribution, each instanton with the $N_f-1$ flavors contracted, will dominate the contribution in the hadronic matrix element. The effective operator produces two types of coupling due to the chirality. The operator produced by $II$ or $AA$ clusters corresponds to a chiral flipping process 
\begin{widetext}
\begin{equation}
\begin{aligned}
    &\frac{1}{V}\int d^4x\mathcal{O}_{2g++,--}(x)e^{-iq \cdot x}=\frac{1}{2N_c(N_c^2-1)}\left(\frac{m^*}{4\pi^2\rho^2}\right)^{2N_f-2}\int d^4z_Id^4z_J\frac{1}{8}\mathrm{tr}\{S(z_I-z_J)\}\\
    &\times\frac{1}{V}\int d^4xe^{-iq\cdot x}F^{(II,AA)}_{2g}(x-z_I,x-z_J)\left[\bar{\psi}(z_I)\frac{1\mp\gamma^5}{2}\psi(z_J)+\bar{\psi}(z_J)\frac{1\mp\gamma^5}{2}\psi(z_I)\right]
\end{aligned}
\end{equation}
where the profile function for the $II$ cluster is defined as
\begin{equation}
\begin{aligned}
\label{eq:2gs_profile_II}
F^{(II)}_{2g}(x-z_I,x-z_J)=&2G^{a}_{\mu \nu}(x-z_I)G^{a}_{\mu\nu}(x-z_J)+4\epsilon^{acd}G_{\mu \nu}^c(x-z_I)A_\mu^d(x-z_I)A_\nu^a(x-z_J)\\
&+4\epsilon^{acd}G_{\mu \nu}^c(x-z_J)A_\mu^d(x-z_J)A_\nu^a(x-z_I)
\end{aligned}
\end{equation}
and the profile function for the $AA$ cluster is equal
\begin{equation}
\begin{aligned}
\label{eq:2gs_profile_II}
   & F^{(II)}_{2g}(x-z_I,x-z_J)=F^{(AA)}_{2g}(x-z_I,x-z_J)
\end{aligned}
\end{equation}
The operator produced by $IA$ molecules corresponds to a chiral-conserving process
\begin{equation}
\begin{aligned}
    &\frac{1}{V}\int d^4x\mathcal{O}_{2g+-}(x)e^{-iq \cdot x}=-\frac{1}{2N_c(N_c^2-1)}\left(\frac{m^*}{4\pi^2\rho^2}\right)^{2N_f-2}\int d^4z_Id^4z_J\frac{1}{4}\mathrm{tr}\{S(z_I-z_J)\gamma_\mu\}\\
    &\times\frac{1}{V}\int d^4xe^{-iq\cdot x}F^{(IA)}_{2g,\mu\nu}(x-z_I,x-z_J)\left[\bar{\psi}(z_I)\gamma_\nu\frac{1+\gamma^5}{2}\psi(z_J)-\bar{\psi}(z_J)\gamma_\nu\frac{1-\gamma^5}{2}\psi(z_I)\right]
\end{aligned}
\end{equation}
where the profile function for the $IA$ molecule is defined as
\bea
\label{eq:2g_profile_IA}
    F^{(IA)}_{2g,\rho\lambda}(x-z_I,x-z_J)=&&\frac{1}{2}\bar{\eta}^a_{\rho\beta}\eta^b_{\lambda\beta}
    \bigg[2G^{a}_{\mu \nu}(x-z_I)G^{b}_{\mu\nu}(x-z_J)
    +4\epsilon^{acd}G_{\mu \nu}^c(x-z_I)A_\mu^d(x-z_I)A_\nu^b(x-z_J) \nonumber\\
    &&+4\epsilon^{bcd}G_{\mu \nu}^c(x-z_J)A_\mu^d(x-z_J)A_\nu^a(x-z_I)\bigg] 
\eea
\end{widetext}

Generally, the operators induced by the instanton clusters are non-local. However, as a result of the diluteness in the instanton ensemble, the highly localized nature of the instantons allows us to approximate the non-local quark operators by local operators,  using the expansion in terms of the relative distance between the instanton pairs $R=z_I-z_J$. The relevant contribution to those matrix elements at higher order of the instanton density, comes from the close clusters. Therefore, we can extract the leading contribution by the $R-$expansion (the local approximation) 
$$
\bar{\psi}(z_I)\psi(z_J)\simeq\bar{\psi}(z)\psi(z)-R_\mu\bar{\psi}(z)\overleftrightarrow{\partial_\mu}\psi(z)+\cdots
$$


Note that the quark line connecting the instanton $I$ and $J$ produces the Euclidean quark propagator with a determinantal mass  $m^*$ in the single instanton approximation (SIA), as opposed to the constituent quark mass in the mean-field approximation (MFA), i.e.
\begin{equation}
    S(x-y)\sim \int \frac{d^4k}{(2\pi)^4}\frac{i\slashed{k}+m^*}{k^2+m^{*2}}\mathcal{F}(\rho k)e^{-ik\cdot(x-y)}
\end{equation}

\subsection{Forward matrix element}
To make the physics more transparent, we properly normalize the hadronic matrix element in the forward limit $q\rightarrow0$. With the consideration of the instanton fluctuations, the hadronic matrix element reads
\begin{widetext}
\begin{equation}
\begin{aligned}
\label{eq:gluonball_scalar}
    &\frac{1}{32\pi^2}\frac{\langle PS| g^2G^2_{\mu \nu}|PS\rangle}{2M_N}
    =-\frac{\sigma_T}{\bar{N}}\frac{1}{4}M^{(0)}_{\mathrm{inv}}\\
    &-\left[\frac{1}{N_c}\left(\frac{2\kappa}{\rho^2 m^{*2}}\right)+\frac{1}{2N_c(N_c^2-1)}\left(\frac{2\kappa}{\rho^2 m^{*2}}\right)^2\rho^2m^2T_{II}\right]\frac{\langle PS|m\bar{\psi}\psi|PS\rangle}{2M_N}
\end{aligned}
\end{equation}
\end{widetext}
The first term comes from the instanton number fluctuation, and the second term comes the quark-instanton interaction in the SIA. The topological compressibility $\sigma_T$ can be estimated by the QCD low-energy theorem \cite{Novikov:1981xi,Diakonov:1995qy}. The value is given by 
\bea
\sigma_T=\langle(N-\bar{N})^2\rangle_{\mathds{P}}=\frac{4}{b}\bar{N}
\eea
where $b=\frac{11}{3}N_c$ is the one-loop beta function in $1/N_c$ limit (quenched QCD). 

In QCD, the gluonic scalar operator is tied to the nucleon mass due by the conformal anomaly. In the chiral limit, the nucleon mass is saturated by the contribution from the anomalous mass (invariant mass $M_{\mathrm{inv}}$) induced by the spontaneous breaking of chiral symmetry,
\begin{equation}
 M_N(m\rightarrow0)\equiv    M_{\mathrm{inv}}=-\frac{b}{32\pi^2} \frac{\langle PS|g^2G^2_{\mu\nu}|PS\rangle}{2M_N}
\end{equation}
In large $N_c$ limit, the invariant mass follows by dimension
\bea
M^{(0)}_{\mathrm{inv}}=C\left(\frac{\bar{N}}{V}\right)^{1/4}
\eea
As in  quenched QCD ($N_c\rightarrow\infty$), the instanton density is the only scale in the QCD instanton vacuum. It is 
the analogue of $\Lambda_{QCD}$ for the gauge configurations in the QCD vacuum, after cooling through  gradient flow.
Beyond the quenched limit, quarks start to contribute. The invariant mass will start to run due to the mixing between the gluon scalar and the quark scalar operator, induced by the quark-instanton interaction. The result to second order in the instanton density is
\begin{widetext}
\begin{equation}
\begin{aligned}
M_{\mathrm{inv}}=M_{\mathrm{inv}}^{(0)}+\left(\frac{b}{N_c}\left(\frac{2\kappa}{\rho^2 m^{*2}}\right)+\frac{b}{2N_c(N_c^2-1)}\left(\frac{2\kappa}{\rho^2 m^{*2}}\right)^23\rho^2m^2T_{II}\right)\frac{\langle PS|m\bar{\psi}\psi|PS\rangle}{2M_N}
\end{aligned}
\end{equation}
\end{widetext}
The  invariant mass is renormalized by the quark-mixing, in the instanton density expansion. With the invariant mass renormalized, the generalization to the off-forward matrix element is now straight-forward, as we now detail.

\subsection{Off-forward matrix element}
In the small size limit with momenta $\rho q\ll1$, the  detailed instanton structure is not probed. In this case, the gluonic operators are reduced to effective quark operators, and the momentum transfer dependence is dominated by quarks. However, when the momentum transfer becomes large enough to probe the instanton size, the momentum transfer dependence will be corrected by the instanton profiles. Hence, the off-forward hadronic matrix element, 
\begin{widetext}
\begin{equation}
    \begin{aligned}
    \label{OFFG2}
    &\frac{1}{32\pi^2}\langle P'S|g^2G^2_{\mu \nu}|PS\rangle=-\left[\frac{1}{4}M^{(0)}_{\mathrm{inv}}\frac{\sigma_T}{\bar{N}}\frac{(2\pi)^4}{V}\delta^4(q)\right]\bar{u}_s(P')u_s(P)\\
    &-\left[\frac{1}{N_c}\left(\frac{2\kappa}{\rho^2 m^{*2}}\right)\beta^{(I)}_{2g}(\rho q)+\frac{1}{2N_c(N_c^2-1)}\left(\frac{2\kappa}{\rho^2 m^{*2}}\right)^23\rho^2m^2 T_{II}\beta^{(II)}_{2g}(\rho q)\right]\langle P'S|m\bar{\psi}\psi|PS\rangle \\
    &+\frac{1}{2N_c(N_c^2-1)}\left(\frac{2\kappa}{\rho^2 m^{*2}}\right)^2\frac{\rho^4m^2}{9}T_{IA}\beta^{(IA)}_{2g}(\rho q)q_\mu q_\nu\langle P'S|\bar{\psi}\left(\gamma_{(\mu}i\overleftrightarrow{\partial}_{\nu)}-\frac{1}{4}g_{\mu\nu}i\overleftrightarrow{\slashed{\partial}}\right)\psi|PS\rangle
    \end{aligned}
\end{equation}

where the non-local form factors induced by the finite instanton size effect, are defined as
\bea
\label{BETAGG}
    \beta^{(I)}_{2g}(q)&=&\frac{1}{q}\int_0^{\infty} dx\frac{24x^2}{(x^2+1)^4}J_1(qx)
= {q^2 \over 2}K_2(q) \nonumber\\
    \beta^{(II)}_{2g}(q)&=&\frac{1}{q}\int_0^\infty dx\frac{8(2-x^2)}{(1+x^2)^4}J_1(qx)
    = \frac{16}{q^2} - \frac 1 {6} q^2\big(K_2(q) + 2 K_4( q)\big) \nonumber\\
    \beta^{(IA)}_{2g}(q)&=&\frac{1}{q}\int_0^\infty dx\frac{576 x^2}{(1+x^2)^4}\frac{J_3(qx)}{q^2x^2}
  =   12 \big[48 (-32 + q^2) +  \nonumber \\
 & &  q^2 (768 + 72 q^2 + q^4) K_0( q)
   + 
   12 q (128 + 28 q^2 + q^4)K_1(q)\big]/q^6
\eea
and are normalized to unity in the forward limit. We plotted them in Fig.\ref{fig_beta_ff}
\end{widetext}

The quark hopping integral between the instanton and anti-instanton $T_{IA}(\rho m^*)$ is defined as
\bea
\label{eq:quark_hopping}
&&T_{II}=\int \frac{d^4R}{4\pi^2R}\int_0^\infty dk\mathcal{F}(k)J_1(kR)\bigg|_{R<\frac{1}{\rho \sqrt[4]{n_{I+A}}}}\nonumber\\
&&T_{IA}=\int \frac{d^4R}{16\pi^2}\int_0^\infty dkk\mathcal{F}(k)J_2(kR)\bigg|_{R<\frac{1}{\rho \sqrt[4]{n_{I+A}}}}\nonumber\\
\eea
With the vacuum parameters, we have $T_{II}=1.666$ and $T_{IA}=0.5834$.

It is clear that the universal fluctuation dominates the gluonic scalar form factor at the leading order of the $1/N_c$ expansion. The $\delta^4(q)$ contribution  reflects on the scalar glueball decorrelator with the glueballs sourced solely by the localized and semi-classical instanton and anti-instanton fields,
\begin{widetext}
\bea
\frac{\sigma_T}{\bar{N}}\frac{(2\pi)^4}{V}\delta^4(q)=
\frac{1}{32\pi^2\langle G^2\rangle}\int d^4xe^{-iq\cdot x}\langle G^2(x)G^2(0)\rangle_c
\eea
\end{widetext}
At non-zero momentum transfer, (\ref{OFFG2}) shows that the gluonic $G^2$ in a hadron,  mixes with the scalar meson ($\sigma$) and tensor meson ($t$) in the hadron. The mixing with the quark traceless tensor operator is penalized further by the instanton size.

\section{Gluonic Pseudoscalar Form factor}
\label{sec_pseudoscalar}
The preceding calculation can be straightforwardly extended to the gluonic pseudoscalar operator,  which enters the spin sum rule in  hadrons through the U(1) anomaly (see below),
\begin{equation}
\label{eq:O2gp}
   \tilde{\mathcal{O}}_{2g}[A]=G^a_{\mu \nu}(x)\tilde{G}^a_{\mu \nu}(x)
\end{equation}
where the dual field strength is defined as
\begin{equation}
    \tilde{G}^a_{\mu \nu}(x)=\frac{1}{2}\epsilon_{\mu\nu\rho\lambda}G^a_{\rho\lambda}(x)
\end{equation}

\subsection{One-instanton contribution}
At leading order in the instanton density, the effective quark operator for $\tilde{\mathcal{O}}_{2g}$ can be obtained using
\begin{equation}
\label{eq:O2gp}
    \tilde{\mathcal{O}}_{2g\pm}(x)=\int d^4z_{J}dU_{J}\tilde{\mathcal{O}}_{2g}[A_J]\Theta_{J}
\end{equation}
where 
\begin{equation}
\begin{aligned}
    &\tilde{\mathcal{O}}_{2g}[A_J]=G^a_{\mu \nu}[A_J]\tilde{G}^a_{\mu \nu}[A_J]
\end{aligned}
\end{equation}
The evaluation of  (\ref{eq:O2gp}) is also illustrated by the same diagrammatic contributions in Fig.\ref{fig:inst_2g-1}. The profile function for the single (anti)-instanton $I$ ($A$) in the gluonic pseudoscalar operator can be deduced from the scalar by using the identity $\tilde{G}^a_{\mu \nu}[I,A]=\pm G^a_{\mu \nu}[I,A]$. More specifically, we have
\begin{widetext}
\bea
\label{eq:O_2gp}
    &&\frac{1}{V}\int d^4x\mathcal{O}_{2g\pm}(x)e^{-iq \cdot x}=\nonumber\\
    &&\frac{1}{V}\int d^4x\tilde{F}^{(I,A)}_{2g}(x)e^{-i\rho q \cdot x}\left[\left(\frac{m^*}{4\pi^2\rho^2}\right)^{N_f}(2\pi)^4\delta^4(q)-\frac{1}{N_c}\left(\frac{m^*}{4\pi^2\rho^2}\right)^{N_f-1}\int d^4z\bar{\psi}(z)\frac{1\mp\gamma^5}{2}\psi(z)e^{-iq \cdot z}\right]\nonumber\\
\eea
\end{widetext}
where the profile function for the single (anti)-instanton $I$ ($A$) is defined as
\begin{equation}
    \tilde{F}^{(I,A)}_{2g}(x)=\pm\frac{192}{(x^2+1)^4}
\end{equation}

Again, the contribution is dominated by the disconnected diagrams where all of the $N_f$ flavors looped up. The leading order (LO)  contribution for the pseudoscalar  follows from   the instanton vacuum fluctuations in the forward limit,

\subsection{Two-instanton contribution}
The contribution at the NLO instanton density c is given as
\bea
    &&\tilde{\mathcal{O}}_{2g\pm\pm}(x)=\nonumber\\
    &&\int d^4z_{I} d^4z_{J} dU_{I}dU_{J}\tilde{\mathcal{O}}_{2g}[A_I,A_J]\Theta_{I}\Theta_{J}
\eea
with
\begin{widetext}
\begin{equation}
\begin{aligned}
\label{eq:O2g_2}
    \tilde{\mathcal{O}}_{2g}[A_I,A_J]=&2G^{a}_{\mu \nu}[I]\tilde{G}^{a}_{\mu \nu}[J]+2G^{a}_{\mu \nu}[I]\tilde{G}^{a}_{\mu \nu}[I,J]+2G^{a}_{\mu \nu}[J]\tilde{G}^{a}_{\mu \nu}[I,J]+G^{a}_{\mu \nu}[I,J]\tilde{G}^{a}_{\mu \nu}[I,J]
\end{aligned}
\end{equation}
The instanton pair contribution is similar to the one illustrated in Fig.~\ref{fig:inst_op_2}. The effective operator produces two types of coupling due to the chirality. The one  produced by the $II$ or $AA$ clusters corresponds to a chiral flipping process 
\begin{equation}
\begin{aligned}
    &\frac{1}{V}\int d^4x\tilde{\mathcal{O}}_{2g++,--}(x)e^{-iq \cdot x}=\frac{1}{2N_c(N_c^2-1)}\left(\frac{m^*}{4\pi^2\rho^2}\right)^{2N_f-2}\int d^4z_Id^4z_J\frac{1}{8}\mathrm{tr}\{S(z_I-z_J)\}\\
    &\times\frac{1}{V}\int d^4xe^{-iq\cdot x}\tilde{F}^{(II,AA)}_{2g}(x-z_I,x-z_J)\left[\bar{\psi}(z_I)\frac{1\mp\gamma^5}{2}\psi(z_J)+\bar{\psi}(z_J)\frac{1\mp\gamma^5}{2}\psi(z_I)\right]
\end{aligned}
\end{equation}
where the profile functions for the $II$ and $AA$ clusters are
\bea
\label{eq:2gp_profile_II}
   && \tilde{F}^{(II)}_{2g}(x-z_I,x-z_J)=F^{(II)}_{2g}(x-z_I,x-z_J)
\eea
and
\bea
\label{eq:2gp_profile_AA}
   && \tilde{F}^{(AA)}_{2g}(x-z_I,x-z_J)=-F^{(II)}_{2g}(x-z_I,x-z_J)
\eea
The operator produced by the $IA$ molecules correspond to a chiral-conserving process
\begin{equation}
\begin{aligned}
    &\frac{1}{V}\int d^4x\tilde{\mathcal{O}}_{2g+-}(x)e^{-iq \cdot x}=-\frac{1}{2N_c(N_c^2-1)}\left(\frac{m^*}{4\pi^2\rho^2}\right)^{2N_f-2}\int d^4z_Id^4z_J\frac{1}{4}\mathrm{tr}\{S(z_I-z_J)\gamma_\mu\}\\
    &\times\frac{1}{V}\int d^4xe^{-iq\cdot x}\tilde{F}^{(IA)}_{2g,\mu\nu}(x-z_I,x-z_J)\left[\bar{\psi}(z_I)\gamma_\nu\frac{1+\gamma^5}{2}\psi(z_J)-\bar{\psi}(z_J)\gamma_\nu\frac{1-\gamma^5}{2}\psi(z_I)\right]
\end{aligned}
\end{equation}
where the profile function for the $IA$ molecule is defined as
\begin{equation}
\begin{aligned}
\label{eq:2gp_profile_IA}
   \tilde{F}^{(IA)}_{2g,\rho\lambda}(x-z_I,x-z_J)=2\bar{\eta}^a_{\rho\beta}\eta^b_{\lambda\beta}
    &\bigg[\epsilon^{acd}G_{\mu \nu}^c(x-z_I)A_\mu^d(x-z_I)A_\nu^b(x-z_J)\\
    &-\epsilon^{bcd}G_{\mu \nu}^c(x-z_J)A_\mu^d(x-z_J)A_\nu^a(x-z_I)\bigg]
\end{aligned}
\end{equation}
\end{widetext} 
Again, we approximate the non-local quark operators by local operators,  by expanding  in terms of the relative distance between the instanton pairs $R=z_I-z_J$. 

\subsection{Forward matrix element}
For the gluonic pseudoscalar, the naive forward limit of the hadronic matrix element vanishes due to the parity selection rule. To extract the forward form facrtor, we choose a different normalization for the forward matrix element by tilting the hadron slightly off-forward and properly normalizing the matrix element with spin, before taking the forward limit. Without loss of generality, we assume that the initial state of the nucleon is in the rest frame. Now, we can rewrite the nucleon spinor product as $$\bar{u}_s(P')i\gamma^5u_s(P)\simeq2M_Ns_v$$ with the helicity defined as $s_v=-\chi_s^\dagger\frac{i}{2}\vec{\sigma}\cdot\vec{v}\chi_s$, where the final state nucleon carries a small recoiled velocity $\vec{v}\simeq \vec{q}/M_N$. Thus, the forward limit of the matrix element can be defined as
\begin{equation}
\frac{\langle PS|\mathcal{O}|PS\rangle}{2M_Ns_v}=\lim_{P'\rightarrow P}\frac{\langle P'S|\mathcal{O}|PS\rangle}{2M_Ns_v}
\end{equation}
With the proper normalization to the matrix element, the result of the hadronic matrix element in the forward limit reads
\begin{widetext}
\begin{equation}
\begin{aligned}
\label{eq:gluonball_pseudoscalar}
    &\frac{1}{32\pi^2}\frac{\langle PS| g^2G^a_{\mu \nu}\tilde{G}^a_{\mu \nu}|PS\rangle}{2M_Ns_v}
    =M_q(0)\frac{\chi_t}{\bar{N}}\\
    &+\left[\frac{1}{N_c}\left(\frac{2\kappa}{\rho^2 m^{*2}}\right)+\frac{1}{2N_c(N_c^2-1)}\left(\frac{2\kappa}{\rho^2 m^{*2}}\right)^23\rho^2m^2T_{II}\right]\frac{\langle PS|m\bar{\psi}i\gamma^5\psi|PS\rangle}{2M_Ns_v}\\
    &-\frac{1}{16N_c(N_c^2-1)}\left(\frac{2\kappa}{\rho^2 m^{*2}}\right)^2\rho^2m^2T_{IA}iq_\mu\frac{\langle P'S|\bar{\psi}\gamma_\mu\gamma^5\psi|PS\rangle}{2M_Ns_v}
\end{aligned}
\end{equation}
\end{widetext}
where $M_q(0)\approx 395\,{\rm MeV}$ is the constituent quark mass in (\ref{VER4}) and the value is estimated in \cite{Liu:2023fpj,Liu:2023yuj}.
Here we assumed the nucleon as a quark–scalar-diquark 
composite,  to capture part of the correlations in the nucleon, but not all \cite{Schafer:1995pz,Zahed:2022wae}, e.g.  $(uud)_{\uparrow} \approx u_{\uparrow}[ud]_0$
for a spin-up proton. It follows that the mixing of the proton spin with the axial charge, is mostly through the unpaired $u$-quark.

The result parallels the gluonic scalar matrix element. The first term stems from the fluctuation of the number difference in the ensemble, and the second term stems from the quark-instanton interaction. The topological susceptibility $\chi_t$ can be estimated by the QCD low-energy theorem \cite{Novikov:1981xi,Kacir:1996qn, Diakonov:1995qy}. When recast in terms of the determinantal mass, the result is
\bea
\label{chi_t}
\chi_t=\langle\Delta N^2\rangle_{\mathds{P}}\sim \bar{N}\left(1+N_f\frac{m^*}{m}\right)^{-1}
\eea
In quenched QCD or gluodynamics, the value in \eqref{chi_t} is about $\bar{N}$ (Poisson), a result supported by quenched lattice simulations~\cite{Luscher:2010ik}. A similar estimate follows from 
the Witten-Veneziano formula in (\ref{CHI}, \ref{CHI1}) with $\chi^{(0)}_t/\bar{N}\sim1.95/N_f$ for a singlet mass $M_1\approx0.85$ GeV. 
In unquenched QCD, (\ref{chi_t}) is significantly screened by the quarks and vanishes in the chiral limit 
$\chi_t/\bar N\sim m/N_f m^*$.
 This result holds in the QCD instanton vacuum~\cite{Zahed:1994qh,Shuryak:1994rr,Diakonov:1995qy,Kacir:1996qn}, and is in agreement with Chiral Perturbation Theory (ChPT)~\cite{DiVecchia:1980yfw,Gasser:1984gg, Leutwyler:1992yt}, and random matrix theory~\cite{Janik:1999ps}.

The gluonic pseudoscalar operator is tied to the quark intrinsic spin  by  the Adler-Bell-Jackiw (ABJ) anomaly. In the chiral limit, the quark intrinsic spin in the nucleon, is saturated by the gluonic helicity induced by the ABJ  anomaly. 
This point will detailed below.

\subsection{Off-Forward matrix element}
The generalization to the off-forward matrix element is straight-forward, with the result
\begin{widetext}
\begin{equation}
\begin{aligned}
\label{HADGGT}
    &\frac{1}{32\pi^2}\langle P'S|g^2G^a_{\mu \nu}\tilde{G}^a_{\mu \nu}|PS\rangle
    =\left[M_q(0)\frac{\chi_t}{\bar{N}}\frac{(2\pi)^4}{V}\delta^4(q)\right]\bar{u}_{s'}(P')i\gamma^5u_s(P)\\
    &+\left[\frac{1}{N_c}\left(\frac{2\kappa}{\rho^2 m^{*2}}\right)\tilde{\beta}^{(I)}_{2g}(\rho q)+\frac{1}{2N_c(N_c^2-1)}\left(\frac{2\kappa}{\rho^2 m^{*2}}\right)^23\rho^2m^2T_{II}\tilde{\beta}^{(II)}_{2g}(\rho q)\right]\langle P'S|m\bar{\psi}i\gamma^5\psi|PS\rangle\\
    &-\frac{1}{16N_c(N_c^2-1)}\left(\frac{2\kappa}{\rho^2 m^{*2}}\right)^2\rho^2m^2T_{IA}\tilde{\beta}^{(IA)}_{2g}(\rho q)iq_\mu\langle P'S|\bar{\psi}\gamma_\mu\gamma^5\psi|PS\rangle
\end{aligned}
\end{equation}
\end{widetext}
where the non-local form factors induced by the finite instanton size effect are defined as
\bea
\label{BETAGGTILDE}
    \tilde{\beta}^{(I)}_{2g}(q)&=&\frac{1}{q}\int_0^{\infty} dx\frac{24x^2}{(x^2+1)^4}J_1(qx)\nonumber\\
    \tilde{\beta}^{(II)}_{2g}(q)&=&\frac{1}{q}\int_0^\infty dx\frac{8(2-x^2)}{(1+x^2)^4}J_1(qx)\nonumber\\
    \tilde{\beta}^{(IA)}_{2g}(q)&=&\frac{1}{q}\int_0^\infty dx\frac{192x^2}{(1+x^2)^5}\frac{J_2(qx)}{qx}\nonumber\\
\eea
and are normalized to unity in the forward limit. The quark hopping integral between the instanton and anti-instanton $T_{IA}(\rho m^*)$ is defined in \eqref{eq:quark_hopping}. 
Note that the ultra-local  $\delta^4(q)$ contribution can
be written as
\begin{widetext}
\bea
\frac{\chi_t}{\bar{N}}\frac{(2\pi)^4}{V}\delta^4(q)
=\frac{1}{32\pi^2\langle G^2\rangle}\int d^4xe^{-iq\cdot x}\langle G\tilde{G}(x)G\tilde{G}(0)\rangle
\eea
\end{widetext}
The non-zero momentum transfer induces  emergent  meson-nucleon couplings in (\ref{HADGGT}), with the pseudoscalar mesons ($\eta'$) and axial vector mesons ($f_1$).

\section{C-odd 3-gluon Form factor}
\label{sec_Codd}
Another class of gluonic operator of interest, is the C-odd  three gluon operator, 
\bea
\label{eq:op_3g}
    \mathcal{O}_{3g}[A]=d^{abc}G^a_{\mu\nu}(x)G^b_{\rho \alpha}(x)G^{c}_{\lambda \alpha}(x)
\eea
which is found to be the leading operator in the photoproduction process of heavy pseudoscalar mesons~\cite{Ma:2003py}.
A similar but not identical operator was considered recently in~\cite{Weiss:2021kpt}, to estimate Weinberg's CP-odd contribution to the nucleon matrix element. 
We now proceed to evaluate it in the QCD 
instanton vacuum, through the substitution of the sum ansatz 
(\ref{eq:gluon_field}), and the ensuing averaging over the moduli.

\subsection{One-instanton contribution}
Since $d^{abc}$ is a symmetric $SU(3)$ structure constant which with no support on the $SU(2)$ subgroup, in LO  of the instanton density, it follows that all associated color orientations  of the gluonic fields rotate congruently in the moduli. Therefore, the structure constant $d^{abc}$ would reduce to $\mathrm{tr}(\tau^a\{\tau^b,\tau^c\})=0$, with zero contribution.

\subsection{Two-instanton contribution}
In NLO in the instanton density expansion, some of the two-instanton terms in $\mathcal{O}_{3g}[A]$ involve the nontrivial relative color rotation $U_I^{\dagger}U_J=U_{IJ}$ between $\tau^a$ and $\mathds{1}_2$ in the $SU(2)$ subgroup, with non-zero net  color structure 
\bea
&&\mathrm{tr}(U_I\tau^aU^\dagger_I\{U_J\tau^bU^\dagger_J,U_J\tau^cU^\dagger_J\})\nonumber\\
&&=\mathrm{tr}(\tau^aU^\dagger_IU_J\mathds{1}_2U^\dagger_JU_I)\delta^{bc}
\eea
This means contributions from the single instanton fields $G_{\mu\nu}[A_I]$ and $G_{\mu\nu}[A_J]$, as well as the overlap between pairs of instanton fields $G_{\mu\nu}[A_I,A_J]$,  a consequence of the non-Abelian gauge nature. (See Appendix \ref{App:sing_inst}). The evaluation of the latters appear involved. 
Fortunately, the calculation can be simplified as the $d^{abc}$ color structure has support only when two different $SU(2)$ subgroups in $SU(N_c)$ overlap through their relative color orientations. To obtain a non-trivial relative color rotation, the crossing term $G_{\mu\nu}[I,J]$ has to be combined with both $G_{\mu\nu}[I]$ and $G_{\mu\nu}[J]$, to carry  the same  color rotation along either $I$ or $J$,
\begin{equation}
    \mathcal{O}_{3g\pm\pm}(x)=\int d^4z_{I}dU_{I}d^4z_{J}dU_{J}\mathcal{O}_{3g}[A_I,A_J]\Theta_{I}\Theta_{J}
\end{equation}
where
\begin{widetext}
\begin{equation}
\begin{aligned}
\label{eq:O3g}
    &\mathcal{O}_{3g}[A_I,A_J]\\
    =&d^{abc}\bigg(G^{a}_{\mu \nu}[I]G^{b}_{\rho \alpha}[I]G^{c}_{\lambda \alpha}[J]+G^{a}_{\mu \nu}[I]G^{b}_{\rho \alpha}[J]G^{c}_{\lambda \alpha}[I]+G^{a}_{\mu\nu}[J]G^{b}_{\rho \alpha}[I]G^{c}_{\lambda \alpha}[I]+G^{a}_{\mu \nu}[J]G^{b}_{\rho \alpha}[J]G^{c}_{\lambda \alpha}[I]\\
    &+G^{a}_{\mu \nu}[J]G^{b}_{\rho \alpha}[I]G^{c}_{\lambda \alpha}[J]+G^{a}_{\mu\nu}[J]G^{b}_{\rho \alpha}[I]G^{c}_{\lambda \alpha}[I]+G^{a}_{\mu\nu}[I]G^{b}_{\rho \alpha}[J]G^{c}_{\lambda \alpha}[I,J]+G^{a}_{\mu\nu}[J]G^{b}_{\rho \alpha}[I]G^{c}_{\lambda \alpha}[I,J]\\
    &+G^{a}_{\mu\nu}[I]G^{b}_{\rho \alpha}[I,J]G^{c}_{\lambda \alpha}[J]+G^{a}_{\mu\nu}[J]G^{b}_{\rho \alpha}[I,J]G^{c}_{\lambda \alpha}[I]+G^{a}_{\mu\nu}[I,J]G^{b}_{\rho \alpha}[I]G^{c}_{\lambda \alpha}[J]+G^{a}_{\mu\nu}[I,J]G^{b}_{\rho \alpha}[J]G^{c}_{\lambda \alpha}[I]\bigg)
\end{aligned}
\end{equation}
\end{widetext}
Here $G_{\mu\nu}[I]$ is short for $G_{\mu\nu}[A_I]$. In the case of a two-instanton cluster and in LO in $1/N_c$ counting, each instanton with $N_f-1$ flavors looped up,  dominates the contribution in the hadronic matrix element. The effective operator produces two types of coupling due to the chirality. The operator produced by $II$ or $AA$ clusters corresponds to a chiral flipping process:
\begin{widetext}
\begin{equation}
\begin{aligned}
    &\frac{1}{V}\int d^4x e^{-iq\cdot x}\mathcal{O}_{3g++,--}(x)=-\frac{N_c-2}{2N_c^2(N_c^2-1)}\left(\frac{m^*}{4\pi^2\rho^2}\right)^{2N_f-2}\int d^4z_Id^4z_J\frac{1}{8}\mathrm{tr}\{S(z_I-z_J)\}\\
    &\times\frac{1}{V}\int d^4x e^{-iq\cdot x}F^{(II)}_{3g,\mu\nu\rho\lambda\alpha\beta}(x-z_I,x-z_J)\left[\bar{\psi}(z_I)\frac{1\mp\gamma^5}{2}\sigma_{\alpha\beta}\psi(z_J)+\bar{\psi}(z_J)\frac{1\mp\gamma^5}{2}\sigma_{\alpha\beta}\psi(z_I)\right]
\end{aligned}
\end{equation}
where the profile function for the $II$ or $AA$ pairs is defined as 
\begin{equation}
\begin{aligned}
\label{eq:3g_profile_II}
    &F^{(II)}_{3g,\mu\nu\rho\lambda,\beta\gamma}(x-z_I,x-z_J)\\
    =&\frac{1}{2}\bar{\eta}^a_{\beta\gamma}\bigg[G^{b}_{\mu \nu}(x-z_I)G^{b}_{\rho \alpha}(x-z_I)G^{a}_{\lambda \alpha}(x-z_J)+G^{b}_{\mu \nu}(x-z_I)G^{a}_{\rho \alpha}(x-z_J)G^{b}_{\lambda \alpha}(x-z_I)\\
    &+G^{a}_{\mu \nu}(x-z_J)G^{b}_{\rho \alpha}(x-z_I)G^{b}_{\lambda \alpha}(x-z_I)\bigg]+(I\leftrightarrow J)\\
    &+\frac{1}{2}\bar{\eta}^a_{\beta\gamma}\frac{1}{4(N_c+2)}\bigg[\epsilon^{acd}G^{c}_{\mu \nu}(x-z_I)\left[A^{d}_{\rho}(x-z_I)A_\alpha^b(x-z_J)-A^{d}_{\alpha}(x-z_I)A_\rho^b(x-z_J)\right]G^{b}_{\lambda \alpha}(x-z_J)\\
    &+\epsilon^{acd}G^{c}_{\rho \alpha}(x-z_I)\left[A^{d}_{\lambda}(x-z_I)A_\alpha^b(x-z_J)-A^{d}_{\alpha}(x-z_I)A_\lambda^b(x-z_J)\right]G^{b}_{\mu\nu}(x-z_J)\\
    &+\epsilon^{acd}G^{c}_{\lambda\alpha}(x-z_I)\left[A^{d}_{\mu}(x-z_I)A_\nu^b(x-z_J)-A^{d}_{\nu}(x-z_I)A_\mu^b(x-z_J)\right]G^{b}_{\rho\alpha}(x-z_J)+(\rho\leftrightarrow\lambda)\bigg]
\end{aligned}
\end{equation}
The  operator produced by $IA$ clusters corresponds to a chiral-conserving process  
\begin{equation}
\begin{aligned}
    &\frac{1}{V}\int d^4x e^{-iq\cdot x}\mathcal{O}_{3g+-}(x)=-\frac{N_c-2}{2N_c^2(N_c^2-1)}\left(\frac{m^*}{4\pi^2\rho^2}\right)^{2N_f-2}\int d^4z_Id^4z_J\frac{i}{4}\mathrm{tr}\{S(z_I-z_J)\gamma_\alpha\}\\
    &\times\frac{1}{V}\int d^4x e^{-iq\cdot x}F^{(IA)}_{3g,\mu\nu\rho\lambda\alpha\beta}(x-z_I,x-z_J)\left[\bar{\psi}(z_I)\gamma_\beta\frac{1+\gamma^5}{2}\psi(z_J)+\bar{\psi}(z_J)\gamma_\beta\frac{1-\gamma^5}{2}\psi(z_I)\right]
\end{aligned}
\end{equation}
where the profile function for the intanton clusters is defined as 
\begin{equation}
\begin{aligned}
\label{eq:3g_profile_IA}
    &F^{(IA)}_{3g,\mu\nu\rho\lambda,\beta\gamma}(x-z_I,x-z_J)\\
    =&\frac{1}{2}\bar{\eta}^a_{\beta\gamma}\bigg[G^{b}_{\mu \nu}(x-z_I)G^{b}_{\rho \alpha}(x-z_I)G^{a}_{\lambda \alpha}(x-z_J)+G^{b}_{\mu \nu}(x-z_I)G^{a}_{\rho \alpha}(x-z_J)G^{b}_{\lambda \alpha}(x-z_I)\\
    &+G^{a}_{\mu \nu}(x-z_J)G^{b}_{\rho \alpha}(x-z_I)G^{b}_{\lambda \alpha}(x-z_I)\bigg]-(I\leftrightarrow J)\\
    &+\frac{1}{2}\bar{\eta}^a_{\beta\gamma}\frac{1}{4(N_c+2)}\bigg[\epsilon^{acd}G^{c}_{\mu \nu}(x-z_I)\left[A^{d}_{\rho}(x-z_I)A_\alpha^b(x-z_J)-A^{d}_{\alpha}(x-z_I)A_\rho^b(x-z_J)\right]G^{b}_{\lambda \alpha}(x-z_J)\\
    &+\epsilon^{acd}G^{c}_{\rho \alpha}(x-z_I)\left[A^{d}_{\lambda}(x-z_I)A_\alpha^b(x-z_J)-A^{d}_{\alpha}(x-z_I)A_\lambda^b(x-z_J)\right]G^{b}_{\mu\nu}(x-z_J)\\
    &+\epsilon^{acd}G^{c}_{\lambda\alpha}(x-z_I)\left[A^{d}_{\mu}(x-z_I)A_\nu^b(x-z_J)-A^{d}_{\nu}(x-z_I)A_\mu^b(x-z_J)\right]G^{b}_{\rho\alpha}(x-z_J)+(\rho\leftrightarrow\lambda)\bigg]
\end{aligned}
\end{equation}
\end{widetext}
In the SIA, these correlation functions can be simplied as single-point profile functions. The renormalization-group (RG) invariant result for the hadronic matrix element is 
\begin{widetext}
\begin{equation}
\begin{aligned}
\label{eq:had_3g_2}
    &\langle P'|g^3d^{abc}G^a_{\mu\nu}G^b_{\rho \alpha}G^{c}_{\lambda \alpha}|P\rangle=-\frac{N_c-2}{2N_c^2(N_c^2-1)}\left(\frac{2\kappa}{\rho^2 m^{*2}}\right)^2\frac{8\pi^2m}{9}\rho^2m^2 T_{II}\beta^{(II)}_{3g}(\rho q)\frac{1}{2}q^2\delta_{\rho\lambda}\langle P'|\bar{\psi}\sigma_{\mu\nu}\psi|P\rangle\\
    &+\frac{N_c-2}{2N_c^2(N_c^2-1)}\left(\frac{2\kappa}{\rho^2 m^{*2}}\right)^2\frac{8\pi^2m}{9}\rho^2m^2 T_{II}\beta^{(II)}_{3g}(\rho q)\left(\delta_{\mu\alpha}q_\beta q_\nu-\delta_{\nu\alpha}q_\beta q_\mu\right)\delta_{\rho\lambda}\langle P'|\bar{\psi}\sigma_{\alpha\beta}\psi|P\rangle\\
    &-\frac{N_c-2}{2N_c^2(N_c^2-1)}\left(\frac{2\kappa}{\rho^2 m^{*2}}\right)^2\rho^2m^2T_{IA}\frac{4\pi^2}{45}\beta^{(IA)}_{3g}(\rho q)\\
    &\times\left[\epsilon_{\beta\gamma\lambda\sigma}q_\sigma q_\nu\delta_{\mu\rho}-\epsilon_{\beta\gamma\nu\sigma}q_\sigma q_\lambda\delta_{\mu\rho}-\frac{1}{2}\epsilon_{\beta\gamma\lambda\nu}q^2\delta_{\mu\rho}+(\rho\leftrightarrow\lambda)-(\mu\leftrightarrow\nu)\right]\langle P'|\bar{\psi}\gamma_{[\beta}i\overleftrightarrow{\partial}_{\gamma]}\gamma^5\psi|P\rangle\\
    &-\frac{N_c-2}{2N_c^2(N_c^2-1)}\left(\frac{2\kappa}{\rho^2 m^{*2}}\right)^2\rho^2m^2T_{IA}\frac{4\pi^2}{15}\beta^{(IA)}_{3g}(\rho q)\delta_{\rho\lambda}\left(\epsilon_{\beta\gamma\mu\sigma}q_\sigma q_\nu-\epsilon_{\beta\gamma\nu\sigma}q_\sigma q_\mu-\frac{1}{2}\epsilon_{\beta\gamma\mu\nu}q^2\right)\langle P'|\bar{\psi}\gamma_{[\beta}i\overleftrightarrow{\partial}_{\gamma]}\gamma^5\psi|P\rangle\\
    &-\frac{N_c-2}{2N_c^2(N_c^2-1)}\left(\frac{2\kappa}{\rho^2 m^{*2}}\right)^2\rho^2m^2T_{IA}\frac{8\pi^2}{45}\beta^{(IA)}_{3g}(\rho q)\\
    &\times\left[\epsilon_{\mu\nu\rho\alpha}\left(\delta_{\beta\lambda}q_\gamma q_\alpha-\delta_{\beta\alpha}q_\gamma q_\lambda\right)-\frac{1}{2}\epsilon_{\mu\nu\rho\gamma}q^2\delta_{\beta\lambda}+(\rho\leftrightarrow\lambda)\right]\langle P'|\bar{\psi}\gamma_{[\beta}i\overleftrightarrow{\partial}_{\gamma]}\gamma^5\psi|P\rangle
\end{aligned}
\end{equation}
where
\bea
\label{BETACODD}
    \beta^{(II)}_{3g}(\rho q)=\frac{1}{q}\int_0^\infty dx\frac{9}{2}\frac{640x^4}{(1+x^2)^6}\frac{J_3(qx)}{q^2x^2}\nonumber\\
    \beta^{(IA)}_{3g}(\rho q)=\frac{1}{q}\int_0^\infty dx\frac{2880x^4}{(1+x^2)^6}\frac{J_3(qx)}{q^2x^2}
\eea
which is seen to vanish in the forward direction.
\\
\\
{\bf Light front twist-3:}
\\
 This operator is of special interest on the light front as a leading  twist-$3$, that sources the coherent photoproduction 
 of $1^{+-}$ heavy mesons off the nucleon state~\cite{Ma:2003py}. In light front signature, it contributes
\begin{equation}
\begin{aligned}
\label{eq:had_2g_2}
    &\langle P'|g^3d^{abc}G^{a+i}(x)G^{b+j}(x)G^{c+}{}_{j}(x)|P\rangle=\frac{N_c-2}{8N_c^2(N_c^2-1)(N_c+2)}\left(\frac{2\kappa}{\rho^2 m^{*2}}\right)^2\\
    &\times\rho^4m^3T_{II}(q^+)^2\beta^+_{3g}(\rho q)\langle P'S|\bar{\psi}\left(\sigma^{i\alpha}q^+q_\alpha-\sigma^{+\alpha}q^iq_\alpha-\frac{1}{2}q^2\sigma^{+i}\right)\psi|PS\rangle\\
\end{aligned}
\end{equation}
where
\begin{equation}
    \beta^+_{3g}(\rho q)=\frac{4\pi^2}{q}\int_0^\infty dx\frac{512x^4}{(1+x^2)^6}\frac{J_5(qx)}{q^4x^4}
\end{equation}
\\
\\
{\bf Gravity dual twist-5:}
\\
The gluonic C-odd operator $d^{abc}G^{a}_{\mu\nu}G^{b}_{\rho\lambda}G^{c}_{\rho\lambda}$
as a twist-5 boundary operator, is argued to be dual of the supergravity  B-field in diffractive hadron-hadron scattering
with Odderon exchange~\cite{Hechenberger:2024abg}. In particular, its off-forward matrix element is
\begin{equation}
\begin{aligned}
    &\langle P'|g^3d^{abc}G^{a}_{\mu\nu}G^{b}_{\rho\lambda}G^{c}_{\rho\lambda}|P\rangle=-\frac{N_c-2}{2N_c^2(N_c^2-1)}\left(\frac{2\kappa}{\rho^2 m^{*2}}\right)^2\rho^2m^2T_{II}\frac{16\pi^2}{9}q^2\langle P'|m\bar{\psi}\sigma_{\mu\nu}\psi|P\rangle\\
    &+\frac{N_c-2}{2N_c^2(N_c^2-1)}\left(\frac{2\kappa}{\rho^2 m^{*2}}\right)^2\rho^2m^2T_{II}\frac{32\pi^2}{9}\left(\delta_{\mu\alpha}q_\beta q_\nu-\delta_{\nu\alpha}q_\beta q_\mu\right)\langle P'|m\bar{\psi}\sigma_{\alpha\beta}\psi|P\rangle\\
    &-\frac{N_c-2}{2N_c^2(N_c^2-1)}\left(\frac{2\kappa}{\rho^2 m^{*2}}\right)^2\rho^2m^2T_{IA}\frac{32\pi^2}{45}\beta^{(IA)}_{3g}(\rho q)\left(\epsilon_{\beta\gamma\mu\sigma}q_\sigma q_\nu-\epsilon_{\beta\gamma\nu\sigma}q_\sigma q_\mu-\frac{1}{2}\epsilon_{\beta\gamma\mu\nu}q^2\right)\langle P'|m\bar{\psi}\gamma_{[\beta}i\overleftrightarrow{\partial}_{\gamma]}\gamma^5\psi|P\rangle\\
    &-\frac{N_c-2}{2N_c^2(N_c^2-1)}\left(\frac{2\kappa}{\rho^2 m^{*2}}\right)^2\rho^2m^2T_{IA}\frac{32\pi^2}{45}\beta^{(IA)}_{3g}(\rho q)\left(\epsilon_{\mu\nu\beta\alpha}q_\gamma q_\alpha-\frac{1}{4}\epsilon_{\mu\nu\beta\gamma}q^2\right)\langle P'|m\bar{\psi}\gamma_{[\beta}i\overleftrightarrow{\partial}_{\gamma]}\gamma^5\psi|P\rangle
\end{aligned}
\end{equation}
\end{widetext}

\section{C-even 3-gluon Form factor}
\label{sec_Ceven}
As a parity check on the preceding calculation, we now consider the C-odd analogue of (\ref{eq:op_3g})
\begin{equation}
\label{eq:op_3gf}
    \tilde{\mathcal{O}}_{3g}[A]=f^{abc}G^a_{\mu\nu}(x)G^b_{\nu\rho}(x)G^{c}_{\rho \mu}(x)
\end{equation}
(\ref{eq:op_3gf}) can again be evaluated semiclassically through  the substitution of the gluonic field in~\eqref{eq:gluon_field}.

\subsection{One-instanton contribution}
In  LO in the density expansion, the effective fermionic operator for $\tilde{\mathcal{O}}_{3g}$ follows from the modular averaging
\begin{equation}
\label{eq:O2g}
    \tilde{\mathcal{O}}_{3g\pm}(x)=\int d^4z_{I}dU_{I}\tilde{\mathcal{O}}_{3g}[A_I]\Theta_{I}
\end{equation}
with
\begin{equation}
\begin{aligned}
    &\tilde{\mathcal{O}}_{3g}[A_I]=\epsilon^{abc}G^a_{\mu\nu}[A_I]G^b_{\nu\rho}[A_I]G^{c}_{\rho \mu}[A_I]
\end{aligned}
\end{equation}
A rerun of the preceding steps give
\begin{widetext}
\bea
\label{eq:O3gf}
    &&\frac{1}{V}\int d^4x\tilde{\mathcal{O}}_{3g\pm}(x)e^{-iq \cdot x}=\nonumber\\
    &&\frac{1}{\rho^2V}\int d^4x\tilde{F}^{(I,A)}_{3g}(x)e^{-i\rho q \cdot x}\left[\left(\frac{m^*}{4\pi^2\rho^2}\right)^{N_f}(2\pi)^4\delta^4(q)-\frac{1}{N_c}\left(\frac{m^*}{4\pi^2\rho^2}\right)^{N_f-1}\int d^4z\bar{\psi}(z)\frac{1\mp\gamma^5}{2}\psi(z)e^{-iq \cdot z}\right]\nonumber\\
\eea
\end{widetext}
where the profile function for the single (anti)-instanton $I$ ($A$) is defined as
\begin{equation}
    \tilde{F}^{(I,A)}_{3g}(x)=\frac{1536}{(x^2+1)^6}
\end{equation}

 \subsection{Two-instanton contribution}
At NLO in the density expansion, the effective fermionic operator for $\tilde{\mathcal{O}}_{3g}$ is obtaied by averaging the leading expansion of $\tilde{\mathcal{O}}_{3g}$ with a two-instanton vertex
\bea
    &&\tilde{\mathcal{O}}_{3g\pm\pm}(x)=\nonumber\\
    &&\int d^4z_{I} d^4z_{J} dU_{I}dU_{J}\tilde{\mathcal{O}}_{3g}[A_I,A_J]\Theta_{I}\Theta_{J}
\eea
with
\begin{widetext}
\begin{equation}
\begin{aligned}
\label{eq:O3g}
    &\tilde{\mathcal{O}}_{3g}[A_I,A_J]\\
    \simeq&\epsilon^{abc}\bigg(G^{a}_{\mu \nu}[I]G^{b}_{\rho \alpha}[I]G^{c}_{\lambda \alpha}[J]+G^{a}_{\mu \nu}[I]G^{b}_{\rho \alpha}[J]G^{c}_{\lambda \alpha}[I]+G^{a}_{\mu\nu}[J]G^{b}_{\rho \alpha}[I]G^{c}_{\lambda \alpha}[I]+G^{a}_{\mu \nu}[J]G^{b}_{\rho \alpha}[J]G^{c}_{\lambda \alpha}[I]\\
    &+G^{a}_{\mu \nu}[J]G^{b}_{\rho \alpha}[I]G^{c}_{\lambda \alpha}[J]+G^{a}_{\mu\nu}[J]G^{b}_{\rho \alpha}[I]G^{c}_{\lambda \alpha}[I]\bigg)
\end{aligned}
\end{equation}
\end{widetext}
Here we dropped the contributions associated to  $G_{\mu\nu}[I,J]$ as they are subleading in power counting.

The emerging fermionic operators are two-fold: chiral conserving and flipping.  The chiral flipping operators are produced by $II$ or $AA$ clusters,
\begin{widetext}
\begin{equation}
\begin{aligned}
    &\frac{1}{V}\int d^4x\tilde{\mathcal{O}}_{3g++,--}(x)e^{-iq \cdot x}=\frac{1}{2N_c(N_c^2-1)}\left(\frac{m^*}{4\pi^2\rho^2}\right)^{2N_f-2}\int d^4z_Id^4z_J\frac{1}{8}\mathrm{tr}\{S(z_I-z_J)\}\\
    &\times\frac{1}{V}\int d^4xe^{-iq\cdot x}\tilde{F}^{(II,AA)}_{3g}(x-z_I,x-z_J)\left[\bar{\psi}(z_I)\frac{1\mp\gamma^5}{2}\psi(z_J)+\bar{\psi}(z_J)\frac{1\mp\gamma^5}{2}\psi(z_I)\right]
\end{aligned}
\end{equation}
where the profile function for the $II$ cluster is defined as
\begin{equation}
\begin{aligned}
\label{eq:3gf_profile_II}
\tilde{F}^{(II)}_{3g}(x-z_I,x-z_J)=&3\epsilon^{abc}G^{a}_{\mu\nu}(x-z_J)G^{b}_{\nu \rho}(x-z_I)G^{c}_{\rho\mu}(x-z_I)+3\epsilon^{abc}G^{a}_{\mu\nu}(x-z_I)G^{b}_{\nu \rho}(x-z_J)G^{c}_{\rho\mu}(x-z_J)
\end{aligned}
\end{equation}
and the profile function for the $AA$ cluster is 
\begin{equation}
\begin{aligned}
\label{eq:3gf_profile_AA}
   & \tilde{F}^{(II)}_{3g}(x-z_I,x-z_J)=\tilde{F}^{(AA)}_{3g}(x-z_I,x-z_J)
\end{aligned}
\end{equation}
The chiral preserving operators are produced by the $IA$ molecules, 
\begin{equation}
\begin{aligned}
    &\frac{1}{V}\int d^4x\tilde{\mathcal{O}}_{3g+-}(x)e^{-iq \cdot x}=-\frac{1}{2N_c(N_c^2-1)}\left(\frac{m^*}{4\pi^2\rho^2}\right)^{2N_f-2}\int d^4z_Id^4z_J\frac{1}{4}\mathrm{tr}\{S(z_I-z_J)\gamma_\mu\}\\
    &\times\frac{1}{V}\int d^4xe^{-iq\cdot x}\tilde{F}^{(IA)}_{3g,\mu\nu}(x-z_I,x-z_J)\left[\bar{\psi}(z_I)\gamma_\nu\frac{1+\gamma^5}{2}\psi(z_J)-\bar{\psi}(z_J)\gamma_\nu\frac{1-\gamma^5}{2}\psi(z_I)\right]
\end{aligned}
\end{equation}
where the profile function for the $IA$ molecule is defined as
\begin{equation}
    \begin{aligned}
     \label{eq:2g_profile_IA}
    \tilde{F}^{(IA)}_{3g,\rho\lambda}(x-z_I,x-z_J)=&\frac{1}{2}\bar{\eta}^a_{\rho\beta}\eta^b_{\lambda\beta}
    \bigg[3\epsilon^{acd}G^{b}_{\mu\nu}(x-z_J)G^{c}_{\nu \rho}(x-z_I)G^{d}_{\rho\mu}(x-z_I)\\
    &+3\epsilon^{bcd}G^{a}_{\mu\nu}(x-z_I)G^{c}_{\nu \rho}(x-z_J)G^{d}_{\rho\mu}(x-z_J)\bigg]  \\
    =&0
    \end{aligned}
\end{equation}
\end{widetext}
and is seen to vanish. 
We again approximated the non-local quark operators by local operators, by expanding in the relative distance between close instanton pairs $R=z_I-z_J$.  

\subsection{Off-forward matrix element}
The general off-forward  hadronic matrix element of the C-even three gluon operator is
\begin{widetext}
\bea
    &&\frac{5\rho^2}{384\pi^2}\langle P'S|g^3f^{abc}G^a_{\mu\nu}G^b_{\nu\rho}G^{c}_{\rho \mu}|PS\rangle=-\left[\frac{1}{4}M^{(0)}_{\mathrm{inv}}\frac{\sigma_T}{\bar{N}}\frac{(2\pi)^4}{V}\delta^4(q)\right]\bar{u}_{s'}(P')u_s(P)\nonumber\\
    &&-\left[\frac{1}{N_c}\left(\frac{2\kappa}{\rho^2 m^{*2}}\right)-\frac{1}{2N_c(N_c^2-1)}\left(\frac{2\kappa}{\rho^2 m^{*2}}\right)^23\rho^2m^2T_{II}\right]\tilde{\beta}_{3g}(\rho q)\langle P'S|m\bar{\psi}\psi|PS\rangle\nonumber \\
\eea
\end{widetext}
where the non-local form factors induced by the finite instanton size effect are defined as
\bea
\label{BETACEVEN}
    \tilde{\beta}_{3g}(q)&=&\frac{1}{q}\int_0^{\infty} dx\frac{80x^2}{(x^2+1)^6}J_1(qx)
\eea
with the normalization to 1 in the forward limit.

\section{Gluonic Gravitational Form factors}
\label{sec_gravitationalff}
Another gluonic operator of interest is the gluonic tensor tied to the QCD energy-momentum tensor (EMT).  To evaluate the QCD energy-momentum tensor using local or non-local  effective 
formulations is subtle, for a recent discussion see~\cite{Freese:2019bhb}. In the present approach, it follows the same reasoning as that for the scalar and pseudoscalar 
operators discussed above.

With this in mind, the EMT in QCD is given by
\begin{equation}
    T_{\mu\nu}=T^g_{\mu\nu}+T^q_{\mu\nu}
\end{equation}
where $T^g_{\mu\nu}$ is the gluonic EMT and $T^q_{\mu\nu}$ is the quark EMT. All issues of operator renormalization
are understood in the sense of a gradient flow cooling to the
semiclassical point with fixed topological charge per 4-volume, with the instanton density as the sole scale.
The energy-momentum tensor can be decomposed as the sum of a traceless and traceful part~\cite{Ji:2021mtz,Ji:1995sv},
\begin{equation}
\label{EMT_decop}
T_{\mu\nu}=\bar{T}_{\mu\nu}+\frac{1}{4}g_{\mu\nu}T_{\alpha\alpha}
\end{equation}
where the traceless part includes the gluonic tensor
\begin{equation}
\begin{aligned}
\bar{T}^g_{\mu\nu}=\frac{1}{4}g_{\mu\nu}G^2_{\lambda\rho}-G^{a}_{\mu\lambda}G^{a}_{\nu\lambda}
\end{aligned}
\end{equation}
and
\begin{equation}
    \bar{T}^q_{\mu\nu}=\bar{\psi}\left(\gamma_{(\mu} i\overleftrightarrow{D}_{\nu)}-\frac{1}{4}g_{\mu\nu}m\right)\psi
\end{equation}
and the traceful part $T_{\alpha\alpha}$ is given by the trace anomaly in \eqref{3X}. Both contributions belong to different irreducible representations of the Lorentz group. Thus, they do not mix in the renormalization.

In the chiral limit, the traceful part is  related to the gluonic scalar by the conformal anomaly and in $1/N_c$ counting rule, it is  independent of the quarks.
We note that this decomposition is commensurate with the analysis of the energy momentum tensor in holographic QCD, through dual gravitons in bulk \cite{Mamo:2019mka}. (Holography is a good example of a strong coupling description of a gauge theory via its gravity dual, where the partonic structure is elusive). The traceful and traceless part of the energy momentum tensor correspond to the spin-2 and spin-0 representations of the Lorentz group, and do not mix under renormalization by symmetry.


The calculation of the gluonic scalar form factor can be easily generalized to the gluonic gravitational form factor. The matrix element vanishes in leading order of the instanton density expansion,
due to  self-duality. Purely tunneling vacuum configurations do not carry energy-momentum. This is best seen in light front signature with
\bea
   &&G^{a+i}G^{a+}{}_{i}=\nonumber\\
    &&-\frac{1}{2}\left(\vec{E}_\perp^a\cdot \vec{E}_\perp^a+2\hat{z}\cdot(\vec{E}_\perp^a\times \vec{B}_\perp^a)+\vec{B}_\perp^a\cdot \vec{B}_\perp^a\right)
\nonumber\\
\eea
Since the instanton and anti-instanton are self-dual, it means the tunneling configuration of the field strength in Minkowski space satisfies $\vec{E}^a=\mp i\vec{B}^a$ for all $I=1,\cdots,N_\pm$. Therefore, the instanton contribution at the leading order of the instanton density in (\ref{eq:gluo_op}) vanishes. The nontrivial instanton contribution starts from the higher orders of instanton density and thus are penalized by an extra factor denoted by $\kappa=\pi^2\rho^4n_{I+A}$, the packing fraction of the instanton vacuum.

\subsection{Two-instanton contribution}
At NLO in the instanton density expansion, the effective fermionic operator for $\mathcal{O}_{2g}$ can be obtained by averaging the leading expansion of $\mathcal{O}_{2g}$ with a two-instanton vertex
\bea
    &&\bar{T}^g_{\mu\nu\pm\pm}(x)=\nonumber\\
    &&\int d^4z_{I} d^4z_{J} dU_{I}dU_{J}\bar{T}^g_{\mu\nu}[A_I,A_J]\Theta_{I}\Theta_{J}
\eea
with
\begin{widetext}
\begin{equation}
\begin{aligned}
\label{eq:O2g_2}
    \bar{T}^g_{\mu\nu}[A_I,A_J]&=G^{a}_{\mu \alpha}[I]G^{a}_{\nu\alpha}[J]+G^{a}_{\mu \alpha}[I]G^{a}_{\nu\alpha}[I,J]+G^{a}_{\mu \alpha}[I,J]G^{a}_{\nu\alpha}[I]+(I\leftrightarrow J)\\
    &+G^{a}_{\mu \alpha}[I,J]G^{a}_{\nu\alpha}[I,J]-\frac{1}{4}\delta_{\mu\nu}\mathcal{O}_{2g}[A_I,A_J]
\end{aligned}
\end{equation}
\end{widetext}
The effective operator produces two types of coupling due to the chirality. The operator produced by $II$ or $AA$ clusters corresponds to a chiral flipping process 
\begin{widetext}
\begin{equation}
\begin{aligned}
    &\frac{1}{V}\int d^4x\bar{T}^g_{\mu\nu++,--}(x)e^{-iq \cdot x}=\frac{1}{2N_c(N_c^2-1)}\left(\frac{m^*}{4\pi^2\rho^2}\right)^{2N_f-2}\int d^4z_Id^4z_J\frac{1}{8}\mathrm{tr}\{S(z_I-z_J)\}\\
    &\times\frac{1}{V}\int d^4xe^{-iq\cdot x}F^{(II,AA)}_{T_g,\mu\nu}(x-z_I,x-z_J)\left[\bar{\psi}(z_I)\frac{1\mp\gamma^5}{2}\psi(z_J)+\bar{\psi}(z_J)\frac{1\mp\gamma^5}{2}\psi(z_I)\right]
\end{aligned}
\end{equation}
where the profile function for $II$ cluster is defined as
\begin{equation}
\begin{aligned}
\label{eq:2gs_profile_II}
F^{(II)}_{T_g,\mu\nu}(x-z_I,x-z_J)=&G^{a}_{\mu \alpha}(x-z_I)G^{a}_{\nu\alpha}(x-z_J)-\frac{1}{4}\delta_{\mu\nu}G^{a}_{\alpha\beta}(x-z_I)G^{a}_{\alpha\beta}(x-z_J)\\
&+\epsilon^{acd}G_{\mu \alpha}^c(x-z_I)\left[A_\nu^d(x-z_I)A_\alpha^a(x-z_J)-A_\alpha^d(x-z_I)A_\nu^a(x-z_J)\right]\\
&+\epsilon^{acd}G_{\nu \alpha}^c(x-z_I)\left[A_\mu^d(x-z_I)A_\alpha^a(x-z_J)-A_\alpha^d(x-z_I)A_\mu^a(x-z_J)\right]\\
&-\delta_{\mu\nu}\epsilon^{acd}G_{\alpha\beta}^c(x-z_I)\left[A_\alpha^d(x-z_I)A_\beta^b(x-z_J)\right]+(I\leftrightarrow J)+\mathcal{O}(1/N_c)
\end{aligned}
\end{equation}
and the profile function for $AA$ cluster is equal
\begin{equation}
\begin{aligned}
\label{eq:2gs_profile_II}
   & F^{(II)}_{T_g,\mu\nu}(x-z_I,x-z_J)=F^{(AA)}_{T_g,\mu\nu}(x-z_I,x-z_J)
\end{aligned}
\end{equation}
Here we dropped the contributions associated to  $G^a_{\mu\alpha}[I,J]G^a_{\nu\alpha}[I,J]$ as they are subleading in power counting.

The operator produced by $IA$ molecules corresponds to a chiral-conserving process
\begin{equation}
\begin{aligned}
    &\frac{1}{V}\int d^4x\bar{T}^g_{\mu\nu+-}(x)e^{-iq \cdot x}=-\frac{1}{2N_c(N_c^2-1)}\left(\frac{m^*}{4\pi^2\rho^2}\right)^{2N_f-2}\int d^4z_Id^4z_J\frac{1}{4}\mathrm{tr}\{S(z_I-z_J)\gamma_\rho\}\\
    &\times\frac{1}{V}\int d^4xe^{-iq\cdot x}F^{(IA)}_{T_g,\mu\nu\rho\lambda}(x-z_I,x-z_J)\left[\bar{\psi}(z_I)\gamma_\lambda\frac{1+\gamma^5}{2}\psi(z_J)-\bar{\psi}(z_J)\gamma_\lambda\frac{1-\gamma^5}{2}\psi(z_I)\right]
\end{aligned}
\end{equation}
where the profile function for $IA$ molecule is defined as
\begin{equation}
\begin{aligned}
\label{eq:2g_profile_IA}
    F^{(IA)}_{T_g,\mu\nu\rho\lambda}(x-z_I,x-z_J)=&\frac{1}{2}\bar{\eta}^a_{\rho\beta}\eta^b_{\lambda\beta}
    \bigg[G^{a}_{\mu \alpha}(x-z_I)G^{b}_{\nu\alpha}(x-z_J)-\frac{1}{4}\delta_{\mu\nu}G^{a}_{\alpha\beta}(x-z_I)G^{b}_{\alpha\beta}(x-z_J)\\
&+\epsilon^{acd}G_{\mu \alpha}^c(x-z_I)\left[A_\nu^d(x-z_I)A_\alpha^b(x-z_J)-A_\alpha^d(x-z_I)A_\nu^b(x-z_J)\right]\\
&+\epsilon^{acd}G_{\nu \alpha}^c(x-z_I)\left[A_\mu^d(x-z_I)A_\alpha^b(x-z_J)-A_\alpha^d(x-z_I)A_\mu^b(x-z_J)\right]\\
&-\delta_{\mu\nu}\epsilon^{acd}G_{\alpha\beta}^c(x-z_I)A_\alpha^d(x-z_I)A_\beta^b(x-z_J)
+\left( \begin{array}{l}
     I\leftrightarrow J \\
     a\leftrightarrow b
 \end{array} \right) \bigg]+\mathcal{O}(1/N_c)
\end{aligned}
\end{equation}

\end{widetext}

\subsection{Off-forward matrix element}
The off-forward hadronic matrix element of the traceless gluonic energy-momentum tensor reads

\begin{widetext}
\begin{equation}
\begin{aligned}
    \langle P'S|g^2\bar{T}^g_{\mu\nu}|PS\rangle=&\frac{1}{2N_c(N_c^2-1)}\left(\frac{2\kappa}{\rho^2 m^{*2}}\right)^2\rho^2m^2T_{IA}\\
    &\times\Bigg\{\frac{16\pi^2}{3}\beta^{(IA)}_{T_g,1}(\rho q)\langle P'S|\bar{\psi}\left(\gamma_{(\mu} i\overleftrightarrow{\partial}_{\nu)}-\frac{1}{4}g_{\mu\nu}i\overleftrightarrow{\slashed{\partial}}\right)\psi|PS\rangle\\
    &-\frac{4\pi^2\rho^2}{9}\beta^{(IA)}_{T_g,2}(\rho q)\left(q_\mu q_\rho g_{\nu\lambda}+q_\nu q_\rho g_{\mu\lambda}-\frac{1}{2} g_{\mu\nu}q_\rho q_\lambda\right)\langle P'S|\bar{\psi}\left(\gamma_{(\rho}i\overleftrightarrow{\partial}_{\lambda)}-\frac{1}{4}g_{\rho\lambda}i\overleftrightarrow{\slashed{\partial}}\right)\psi|PS\rangle\\
    &-4\pi^2\rho^4\beta^{(IA)}_{T_g,3}(\rho q)\left(q_\mu q_\nu-\frac{1}{4}q^2g_{\mu\nu}\right)q_\rho q_\lambda\langle P'S|\bar{\psi}\left(\gamma_{(\rho}i\overleftrightarrow{\partial}_{\lambda)}-\frac{1}{4}g_{\rho\lambda}i\overleftrightarrow{\slashed{\partial}}\right)\psi|PS\rangle\Bigg\}\\
    &+\frac{1}{2N_c(N_c^2-1)}\left(\frac{2\kappa}{\rho^2 m^{*2}}\right)^2\rho^2m^2 T_{II}\frac{8\pi^2\rho^2}{9}\beta^{(II)}_{T_g}(\rho q)\left(q_\mu q_\nu-\frac{1}{4}q^2g_{\mu\nu}\right)\langle P'S|m\bar{\psi}\psi|PS\rangle
\end{aligned}
\end{equation}
where the non-local form factors are defined as

\begin{equation}
\label{BETAEMT}
    \beta^{(IA)}_{T_g,1}(q)=\frac{1}{q}\int_0^\infty dx\left[\frac{24}{(1+x^2)^4}J_1(qx)+\frac{24x^2}{(1+x^2)^4}J_3(qx)-\frac{192}{(1+x^2)^3}\frac{J_3(qx)}{q^2x^2}\right]
\end{equation}

\begin{equation}
    \beta^{(IA)}_{T_g,2}(q)=\frac{1}{q}\int_0^\infty dx9x^2\left[\frac{128x^2}{(1+x^2)^4}\frac{J_3(qx)}{q^2x^2}-\frac{512}{(1+x^2)^3}\frac{J_4(qx)}{q^3x^3}\right]
\end{equation}

\begin{equation}
    \beta^{(IA)}_{T_g,3}(q)=\frac{1}{q}\int_0^\infty dx\frac{256x^4}{(1+x^2)^3}\frac{J_5(qx)}{q^4x^4}
\end{equation}

\begin{equation}
    \beta^{(II)}_{T_g}(q)=\frac{1}{q}\int_0^\infty dx\frac{576x^2}{(1+x^2)^4}\frac{J_3(qx)}{q^2x^2}
\end{equation}
and are normalized to 1.
For the quark traceless EMT, the matching is even simpler and straightforward at the leading order of the instanton density expansion
\begin{equation}
    \langle P'S|\bar{T}^q_{\mu\nu}|PS\rangle=\langle P'S|\bar{\psi}\left(\gamma_{(\mu} i\overleftrightarrow{\partial}_{\nu)}-\frac{1}{4}g_{\mu\nu}i\overleftrightarrow{\slashed{\partial}}\right)\psi|PS\rangle+\mathcal{O}(n_{I+A})
\end{equation}
\end{widetext}
The traceless gluonic and quark EMT now is matched to the effective traceless quark energy-momentum tensor derived from the effective field theory \eqref{eq:effective_action}. It is expected as they are under the same Lorentz irreducible representation. The finite-sized instanton profile also induces the mixing of the scalar quark operator.

In  the nucleon states, we can parameterize the matrix element for the effective quark traceless EMT operator and the scalar operator
\begin{widetext}
\begin{equation}
\begin{aligned}
\label{T_traceless}
    &\langle P'S|\bar{T}^{q,g}_{\mu\nu}|PS\rangle\\
    &=\bar{u}_{s'}(P')\bigg(A_{q,g}(q)\left(\gamma^{(\mu} \bar{P}^{\nu)}-\frac{1}{4}g^{\mu\nu}M_N\right)\\
    &+B_{q,g}(q)\left(\frac{i\bar{P}^{(\mu}\sigma^{\nu)\alpha}q_\alpha}{2M_N}-g^{\mu\nu}\frac{q^2}{16M_N}\right)+C_{q,g}(q)\frac{1}{M_N}\left(q^{\mu}q^{\nu}-\frac{1}{4}g^{\mu\nu}q^2\right)\bigg)u_{s}(P)
    \end{aligned}
\end{equation}
\end{widetext}
and the scalar form factor
\begin{equation}
\begin{aligned}
    &\langle P'S|m\bar{\psi}\psi|PS\rangle=\sigma(q)\bar{u}_{s'}(P')u_{s}(P)
\end{aligned}
\end{equation}

The gluonic gravitational form factors are related to the effective quark form factors as
\begin{widetext}
\begin{equation}
\begin{aligned}
\label{eq:A_g}
    A_g(q)=&\frac{1}{2N_c(N_c^2-1)}\left(\frac{2\kappa}{\rho^2 m^{*2}}\right)^2\rho^2m^2T_{IA}(\rho m^*)\frac{16\pi^2}{3}\beta^{(IA)}_{T_g,1}(\rho q)A_q(q)
\end{aligned}
\end{equation}

\begin{equation}
\begin{aligned}
\label{eq:B_g}
    B_g(q)=&\frac{1}{2N_c(N_c^2-1)}\left(\frac{2\kappa}{\rho^2 m^{*2}}\right)^2\rho^2m^2T_{IA}(\rho m^*)\frac{16\pi^2}{3}\beta^{(IA)}_{T_g,1}(\rho q)B_q(q)
\end{aligned}
\end{equation}

\begin{equation}
\begin{aligned}
    C_g(q)=&\frac{1}{2N_c(N_c^2-1)}\left(\frac{2\kappa}{\rho^2 m^{*2}}\right)^2\rho^2m^2T_{IA}(\rho m^*)\\
    &\times\Bigg\{\frac{16\pi^2}{3}\beta^{(IA)}_{T_g,1}(\rho q)C_q(q)+\frac{2\pi^2\rho^2}{9}M_N^2\left(\beta^{(IA)}_{T_g,2}(\rho q)+\frac{9}{2}\rho^2q^2\beta^{(IA)}_{T_g,3}(\rho q)\right)\left(A_q+B_q\frac{q^2}{4M^2_N}-C_q\frac{3q^2}{M^2_N}\right)\Bigg\}\\
    &+\frac{1}{2N_c(N_c^2-1)}\left(\frac{2\kappa}{\rho^2 m^{*2}}\right)^2\rho^2m^2T_{II}\frac{4\pi^2\rho^2M_N^2}{9}\beta^{(II)}_{T_g}(\rho q)\frac{\sigma(q)}{M_N}
\end{aligned}
\end{equation}


\subsection{Gravitational charges: $A_{q,g}(0)$}

In the forward limit $q\rightarrow0$, the gluonic traceless part \eqref{T_traceless} is
\begin{equation}
\langle PS|g^2\bar{T}^g_{\mu\nu}|PS\rangle=\frac{1}{2N_c(N_c^2-1)}\left(\frac{n_{I+A}}{2}\frac{4\pi^2\rho^2}{m^*}\right)^2\rho^2T_{IA}(\rho m^*)\frac{16\pi^2}{3}\langle PS|\bar{T}^q_{\mu\nu}|PS\rangle
\end{equation}
\end{widetext}
and the quark traceless part dominates the matrix element at the LO in the density expansion,
\begin{equation}
    \langle PS|\bar{T}^q_{\mu\nu}|PS\rangle=2\left(P_\mu P_\nu-\frac{1}{4}g_{\mu\nu}M_N^2\right) A_q(0)
\end{equation}
 Poincare symmetry guarantees the energy momentum conservation of the $A$-form factor in the zero momentum limit $$A_q(0)+A_g(0)=1$$, with
 \begin{widetext}
 $$A_g(0)=\frac{1}{2N_c(N_c^2-1)}\left(\frac{2\kappa}{\rho^2 m^{*2}}\right)^2\frac{16\pi^2\rho^2m^2}{3}T_{IA}(\rho m^*)\left[1+\frac{1}{2N_c(N_c^2-1)}\left(\frac{2\kappa}{\rho^2 m^{*2}}\right)^2\frac{16\pi^2\rho^2m^2}{3}T_{IA}(\rho m^*)\right]^{-1}$$ 
\end{widetext}
The gluonic contribution to the nucleon gravitational charge $A_g(0)$ receives a contribution from the instanton-antiinstanton molecules at NLO. Using the QCD instanton vacuum parameters
\cite{Liu:2023fpj,Liu:2023yuj} (and references therein), we have
at the resolution $\mu\approx1/\rho$
$$
A_q(0)=0.9865\qquad
A_g(0)=0.0135
$$
These estimates are consistent with those given in~\cite{Zahed:2021fxk}. 

To compare with other results, we can evolve the result from $\mu_0=0.56~\mathrm{GeV}\approx1/\rho$ where the evolution starts to resolve the instanton vacuum \cite{Liu:2023fpj,Liu:2023yuj} with DGLAP equation~\cite{Hatta:2018sqd}
\bea
\label{DGLAP}
   && \mu\frac{d}{d\mu}\begin{pmatrix}
    A_q(0) \\
    A_g(0)
    \end{pmatrix}_\mu=\nonumber\\
    &&\frac{\alpha_s}{4\pi}\begin{pmatrix}
    -\frac{16}{3}C_F & \frac{4}{3}N_f\\
    \frac{16}{3}C_F & -\frac{4}{3}N_f
    \end{pmatrix}
    \begin{pmatrix}
    A_q(0) \\
    A_g(0)
    \end{pmatrix}_\mu
\eea
where $C_F=\frac{N_c^2-1}{2N_c}$. The asymptotic form thus is given by the solution at $\mu\rightarrow\infty$,

$$
A_q(0)=\frac{N_f}{4C_F+N_f}
$$
and
$$
A_g(0)=\frac{4C_F}{4C_F+N_f}
$$
At $\mu=2$ GeV, the results are shown in the table
\begin{widetext}
\begin{center}
\begin{tabular}{|l|c|c|}
    \hline
    & $A_q(0)$ & $A_g(0)$ \\
    \hline
    ILM (this work) & $0.560$ & $0.364$ \\
    Asymptotic \cite{Gross:1974cs,Politzer:1974sm} & $0.529$ & $0.471$ \\
    Skyrmion \cite{Jung:2013bya} & $0.5$ & $0.5$ \\
    Lattice (dipole, 2018) \cite{Shanahan:2018pib} & 0.46(8) & 0.54(8) \\
    Lattice (dipole, 2023) \cite{Hackett:2023rif} & $0.510(25)$ & $0.501(27)$ \\
    Lattice (tripole) \cite{Pefkou:2021fni} & 0.57(4) & 0.429(39) \\
    Lattice (Extended Twisted Mass Collaboration) \cite{Alexandrou:2020sml} & 0.618(60) & 0.427(92) \\
    Global Analysis \cite{Hou:2019efy} & $0.58(1)$ & $0.414(8)$ \\
    \hline
\end{tabular}
\end{center}
In the second and forth lattice result, $A_q(0)$ is obtained by imposing the momentum conservation $A_q(0)=1-A_g(0)$. Extended Twisted Mass Collaboration is using physical pion mass.

\subsection{Gravitational charges: $C_{q,g}(0)$}

The $C$-form factor at zero momentum transfer is given by
\begin{equation}
\begin{aligned}
    C_g(0)=&\frac{1}{2N_c(N_c^2-1)}\left(\frac{2\kappa}{\rho^2 m^{*2}}\right)^2\rho^2m^2T_{IA}\Bigg\{\frac{16\pi^2}{3}C_q(0)+\frac{2\pi^2\rho^2}{9}M_N^2A_q(0)\Bigg\}\\
    &+\frac{1}{2N_c(N_c^2-1)}\left(\frac{2\kappa}{\rho^2 m^{*2}}\right)^2\rho^2m^2T_{II}\frac{4\pi^2\rho^2}{9}M_N\sigma_{\pi N}
\end{aligned}
\end{equation}
\end{widetext}
The estimate follows by saturating the scalar and flavor-singlet nucleon form factor,  by the  $\sigma$ exchange, with the result 
\begin{widetext}
\begin{equation}
\label{SIGMAXX}
\sigma_{\pi N}\equiv \frac{\langle PS|m\bar{\psi}\psi|PS\rangle}{2M_N}= 3m\times\frac{M_q(0)}{M_N}\frac{g_{\sigma qq}^2/G_\sigma}{m^2_\sigma}
\end{equation}
\end{widetext}
The overall factor of 3 counts the 3 constituent quarks in the nucleon. (\ref{SIGMAXX}) yields a nucleon sigma-term of about $\sigma_{\pi N}= 11.5\, \rm MeV$ with our value of $m=12.2\,\rm MeV$ in Table \ref{tab:ILM} and consistently the model estimation on the parameters $m_{\sigma}=683.1$ MeV and $g_{\sigma qq}=3.841$ with $G_{\sigma}=42.32$ GeV${}^{-2}$ in Appendix \ref{App:tHooft}. This is small in comparison to the empirical value 
$\sigma_{\pi N}= 60\, \rm MeV$~\cite{Schweitzer:2003sb}
(and references therein). Additional contributions
to the simple mean-field estimate with only $\sigma$-exchange, and beyond the constituent quark description for the nucleon, are expected to narrow the difference. For the mass budget to follow, we will make use of the empirical value

The  gluon $C$-form factor at zero momentum, requires the explicit
nucleon wavefunction in the QCD instanton vacuum which we plan to construct. 
For an estimate, we will use the results from topological models of the nucleon~\cite{Wakamatsu:2007uc,Cebulla:2007ei,Jung:2013bya} which are
about $-1$ (see table) at the low resolution of $1/\rho$. The chiral quark soliton model ($\chi$QM) in~\cite{Wakamatsu:2007uc}, is usually argued to emerge from the QCD instanton vacuum in the large $N_c$ limit~\cite{Diakonov:2002mb}, at the same resolution.

\begin{center}
\begin{tabular}{|l|c|c|c|c|}
    \hline
    & $C_q(0)$ & $C^{\mathrm{ILM}}_g(0)$ (this work)  \\
    \hline
    $\pi\rho\omega$ soliton \cite{Jung:2013bya} & $-1.006$ & $-0.0117$  \\
    Skyrme \cite{Cebulla:2007ei} & $-0.896$ & $-0.0102$  \\
    $\chi$QSM \cite{Wakamatsu:2007uc} & $-0.97$ & $-0.0112$  \\
    \hline
\end{tabular}
\end{center}

\begin{widetext}
\begin{center}
\begin{tabular}{|l|c|c|c|c|}
    \hline
    & $C_q(0)$ & $C_g(0)$ \\
    \hline
    ILM (this work) & -0.617 & -0.365  \\
    Lattice (dipole, 2018) \cite{Shanahan:2018pib} & - & $-2.5$  \\
    Lattice (dipole, 2023) \cite{Hackett:2023rif} & $-0.325(12)$ & $-0.643(21)$ \\
    Lattice (tripole) \cite{Pefkou:2021fni} & - & $-0.485$ \\
    Measurement (6 GeV) \cite{Burkert:2018bqq} & $-0.408\pm0.028\pm0.033$ & -  \\
    \hline
\end{tabular}
\end{center}
\end{widetext}

To compare with other results, again we can evolve the result from $\mu_0=0.56~\mathrm{GeV}\approx1/\rho$. For $C$-form factor, they obey the same DGLAP equation in \eqref{DGLAP}. Thus, their asymptotic form  is given by

$$
C_q(0)=C\frac{N_f}{4C_F+N_f}
$$
and
$$
C_g(0)=C\frac{4C_F}{4C_F+N_f}
$$
where $C\equiv C(0)=D/4$ is the intrinsic charge related to the mechanical property inside the nucleon, also known as $D$-term \cite{Polyakov:2018zvc}. Here we choose the result estimated from $\chi$QSM \cite{Wakamatsu:2007uc} for $C_q(0)$. Evolved to $\mu=2$ GeV, the results are shown in the table.

\begin{figure}
\hfill
\subfloat[\label{fig:mass_330}]{%
    \includegraphics[scale=0.30]{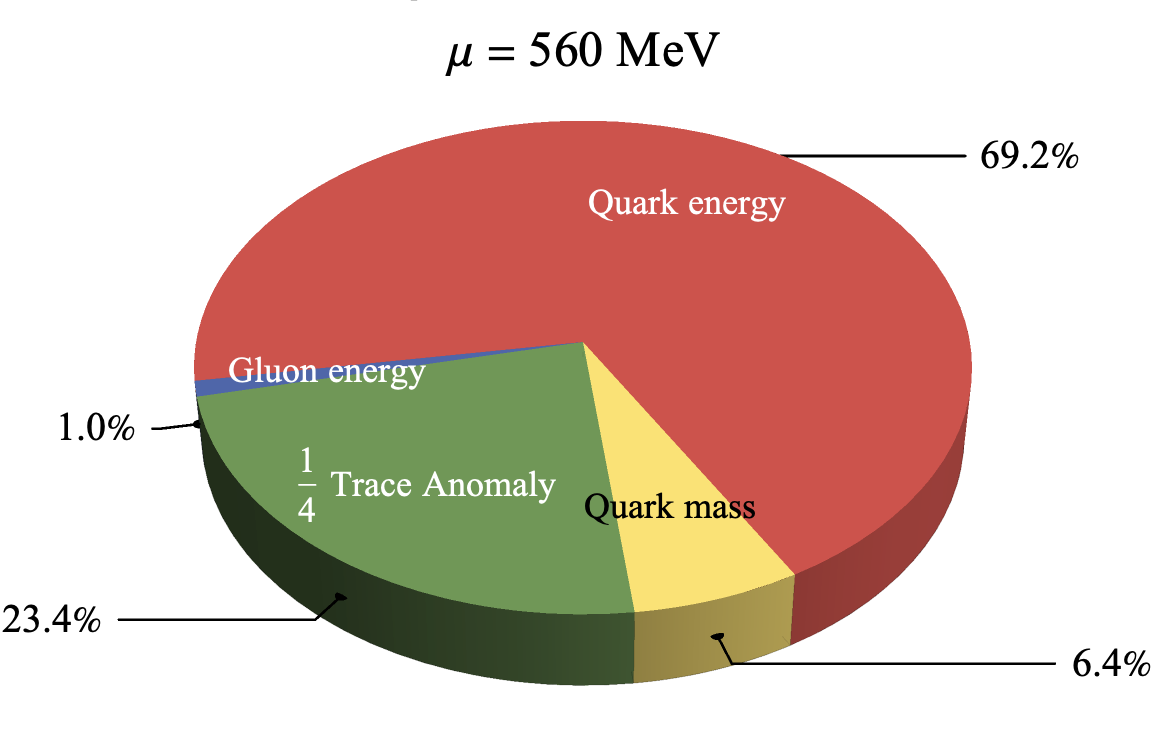}}
\hfill
\subfloat[\label{fig:mass_2000}]{%
    \includegraphics[scale=0.30]{mass_sum_rule_2000.png}}
\hfill
\subfloat[\label{PIELAT}]{%
    \includegraphics[scale=0.30]{lattice_mass.png}}
\caption{ 
Mass decomposition using Ji's nucleon mass sum rule, in the QCD instanton vacuum at the resolution $\mu=560~\mathrm{MeV}
\sim1/\rho$ (a), and after DGLAP evolution at a resolution 
$\mu=2~\mathrm{GeV}$ (b); (c) Mass decomposition using Ji's nucleon mass sum rule, at a resolution of $\mu=2$ GeV from the lattice collaboration~\cite{Yang:2018nqn}.
}
\label{PIEVAC}
\end{figure}

\section{Hadron  mass identity}
\label{sec_mass}
In QCD, the  breaking of conformal symmetry is captured by the trace anomaly
\eqref{3X}.
By Poincare symmetry, all squared hadron masses satisfy the mass identity 
\begin{equation}
    M^2_H= \frac 12 {\langle PS|T_{\mu\mu}|PS\rangle}
\end{equation}
which is distinct from the mass budgeting sum rule to be discussed below. 
In particular, \eqref{3X} gives
\bea
\label{forwardT}
M_H^2
&=&-\frac{b}{64\pi^2} {\langle PS|g^2G^a_{\mu\nu}G^{a}_{\mu\nu}|PS\rangle}
\nonumber\\&&+\frac m2{\langle PS|\bar{\psi}\psi|PS\rangle}
\eea
Thus, all hadron masses are composed of
\begin{equation}
\begin{aligned}
    M_H=M_{\mathrm{inv}}+\sigma_{\pi H}
\end{aligned}
\end{equation}
of the "invariant" mass (fixed by $\Lambda_{\rm QCD}$) and the chiral breaking mass  (sigma term) at some resolution. 
In quenched QCD (gluodynamics), the hadron mass is the quenched invariant mass $M^{(0)}_{\mathrm{inv}}$ induced solely by the spontaneous breaking of chiral symmetry and/or confinement from gluons. In QCD, quarks  contribute even in the chiral limit. Away from the quenched and chiral limit, the hadronic mass receives additional contributions from the  $\sigma_{\pi H}$ term~\cite{Hoferichter:2016ocj,Steele:1995yr}.

In the QCD instanton vacuum $\Lambda_{\rm QCD}$ at low resolution, is played by the mean instanton  density at LO. Recall that even the constituent quark mass is
fixed by this density, as the mean instanton size is also fixed by this density implicitly through modular interactions. In the QCD instanton vacuum, the mass breaking terms  are of order ${\cal O}(mR)$, and small for light quarks. At this low resolution, it is possible to budget the mass carried by the emerging quarks, and the semi-classical  gluons that permeate the vacuum, as we now detail.

\section{Nucleon mass sum rule}
\label{sec_sum}
The trace identity (\ref{forwardT}) reflects on the general fact that all hadron masses in QCD  are tied to the quantum  breaking of conformal symmetry as we noted earlier, and should be enforced by any non-perturbative quantum description, whether numerical such as the lattice,
or semi-classical such as the QCD instanton vacuum. 
However,  it does not specifically budget this mass breaking in terms of  the emerging  constituents in hadrons.

In a strongly interacting  theory,  the concept of constituents is subtle  and resolution dependent. Fortunately, the QCD instanton vacuum emerging from cooled lattice
simulations, allows for a quantitative description, all within the well-defined framework of semi-classics. In this spirit, a  physically  motivated proposal to budget the mass, was 
put forth by Ji in~\cite{Ji:1994av,Ji:1995sv},  and  since revisited by many~\cite{Lorce:2017xzd,Roberts:2021xnz,Metz:2020vxd} (and references therein). The ensuing mass composition
involves the sum of partonic contributions, some of which may be measurable  using DIS experiments.
The proposal relies on an invariant decomposition of the energy momentum tensor (\ref{EMT_decop}).

The tracefull and traceless part of the
energy momentum tensor  (\ref{EMT_decop}) correspond to the spin-2 and spin-0 representations of the Lorentz group, and do not mix  under renormalization by symmetry, as we noted earlier. 
Their renormalization  at  the  instanton size scale $\rho\approx 0.3\,{\rm fm}$ is  achieved by cooling through gradient flow, under the constaint of a fixed topological susceptibility.   Note that this renormalization scale is softer than the one  used in currently  fine lattices with $1/\mu\approx 0.1\,{\rm fm}$ ($\overline{\rm MS}$ scheme)~\cite{Yang:2018nqn}.

 With this in mind, the  corresponding Hamiltonian in Minkowski signature, follows from the 00-component of (\ref{EMT_decop})
\bea
\label{T5X}
H_G&=&\int d^3x\,\bar T_{00}^g=\int d^3x\,\frac 12(E^2+B^2)\nonumber\\
H_Q&=&\int d^3x\,\left(\bar T_{00}^q-\frac{3}{4}m\overline\psi\psi\right)=\int d^3x\, \overline\psi \gamma\cdot  i\overleftrightarrow D\psi\nonumber\\
H_A&=&\int d^3x\,\frac 14 \bigg(\frac{\beta(g^2)}{4g^2}G^2\approx 
-\frac {b}{32\pi^2}g^2G^2\bigg)\nonumber\\
H_m&=&\int d^3x\, m\overline\psi\psi
\eea
We rearranged the quark mass term so that the nucleon mass budget is then
\begin{widetext}
\be
\label{T6}
M_N=\frac{\left<P|H_G+H_Q+H_A+H_m|P\right>}{\left<P|P\right>}\equiv M^N_G+M^N_Q+M^N_A+M^N_m
\ee
\end{widetext}
with the identification
\be
M_{\rm inv}=M^N_G+M^N_Q+M^N_A
\ee

Using (\ref{eq:A_g}), we note that the forward matrix element of the gluonic $H_G$ contribution in (\ref{T5X}) vanishes at NLO in the QCD  instanton vacuum. Hence, at the resolution of the order of the inverse instanton size, we have
\bea
\label{T8}
\frac{M^N_Q}{M_N}&= &\frac 34 \left(A_q(0)-\frac{\sigma_{\pi N}}{M_N}\right)\approx   69.19\%\nonumber\\
\frac{M^N_G}{M_N}&= &\frac 34 A_g(0)\approx   1.01\%\nonumber\\
\frac{M^N_A}{M_N}&=&\frac {1}4\bigg(1-\frac{\sigma_{\pi N}}{M_N}\bigg) \approx 23.40\% \nonumber\\
\frac{M^N_m}{M_M}&=&\frac{\sigma_{\pi N}}{M_N}\approx 6.39\%
\eea
using the empirical pion-nucleon sigma term~\cite{Hoferichter:2016ocj,Alarcon:2021dlz}.
(\ref{T8}) shows that in the QCD instanton vacuum, about 69\%  of the nucleon mass
stems from the valence quarks (hopping zero modes),  23.4\% from the gluon condensate (displaced vacuum instanton field), 
and 1\% from emerging  valence gluons. This is illustrated in~\ref{PIEVAC}a. 
The nucleon is composed mostly of quark constituents hopping and scooping the vacuum gluon fields.  This result is consistent with the one observed in~\cite{Zahed:2021fxk}.

The budgeting of the nucleon mass evolves as the energy scale varies. At $\mu=2$ GeV, the valence quark and gluon energy contribution redistributes as illustrated in~\ref{PIEVAC}b. The  budgeting of the nucleon mass in
(\ref{T9}) from the QCD instanton vacuum, is consistent with the one reported on the lattice in~\cite{Yang:2018nqn} as illustrated in Fig.~\ref{PIELAT}. Under DGLAP evolution,   the gluons carry a larger energy fraction at the expense of the quarks. 

\bea
\label{T9}
\frac{M^N_Q}{M_N}&\approx   42.91\%\nonumber\\
\frac{M^N_G}{M_N}&\approx   27.29\%\nonumber\\
\eea

\section{Nucleon spin sum rule}
\label{sec_spin}
The spin structure of the nucleon has been addressed both empirically and theoretically by many, and we refer to the review in~\cite{Deur:2018roz} for an exhaustive account and references. Here, we  will address it in the QCD instanton vacuum following  recent estimates by one of us~\cite{Zahed:2021fxk}, using 
Ji's nucleon spin decomposition~\cite{Ji:1996ek},
\begin{equation}
    S_N=\frac{1}{2}\Sigma_q+L_q+J_g
\end{equation}
where $S_N=1/2$ is the nucleon spin.

\subsection{Intrinsic quark spin $\Sigma_q$}
The quark intrinsic spin 
\be
\label{spin_q}
\Sigma_q=\int d^3\vec{x} \bar{\psi}\vec{\gamma}\gamma^5\psi
\ee
captures the isoscalar axial charge inside the nucleon, which is best described by the hadronic matrix element of the flavor-singlet axial current,
\begin{equation}
    \langle PS|\bar{\psi}\gamma_\mu\gamma^5\psi|PS\rangle=2\Sigma_qS_\mu
\end{equation}
Here $S_\mu$ is the spin vector of the nucleon with the normalization $S^2=-M_N^2$ and $P\cdot S=0$.

The quark intrinsic spin $\Sigma_q$ is tied to the pseudoscalar gluon operator by the Adler-Bell-Jackiw (ABJ) anomaly~\cite{Adler:1969gk}
\begin{equation}
\label{chiral_anomaly}
\partial_\mu \bar{\psi}\gamma_\mu\gamma^5\psi = \frac{N_f}{16\pi^2} g^2G^a_{\mu\nu}\tilde{G}^{a}_{\mu\nu}+2m\bar{\psi}i\gamma^5\psi
\end{equation}
The  quark intrinsic spin consists of the anomalous gluonic contribution plus the explicit breaking of the $U(1)$ axial symmetry by the quark current mass $m$, 
\begin{widetext}
\bea
\frac{\langle PS|\partial_\mu\bar{\psi}\gamma_\mu\gamma^5\psi|PS\rangle}{2M_Ns_v}=2M_N\Sigma_q
= \frac{N_f}{16\pi^2} \frac{\langle PS|g^2G^a_{\mu\nu}\tilde{G}^{a}_{\mu\nu}|PS\rangle}{2M_Ns_v}+2m\frac{\langle PS|\bar{\psi}i\gamma^5\psi|PS\rangle}{2M_Ns_v}
\eea
\end{widetext}
Thus the intrinsic quark spin can be decomposed into \cite{Zahed:2022wae} (and references therein)
\begin{equation}
    \Sigma_q=\Delta \tilde{q}-N_f \Delta g
\end{equation}
where 
\be
\frac{\langle PS|m\bar{\psi}i\gamma^5\psi|PS\rangle}{2M_Ns_v}=M_N\Delta\tilde{q}\
\ee
and
\be
\Delta g=-\tilde{G}_N(0)
\ee

\begin{widetext}
\begin{center}
\begin{tabular}{|l|c|}
    \hline
    & $\Sigma_q$ \\
    \hline
    ILM (quenched, 2 GeV) & $0.65$  \\
    ILM (unquenched with 2 flavors, 2 GeV) & $0.105$  \\
    EMC (1987) \cite{EuropeanMuon:1989yki,Brodsky:1988ip,Ellis:1995jx} & $0.12 \pm 0.17$ \\
    COMPASS ($\sqrt{3}$ GeV) \cite{COMPASS:2006mhr} & $0.35 \pm 0.03~ (\mathrm{stat.}) \pm 0.05~ (\mathrm{syst.})$\\
    HERMES ($\sqrt{5}$ GeV) \cite{HERMES:2006jyl} & $0.33\pm0.011(\mathrm{theo})\pm0.025(\mathrm{exp})\pm0.028(\mathrm{evol})$ \\
    Lattice ($2$ GeV, helicity quasi PDF) \cite{Alexandrou:2021oih} & $0.467(58)$\\
    Lattice ($2$ GeV, axial form factor) \cite{Alexandrou:2021wzv} & $0.392(26)$\\
    Lattice ($2$ GeV, spin decomposition) \cite{Alexandrou:2020sml} & $0.382(30)$ \\
    Lattice (\cite{Liu:2015nva}) & $0.30(6)$\\
    \hline
\end{tabular}
\end{center}
\end{widetext}

In the quenched  and LO in the instanton density, the intrinsic spin is saturated by the ABJ  anomaly contribution. This is achieved in the QCD instanton vacuum at LO in the pseudoparticle density, with the intrinsic quark spin  fixed by the quenched topological susceptibility $\chi^{(0)}_t$,
\begin{widetext}
\bea
\Sigma_q^{(0)}= -N_f \Delta g=
N_f\frac{M_q(0)}{M_N}\frac{\chi^{(0)}_t}{\bar N}=
\frac{M_q(0)}{M_N}\frac{f_\pi^2M_1^2}{2n_{I+\overline I}}\rightarrow  0.65
\eea
The singlet squared mass follows from the Witten-Veneziano relation
(\ref{CHI1}). The rightmost result uses ${M_q(0)}/{M_N}\sim 1/3$ \cite{Zahed:2022wae}, 
and the Witten-Veneziano relation (\ref{CHI1}).

In the unquenched and NLO instanton density, there is mixing with the emergent 
constituent quarks. As a result, the intrinsic spin gets modified 
\bea
\label{const_Sigma_q}
\Sigma_q=&&N_f\frac{M_q(0)}{M_N}\frac{\chi_t}{\bar{N}}+\left[1+\frac{N_f}{N_c}\left(\frac{2\kappa}{\rho^2 m^{*2}}\right)+\frac{N_f}{2N_c(N_c^2-1)}\left(\frac{2\kappa}{\rho^2 m^{*2}}\right)^23\rho^2m^2T_{II}\right]\Delta \tilde q
\nonumber\\
&&-\frac{1}{8N_c(N_c^2-1)}\left(\frac{2\kappa}{\rho^2 m^{*2}}\right)^2\rho^2m^2T_{IA}\Sigma_q\rightarrow 0.1048
\eea
with the  unquenched and large volume topological susceptibility $\chi_t$ given in (\ref{chi_t}). 
The rightmost estimate follows by saturating the pseudoscalar and flavor-singlet
nucleon form factor,  by the  $\eta'$ exchange.  In the QCD instanton vacuum, the exchange is given by 
\bea
\label{Delta_q}
M_N\Delta\tilde{q}\equiv\frac{\langle PS|m\bar{\psi}i\gamma^5\psi|PS\rangle}{2M_Ns_v}
=3m\times\frac{M_q(0)}{M_N} \frac{g_{\eta' qq}^2/G_{\eta'}}{m_{\eta'}^2}
\eea
\end{widetext}
The overall factor of 3 counts the 3 constituent quarks in the nucleon. \eqref{Delta_q} yields a value of explicit chiral symmetry breaking about $\Delta\tilde q=0.0119$. Here we used the empirical value $m=0.122$ MeV in Table \ref{tab:ILM} and the consistent  model  parameters $m_{\eta'}=681.6$ MeV and $g_{\eta' qq}=1.686$ with $G_{\eta'}=8.417$ GeV${}^{-2}$ in Appendix \ref{App:tHooft}. 

Note that if we were to use a quark-diquark description of the proton or neutron as a strongly correlated quark-diquark states, with a 
tight scalar-iso-scalar diquark $[ud]_S$ and weaker  axial-vector flavor-triplet  diquark $[ud]_A$~\cite{Schafer:1996wv}, then 
 the proton SU(6) wavefunctions can be repacked in quark-diquark contributions as~\cite{Anselmino:1992vg}
\begin{widetext}
\bea
\label{SU6}
p\uparrow=\frac 1{\sqrt {18}}
 \bigg(3[ud]_S u\uparrow + 2[uu]^+_A d\downarrow 
 -\sqrt 2[uu]^0_A d\uparrow-\sqrt 2[ud]^+_A u\downarrow + [ud]^0_A u\uparrow\bigg)
 \eea
\end{widetext}
This would suggest that for the proton the $u$-quark and $d$-quark spin are opposite, with a ratio
\bea
\left|\frac{\Sigma_d}{\Sigma_u}\right|=\frac 14
\eea
and in particular
$\Sigma_u=0.0286, \,  \Sigma_d=-0.0071$, with $\Sigma_q=\Sigma_u+\Sigma_d=0.0215$. 
In this SU(6) repacking, the discrepancy with \eqref{const_Sigma_q} stems from the left out  diquarks. A more realistic diquark wavefunction for the nucleon in the ILM will be addressed elsewhere.

At this stage it is worth noting that while the topological charge fluctuations
in the 4-volume are screened in the QCD instanton vacuum as per our discussion, the topological charge 
fluctuations in small sub-volumes are finite, see~\cite{Shuryak:1994rr} for a definition. In the QCD instanton vacuum with light quarks,  they were found to be numerically Poissonian~\cite{Shuryak:1994rr}
\bea
\label{POISS}
\lim_{V\to 0}\frac{\langle \Delta N^2(V)\rangle}V=n_{I+A}
\eea
with the unquenched singlet pseudoscalar mass  $M_1$, now given by
\bea
\label{WZX}
M_1^2\rightarrow\frac{2N_f}{f^2}\lim_{V\to 0}\frac{\langle \Delta N^2(V)\rangle}V
=\frac{2N_f}{f^2}n_{I+A}\nonumber\\
\eea
which is finite even  in the chiral limit, and in agreement with the
result  in~\cite{Alkofer:1989uj,Zahed:1994qh}. This is supported by the fact that the  eta$^\prime$ mass is large
in nature, which is unquenched QCD. Since the world-volume of a hadron in the QCD 
instanton vacuum can be considered as a vanishingly small sub-volume, it is plausible to use (\ref{WZX}) in our analysis of the intrinsic spin. This leads to the result, that the quenched and unquenched results for the intrinsic spin are mostly unchanged.


\begin{center}
\begin{tabular}{|l|c|c|c|}
    \hline
     & $J_q$ & $J_g$\\
    \hline
    ILM (this work, 2 GeV) & $0.318$ & $0.182$  \\
    Lattice (ETMC) \cite{Alexandrou:2020sml} & 0.285(45) & 0.187(46) \\
    Lattice ($\chi$QCD) \cite{Wang:2021vqy} & $0.265(32)$ & $ 0.230(25) $\\
    \hline
\end{tabular}
\end{center}

\begin{figure}
    \subfloat[\label{}]{
    \includegraphics[scale=0.46]{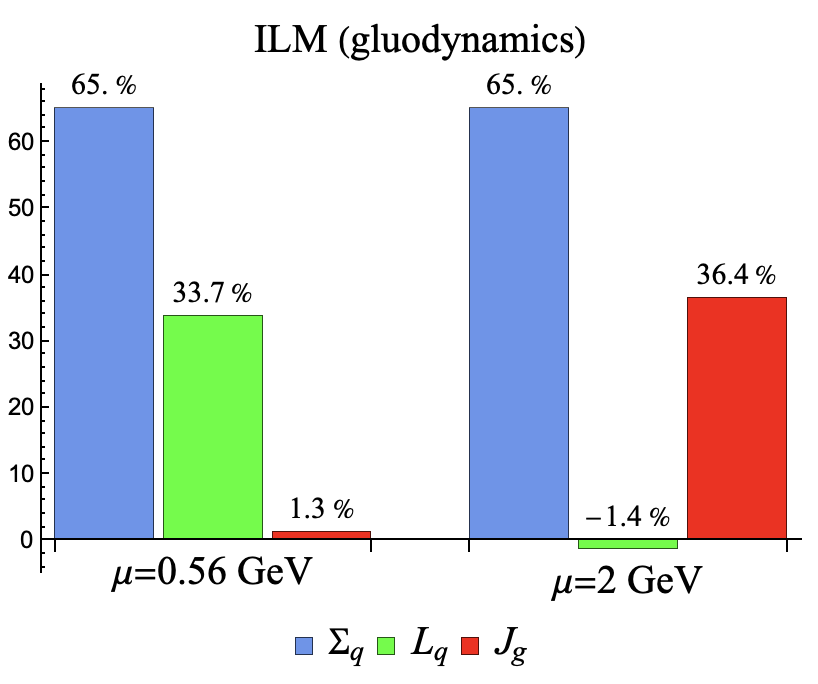}}
    \hfill
    \subfloat[\label{}]{
    \includegraphics[scale=0.46]{spin_sum_unquenched.png}}
    \hfill
    \subfloat[\label{}]{
    \includegraphics[scale=0.46]{spin_sum_lat.png}}
    \caption{a: Spin decomposition using Ji's sum rule, in the quenched QCD instanton vacuum at $\mu=0.56\,{\rm GeV}$ (left),  and $\mu=2\,{\rm GeV}$ (right); b: Spin decomposition using Ji's sum rule, in the $N_f=2$ QCD instanton vacuum at $\mu=0.56\,{\rm GeV}$ (left) and $\mu=2\,{\rm GeV}$ (right); 
    c: Spin decomposition using Ji's sum rule at a resolution of $\mu=2\,{\rm GeV}$, from the lattice collaboration $\chi$QCD ~\cite{Wang:2021vqy} (left) and the ETMC~\cite{Alexandrou:2020sml} (right).}
    \label{fig:spin_sum}
\end{figure}

\subsection{Quark and gluon orbital momenta}
The quark orbital angular momentum (OAM) is given by
\be
L_q=\int d^3\vec{x} \bar{\psi}\gamma^0\vec{x}\times i\vec{D}\psi
\ee
Combined with the intrinsic quark spin \eqref{spin_q}, we have the quark total angular momentum

\be
J_q=\frac{1}{2}\Sigma_q+L_q
\ee

The quark total angular momentum $J_q$ is calculated by the forward hadronic matrix element of the traceless part of the quark $0j$-EMT, 
\bea
\label{SPINQ}
J_q&=&\frac{\langle PS|\int d^3\vec{x}\epsilon^{3ij}x^i\bar{T}_q^{0j}|PS\rangle}{\langle PS|PS\rangle} \nonumber\\
&=&\epsilon^{3ij}i\partial_q^i
\frac{\langle P'S|\bar{T}_q^{0j}|PS\rangle}{\langle PS|PS\rangle}
\eea
where $\partial^i_q$ refers to the derivative with respect to the momentum transfer, followed by the zero momentum transfer limit. 
Similarly, the angular momentum carried by the gluons is
\be
J_g=\int d^3\vec{x} \vec{x}\times(\vec{E}^a\times\vec{B}^a)
\ee
which translates to the forward hadronic matrix element.
\bea
\label{SPING}
J_g&=&\frac{\langle PS|\int d^3\vec{x}\epsilon^{3ij}x^i\bar{T}_g^{0j}|PS\rangle}{\langle PS|PS\rangle} \nonumber\\
&=&\epsilon^{3ij}i\partial_q^i
\frac{\langle P'S|\bar{T}_g^{0j}|PS\rangle}{\langle PS|PS\rangle}
\eea
Using (\ref{eq:A_g}) it follows that (\ref{SPING}) is penalized by the instanton density in the forward limit, at NLO in the instanton density expansion. Hence, at the resolution of the order of the inverse instanton size, Ji's spin sum rule is given by

\bea
\label{S8}
\frac{\frac{1}{2}\Sigma_q}{S_N}&= &\Delta \tilde{q}-\frac{g^2}{8\pi^2}N_f \Delta g\approx 65\%\nonumber\\
\frac{L_q}{S_N}&= &A_q(0)+B_q(0)-\frac{\frac{1}{2}\Sigma_q}{S_N}\approx 33.7\%\nonumber\\
\frac{J_g}{S_N}&=&A_g(0)+B_g(0) \approx 1.3\%
\eea
In the case of the unquenched QCD in the instanton vacuum with 2 flaovrs, we have the topological susceptibility screened. 
\bea
\label{S8-2}
\frac{\frac{1}{2}\Sigma_q}{S_N}&= &\Delta \tilde{q}-\frac{g^2}{8\pi^2}N_f \Delta g\approx 10.5\%\nonumber\\
\frac{L_q}{S_N}&= &A_q(0)+B_q(0)-\frac{\frac{1}{2}\Sigma_q}{S_N}\approx 88.2\%\nonumber\\
\frac{J_g}{S_N}&=&A_g(0)+B_g(0) \approx 1.4\%
\eea
at the low resolution of $1/\rho\approx 560\,{\rm MeV}$.
Here we assumed $B_q(0)=0$ \cite{Mamo:2022eui}, hence $B_g(0)=0$ from \eqref{eq:B_g}.  

\eqref{S8} shows that in the QCD instanton vacuum, about $65\%$  of the nucleon spin stems from the spin of the valence quarks as they hop and mix with the vacuum topological charge fluctuations, $34\%$ stems from their orbital angular motion (OAM), and only $1\%$ from the emerging valence gluons as the topological charge fluctuates in small sub-volumes.

The budgeting of the nucleon spin evolves as the energy scale varies. For the quenched QCD instanton vacuum  at $\mu=2$ GeV, the valence quark OAM and gluon angular momentum redistributes as
\bea
\label{S9}
\frac{\frac{1}{2}\Sigma_q}{S_N}&\approx&   65\%\nonumber\\
\frac{L_q}{S_N}&\approx&   -1.4\%\nonumber\\
\frac{J_g}{S_N}&\approx&   36.4\%
\eea
For the QCD instanton vacuum with 2 flavors, we have 
\bea
\label{S9-2}
\frac{\frac{1}{2}\Sigma_q}{S_N}&\approx& 10.5\%\nonumber\\
\frac{L_q}{S_N}&\approx& 53.1\%\nonumber\\
\frac{J_g}{S_N}&\approx& 36.4\%
\eea

In the QCD instanton vacuum, the intrinsic spin does not renormalize,
as it captures the vacuum topological
susceptibility scooped by the nucleon, in the small volume limit. As a result, DGLAP
evolution enhances the gluon contribution at the sole expense of the quark orbital contribution, both of which are not topological in our analysis.

The results for Ji's spin decomposition in the QCD instanton vacuum, are illustrated in Fig.~\ref{fig:spin_sum}a (quenched) and 
Fig.~\ref{fig:spin_sum}b (unquenched)  at $\mu=0.64$ GeV (left), and at $\mu=2$ GeV after DGLAP evolution (right). 
They are compared to the reported 
results from the $\chi$-QCD collaboration~\cite{Wang:2021vqy} in Fig.~\ref{fig:spin_sum}c (left),
and from the ETMC collaboration~\cite{Alexandrou:2020sml} 
in Fig.~\ref{fig:spin_sum}b (right).

While the gluonic contributions are comparable to the one reported by both lattic collaborations, there is a difference in the way the quarks 
are carrying the spin. In the lattice, the intrinsic spin to OAM ratio is about 3:1 ($\chi$QCD) and 2:1 (ETMC), which is to be compared to 
1:1 in the QCD instanton vacuum with two flavors. The origin, 
is the substantial depletion of the intrinsic spin at low resolution, owing to the strong screening of the large volume topological susceptibility. Although we suggested earlier that the small volume
topological susceptibility remains large, and may cause the intrinsic spin to be larger even in the unquenched case, our analysis shows that it would lead
to a considerably large ratio for the intrinsic spin to OAM as illustrated in Fig.~\ref{fig:spin_sum}a (left). This shows the importance of studying and reporting on both the large and small volume topological susceptibilities, when reporting the
results for the spin composition of hadrons, using QCD lattice simulations.

\section{Conclusion}
\label{sec_conclusion}
The QCD instanton vacuum is well supported by current lattice QCD simulations. When the zero-point
gauge fluctuations are removed  numerically by cooling using the gradient flow method~\cite{Luscher:2009eq,Luscher:2011bx},
the QCD vacuum is found to be populated by topological 
lumps of gauge fields. These lumps are tunneling gauge configurations describing  instantons/antiinstanton pseudoparticles, or failed tunneling gauge molecules made  of instanton-antiinstanton pairs.

Deep in the cooling procedure, the corresponding
gauge fields are strong and localized, with a mean
 size of  $\rho=\frac 13\, {\rm fm}$ and a mean density $n_{I+A}=1\,{\rm fm}^{-4}$.  
Confinement, is likely caused by center-vortices through percolating long $Z_{N_c}$ strings of about $4\,{\rm fm}^{-2}$, with topologically active branch points, the likely anchors of the pseudoparticles.

Hadrons in the QCD vacuum are small ripples propagating in the QCD vacuum. At the resolution 
of $\rho=\frac 13\,{\rm fm}$, the composition and
properties of the low-lying hadrons can be fairly
approximated in the QCD instanton vacuum, where
their bound state structure and small sizes, make them less prone to disordering by the long center-vortices. This is 
less so for their excited states which are larger in size, and more prone to flux piercing and the string tension. Throughout, we have assumed that this is the case, and pursued all the analyses of the gluons in hadrons solely in the context of the QCD instanton vacuum.

At medium resolution  and in LO in the instanton density, the pseudoparticles
strong gauge fields dominate the gluonic content of the forward matrix elements. The scalar and pseudoscalar gluon matrix elements receive contributions from their number fluctuations at LO, and from pairs of pseudoparticles at NLO. We have shown that the off-forward gluonic scalar and pseudoscalar matrix elements are readily expressed
in terms of the pseudoparticle moduli, essentially the fermionic zero modes in the form of multiflavor
effective vertices after averaging over the cor moduli. The same carries to higher dimensional gluonic operators, as we have shown for the C-even and C-odd dimension-6 gluon operators. The latters 
map on leading twist contributions in diffractive C-even and C-odd vector meson production.

The self-dual character of the pseudoparticles, shows that only pairs of pseudoparticles or molecules can contribute at NLO to the traceless part of the QCD energy-momentum tensor. The trace part is anomalous, and receives contributions in LO from the pseudoparticles. We have shown how they contribute to the gluonic form factors
in hadrons, in the form of effective fermionic operators, once the modular integration is carried out. The results allow for a detailed budgeting of the quarks and gluons  contributions to the nucleon mass and spin, at low resolution.
Their evolution at higher resolution, are in good  agreement  with the current lattice simulations. The comparison would be enhanced, if the lattice collaboration~\cite{Wang:2021vqy,Alexandrou:2020sml} could  also
report their large and small volume topological susceptibities,
along with their nucleon spin results.

\begin{acknowledgments}
We thank  Christian Weiss for discussions.
This work is supported by the Office of Science, U.S. Department of Energy under Contract No. DE-FG-88ER40388,
and in part within the framework of the Quark-Gluon Tomography (QGT) Topical Collaboration, under contract no. DE-SC0023646.
\end{acknowledgments}

\appendix

\section{Emergent effective t'Hooft interactions  with gluons in SIA}
\label{App:effective_Langrangian}
The effective Lagrangian following from (\ref{eq:effective_action}) after averaging over the instanton moduli with fermionic zero modes in the SIA,
yield emergent multi-flavor interactions with gluon insertions. 
The latters follow by LSZ reduction of the instanton profile. 
To  characterize these interactions, we use the book-keeping in 
$1/N_c$.

For a single instanton with $N_f$ quarks and $N_g$ gluons, the vertices in (\ref{eq:tHooft}) give rise to an effective 
t's Hooft interaction coupling 
$$
G_{f+g}\sim\frac{n_{I+A}}{2}\frac{1}{N_c^{N_f+N_g}}\left(\frac{4\pi^2\rho^2}{m^*}\right)^{N_f}\left(\frac{2\pi^2\rho^2}{g}\right)^{N_g}
$$
characteristic of a vacuum contribution. Note 
that each gluon insertion from the instanton tail, is further suppressed by the instanton size. In the SIA, the effective interactions are given by
\begin{widetext}
\begin{equation}
\begin{aligned}
\label{LEFFSIA}
&\mathcal{L}_{\mathrm{eff}}=\bar{\psi}(i\slashed{\partial}-m)\psi-\frac{1}{4}(G^a_{\mu\nu})^2+\frac{n_{I+A}}{2}\frac{4\pi^2\rho^2}{m^*}\mathcal{V}_{N_f=1}+\frac{n_{I+A}}{2}\left(\frac{4\pi^2\rho^2}{m^*}\right)^2\mathcal{V}_{N_f=2}+\mathcal{O}\left(\frac{n_{I+A}}{2}\left(\frac{4\pi^2\rho^2}{m^*}\right)^3\right)
\end{aligned}
\end{equation}
where the one-body interaction is defined as
\begin{equation}
\begin{aligned}
   \mathcal{V}_{N_f=1}=& -\frac{1}{N_c}\bar{\psi}\psi+\frac{1}{N^2_c-1}\left(\frac{2\pi^2\rho^2}{g}\right)\bar{\psi}\sigma_{\mu\nu}\frac{\lambda^a}{2}\psi G^a_{\mu\nu}-\frac{1}{N_c(N^2_c-1)}\left(\frac{2\pi^2\rho^2}{g}\right)^2f^{abc}\bar{\psi}\sigma_{\mu\nu}\lambda^a\psi G^b_{\mu\rho}G^c_{\nu\rho}\\
&-\frac{1}{N_c(N^2_c-1)}\left(\frac{2\pi^2\rho^2}{g}\right)^2\left(\delta^{bc}\bar{\psi}\psi+\frac{N_c}{2(N_c+2)}d^{abc}\bar{\psi}\lambda^a\psi\right) G^b_{\mu\nu}G^c_{\mu\nu}\\
&-\frac{1}{N_c(N^2_c-1)}\left(\frac{2\pi^2\rho^2}{g}\right)^2\left(\delta^{bc}\bar{\psi}\gamma^5\psi+\frac{N_c}{2(N_c+2)}d^{abc}\bar{\psi}\lambda^a\gamma^5\psi\right)G^b_{\mu\nu}\tilde{G}^c_{\mu\nu}\\
&+\frac{1}{6(N^2_c-1)}\left(\frac{2\pi^2\rho^2}{g}\right)^3\left(\frac{8}{N_c^2}f^{abc}\bar{\psi}\psi+\frac{12}{N_c^2-4}d^{dae}f^{bce}\bar{\psi}\lambda^d\psi\right) G^a_{\mu\nu}G^b_{\nu\rho}G^c_{\rho\mu}\\
&+\frac{1}{6(N^2_c-1)}\left(\frac{2\pi^2\rho^2}{g}\right)^3\left(\frac{8}{N_c^2}f^{abc}\bar{\psi}\gamma^5\psi+\frac{12}{N_c^2-4}d^{dae}f^{bce}\bar{\psi}\lambda^d\gamma^5\psi\right)\tilde{G}^a_{\mu\nu}G^b_{\nu\rho}G^c_{\rho\mu}\\
    &+\frac{1}{6(N_c^2-1)}\left(\frac{2\pi^2\rho^2}{g}\right)^3\left(\frac{3}{(N_c^2-1)}\delta^{ab}\delta^{cd}+\frac{3N_c}{(N_c^2-4)(N_c+2)}d^{abe}d^{cde}+\frac{6}{N_c^2}f^{ace}f^{bde}\right)\bar{\psi}\sigma_{\mu\nu}\lambda^a\psi G^b_{\mu\nu}G^c_{\rho\lambda}G^d_{\rho\lambda}\\
     &+\frac{1}{6(N_c^2-1)}\left(\frac{2\pi^2\rho^2}{g}\right)^3\left(\frac{3}{(N_c^2-1)}\delta^{ab}\delta^{cd}+\frac{3N_c}{(N_c^2-4)(N_c+2)}d^{abe}d^{cde}+\frac{6}{N_c^2}f^{ace}f^{bde}\right)\bar{\psi}\sigma_{\mu\nu}\lambda^a\gamma^5\psi G^b_{\mu\nu}\tilde{G}^c_{\rho\lambda}G^d_{\rho\lambda}\\
    &+\mathcal{O}\left(\left(\frac{2\pi^2\rho^2}{g}\right)^4\right)\\
\end{aligned}
\end{equation}
and the two-body interaction is defined as
\begin{equation}
\begin{aligned}
    \mathcal{V}_{N_f=2}=&\frac{2N_c-1}{16N_c(N^2_c-1)}\left[(\bar{\psi}\psi)^2-(\bar{\psi}\tau^a\psi)^2-(\bar{\psi}i\gamma^5\psi)^2+(\bar{\psi}i\gamma^5\tau^a\psi)^2\right]+\frac{1}{32N_c(N^2_c-1)}\left[\left(\bar{\psi}\sigma_{\mu\nu}\psi\right)^2-\left(\bar{\psi}\sigma_{\mu\nu}\tau^a\psi\right)^2\right]\\
    &-\frac{1}{N_c(N_c^2-1)}\left(\frac{2\pi^2\rho^2}{g}\right)\left[\bar{u}_Ru_L\bar{d}_R\sigma_{\mu\nu}\frac{\lambda^a}{2}d_L+\bar{u}_R\sigma_{\mu\nu}\frac{\lambda^a}{2}u_L\bar{d}_Rd_L\right]G^a_{\mu\nu}\\
    &-\frac{1}{(N_c+2)(N_c^2-1)}\left(\frac{2\pi^2\rho^2}{g}\right)d^{abc}\left[\bar{u}_R\frac{\lambda^a}{2}u_L\bar{d}_R\sigma_{\mu\nu}\frac{\lambda^b}{2}d_L+\bar{u}_R\sigma_{\mu\nu}\frac{\lambda^a}{2}u_L\bar{d}_R\frac{\lambda^b}{2}d_L\right]G^c_{\mu\nu}\\
    &-\frac{1}{(2N_c)(N_c^2-1)}\left(\frac{2\pi^2\rho^2}{g}\right)f^{abc}\left[\bar{u}_R\sigma_{\mu\rho}\frac{\lambda^a}{2}u_L\bar{d}_R\sigma_{\nu\rho}\frac{\lambda^b}{2}d_L+\bar{u}_R\sigma_{\mu\rho}\frac{\lambda^a}{2}u_L\bar{d}_R\sigma_{\nu\rho}\frac{\lambda^b}{2}d_L\right](G^c_{\mu\nu}-\tilde{G}^c_{\mu\nu})\\
    &+\mathcal{O}\left(\left(\frac{2\pi^2\rho^2}{g}\right)^2\right)\\
\end{aligned}
\end{equation}
\end{widetext}

 The emergent couplings with constitutive quarks and gluons, follow by color averaging, after pertinent fermionic lines closing in the SIA. More specifically, in the color average, each of  the $UU^\dagger$ pair gives a $1/N_c$ factor in the large $N_c$ limit. Therefore, the power counting of each vertex $1/N_c^{N_f+N_g}$ is given by the flavor number $N_f$ and the gluon number $N_g$. That is, the more quarks and gluons involved in the instanton, the more $1/N_c$ suppression. Here we show the one-body interaction with one to three gluons involved, and the two-body interaction with one gluon involved. Higher order interactions follow the same reasoning, but are more involved.
Using (\ref{}) in the mean-field approximation , yields the SIA determinantal mass 
\bea
m_f^*=m_f +\frac{n_{I+A}}{2}\frac{4\pi^2\rho^2}{m_f^*}\langle \mathcal{V}_{N_f=1}\rangle
\eea

\section{$\sigma$ and $\eta'$ in hadronic matrix elements}
\label{App:tHooft}

\begin{figure*}
\includegraphics[scale=0.75]{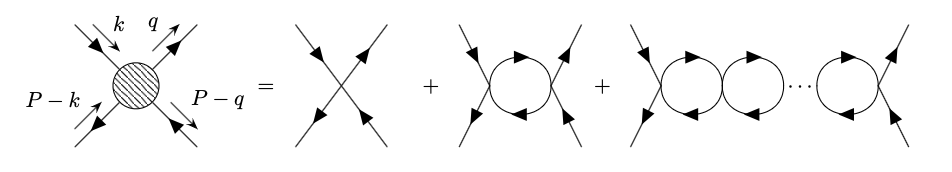}
\label{fig:BS}
\caption{$\sigma, \eta'$ Bethe-Salpeter kernels in the QCD instanton vacuum, in leading order in the $1/N_c$ book-keeping approximation.}
\end{figure*}

In this Appendix, we will estimate 
the meson-quark effective couplings in the $\sigma$ and $\eta'$ channels for the hadronic matrix elements of $\bar{\psi}\psi$ and $\bar{\psi}i\gamma^5\psi$, appearing in the gravitational form factors.

The quark part of the effective Lagrangian in the instanton vacuum in \eqref{eq:effective_action} determines the meson mass spectrum and meson-quark effective coupling in the QCD instanton vacuum. For that we need to go beyond the SIA, by resumming the
contributions to the quark propagator that includes both the close and far pseudoparticles.

The correponding  effective Lagrangian with induced vertices from unpaired (I,A) and paired (IA) pseudoparticles, is given byby~\cite{Liu:2023fpj}, 
\begin{widetext}
\begin{equation}
\begin{aligned}
\label{tHooft}
\mathcal{L}_{\mathrm{tHooft}}=&\bar{\psi}(i\slashed{\partial}-M)\psi+\frac{G_\sigma}{2}(\bar{\psi}\psi)^2+\frac{G_{a_0}}{2}(\bar{\psi}\tau^a\psi)^2+\frac{G_{\eta'}}{2}(\bar{\psi}i\gamma^5\psi)^2+\frac{G_\pi}{2}(\bar{\psi}i\gamma^5\tau^a\psi)^2\\
 &-\frac{G_\omega}{2}(\bar{\psi}\gamma_\mu\psi)^2-\frac{G_\rho}{2}(\bar{\psi}\gamma_\mu\tau^a\psi)^2-\frac{G_{f_1}}{2}(\bar{\psi}\gamma_\mu\gamma^5\psi)^2-\frac{G_{a_1}}{2}(\bar{\psi}\gamma_\mu\gamma^5\tau^a\psi)^2 
\end{aligned}
\end{equation}
As we noted earlier, it is the the constituent mass $M\equiv M_q(0)=395.17~\mathrm{MeV}$ in \eqref{VER4}, as opposed to the determinantal mass $m^*$, that enters the analysis of the long range hadronic correlations. The induced ${}^\prime$t Hooft couplings are
\begin{align*}
G_\sigma&=\frac{G_{I}}{4N_c^2}+\frac{4G_{IA}}{N_c^2}  &  G_{a_0}&=-\frac{G_{I}}{4N_c^2}+\frac{4G_{IA}}{N_c^2}   & 
G_\pi &=\frac{G_{I}}{4N_c^2}+\frac{4G_{IA}}{N_c^2}   &  G_{\eta'}&=-\frac{G_{I}}{4N_c^2}+\frac{4G_{IA}}{N_c^2}\\
G_\omega&=\frac{G_{IA}}{N_c^2}   &  G_\rho&=\frac{G_{IA}}{N_c^2}       & G_{a_1}&=\frac{G_{IA}}{N_c^2}    &  G_{f_1}&=-3\frac{G_{IA}}{N_c^2}
\end{align*}
\end{widetext}
The coupling constant of each channel is fixed by two parameters $G_{I}$ and  $G_{IA}$, directly determined from the QCD instanton vacuum with $\rho=0.31$ fm, $n_{I+A}=1$ fm$^{-4}$. For two flavors, the values are given by
\bea
G_I&=&\frac{n_{I+A}}{2}\left(\frac{4\pi^2\rho^2}{m^*}\right)^2=610.3~\mathrm{GeV}^{-2}\nonumber\\
G_{IA}&=&\frac{G^2_I}{8}\left\langle \left( \frac{T_{IA}}{4\pi^2\rho^2}\right)^{2}\right\rangle=57.08~\mathrm{GeV}^{-2}\nonumber
\eea
Since $n_{I+A}$ is of order $N_c$, both $G_I$ and $G_{IA}$ are of the same order in the $1/N_c$ book-keeping. The quark hopping integral $T_{IA}=\int d^4x\phi_I^\dagger(x-z_I)i\slashed{\partial}\phi_A(x-z_A)$ in the IA-molecule, is defined by the overlap between the quark zero modes in \eqref{VER2}.

With this in mind, the bubble chain in Fig.\ref{fig:BS} with the scalar 't Hooft interaction in \eqref{tHooft}, yields the gap-like equation for the $\sigma$ meson mass 
\begin{align}
\label{sigma_gap}
&1=G_{\sigma}\Pi_{SS}(m_{\sigma}^2)
\end{align}
where the scalar to scalar vacuum polarization function is defined as
\begin{widetext}
\begin{equation}
\begin{aligned}
   \Pi_{SS}(P^2)
   = -2iN_c\int\frac{d^4k}{(2\pi)^4}\mathrm{tr}\left[S_{\mathrm{ILM}}(k)S_{\mathrm{ILM}}(k-P)\right]\mathcal{F}(k)\mathcal{F}(P-k)
\end{aligned}
\end{equation}
\end{widetext}
The  effective Euclidean quark propagator in the QCD instanton vacuum involves the running quark mass (not the determinantal mass)~\cite{Liu:2023fpj,Liu:2023yuj}
\begin{equation}
    S_{\mathrm{ILM}}(k)=\frac{i\slashed{k}+M_q(k)}{k^2+M_q(k)^2}
\end{equation}
With \eqref{sigma_gap}, the sigma mass is obtained
\be
\label{sigmaX}
m_\sigma=683.1~\mathrm{MeV}
\ee
The sigma meson-quark effective coupling can also be determined by
\begin{align}
&g^2_{\sigma qq}=\left(\frac{\partial \Pi_{SS}(P^2)}{\partial P^2}\right)^{-1}\Big|_{P^2=m_{\sigma}^2}\rightarrow3.841
\end{align}

In the pseudoscalar-flavor-singlet channel, the chiral Ward identity is violated, the pseudoscalar and axial vector mix. 
The $\eta'$ mass follows from the resummation in Fig.\ref{fig:BS}, in leading order in the $1/N_c$ book-keeping
Using
the emerging 't Hooft interaction in the pseudoscalar and axial vector channels, the mass of the two-flavor  meson follows from the gap-like equation with $\eta'$-$f_1$ mixing
\begin{widetext}
\begin{align}
1=G_{\eta'}\Pi_{PP}(m_{\eta'}^2)+G_{f_1}\Pi^{(l)}_{AA}(m_{\eta'}^2)-G_{\eta',\pi}G_{f_1}\Pi_{PA}(m_{\eta'}^2)\Pi_{AP}(m_{\eta',\pi}^2)
\end{align}
\end{widetext}
The corresponding $\eta'$ mass   is
\be
\label{MASSETAPRIME}
m_{\eta'}=681.7~\mathrm{MeV}
\ee
which is comparable to the scalar mass (\ref{sigmaX}).  This is to be compared to the two-flavor $m_{\eta'}=772$ MeV reported on the lattice~\cite{Dimopoulos:2018xkm}.
The interactions in (\ref{tHooft}) are mostly repulsive in the scalar and eta' channels, and attractive in the pion and $a_0$ channels, hence the  heavy eta'.

The meson-quark effective coupling in the pseudoscalar channel is adjusted by the mixing from the axial vectors,
\begin{widetext}
\begin{equation}
\begin{aligned}
g^2_{\eta' qq}=\frac{G_{\eta'}-G_{f_1}+G_{\eta'}G_{f_1}(\Pi_{PP}-\Pi^{(l)}_{AA})}{G_{\eta'}(1-G_{f_1}\Pi^{(l)}_{AA})\frac{\partial \Pi_{PP}}{\partial P^2}+G_{f_1}(1-G_{\eta'}\Pi_{PP})\frac{\partial \Pi^{(l)}_{AA}}{\partial P^2}-2G_{\eta'} G_{f_1}\Pi_{PA}\frac{\partial\Pi_{PA}}{\partial P^2}}\Bigg|_{P^2=m_{\eta'}^2}
\rightarrow ~1.686
\end{aligned}
\end{equation}
The vacuum polarization functions regarding the mixing are defined as
\begin{equation}
\begin{aligned}
   \Pi_{PP}(P^2)=& -2iN_c\int\frac{d^4k}{(2\pi)^4}\mathrm{tr}\left[S_{\mathrm{ILM}}(k)i\gamma^5S_{\mathrm{ILM}}(k-P)i\gamma^5\right]\mathcal{F}(k)\mathcal{F}(P-k)
\end{aligned}
\end{equation}

\begin{equation}
\begin{aligned}
   \Pi^{\mu}_{PA}(P^2)= -2iN_c\int\frac{d^4k}{(2\pi)^4}\mathrm{tr}\left[S_{\mathrm{ILM}}(k)i\gamma^5S_{\mathrm{ILM}}(k-P)\gamma^\mu\gamma^5\right]\mathcal{F}(k)\mathcal{F}(P-k)
   =i\Pi_{PA}(P^2)\frac{P^\mu}{\sqrt{P^2}}
\end{aligned}
\end{equation}
with $\Pi^{\mu}_{AP}=\left(\Pi^{\mu}_{PA}\right)^*$ its complex conjugate, and 
\begin{equation}
\begin{aligned}
   \Pi^{\mu\nu}_{AA}(P^2)=& -2iN_c\int\frac{d^4k}{(2\pi)^4}\mathrm{tr}\left[S_{\mathrm{ILM}}(k)\gamma^\mu\gamma^5S_{\mathrm{ILM}}(k-P)\gamma^\nu\gamma^5\right]\mathcal{F}(k)\mathcal{F}(P-k)\\
   =&-\Pi^{(t)}_{AA}(P^2)\left(g^{\mu\nu}-\frac{P^\mu P^\nu}{P^2}\right)-\Pi^{(l)}_{AA}(P^2)\frac{P^\mu P^\nu}{P^2}
\end{aligned}
\end{equation}
\end{widetext}

\section{Instanton field in singular gauge}
\label{App:sing_inst}
In singular gauge, the instanton gauge field $A^a_\mu(x)$ 
is given by 
\begin{equation}
\label{FSTX}
A^a_{I\mu}(x-z_I,\rho,U_I)=R^{ab}(U_I)A^b_\mu(x-z_I)
\end{equation}
The  gauge profile is defined in  (\ref{INSINGULAR}),
and $R^{ab}(U_I)=\frac{1}{2}\mathrm{Tr}(\tau^aU_I\tau^bU_I^\dagger)$,  where $\tau^a$ is a $N_c\times N_c$ matrices with the $2\times2$ Pauli matrices embedded in the upper left corner. For the anti-instanton field, we  substitute  $\bar{\eta}^a_{\mu\nu}$ by $\eta^a_{\mu\nu}$ and flip the sign in front of Levi-Cevita tensor, $\epsilon_{\mu\nu\rho\lambda}\rightarrow-\epsilon_{\mu\nu\rho\lambda}$. The instanton moduli  is captured by the rigid color rotation, instanton 
location and size. 

The field strength associated to (\ref{FSTX}) 
reads
\begin{equation}
    G^a_{\mu\nu}[A_I]=R^{ab}(U_I)G^a_{\mu\nu}(x-z_I)
\end{equation}
with the corresponding field strength profile
\begin{widetext}
\begin{equation}
    G^a_{\mu\nu}(x)=
        \frac{1}{g}\frac{8\rho^2}{x^2(x^2+\rho^2)^2}\left[\bar{\eta}^a_{\mu\rho}\left(\frac{x_\rho x_\nu}{x^2}-\frac{1}{4}\delta_{\rho\nu}\right)-\bar{\eta}^a_{\nu\rho}\left(\frac{x_\rho x_\mu}{x^2}-\frac{1}{4}\delta_{\rho\mu}\right)\right]
\end{equation}
\end{widetext}
In the sum ansatz, the gluonic field strength for the multi-instanton configuration can be split  into single instanton fields, and crossing terms typical of non-Abelian fields
\begin{equation}
G^a_{\mu\nu}[A_{\mathrm{inst}}]=\sum_{I}G^a_{\mu\nu}[A_I]+\sum_{I\neq J}G^a_{\mu\nu}[A_I,A_J]
\end{equation}
The non-Abelian gauge crossing term between two instantons can be expressed as
\begin{widetext}
\begin{equation}
\begin{aligned}
\label{eq:cross_term}
G^a_{\mu\nu}[A_I,A_J]&=gR^{aa'}(U_I)\epsilon^{a'bc}R^{cd}(U_I^\dagger U_J)\left[A^b_{\mu}(x-z_I) A^d_{\nu}(x-z_J)-A^b_{\nu}(x-z_I) A^d_{\mu}(x-z_J)\right]\\ 
\end{aligned}
\end{equation}
\end{widetext}
The 't Hooft symbol in Euclidean space is defined as
\begin{equation}
    \bar{\eta}^a_{\mu\nu}=\frac{1}{4i}\mathrm{Tr}\left[\tau^a(\tau_\mu^-\tau_\nu^+-\tau_\nu^-\tau_\mu^+)\right]
\end{equation}

\begin{equation}
    \eta^a_{\mu\nu}=\frac{1}{4i}\mathrm{Tr}\left[\tau^a(\tau_\mu^+\tau_\nu^--\tau_\nu^+\tau_\mu^-)\right]
\end{equation}
where the Pauli 4-vector is defined as $\tau^\pm_\mu=(\vec{\tau},\mp i)$. The conversion to  Minkowski space is follows by adding extra $i$'s to each $4$th component of the Lorentz indices. Thus, the 't Hooft symbol in Minkowski space is defined as \cite{Vainshtein:1981wh}
\begin{equation}
    \eta^{a}_{\mu\nu}=\begin{cases}
        \epsilon^{a}{}_{\mu\nu} ,\ & \mu\neq0,\ \nu\neq0\\
        i\delta^{a}_{\mu} ,\ & \mu\neq0,\ \nu=0 \\
        -i\delta^{a}_\nu ,\ & \mu=0,\ \nu\neq0
    \end{cases}
\end{equation}
and its  conjugate,
\begin{equation}
    \bar{\eta}^{a}_{\mu\nu}=\begin{cases}
        \epsilon^{a}{}_{\mu\nu} ,\ & \mu\neq0,\ \nu\neq0\\
        -i\delta^{a}_\mu ,\ & \mu\neq0,\ \nu=0 \\
        i\delta^{a}_\nu ,\ & \mu=0,\ \nu\neq0
    \end{cases}
\end{equation}

\section{Two-instanton Configurations on the Gluonic Operators}
\label{App:pairs}
The gluonic operators for the multi-instanton configuration can be  constructed using  the gluonic field strength following from the sum ansatz. Throughout, we will limit the discussion to instanton pairs. The consideration of higher clusters goes beyond the scope of this work.
\begin{widetext}

\subsection{Two-gluon operators}
In the case of the two-gluon operators, we have three typical instanton pair configurations,
\begin{equation}
    G^{a}_{\mu \alpha}[A_J]G^{a}_{\nu \alpha}[A_K]=\frac{1}{2}\mathrm{tr}(\tau^aU_{JK}\tau^bU^{\dagger}_{JK})G^{a}_{\mu \alpha}(x-z_J)G^{b}_{\nu \alpha}(x-z_K)
\end{equation}
\begin{equation}
\begin{aligned}
    G^{a}_{\mu \alpha}[A_J]G^{a}_{\nu \alpha}[A_J,A_K]=&\frac{1}{2}\mathrm{tr}(\tau^aU_{JK}\tau^bU^{\dagger}_{JK})\\
    &\times\epsilon^{acd}G_{\mu \alpha}^c(x-z_J)[A_\nu^d(x-z_J)A_\alpha^b(x-z_K)-A_\alpha^d(x-z_J)A_\nu^b(x-z_K)]
\end{aligned}
\end{equation}
\bea
   G^{a}_{\mu \alpha}[A_J,A_K]G^{a}_{\nu \alpha}[A_J,A_K]=&&\frac{1}{2}\mathrm{tr}(\tau^aU_{JK}\tau^bU^{\dagger}_{JK})\frac{1}{2}\mathrm{tr}(\tau^cU_{JK}\tau^dU^{\dagger}_{JK})\nonumber\\
   &&\times[A_\mu^a(x-z_J)A_\alpha^b(x-z_K)-A_\alpha^a(x-z_J)A_\mu^b(x-z_K)]\nonumber\\
   &&\times[A_\nu^{c}(x-z_J)A_\alpha^{d}(x-z_K)-A_\alpha^{c}(x-z_J)A_\nu^{d}(x-z_K)]
\eea

\subsection{$C$-odd three-gluon operators}
In the case of the $C$-odd three-gluon operators, we have two typical instanton pair configurations,
\begin{equation}
    d^{abc}G^{a}_{\mu \nu}[A_J]G^{b}_{\rho \alpha}[A_J]G^{c}_{\lambda \alpha}[A_K]=\frac{1}{2}\mathrm{tr}(\mathds{1}_2U_{JK}\tau^aU^\dagger_{JK})G^{b}_{\mu \nu}(x-z_J)G^{b}_{\rho \alpha}(x-z_J)G^{a}_{\lambda \alpha}(x-z_K)
\end{equation}
\bea
   && d^{abc}G^{a}_{\mu \nu}[A_J]G^{b}_{\rho \alpha}[A_J,A_K]G^{c}_{\lambda \alpha}[A_K]=\frac{1}{2}\mathrm{tr}(\mathds{1}_2U_{JK}\tau^aU^\dagger_{JK})\frac{1}{2}\mathrm{tr}(\tau^bU_{JK}\tau^cU^\dagger_{JK})\nonumber\\
    &&\times\epsilon^{ade}G^{d}_{\mu \nu}(x-z_J)\left[A^{e}_{\rho}(x-z_J)A_\alpha^b(x-z_K)-A^{e}_{\alpha}(x-z_J)A_\rho^b(x-z_K)\right]G^{c}_{\lambda \alpha}(x-z_K)
\eea

\subsection{$C$-even three-gluon operators}
In the case of the $C$-even three-gluon operators, we have two typical instanton pair configurations. 

\begin{equation}
    f^{abc}G^{a}_{\mu \nu}[A_J]G^{b}_{\nu\lambda}[A_J]G^{c}_{\lambda \mu}[A_K]=\frac{1}{2}\mathrm{tr}(\tau^aU_{JK}\tau^bU^\dagger_{JK})\epsilon^{acd}G^{c}_{\mu \nu}(x-z_J)G^{d}_{\nu\lambda}(x-z_J)G^{b}_{\lambda \mu}(x-z_K)
\end{equation}
\bea
    &&f^{abc}G^{a}_{\mu \nu}[A_J]G^{b}_{\nu\lambda }[A_J,A_K]G^{c}_{\lambda \mu}[A_K]=\frac{1}{2}\mathrm{tr}(\tau^aU_{JK}\tau^bU^\dagger_{JK})\frac{1}{2}\mathrm{tr}(\tau^cU_{JK}\tau^dU^\dagger_{JK})\nonumber\\
    &&\times\{\left[G^{a}_{\mu \nu}(x-z_J)A^{c}_{\nu}(x-z_J)-\delta^{ac}G^{e}_{\mu \nu}(x-z_J)A^{e}_{\nu}(x-z_J)\right]A^b_\lambda(x-z_K)G^{d}_{\lambda \mu}(x-z_K)\nonumber\\
    &&-\left[G^{a}_{\mu \nu}(x-z_J)A^{c}_{\lambda}(x-z_J)-\delta^{ac}G^{e}_{\mu \nu}(x-z_J)A^{e}_{\lambda}(x-z_J)\right]A^b_\nu(x-z_K)G^{d}_{\lambda \mu}(x-z_K)\}
\eea
\end{widetext}

We now  detail the steps in the calculation of the gluonic operators in the instanton ensemble,  to include correlations between instantons.

\subsection{Color averages}
In the case of a two-instanton configuration $J, K$, the color structure of the gluonic operators only depends on the relative color rotation $U_{JK}=U_J^\dagger U_K$. Only few color structures are involved in the calculation of the color averages, which we now list.

\begin{widetext}
\subsubsection{$\frac{1}{2}\mathrm{tr}(\tau^aU_{JK}\tau^bU^{\dagger}_{JK})$}

\begin{equation}
\begin{aligned}
    &\int dU_JdU_K\frac{1}{2}\mathrm{tr}(\tau^aU_{JK}\tau^bU^{\dagger}_{JK})\Theta_J\Theta_K\\
    =&\frac{1}{2N_c(N_c^2-1)}\left(\frac{m^*}{4\pi^2\rho^2}\right)^{2N_f-2}\Bigg[\bar{\psi}(z_J)\frac{1\mp\gamma^5}{2}\frac{1}{4}\bar{\eta}_{\alpha\beta}^a(J)\sigma_{\alpha\beta} S(z_J-z_K)\frac{1}{4}\bar{\eta}_{\rho\lambda}^b(K)\sigma_{\rho\lambda} \frac{1\mp\gamma^5}{2}\psi(z_K)\\
    &+\bar{\psi}(z_K)\frac{1\mp\gamma^5}{2}\frac{1}{4}\bar{\eta}_{\rho\lambda}^b(K)\sigma_{\rho\lambda} S(z_K-z_J)\frac{1}{4}\bar{\eta}_{\alpha\beta}^a(J)\sigma_{\alpha\beta} \frac{1\mp\gamma^5}{2}\psi(z_J)\Bigg]\\
\end{aligned}
\end{equation}
where the 't-Hooft symbol is defined as $\eta_{\mu\nu}^a(J)=\bar{\eta}_{\mu\nu}^a$ if $J\in I$ (instanton) and $\eta_{\mu\nu}^a(J)=\eta_{\mu\nu}^a$ if $J\in A$ (anti-instanton). If the instanton pair is an instanton-antiinstanton molecule, the averaging produces a chiral conserving quark operator, and a chiral flipping quark operator
otherwise.
\\
\begin{itemize}
    \item \textbf{$II$ cluster}
    
When the pair carries like-topological-charge, i.e. $JK=II$ or $AA$
\begin{equation}
\begin{aligned}
    &\int dU_JdU_K\frac{1}{2}\mathrm{tr}(\tau^aU_{JK}\tau^bU^{\dagger}_{JK})\Theta_I\Theta_J\\
    =&\frac{1}{2N_c(N_c^2-1)}\left(\frac{m^*}{4\pi^2\rho^2}\right)^{2N_f-2}\frac{1}{8}\mathrm{tr}\{S(z_J-z_K)\}\delta^{ab}\left[\bar{\psi}(z_J)\frac{1\mp\gamma^5}{2}\psi(z_K)+\bar{\psi}(z_K)\frac{1\mp\gamma^5}{2}\psi(z_J)\right]
\end{aligned}
\end{equation}

    \item \textbf{$IA$ cluster}

When the pair or  molecule carries unlike-topological-charge, i.e. $JK=IA$ or $AI$
\begin{equation}
\begin{aligned}
    &\int dU_JdU_K\frac{1}{2}\mathrm{tr}(\tau^aU_{JK}\tau^bU^{\dagger}_{JK})\Theta_J\Theta_K\\
    =&-\frac{1}{2N_c(N_c^2-1)}\left(\frac{m^*}{4\pi^2\rho^2}\right)^{2N_f-2}\frac{1}{8}\mathrm{tr}\{S(z_J-z_K)\gamma_\rho\}\bar{\eta}^a_{\rho\beta}\eta^b_{\lambda\beta}\left[\bar{\psi}(z_J)\gamma_\lambda\frac{1\pm\gamma^5}{2}\psi(z_K)-\bar{\psi}(z_K)\gamma_\lambda\frac{1\mp\gamma^5}{2}\psi(z_J)\right]
\end{aligned}
\end{equation}

\subsubsection{$\frac{1}{2}\mathrm{tr}(\tau^aU_{JK}\tau^bU^{\dagger}_{JK})\frac{1}{2}\mathrm{tr}(\tau^cU_{JK}\tau^dU^{\dagger}_{JK})$}

\begin{equation}
\begin{aligned}
    &\int dU_JdU_K\frac{1}{2}\mathrm{tr}(\tau^aU_{JK}\tau^bU^{\dagger}_{JK})\frac{1}{2}\mathrm{tr}(\tau^cU_{JK}\tau^dU^{\dagger}_{JK})\Theta_J\Theta_K\\
    =&\frac{1}{2N_c^2(N_c^2-1)}\left(\frac{m^*}{4\pi^2\rho^2}\right)^{2N_f-2}\epsilon^{ace}\epsilon^{bdf}\Bigg[\bar{\psi}(z_J)\frac{1\mp\gamma^5}{2}\frac{1}{4}\bar{\eta}_{\alpha\beta}^e(J)\sigma_{\alpha\beta} S(z_J-z_K)\frac{1}{4}\bar{\eta}_{\rho\lambda}^f(K)\sigma_{\rho\lambda} \frac{1\mp\gamma^5}{2}\psi(z_K)\\
    &+\bar{\psi}(z_K)\frac{1\mp\gamma^5}{2}\frac{1}{4}\bar{\eta}_{\rho\lambda}^f(K)\sigma_{\rho\lambda} S(z_K-z_J)\frac{1}{4}\bar{\eta}_{\alpha\beta}^e(J)\sigma_{\alpha\beta} \frac{1\mp\gamma^5}{2}\psi(z_J)\Bigg]\\
    +&\frac{1}{2N_c^2(N_c^2-1)}\left(\frac{m^*}{4\pi^2\rho^2}\right)^{2N_f-2}\left(2+\frac{N_c-2}{4(N_c+2)}\right)\delta^{ac}\delta^{bd}\\
    &\times\left[\bar{\psi}(z_J)\frac{1\mp\gamma^5}{2} S(z_J-z_K)\frac{1\mp\gamma^5}{2}\psi(z_K)+\bar{\psi}(z_K)\frac{1\mp\gamma^5}{2} S(z_K-z_J)\frac{1\mp\gamma^5}{2}\psi(z_J)\right]
\end{aligned}
\end{equation}
\end{itemize}
This type of color structure is down by $1/N_c$ in the two-gluon operators. Therefore, they can be neglected in the $1/N_c$ counting.
\\
\begin{itemize}
    \item \textbf{$II$ cluster}
When the pair carries like-topological-charge, i.e. $JK=II$ or $AA$

\begin{equation}
\begin{aligned}
    &\int dU_JdU_K\frac{1}{2}\mathrm{tr}(\tau^aU_{JK}\tau^bU^{\dagger}_{JK})\frac{1}{2}\mathrm{tr}(\tau^cU_{JK}\tau^dU^{\dagger}_{JK})\Theta_J\Theta_K\\
    =&\frac{1}{2N_c(N_c^2-1)}\left(\frac{m^*}{4\pi^2\rho^2}\right)^{2N_f-2}\frac{1}{8}\mathrm{tr}\{S(z_J-z_K)\}\left[\frac{1}{N_c}\epsilon^{ace}\epsilon^{bde}+\left(\frac{4}{N_c}+\frac{N_c-2}{2N_c(N_c+2)}\right)\delta^{ac}\delta^{bd}\right]\\
    &\times\left[\bar{\psi}(z_I)\frac{1\mp\gamma^5}{2}\psi(z_J)+\bar{\psi}(z_J)\frac{1\mp\gamma^5}{2}\psi(z_I)\right]
\end{aligned}
\end{equation}

    \item \textbf{$IA$ cluster}
    
When the pair or  molecule carries unlike-topological-charge,  i.e. $JK=IA$ or $AI$
\begin{equation}
\begin{aligned}
        &\int dU_JdU_K\frac{1}{2}\mathrm{tr}(\tau^aU_{JK}\tau^bU^{\dagger}_{JK})\frac{1}{2}\mathrm{tr}(\tau^cU_{JK}\tau^dU^{\dagger}_{JK})\Theta_J\Theta_K\\
    &=\frac{1}{2N_c(N_c^2-1)}\left(\frac{m^*}{4\pi^2\rho^2}\right)^{2N_f-2}\frac{1}{8}\mathrm{tr}\{S(z_I-z_J)\gamma_\rho\}\left[\bar{\eta}^e_{\rho\beta}\eta^f_{\lambda\beta}
    \frac{1}{N_c}\epsilon^{ace}\epsilon^{bdf}+\delta_{\rho\lambda}\left(\frac{4}{N_c}+\frac{N_c-2}{2N_c(N_c+2)}\right)\delta^{ac}\delta^{bd}\right]\\
    &\times\left[\bar{\psi}(z_I)\gamma_\lambda\frac{1\pm\gamma^5}{2}\psi(z_J)-\bar{\psi}(z_J)\gamma_\lambda\frac{1\mp\gamma^5}{2}\psi(z_I)\right]
\end{aligned}
\end{equation}

\end{itemize}

\subsubsection{$\frac{1}{2}\mathrm{Tr}(\mathds{1}_2U_{JK}\tau^aU^\dagger_{JK})$}

\begin{equation}
\begin{aligned}
&\int dU_JdU_K\frac{1}{2}\mathrm{Tr}(\mathds{1}_2U_{JK}\tau^aU^\dagger_{JK})\Theta_J\Theta_K\\
=&-\frac{N_c-2}{2N_c^2(N_c^2-1)}\left(\frac{m^*}{4\pi^2\rho^2}\right)^{2N_f-2}\frac{1}{4}\bar{\eta}^a_{\mu\nu}(K)\\
&\times\left[\bar{\psi}(z_J)\frac{1\mp\gamma^5}{2}S(z_J-z_K)\sigma_{\mu\nu}\frac{1\mp\gamma^5}{2}\psi(z_K)+\bar{\psi}(z_K)\frac{1\mp\gamma^5}{2}\sigma_{\mu\nu}S(z_K-z_J)\frac{1\mp\gamma^5}{2}\psi(z_J)\right]
\end{aligned}
\end{equation}

\begin{itemize}
    \item \textbf{$II$ cluster}

When the pair carries like-topological-charge, i.e. $JK=II$ or $AA$

\begin{equation}
\begin{aligned}
    &\int dU_JdU_K\frac{1}{2}\mathrm{Tr}(\mathds{1}_2U_{JK}\tau^aU^\dagger_{JK})\Theta_J\Theta_K\\
    =&-\frac{N_c-2}{2N_c^2(N_c^2-1)}\left(\frac{m^*}{4\pi^2\rho^2}\right)^{2N_f-2}\frac{1}{8}\mathrm{tr}\{S(z_J-z_K)\}\frac{1}{2}\bar{\eta}^a_{\mu\nu}(K)\left[\bar{\psi}(z_J)\sigma_{\mu\nu}\frac{1\mp\gamma^5}{2}\psi(z_K)+\bar{\psi}(z_K)\sigma_{\mu\nu}\frac{1\mp\gamma^5}{2}\psi(z_J)\right]
\end{aligned}
\end{equation}

    \item \textbf{$IA$ cluster}
    
When the pair or  molecule carries unlike-topological-charge,  i.e. $JK=IA$ or $AI$
\begin{equation}
\begin{aligned}
        &\int dU_JdU_K\frac{1}{2}\mathrm{Tr}(\mathds{1}_2U_{JK}\tau^aU^\dagger_{JK})\Theta_J\Theta_K\\
    =&-\frac{N_c-2}{2N_c^2(N_c^2-1)}\left(\frac{m^*}{4\pi^2\rho^2}\right)^{2N_f-2}\frac{i}{4}\mathrm{tr}\{S(z_J-z_K)\gamma_\mu\}\frac{1}{2}\bar{\eta}^a_{\mu\nu}(K)\left[\bar{\psi}(z_J)\gamma_\nu\frac{1\pm\gamma^5}{2}\psi(z_K)+\bar{\psi}(z_K)\gamma_\nu\frac{1\mp\gamma^5}{2}\psi(z_J)\right]
\end{aligned}
\end{equation}

\end{itemize}

\subsubsection{$\frac{1}{2}\mathrm{tr}(\mathds{1}_2U_{JK}\tau^aU^\dagger_{JK})\frac{1}{2}\mathrm{tr}(\tau^bU_{JK}\tau^cU^\dagger_{JK})$}
\begin{equation}
\begin{aligned}
    &\int dU_JdU_K\frac{1}{2}\mathrm{tr}(\mathds{1}_2U_{JK}\tau^aU^\dagger_{JK})\frac{1}{2}\mathrm{tr}(\tau^bU_{JK}\tau^cU^\dagger_{JK})\Theta_J\Theta_K\\
=&-\frac{N_c-2}{2N_c^2(N_c^2-1)}\frac{1}{4(N_c+2)}\frac{1}{4}\bar{\eta}^a_{\mu\nu}(J)\delta^{bc}\\
&\times\left[\bar{\psi}(z_J)\frac{1\mp\gamma^5}{2}\sigma_{\mu\nu}S(z_J-z_K)\frac{1\mp\gamma^5}{2}\psi(z_K)+\bar{\psi}(z_K)\frac{1\mp\gamma^5}{2}S(z_K-z_J)\sigma_{\mu\nu}\frac{1\mp\gamma^5}{2}\psi(z_J)\right]
\end{aligned}
\end{equation}

\begin{itemize}

    \item $II$, $AA$ cluster

When the pair carries like-topological-charge, i.e. $JK=II$ or $AA$

\begin{equation}
\begin{aligned}
    &\int dU_JdU_K\frac{1}{2}\mathrm{tr}(\mathds{1}_2U_{JK}\tau^aU^\dagger_{JK})\frac{1}{2}\mathrm{tr}(\tau^bU_{JK}\tau^cU^\dagger_{JK})\Theta_J\Theta_K\\
    =&-\frac{N_c-2}{2N_c^2(N_c^2-1)}\frac{1}{4(N_c+2)}\left(\frac{m^*}{4\pi^2\rho^2}\right)^{2N_f-2}\frac{1}{8}\mathrm{tr}\{S(z_J-z_K)\}\frac{1}{2}\bar{\eta}^a_{\mu\nu}(J)\delta^{bc}\\
    &\times\left[\bar{\psi}(z_I)\sigma_{\mu\nu}\frac{1\mp\gamma^5}{2}\psi(z_J)+\bar{\psi}(z_J)\sigma_{\mu\nu}\frac{1\mp\gamma^5}{2}\psi(z_I)\right]
\end{aligned}
\end{equation}

    \item $IA$ cluster

When the pair or  molecule carries unlike-topological-charge, i.e. $JK=IA$ or $AI$
\begin{equation}
\begin{aligned}
    &\int dU_JdU_K\frac{1}{2}\mathrm{tr}(\mathds{1}_2U_{JK}\tau^aU^\dagger_{JK})\frac{1}{2}\mathrm{tr}(\tau^bU_{JK}\tau^cU^\dagger_{JK})\Theta_J\Theta_K\\
    &=-\frac{N_c-2}{2N_c^2(N_c^2-1)}\frac{1}{4(N_c+2)}\left(\frac{m^*}{4\pi^2\rho^2}\right)^{2N_f-2}\frac{i}{4}\mathrm{tr}\{S(z_J-z_K)\gamma_\mu\}\frac{1}{2}\bar{\eta}^a_{\mu\nu}(J)\delta^{bc}\\
    &\times\left[\bar{\psi}(z_J)\gamma_\nu\frac{1\pm\gamma^5}{2}\psi(z_K)+\bar{\psi}(z_K)\gamma_\nu\frac{1\mp\gamma^5}{2}\psi(z_J)\right]
\end{aligned}
\end{equation}

\end{itemize}
\end{widetext}

\section{Gluons captured by the instanton tails}
\label{App:tails}
\begin{figure*}
\hfill
\subfloat[\label{fig:inst_2g-3}]{%
    \includegraphics[scale=0.75]{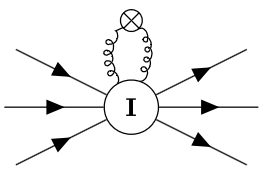}
}\hfill
\subfloat[\label{fig:inst_2g-4}]{%
 \includegraphics[scale=0.75]{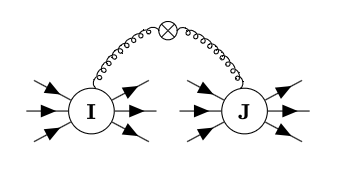}
}\hfill
\subfloat[\label{fig:inst_3g-4}]{%
\includegraphics[scale=0.75]{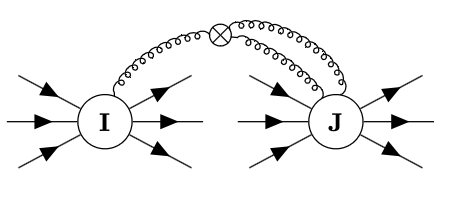}
}
\caption{ 
a: The diagram of $\mathcal{O}[I]$ in the multi-instanton expansion of the two gluon operator $\mathcal{O}_{2g}$. Each dashed line connected to the instanton represents the classical background field;
b: The diagram of $\mathcal{O}[I]$ in the multi-instanton expansion of the two gluon operator $\mathcal{O}_{2g}$. Each dashed line connected to the instanton represents the classical background field.
c: The effective two-(anti)-instanton operators $\mathcal{O}_{3g++,+-,--}$ which represents all of the gluon in the operator $\mathcal{O}_{3g}[A]$ are plane waves. one of the gluon plane waves are connected to the instanton $I$ and the remaining two gluon plane waves are connected to $J$. The conjugate diagram could be obtained by $I\leftrightarrow J$;
}
\end{figure*}

When strict factorization is enforced on hadronic kernels, 
and Feynman-like graphs are used, 
it maybe more appropriate to evaluate the hadronic matrix elements  using  the effective vertices in  $S_{\mathrm{eff}}$ in (\ref{eq:effective_action}), with the LSZ reduced gluons. This is  graphically illustrated by the Feynman diagrams in Fig.\ref{fig:inst_2g-3}, \ref{fig:inst_2g-4}, and \ref{fig:inst_3g-4}, where the gluonic lines denotes the 
LSZ reduced gluons. 

We note that in the  ensemble averaging 
described earlier, the dashed gluons are moduli gluons,
as opposed to the LSZ reduced gluons in this section. The latters are only souced by small size pseudoparticles in the QCD vacuum, while the formers are for any pseudoparticle size in the QCD vacuum.

With this in mind and using the $1/N_c$ book-keeping, the one-body interaction is seen to dominate matrix elements, with the rest of the  $N_f-1$ flavors looped up. Therefore, in the following calculations, we will only retain the one-body induced interaction.

For the gluonic scalar  $G^a_{\mu \nu}G^a_{\mu \nu}$, the result of the Feynman diagram in Fig.\ref{fig:inst_2g-3} reads
\begin{widetext}
\begin{equation}
\begin{aligned}
\label{eq:O_2g}
    \langle P'|G^a_{\mu \nu}G^a_{\mu \nu}|P\rangle=&-\frac{G}{N_c}\left(\frac{m^*}{4\pi^2\rho^2}\right)^{N_f-1}\langle P'|\bar{\psi}\psi|P\rangle \int d^4x\left(\frac{2\pi^2\rho^2}{g}\right)^2D_{\mu \nu\rho\lambda}(x)D_{\mu \nu\rho\lambda}(x)e^{-iq\cdot x}
\end{aligned}
\end{equation}
where the Euclidean gluon propagator connecting the instanton and the effective vertices,  is defined by
\begin{equation}
\begin{aligned}
\label{eq:glu_prop}
    D_{\mu\nu\rho\lambda}(x)=&\int\frac{d^4q}{(2\pi)^4}\mathcal{F}_g(\rho q)\frac{1}{q^2}(q_\mu q_\rho \delta_{\nu\lambda}-q_\nu q_\rho \delta_{\mu\lambda}-q_\mu q_\lambda \delta_{\nu\rho}+q_\nu q_\lambda\delta_{\mu\rho})\theta(1/\rho^2-q^2)e^{iq\cdot x}\\
    =&-\frac{1}{4\pi^2x}\left(\frac{x_\mu x_\rho}{x^2} \delta_{\nu\lambda}-\frac{x_\nu x_\rho}{x^2} \delta_{\mu\lambda}-\frac{x_\mu x_\lambda}{x^2} \delta_{\nu\rho}+\frac{x_\nu x_\lambda}{x^2}\delta_{\mu\rho}\right)\int_0^\infty dqq^2\mathcal{F}_g(\rho q)\theta(1/\rho^2-q^2)J_3(qx)\\
    &+\frac{1}{2\pi^2x}\left(\delta_{\mu\rho} \delta_{\nu\lambda}-\delta_{\mu\lambda} \delta_{\nu\rho}\right)\int_0^\infty dqq^2\mathcal{F}_g(\rho q)\theta(1/\rho^2-q^2)\frac{J_2(qx)}{qx}
\end{aligned}
\end{equation}
\end{widetext}
The  gluon emitted from the instanton carries the instanton gluonic form factor, while the gluon absorbed by the operator is cut by the hard-cutoff $1/\rho$.

The calculation can  be extended to second order in the instanton density. The result of the Feynman diagram in Fig.\ref{fig:inst_2g-4} depends on the topological charges of the $IJ$ pairs. The pairs with the same topological charges such as the $II$ or $AA$ pair yield a chiral conserving quark operator. The result reads
\begin{widetext}
\begin{equation}
\begin{aligned}
&\langle P'|G^a_{\mu \nu}G^a_{\rho\lambda}|P\rangle\bigg|_{II,AA}
=\frac{G^2}{N_c(N_c^2-1)}\left(\frac{m^*}{4\pi^2\rho^2}\right)^{2(N_f-1)}\left(\frac{2\pi^2\rho^2}{g}\right)^2\int d^4z_I d^4z_J\frac{1}{4}\mathrm{tr}\{S(z_I-z_J)\}\\
&\times D_{\mu \nu\rho\lambda}(x-z_I)D_{\mu \nu\rho\lambda}(x-z_J)\langle P'|\left[\bar{\psi}(z_I)\frac{1\mp\gamma^5}{2}\psi(z_J)+\bar{\psi}(z_J)\frac{1\mp\gamma^5}{2}\psi(z_I)\right]|P\rangle
\end{aligned}
\end{equation}
Similarly, the pairs with opposite topological charges in the $IA$ or $AI$ molecules, yield a chiral flipping quark operator
given by
\begin{equation}
\begin{aligned}
&\langle P'|G^a_{\mu \nu}G^a_{\rho\lambda}|P\rangle\bigg|_{IA,AI}
=-\frac{G^2}{N_c(N_c^2-1)}\left(\frac{m^*}{4\pi^2\rho^2}\right)^{2(N_f-1)}\left(\frac{2\pi^2\rho^2}{g}\right)^2\int d^4z_I d^4z_J\frac{1}{4}\mathrm{tr}\{S(z_I-z_J)\gamma_\rho\}\\
&\times \left[D_{\mu \nu\rho\alpha}(x-z_I)D_{\mu \nu\lambda\alpha}(x-z_J)+D_{\mu \nu\lambda\alpha}(x-z_I)D_{\mu \nu\rho\alpha}(x-z_J)\right]\\
&\times\langle P'|\left[\bar{\psi}(z_I)\gamma_\lambda\frac{1\pm\gamma^5}{2}\psi(z_J)-\bar{\psi}(z_J)\gamma_\lambda\frac{1\mp\gamma^5}{2}\psi(z_I)\right]|P\rangle
\end{aligned}
\end{equation}
\end{widetext}

The localized form of the pseudoparticles in the QCD instanton vacuum, allows for the use of the R-expansion or local approximation, in the evaluation of the double integral over the
position of the pseudoparticles, with the result to second order in the instanton density
\begin{widetext}
\begin{equation}
\begin{aligned}
\langle P'|g^2G^a_{\mu \nu}G^a_{\mu \nu}|P\rangle
=&\left[-\frac{G}{N_c}\left(\frac{m^*}{4\pi^2\rho^2}\right)^{N_f-1}+\frac{G^2}{N_c(N_c^2-1)}\left(\frac{m^*}{4\pi^2\rho^2}\right)^{2(N_f-1)}\frac{1}{m^*}\right]\langle P'|\bar{\psi}\psi|P\rangle\\
&\times \left(2\pi^2\right)^2\int d^4x D_{\mu \nu\rho\lambda}(x)D_{\mu \nu\rho\lambda}(x)e^{-i\rho q\cdot x}\\
&-\frac{G^2}{N_c(N_c^2-1)}\left(\frac{m^*}{4\pi^2\rho^2}\right)^{2(N_f-1)}\rho^2 T_{IA}(\rho m^*)\langle P'|\bar{\psi}\gamma_\rho\overleftrightarrow{\partial}_\lambda\psi|P\rangle\\
&\times \left(2\pi^2\right)^2\int d^4x\left[D_{\mu \nu\rho\alpha}(x)D_{\mu \nu\lambda\alpha}(x)+D_{\mu \nu\lambda\alpha}(x)D_{\mu \nu\rho\alpha}(x)\right]e^{-i\rho q\cdot x}\\
&+\mathcal{O}(G^3)
\end{aligned}
\end{equation}
\end{widetext}
The  quark hopping interal is given in (\eqref{eq:quark_hopping}).


The calculation can be extended to the three gluon operators. Here we detail it for the $C$-even operator. As expected, there  is no contribution from a single pseudoparticle.
The Feynman diagram in Fig.~\ref{fig:inst_3g-4} with 
like-pairs the pairs $II$ or $AA$, yields the chiral conserving quark operator,
\begin{widetext}
\begin{equation}
\begin{aligned}
&\langle P'|d^{abc}G^a_{\mu \nu}G^b_{\rho \alpha}G^c_{\lambda \alpha}|P\rangle\bigg|_{II,AA}=-\frac{N_c-2}{2N_c^2(N^2_c-1)}G^2\left(\frac{m^*}{4\pi^2\rho^2}\right)^{2(N_f-1)}\left(\frac{2\pi^2\rho^2}{g}\right)^3\int d^4z_Id^4z_J\frac{1}{4}\mathrm{tr}\{S(z_I-z_J)\}\\
&\times \bigg[D_{\mu \nu\beta\gamma}(x-z_J)D_{\rho \alpha\delta\sigma}(x-z_I)D_{\lambda\alpha\delta\sigma}(x-z_I)+2D_{\rho \alpha\beta\gamma}(x-z_J)D_{\mu \nu\delta\sigma}(x-z_I)D_{\lambda \alpha\delta\sigma}(x-z_I)+(\rho\leftrightarrow\lambda)+(I\leftrightarrow J)\bigg]\\
&\times\langle P'|\left[\bar{\psi}(z_I)\sigma_{\beta\gamma}\frac{1\mp\gamma^5}{2}\psi(z_J)+\bar{\psi}(z_J)\sigma_{\beta\gamma}\frac{1\mp\gamma^5}{2}\psi(z_I)\right]|P\rangle\\
\end{aligned}
\end{equation}
For the unlike-pairs, 
 the result gives the chiral flipping quark operator,
\begin{equation}
\begin{aligned}
&\langle P'|d^{abc}G^a_{\mu \nu}G^b_{\rho \alpha}G^c_{\lambda \alpha}|P\rangle\bigg|_{IA,AI}=-\frac{N_c-2}{2N_c^2(N^2_c-1)}G^2\left(\frac{m^*}{4\pi^2\rho^2}\right)^{2(N_f-1)}\left(\frac{2\pi^2\rho^2}{g}\right)^3\int d^4z_Id^4z_J\frac{i}{2}\mathrm{tr}\{S(z_I-z_J)\gamma_\beta\}\\
&\times \bigg[D_{\mu \nu\beta\gamma}(x-z_J)D_{\delta\sigma\rho \alpha}(x-z_I)D_{\delta\sigma\lambda\alpha}(x-z_I)+2D_{\rho \alpha\beta\gamma}(x-z_J)D_{\mu \nu\delta\sigma}(x-z_I)D_{\lambda \alpha\delta\sigma}(x-z_I)+(\rho\leftrightarrow\lambda)-(I\leftrightarrow J)\bigg]\\
&\times\langle P'|\left[\bar{\psi}(z_I)\gamma_{\gamma}\frac{1\pm\gamma^5}{2}\psi(z_J)+\bar{\psi}(z_J)\gamma_{\gamma}\frac{1\mp\gamma^5}{2}\psi(z_I)\right]|P\rangle\\
\end{aligned}
\end{equation}
\end{widetext}
The  Euclidean gluon propagator connecting the instanton and the operators,  is given in (\eqref{eq:glu_prop}). 

Similarly, in the forward limit, we can average over the local operator around the neighborhood of one of the instanton field shared by the operator. Thereforw, after averaging the gluonic operator over 4-dimensional Euclidean space, the profile functions only depends on the relative distance between two instantons  Again,  the effective fermionic operator is non-local. Using the R-expansion or local approximation, the result is
\begin{widetext}
\begin{equation}
\begin{aligned}
\label{eq:had_3g}
    &\langle P'|g^3d^{abc}G^a_{\mu \nu}G^b_{\rho \alpha}G^c_{\lambda \alpha}|P\rangle=-\frac{N_c-2}{N_c^2(N^2_c-1)}G^2\left(\frac{m^*}{4\pi^2\rho^2}\right)^{2(N_f-1)}\frac{2}{\rho^2 m^*}\langle P'|\bar{\psi}\sigma_{\beta\gamma}\psi|P\rangle
    \\
&\times \left(2\pi^2\right)^3\int d^4x e^{-i\rho q\cdot x}\bigg[D_{\mu \nu\beta\gamma}(x)D_{\rho \alpha\delta\sigma}(x)D_{\lambda\alpha\delta\sigma}(x)+2D_{\rho\alpha\beta\gamma}(x)D_{\mu \nu\delta\sigma}(x)D_{\lambda \alpha\delta\sigma}(x)+(\rho\leftrightarrow\lambda)\bigg]\\
&-\frac{N_c-2}{N_c^2(N^2_c-1)}G^2\left(\frac{m^*}{4\pi^2\rho^2}\right)^{2(N_f-1)}
\rho^2T_{IA}(\rho m^*)\langle P'|\bar{\psi}\gamma_{\gamma}\psi|P\rangle\\
&\times \left(2\pi^2\right)^3\int d^4x e^{-i\rho q\cdot x}\frac{\partial}{\partial x'_\beta}\bigg[D_{\mu \nu\beta\gamma}(x)D_{\rho \alpha\delta\sigma}(x')D_{\lambda\alpha\delta\sigma}(x')+2D_{\rho \alpha\beta\gamma}(x)D_{\mu \nu\delta\sigma}(x')D_{\lambda \alpha\delta\sigma}(x')\\
&-D_{\mu \nu\beta\gamma}(x')D_{\delta\sigma\rho \alpha}(x)D_{\delta\sigma\lambda\alpha}(x)-2D_{\rho \alpha\beta\gamma}(x')D_{\mu \nu\delta\sigma}(x)D_{\lambda \alpha\delta\sigma}(x)+(\rho\leftrightarrow\lambda)\bigg]\bigg|_{x=x'}
\end{aligned}
\end{equation}
\end{widetext}
The quark hopping interal $T_{IA}(\rho m^*)$ given in (\eqref{eq:quark_hopping}). 

\section{Grand canonical ensemble}
\label{App:grand}
To account for the fluctuations of the topological charges
in the present  description using an ensemble of pseudoparticles, we need to extend it to a grand canonical ensemble. For that, 
we will allow for the number sum $N=N_++N_-$ and the number difference $\Delta N=N_+-N_-$ to fluctuate, with a universal distribution $\mathds{P}(N_+,N_-)$ fixed by low-energy theorems for $N$ and a topological variance $\chi_t$ for $\Delta N$\cite{Zahed:2021fxk,Zahed:2022wae,Diakonov:1995qy,Schafer:1996wv}
\begin{equation}
\label{DISTX}
    \mathds{P}(N_+,N_-)\propto\left(\frac{\bar{N}^{N}}{N!}\right)^{b/4}\frac{1}{\sqrt{2\pi\chi_t}}\exp\left(-\frac{\Delta N^2}{2\chi_t}\right)
\end{equation}
with $\chi_t=\langle \Delta N^2\rangle$ and $b=\frac{11}{3}N_c-\frac{2}{3}N_f$. The average instanton number is consistent with the parameters $\bar{N}=\langle N\rangle$, and the mean topological charge is null $Q_t=\langle\Delta N\rangle=0$.
As a result, most of the operators we encountered earlier 
can be further averaged,
\begin{equation}
\label{grand_canonical}
    \langle \mathcal{O}\rangle=\sum_{N_+,N_-}\mathds{P}(N_+,N_-)\langle \mathcal{O}\rangle_{N_\pm}
    \equiv \overline{\langle\mathcal{O}\rangle}_{N_\pm} 
\end{equation}
The  averaging is carried  over the configurations with fixed $N_\pm$ (canonical ensemble average),  followed by an averaging over the distribution (\ref{DISTX}).

Since the multiflavor emergent coupling $G(1\pm\delta)=N_\pm/\langle\theta_\pm\rangle_{\mathrm{eff}}$ is fixed by the saddle point approximation in the canonical ensemble ensemble, we can expand $\langle\mathcal{O}\rangle_{\mathrm{eff}}$ in \eqref{eq:inst_contribution}, in terms of connected diagrams with different orders of instanton numbers $N_\pm$. Therefore, the total fixed-$N_\pm$ ensemble average can be written as a certain function expanded in terms of instanton numbers $N_\pm$,
\begin{equation}
\begin{aligned}
    \langle \mathcal{O}[A]\rangle_{N_\pm}=\mathcal{O}(N_+,N_-)
\end{aligned}
\end{equation}
All vacuum expectation values $\langle \mathcal{O}[A]\rangle$ in the grand canonical ensemble, can be expressed as
\begin{equation}
    \langle \mathcal{O}[A]\rangle=\overline{\mathcal{O}(N_+,N_-)}
\end{equation}

With this in mind, the evaluation of the hadronic matrix elements
can be formally written as a large-T reduction of a  3-point function
\begin{equation}
    \frac{\langle h|\mathcal{O}|h\rangle}{\langle h|h\rangle}=\lim_{T\rightarrow\infty}\frac{\langle J^\dagger_h(T/2)\mathcal{O} J_h(-T/2)\rangle_{\mathrm{con}}}{\langle J^\dagger_h(T/2)J_h(-T/2)\rangle}
\end{equation}
where $J_h(t)$ is a pertinent source for  the hadronic state $h$.
The averaging of the connected 3-point function, reads
\begin{widetext}
\begin{equation}
\begin{aligned}
\langle J^\dagger_h(T/2)\mathcal{O}[A] J_h(-T/2)\rangle_{\mathrm{con}}=&\langle J^\dagger_h(T/2)\mathcal{O}[A] J_h(-T/2)\rangle-\langle\mathcal{O}[A]\rangle\langle J^\dagger_h(T/2) J_h(-T/2)\rangle_{\mathrm{con}}\\
=&\sum_{N_+,N_-}\mathds{P}(N_+,N_-)\left(\mathcal{O}(N_+,N_-)-\overline{\mathcal{O}(N_+,N_-)}\right)\langle J^\dagger_h(T/2)J_h(-T/2)\rangle_{N_\pm}\\
&+\langle J^\dagger_h(T/2) :\mathcal{O}[\psi,\bar{\psi}]: J_h(-T/2)\rangle_{\mathrm{eff}}
\end{aligned}    
\end{equation}
\end{widetext}
where $:\mathcal{O}[\psi,\bar{\psi}]:$ denotes the effective quark operator connected to the hadronic sources. The quark contribution are usually penalized by $1/N_c$-counting, as they are rooted in the quark-instanton interaction.

To extract the nontrivial contribution from the disconnected diagrams, we need to consider the fluctuations.
In the $1/N_c$ book-keeping, the dominant contributions are given by
\begin{widetext}
\begin{equation}
\begin{aligned}
    &\frac{\langle P|\mathcal{O}[A]|P\rangle}{\langle P|P\rangle}
    =\sum_{N_+,N_-}\mathds{P}(N_+,N_-)\left(\mathcal{O}(N_+,N_-)-\overline{\mathcal{O}(N_+,N_-)}\right)\\
    \times&\lim_{T\rightarrow\infty}\left[(N-\bar{N})\left(\frac{\partial}{\partial N}\ln\left\langle J^\dagger_h(T/2)J_h(-T/2)\right\rangle_{\mathrm{eff}}\right)_{N=\bar{N}}+\Delta N\left(\frac{\partial}{\partial \Delta N}\ln\left\langle J^\dagger_h(T/2)J_h(-T/2)\right\rangle_{\mathrm{eff}}\right)_{\Delta N=0}\right]
\end{aligned}
\end{equation}
with the number sum  $N=N_++N_-$, the number difference $\Delta N=N_+-N_-$, the tmean number $\bar{N}=\langle N\rangle$, and the mean topological charge $Q_t=\langle\Delta N\rangle=0$. 
\end{widetext}

To show how the 3-point function is determined, we will consider
few examples. When the gluonic operator is proportional to the total instanton number $$\mathcal{O}(N_+,N_-)=\alpha N/V$$.
For asymptotic Euclidean times
$$\left\langle J^\dagger_h(T/2)J_h(-T/2)\right\rangle_{\mathrm{eff}}\rightarrow e^{-M_h(N_+,N_-)T}$$ the matrix element at the leading $1/N_c$ is tied to 
the topological compressibility. 
\begin{equation}
\begin{aligned}
    &\langle P|\mathcal{O}[A]|P\rangle\\
    =&-2M_h^2\alpha\left[\frac{\langle(N-\bar{N})^2\rangle_{\mathds{P}}}{\bar{N}}\right]\bar{N}\frac{\partial\ln M_h}{\partial \bar{N}}
\end{aligned}
\end{equation}
In particular, for the nucleon the mass is related to the instanton density by the scaling relation
$$M_N=C\left(\frac{\bar{N}}{V}\right)^{1/4}+\sigma_{\pi N}$$
so that
\begin{equation}
\begin{aligned}
    &\langle P|\mathcal{O}[A]|P\rangle\\
    =&-2M_N^2\alpha\frac{1}{4}\left[\frac{\langle(N-\bar{N})^2\rangle_{\mathds{P}}}{\bar{N}}\right]\left(1-\frac{\sigma_{\pi N}}{M_N}\right)
\end{aligned}
\end{equation}

When the gluonic operator is proportianal to the number difference say $\mathcal{O}(N_+,N_-)=\alpha \Delta N/V$,
a rerun of the preceding reasoning gives
\begin{equation}
\begin{aligned}
    \langle P|\mathcal{O}[A]|P\rangle=-2M_h^2\alpha\chi_t\frac{\partial \ln M_h}{\partial\Delta N}
\end{aligned}
\end{equation}
For a polarized nucleon consisting of quark-scalar-diquark~\cite{Zahed:2022wae}, $$M_N(\Delta N)=M_N-M_Ns_v\frac{\Delta N}{\bar{N}}$$
so that
\begin{equation}
\begin{aligned}
    \langle P|\mathcal{O}[A]|P\rangle=2M_N^2\alpha\frac{\chi_t}{\bar{N}}s_v
\end{aligned}
\end{equation}

This calculation can be extended to constituent quark model and $SU(6)$ quark-diquark model.

\section{Averaging the instanton interactions over the $SU(3)$ rotations}
\label{App:average}
One way to carry out the color averaging in the effective instanton interaction $\mathrm{det}_\pm$ is by determinantal reduction~\cite{creutz1978invariant}
\begin{equation}
    \int dU\prod_{i=1}^{N_c}U_{a_ib_i}=\frac{1}{N_c!}\epsilon_{a_1\cdots a_{N_c}}\epsilon_{b_1\cdots b_{N_c}}
\end{equation}
and
\begin{equation}
\begin{aligned}
U^\dagger_{ba}=&\frac{1}{(N_c-1)!}\\
    &\times\epsilon_{aa_1\cdots a_{N_c-1}}\epsilon_{bb_1\cdots b_{N_c-1}}U_{a_1b_1}\cdots U_{a_{N_c-1}b_{N_c-1}}    
\end{aligned}
\end{equation}
where $\epsilon_{a_1\cdots a_{N_c}}$ is the Levi-Civita tensor of rank-$N_c$ with $\epsilon_{12\cdots N_c}=1$

With these two identities, the color averagings of $(UU^\dagger)^p$ with small $p$, are

\begin{enumerate}
    \item $p=1$
      \begin{equation}
      \int dU U_{ab}U^{\dagger}_{cd}=\frac{1}{N_c}\delta_{ad}\delta_{cb}
      \end{equation}
    \item $p=2$
      \begin{equation}
      \begin{aligned}
      &\int dU U_{a_1b_1}U^{\dagger}_{c_1d_1}U_{a_2b_2}U^{\dagger}_{c_2d_2}
      =\\
      &\frac{1}{N_c^2-1}\left(\delta_{a_1d_1}\delta_{a_2d_2}\delta_{c_1b_1}\delta_{c_2b_2}+\delta_{a_1d_2}\delta_{a_2d_1}\delta_{c_1b_2}\delta_{c_2b_1}\right)\\
      &-\frac{1}{N_c(N_c^2-1)}\left(\delta_{a_1d_1}\delta_{a_2d_2}\delta_{c_1b_2}\delta_{c_2b_1}+\delta_{a_1d_2}\delta_{a_2d_1}\delta_{c_1b_1}\delta_{c_2b_2}\right)
      \end{aligned}
      \end{equation}
      
    \begin{widetext}
    \item $p=3$
      \begin{equation}
      \begin{aligned}
      &\int dU U_{a_1b_1}U^{\dagger}_{c_1d_1}U_{a_2b_2}U^{\dagger}_{c_2d_2}U_{a_3b_3}U^{\dagger}_{c_3d_3}\\
      =&\frac{N_c^2-2}{N_c(N^2_c-4)(N_c^2-1)}\\
      &\times(\delta_{a_1d_1}\delta_{a_2d_2}\delta_{a_3d_3}\delta_{c_1b_1}\delta_{c_2b_2}\delta_{c_3b_3}
      +\delta_{a_1d_2}\delta_{a_2d_1}\delta_{a_3d_3}\delta_{c_1b_2}\delta_{c_2b_1}\delta_{c_3b_3}\\
      &+\delta_{a_1d_3}\delta_{a_2d_2}\delta_{a_3d_1}\delta_{c_1b_3}\delta_{c_2b_2}\delta_{c_3b_1}
      +\delta_{a_1d_1}\delta_{a_3d_2}\delta_{a_2d_3}\delta_{c_1b_1}\delta_{c_3b_2}\delta_{c_2b_3}\\
      &+\delta_{a_1d_3}\delta_{a_3d_2}\delta_{a_2d_1}\delta_{c_1b_3}\delta_{c_3b_2}\delta_{c_2b_1}
      +\delta_{a_1d_2}\delta_{a_2d_3}\delta_{a_3d_1}\delta_{c_1b_2}\delta_{c_2b_3}\delta_{c_3b_1})\\
      &-\frac{1}{(N_c^2-4)(N_c^2-1)}\\
      &\times(\delta_{a_1d_1}\delta_{a_2d_2}\delta_{a_3d_3}\delta_{c_1b_2}\delta_{c_2b_1}\delta_{c_3b_3}
      +\delta_{a_1d_2}\delta_{a_2d_1}\delta_{a_3d_3}\delta_{c_1b_1}\delta_{c_2b_2}\delta_{c_3b_3}\\
      &+\delta_{a_1d_1}\delta_{a_2d_2}\delta_{a_3d_3}\delta_{c_1b_3}\delta_{c_2b_2}\delta_{c_3b_1}
      +\delta_{a_1d_3}\delta_{a_2d_2}\delta_{a_3d_1}\delta_{c_1b_1}\delta_{c_2b_2}\delta_{c_3b_3}\\
      &+\delta_{a_1d_1}\delta_{a_2d_2}\delta_{a_3d_3}\delta_{c_1b_1}\delta_{c_3b_2}\delta_{c_2b_3}
      +\delta_{a_1d_1}\delta_{a_3d_2}\delta_{a_2d_3}\delta_{c_1b_1}\delta_{c_2b_2}\delta_{c_3b_3}\\
      &+\delta_{a_1d_3}\delta_{a_3d_2}\delta_{a_2d_1}\delta_{c_1b_1}\delta_{c_3b_2}\delta_{c_2b_3}
      +\delta_{a_1d_3}\delta_{a_3d_2}\delta_{a_2d_1}\delta_{c_3b_1}\delta_{c_2b_2}\delta_{c_1b_3}\\
      &+\delta_{a_1d_3}\delta_{a_3d_2}\delta_{a_2d_1}\delta_{c_1b_2}\delta_{c_2b_1}\delta_{c_3b_3}
      +\delta_{a_1d_1}\delta_{a_3d_2}\delta_{a_2d_3}\delta_{c_1b_3}\delta_{c_3b_2}\delta_{c_2b_1}\\
      &+\delta_{a_1d_1}\delta_{a_3d_2}\delta_{a_2d_3}\delta_{c_1b_3}\delta_{c_3b_2}\delta_{c_2b_1}
      +\delta_{a_1d_1}\delta_{a_3d_2}\delta_{a_2d_3}\delta_{c_1b_3}\delta_{c_3b_2}\delta_{c_2b_1}\\
      &+\delta_{a_1d_2}\delta_{a_2d_3}\delta_{a_3d_1}\delta_{c_1b_1}\delta_{c_3b_2}\delta_{c_2b_3}
      +\delta_{a_1d_2}\delta_{a_2d_3}\delta_{a_3d_1}\delta_{c_3b_1}\delta_{c_2b_2}\delta_{c_1b_3}\\
      &+\delta_{a_1d_2}\delta_{a_2d_3}\delta_{a_3d_1}\delta_{c_1b_2}\delta_{c_2b_1}\delta_{c_3b_3}
      +\delta_{a_1d_1}\delta_{a_3d_2}\delta_{a_2d_3}\delta_{c_1b_2}\delta_{c_2b_3}\delta_{c_3b_1}\\
      &+\delta_{a_1d_1}\delta_{a_3d_2}\delta_{a_2d_3}\delta_{c_1b_2}\delta_{c_2b_3}\delta_{c_3b_1}
      +\delta_{a_1d_1}\delta_{a_3d_2}\delta_{a_2d_3}\delta_{c_1b_2}\delta_{c_2b_3}\delta_{c_3b_1})\\
      &+\frac{2}{N_c(N_c^2-4)(N_c^2-1)}\\
      &\times(\delta_{a_1d_2}\delta_{a_2d_3}\delta_{a_3d_1}\delta_{c_1b_1}\delta_{c_2b_2}\delta_{c_3b_3}
      +\delta_{a_1d_1}\delta_{a_2d_2}\delta_{a_3d_3}\delta_{c_1b_2}\delta_{c_2b_3}\delta_{c_3b_1}\\
      &+\delta_{a_1d_3}\delta_{a_3d_2}\delta_{a_2d_1}\delta_{c_1b_1}\delta_{c_2b_2}\delta_{c_3b_3}
      +\delta_{a_1d_1}\delta_{a_2d_2}\delta_{a_3d_3}\delta_{c_1b_3}\delta_{c_3b_2}\delta_{c_2b_1}\\
      &+\delta_{a_1d_2}\delta_{a_2d_3}\delta_{a_3d_1}\delta_{c_1b_3}\delta_{c_3b_2}\delta_{c_2b_1}
      +\delta_{a_1d_3}\delta_{a_3d_2}\delta_{a_2d_1}\delta_{c_1b_2}\delta_{c_2b_3}\delta_{c_3b_1}\\
      &+\delta_{a_1d_1}\delta_{a_3d_2}\delta_{a_2d_3}\delta_{c_3b_2}\delta_{c_2b_2}\delta_{c_2b_3}
      +\delta_{a_1d_1}\delta_{a_3d_2}\delta_{a_2d_3}\delta_{c_1b_2}\delta_{c_2b_1}\delta_{c_3b_3}\\
      &+\delta_{a_1d_3}\delta_{a_2d_2}\delta_{a_3d_1}\delta_{c_1b_2}\delta_{c_2b_1}\delta_{c_3b_3}
      +\delta_{a_1d_3}\delta_{a_2d_2}\delta_{a_3d_1}\delta_{c_1b_1}\delta_{c_3b_2}\delta_{c_2b_3}\\
      &+\delta_{a_1d_2}\delta_{a_2d_1}\delta_{a_3d_3}\delta_{c_1b_1}\delta_{c_3b_2}\delta_{c_2b_3}
      +\delta_{a_1d_2}\delta_{a_2d_1}\delta_{a_3d_3}\delta_{c_3b_2}\delta_{c_2b_2}\delta_{c_2b_3})
      \end{aligned}
      \end{equation}
    \end{widetext}
\end{enumerate}

However, For large values of $p$, this averaging method is tedious.  Since $N_c \otimes N_c = 1 \oplus (N_c^2-1)$, the group integral practically reduces to finding all projections of the product of adjoint representations onto the singlet for $SU(N_c)$. The result can be obtained by using the graphical color projection rules~\cite{Chernyshev:1995gj,nowak1989flavor,Miesch:2023hjt}, 
with the following results
\begin{enumerate}
    \item $p=2$
      \begin{equation}
      \begin{aligned}
      &\int dU U_{a_1b_1}U^{\dagger}_{c_1d_1}U_{a_2b_2}U^{\dagger}_{c_2d_2}=\frac{1}{N_c^2}\delta_{a_1d_1}\delta_{a_2d_2}\delta_{c_1b_1}\delta_{c_2b_2}\\
      &+\frac{1}{4(N_c^2-1)}\lambda^\alpha_{a_1d_1}\lambda^\alpha_{a_2d_2}\lambda^\beta_{c_1b_1}\lambda^\beta_{c_2b_2}
      \end{aligned}
      \end{equation}
\begin{widetext}
    \item $p=3$
      \begin{equation}
      \begin{aligned}
      &\int dU U_{a_1b_1}U^{\dagger}_{c_1d_1}U_{a_2b_2}U^{\dagger}_{c_2d_2}U_{a_3b_3}U^{\dagger}_{c_3d_3}=\frac{1}{N_c^3}\delta_{a_1d_1}\delta_{a_2d_2}\delta_{a_3d_3}\delta_{c_1b_1}\delta_{c_2b_2}\delta_{c_3b_3}\\
      &+\frac{1}{4N_c(N_c^2-1)}\left(\lambda^\alpha_{a_1d_1}\lambda^\alpha_{a_2d_2}\delta_{a_3d_3}\lambda^\beta_{c_1b_2}\lambda^\beta_{c_2b_1}\delta_{c_3b_3}+\delta_{a_1d_1}\lambda^\alpha_{a_2d_2}\lambda^\alpha_{a_3d_3}\delta_{c_1b_1}\lambda^\beta_{c_2b_2}\lambda^\beta_{c_3b_3}+\lambda^\alpha_{a_1d_1}\delta_{a_2d_2}\lambda^\alpha_{a_3d_3}\lambda^\beta_{c_1b_1}\delta_{c_2b_2}\lambda^\beta_{c_3b_3}\right)\\
       &+\frac{1}{4(N_c^2-1)}\left(\frac{N_c}{2(N_c^2-4)}d^{\alpha\beta\gamma}d^{\alpha'\beta'\gamma'}\lambda^\alpha_{a_1d_1}\lambda^\beta_{a_2d_2}\lambda^\gamma_{a_3d_3}\lambda^{\alpha'}_{c_1b_2}\lambda^{\beta'}_{c_2b_1}\lambda^{\gamma'}_{c_3b_3}+\frac{1}{2N_c}f^{\alpha\beta\gamma}f^{\alpha'\beta'\gamma'}\lambda^\alpha_{a_1d_1}\lambda^\beta_{a_2d_2}\lambda^\gamma_{a_3d_3}\lambda^{\alpha'}_{c_1b_2}\lambda^{\beta'}_{c_2b_1}\lambda^{\gamma'}_{c_3b_3}\right)
      \end{aligned}
      \end{equation}
    \item $p=4$
      \begin{equation}
      \begin{aligned}
      &\int dU U_{a_1b_1}U^{\dagger}_{c_1d_1}U_{a_2b_2}U^{\dagger}_{c_2d_2}U_{a_3b_3}U^{\dagger}_{c_3d_3}U_{a_4b_4}U^{\dagger}_{c_4d_4}=\frac{1}{N_c^4}\delta_{a_1d_1}\delta_{a_2d_2}\delta_{a_3d_3}\delta_{a_4d_4}\delta_{c_1b_1}\delta_{c_2b_2}\delta_{c_3b_3}\delta_{c_4b_4}\\
      &+\left[\frac{1}{N_c}\delta_{a_4d_4}\delta_{c_4b_4}\left(\int dU U_{a_1b_1}U^{\dagger}_{c_1d_1}U_{a_2b_2}U^{\dagger}_{c_2d_2}U_{a_3b_3}U^{\dagger}_{c_3d_3}-\frac{1}{N_c^3}\delta_{a_1d_1}\delta_{a_2d_2}\delta_{a_3d_3}\delta_{c_1b_1}\delta_{c_2b_2}\delta_{c_3b_3}\right)+\mathrm{permutations}\right]\\
      &+\lambda^\alpha_{a_1d_1}\lambda^\beta_{a_2d_2}\lambda^\gamma_{a_3d_3}\lambda^\delta_{a_4d_4}\lambda^{\alpha'}_{c_1b_1}\lambda^{\beta'}_{c_2b_2}\lambda^{\gamma'}_{c_3b_3}\lambda^{\delta'}_{c_4b_4}\bigg[\frac{1}{16(N_c^2-1)^2}\left(\delta^{\alpha\beta}\delta^{\gamma\delta}\delta^{\alpha'\beta'}\delta^{\gamma'\delta'}+\delta^{\alpha\gamma}\delta^{\beta\delta}\delta^{\alpha'\gamma'}\delta^{\beta'\delta'}+\delta^{\alpha\delta}\delta^{\beta\gamma}\delta^{\alpha'\delta'}\delta^{\beta'\gamma'}\right)\\
       &+\frac{1}{4(N_c^2-1)}\bigg(\frac{N_c^2}{4(N_c^2-4)^2}d^{\alpha\beta\epsilon}d^{\gamma\delta\epsilon}d^{\alpha'\beta'\rho}d^{\gamma'\delta'\rho}+\frac{1}{4(N_c^2-4)}d^{\alpha\beta\epsilon}f^{\gamma\delta\epsilon}d^{\alpha'\beta'\epsilon'}f^{\gamma'\delta'\epsilon'}+\frac{1}{4(N_c^2-4)}f^{\alpha\beta\epsilon}d^{\gamma\delta\epsilon}f^{\alpha'\beta'\epsilon'}d^{\gamma'\delta'\epsilon'}\\
       &+\frac{1}{4N_c^2}f^{\alpha\beta\epsilon}f^{\gamma\delta\epsilon}f^{\alpha'\beta'\epsilon'}f^{\gamma'\delta'\epsilon}\bigg)\bigg]
      \end{aligned}
      \end{equation}
\end{widetext}
\end{enumerate}

\bibliography{ref}

\begin{thebibliography}{119}%
\makeatletter
\providecommand \@ifxundefined [1]{%
 \@ifx{#1\undefined}
}%
\providecommand \@ifnum [1]{%
 \ifnum #1\expandafter \@firstoftwo
 \else \expandafter \@secondoftwo
 \fi
}%
\providecommand \@ifx [1]{%
 \ifx #1\expandafter \@firstoftwo
 \else \expandafter \@secondoftwo
 \fi
}%
\providecommand \natexlab [1]{#1}%
\providecommand \enquote  [1]{``#1''}%
\providecommand \bibnamefont  [1]{#1}%
\providecommand \bibfnamefont [1]{#1}%
\providecommand \citenamefont [1]{#1}%
\providecommand \href@noop [0]{\@secondoftwo}%
\providecommand \href [0]{\begingroup \@sanitize@url \@href}%
\providecommand \@href[1]{\@@startlink{#1}\@@href}%
\providecommand \@@href[1]{\endgroup#1\@@endlink}%
\providecommand \@sanitize@url [0]{\catcode `\\12\catcode `\$12\catcode
  `\&12\catcode `\#12\catcode `\^12\catcode `\_12\catcode `\%12\relax}%
\providecommand \@@startlink[1]{}%
\providecommand \@@endlink[0]{}%
\providecommand \url  [0]{\begingroup\@sanitize@url \@url }%
\providecommand \@url [1]{\endgroup\@href {#1}{\urlprefix }}%
\providecommand \urlprefix  [0]{URL }%
\providecommand \Eprint [0]{\href }%
\providecommand \doibase [0]{http://dx.doi.org/}%
\providecommand \selectlanguage [0]{\@gobble}%
\providecommand \bibinfo  [0]{\@secondoftwo}%
\providecommand \bibfield  [0]{\@secondoftwo}%
\providecommand \translation [1]{[#1]}%
\providecommand \BibitemOpen [0]{}%
\providecommand \bibitemStop [0]{}%
\providecommand \bibitemNoStop [0]{.\EOS\space}%
\providecommand \EOS [0]{\spacefactor3000\relax}%
\providecommand \BibitemShut  [1]{\csname bibitem#1\endcsname}%
\let\auto@bib@innerbib\@empty
\bibitem [{\citenamefont {Athenodorou}\ \emph {et~al.}(2018)\citenamefont
  {Athenodorou}, \citenamefont {Boucaud}, \citenamefont {De~Soto},
  \citenamefont {Rodr\'\i{}guez-Quintero},\ and\ \citenamefont
  {Zafeiropoulos}}]{Athenodorou:2018jwu}%
  \BibitemOpen
  \bibfield  {author} {\bibinfo {author} {\bibfnamefont {A.}~\bibnamefont
  {Athenodorou}}, \bibinfo {author} {\bibfnamefont {Ph.}\ \bibnamefont
  {Boucaud}}, \bibinfo {author} {\bibfnamefont {F.}~\bibnamefont {De~Soto}},
  \bibinfo {author} {\bibfnamefont {J.}~\bibnamefont
  {Rodr\'\i{}guez-Quintero}}, \ and\ \bibinfo {author} {\bibfnamefont
  {S.}~\bibnamefont {Zafeiropoulos}},\ }\bibfield  {title} {\enquote {\bibinfo
  {title} {{Instanton liquid properties from lattice QCD}},}\ }\href {\doibase
  10.1007/JHEP02(2018)140} {\bibfield  {journal} {\bibinfo  {journal} {JHEP}\
  }\textbf {\bibinfo {volume} {02}},\ \bibinfo {pages} {140} (\bibinfo {year}
  {2018})},\ \Eprint {http://arxiv.org/abs/1801.10155} {arXiv:1801.10155
  [hep-lat]} \BibitemShut {NoStop}%
\bibitem [{\citenamefont {Belavin}\ \emph {et~al.}(1975)\citenamefont
  {Belavin}, \citenamefont {Polyakov}, \citenamefont {Schwartz},\ and\
  \citenamefont {Tyupkin}}]{Belavin:1975fg}%
  \BibitemOpen
  \bibfield  {author} {\bibinfo {author} {\bibfnamefont {A.~A.}\ \bibnamefont
  {Belavin}}, \bibinfo {author} {\bibfnamefont {Alexander~M.}\ \bibnamefont
  {Polyakov}}, \bibinfo {author} {\bibfnamefont {A.~S.}\ \bibnamefont
  {Schwartz}}, \ and\ \bibinfo {author} {\bibfnamefont {Yu.~S.}\ \bibnamefont
  {Tyupkin}},\ }\bibfield  {title} {\enquote {\bibinfo {title} {{Pseudoparticle
  Solutions of the Yang-Mills Equations}},}\ }\href {\doibase
  10.1016/0370-2693(75)90163-X} {\bibfield  {journal} {\bibinfo  {journal}
  {Phys. Lett. B}\ }\textbf {\bibinfo {volume} {59}},\ \bibinfo {pages}
  {85--87} (\bibinfo {year} {1975})}\BibitemShut {NoStop}%
\bibitem [{\citenamefont {Callan}\ \emph {et~al.}(1976)\citenamefont {Callan},
  \citenamefont {Dashen},\ and\ \citenamefont {Gross}}]{Callan:1976je}%
  \BibitemOpen
  \bibfield  {author} {\bibinfo {author} {\bibfnamefont {Curtis~G.}\
  \bibnamefont {Callan}, \bibfnamefont {Jr.}}, \bibinfo {author} {\bibfnamefont
  {R.~F.}\ \bibnamefont {Dashen}}, \ and\ \bibinfo {author} {\bibfnamefont
  {David~J.}\ \bibnamefont {Gross}},\ }\bibfield  {title} {\enquote {\bibinfo
  {title} {{The Structure of the Gauge Theory Vacuum}},}\ }\href {\doibase
  10.1016/0370-2693(76)90277-X} {\bibfield  {journal} {\bibinfo  {journal}
  {Phys. Lett. B}\ }\textbf {\bibinfo {volume} {63}},\ \bibinfo {pages}
  {334--340} (\bibinfo {year} {1976})}\BibitemShut {NoStop}%
\bibitem [{\citenamefont {Callan}\ \emph {et~al.}(1978)\citenamefont {Callan},
  \citenamefont {Dashen},\ and\ \citenamefont {Gross}}]{Callan:1977gz}%
  \BibitemOpen
  \bibfield  {author} {\bibinfo {author} {\bibfnamefont {Curtis~G.}\
  \bibnamefont {Callan}, \bibfnamefont {Jr.}}, \bibinfo {author} {\bibfnamefont
  {Roger~F.}\ \bibnamefont {Dashen}}, \ and\ \bibinfo {author} {\bibfnamefont
  {David~J.}\ \bibnamefont {Gross}},\ }\bibfield  {title} {\enquote {\bibinfo
  {title} {{Toward a Theory of the Strong Interactions}},}\ }\href {\doibase
  10.1103/PhysRevD.17.2717} {\bibfield  {journal} {\bibinfo  {journal} {Phys.
  Rev. D}\ }\textbf {\bibinfo {volume} {17}},\ \bibinfo {pages} {2717}
  (\bibinfo {year} {1978})}\BibitemShut {NoStop}%
\bibitem [{\citenamefont {Shuryak}(1982)}]{Shuryak:1981ff}%
  \BibitemOpen
  \bibfield  {author} {\bibinfo {author} {\bibfnamefont {Edward~V.}\
  \bibnamefont {Shuryak}},\ }\bibfield  {title} {\enquote {\bibinfo {title}
  {{The Role of Instantons in Quantum Chromodynamics. 1. Physical Vacuum}},}\
  }\href {\doibase 10.1016/0550-3213(82)90478-3} {\bibfield  {journal}
  {\bibinfo  {journal} {Nucl. Phys. B}\ }\textbf {\bibinfo {volume} {203}},\
  \bibinfo {pages} {93} (\bibinfo {year} {1982})}\BibitemShut {NoStop}%
\bibitem [{\citenamefont {Sch\"afer}\ and\ \citenamefont
  {Shuryak}(1998)}]{Schafer:1996wv}%
  \BibitemOpen
  \bibfield  {author} {\bibinfo {author} {\bibfnamefont {Thomas}\ \bibnamefont
  {Sch\"afer}}\ and\ \bibinfo {author} {\bibfnamefont {Edward~V.}\ \bibnamefont
  {Shuryak}},\ }\bibfield  {title} {\enquote {\bibinfo {title} {{Instantons in
  QCD}},}\ }\href {\doibase 10.1103/RevModPhys.70.323} {\bibfield  {journal}
  {\bibinfo  {journal} {Rev. Mod. Phys.}\ }\textbf {\bibinfo {volume} {70}},\
  \bibinfo {pages} {323--426} (\bibinfo {year} {1998})},\ \Eprint
  {http://arxiv.org/abs/hep-ph/9610451} {arXiv:hep-ph/9610451} \BibitemShut
  {NoStop}%
\bibitem [{\citenamefont {Shuryak}\ and\ \citenamefont
  {Zahed}(2023)}]{Shuryak:2021fsu}%
  \BibitemOpen
  \bibfield  {author} {\bibinfo {author} {\bibfnamefont {Edward}\ \bibnamefont
  {Shuryak}}\ and\ \bibinfo {author} {\bibfnamefont {Ismail}\ \bibnamefont
  {Zahed}},\ }\bibfield  {title} {\enquote {\bibinfo {title} {{Hadronic
  structure on the light front. I. Instanton effects and quark-antiquark
  effective potentials}},}\ }\href {\doibase 10.1103/PhysRevD.107.034023}
  {\bibfield  {journal} {\bibinfo  {journal} {Phys. Rev. D}\ }\textbf {\bibinfo
  {volume} {107}},\ \bibinfo {pages} {034023} (\bibinfo {year} {2023})},\
  \Eprint {http://arxiv.org/abs/2110.15927} {arXiv:2110.15927 [hep-ph]}
  \BibitemShut {NoStop}%
\bibitem [{\citenamefont {Sch\"afer}\ and\ \citenamefont
  {Shuryak}(1995)}]{Schafer:1994fd}%
  \BibitemOpen
  \bibfield  {author} {\bibinfo {author} {\bibfnamefont {Thomas}\ \bibnamefont
  {Sch\"afer}}\ and\ \bibinfo {author} {\bibfnamefont {Edward~V.}\ \bibnamefont
  {Shuryak}},\ }\bibfield  {title} {\enquote {\bibinfo {title} {{Glueballs and
  instantons}},}\ }\href {\doibase 10.1103/PhysRevLett.75.1707} {\bibfield
  {journal} {\bibinfo  {journal} {Phys. Rev. Lett.}\ }\textbf {\bibinfo
  {volume} {75}},\ \bibinfo {pages} {1707--1710} (\bibinfo {year} {1995})},\
  \Eprint {http://arxiv.org/abs/hep-ph/9410372} {arXiv:hep-ph/9410372}
  \BibitemShut {NoStop}%
\bibitem [{\citenamefont {Kacir}\ \emph {et~al.}(1999)\citenamefont {Kacir},
  \citenamefont {Prakash},\ and\ \citenamefont {Zahed}}]{Kacir:1996qn}%
  \BibitemOpen
  \bibfield  {author} {\bibinfo {author} {\bibfnamefont {M.}~\bibnamefont
  {Kacir}}, \bibinfo {author} {\bibfnamefont {M.}~\bibnamefont {Prakash}}, \
  and\ \bibinfo {author} {\bibfnamefont {I.}~\bibnamefont {Zahed}},\ }\bibfield
   {title} {\enquote {\bibinfo {title} {{Hadrons and QCD instantons: A
  Bosonized view}},}\ }\href@noop {} {\bibfield  {journal} {\bibinfo  {journal}
  {Acta Phys. Polon. B}\ }\textbf {\bibinfo {volume} {30}},\ \bibinfo {pages}
  {287--348} (\bibinfo {year} {1999})},\ \Eprint
  {http://arxiv.org/abs/hep-ph/9602314} {arXiv:hep-ph/9602314} \BibitemShut
  {NoStop}%
\bibitem [{\citenamefont {Sch\"afer}\ and\ \citenamefont
  {Zetocha}(2004)}]{Schafer:2004ke}%
  \BibitemOpen
  \bibfield  {author} {\bibinfo {author} {\bibfnamefont {T.}~\bibnamefont
  {Sch\"afer}}\ and\ \bibinfo {author} {\bibfnamefont {V.}~\bibnamefont
  {Zetocha}},\ }\bibfield  {title} {\enquote {\bibinfo {title} {{Instantons and
  the spin of the nucleon}},}\ }\href {\doibase 10.1103/PhysRevD.69.094028}
  {\bibfield  {journal} {\bibinfo  {journal} {Phys. Rev. D}\ }\textbf {\bibinfo
  {volume} {69}},\ \bibinfo {pages} {094028} (\bibinfo {year} {2004})},\
  \Eprint {http://arxiv.org/abs/hep-ph/0401165} {arXiv:hep-ph/0401165}
  \BibitemShut {NoStop}%
\bibitem [{\citenamefont {Iatrakis}\ \emph {et~al.}(2015)\citenamefont
  {Iatrakis}, \citenamefont {Ramamurti},\ and\ \citenamefont
  {Shuryak}}]{Iatrakis:2015rga}%
  \BibitemOpen
  \bibfield  {author} {\bibinfo {author} {\bibfnamefont {Ioannis}\ \bibnamefont
  {Iatrakis}}, \bibinfo {author} {\bibfnamefont {Adith}\ \bibnamefont
  {Ramamurti}}, \ and\ \bibinfo {author} {\bibfnamefont {Edward}\ \bibnamefont
  {Shuryak}},\ }\bibfield  {title} {\enquote {\bibinfo {title} {{Collective
  String Interactions in AdS/QCD and High-Multiplicity pA Collisions}},}\
  }\href {\doibase 10.1103/PhysRevD.92.014011} {\bibfield  {journal} {\bibinfo
  {journal} {Phys. Rev. D}\ }\textbf {\bibinfo {volume} {92}},\ \bibinfo
  {pages} {014011} (\bibinfo {year} {2015})},\ \Eprint
  {http://arxiv.org/abs/1503.04759} {arXiv:1503.04759 [hep-ph]} \BibitemShut
  {NoStop}%
\bibitem [{\citenamefont {Shuryak}(2018)}]{Shuryak:2018fjr}%
  \BibitemOpen
  \bibfield  {author} {\bibinfo {author} {\bibfnamefont {Edward}\ \bibnamefont
  {Shuryak}},\ }\bibfield  {title} {\enquote {\bibinfo {title} {{Lectures on
  nonperturbative QCD ( Nonperturbative Topological Phenomena in QCD and
  Related Theories)}},}\ }\href@noop {} {\  (\bibinfo {year} {2018})},\ \Eprint
  {http://arxiv.org/abs/1812.01509} {arXiv:1812.01509 [hep-ph]} \BibitemShut
  {NoStop}%
\bibitem [{\citenamefont {Duran}\ \emph {et~al.}(2023)\citenamefont {Duran}
  \emph {et~al.}}]{Duran:2022xag}%
  \BibitemOpen
  \bibfield  {author} {\bibinfo {author} {\bibfnamefont {B.}~\bibnamefont
  {Duran}} \emph {et~al.},\ }\bibfield  {title} {\enquote {\bibinfo {title}
  {{Determining the gluonic gravitational form factors of the proton}},}\
  }\href {\doibase 10.1038/s41586-023-05730-4} {\bibfield  {journal} {\bibinfo
  {journal} {Nature}\ }\textbf {\bibinfo {volume} {615}},\ \bibinfo {pages}
  {813--816} (\bibinfo {year} {2023})},\ \Eprint
  {http://arxiv.org/abs/2207.05212} {arXiv:2207.05212 [nucl-ex]} \BibitemShut
  {NoStop}%
\bibitem [{\citenamefont {Meziani}(2024)}]{Meziani:2024cke}%
  \BibitemOpen
  \bibfield  {author} {\bibinfo {author} {\bibfnamefont {Zein-Eddine}\
  \bibnamefont {Meziani}},\ }\bibfield  {title} {\enquote {\bibinfo {title}
  {{Gluonic gravitational form factors of the proton}},}\ }in\ \href@noop {}
  {\emph {\bibinfo {booktitle} {{25th International Spin Symposium}}}}\
  (\bibinfo {year} {2024})\ \Eprint {http://arxiv.org/abs/2403.08423}
  {arXiv:2403.08423 [nucl-ex]} \BibitemShut {NoStop}%
\bibitem [{\citenamefont {Diakonov}\ \emph {et~al.}(1996)\citenamefont
  {Diakonov}, \citenamefont {Polyakov},\ and\ \citenamefont
  {Weiss}}]{Diakonov:1995qy}%
  \BibitemOpen
  \bibfield  {author} {\bibinfo {author} {\bibfnamefont {Dmitri}\ \bibnamefont
  {Diakonov}}, \bibinfo {author} {\bibfnamefont {Maxim~V.}\ \bibnamefont
  {Polyakov}}, \ and\ \bibinfo {author} {\bibfnamefont {C.}~\bibnamefont
  {Weiss}},\ }\bibfield  {title} {\enquote {\bibinfo {title} {{Hadronic matrix
  elements of gluon operators in the instanton vacuum}},}\ }\href {\doibase
  10.1016/0550-3213(95)00675-3} {\bibfield  {journal} {\bibinfo  {journal}
  {Nucl. Phys. B}\ }\textbf {\bibinfo {volume} {461}},\ \bibinfo {pages}
  {539--580} (\bibinfo {year} {1996})},\ \Eprint
  {http://arxiv.org/abs/hep-ph/9510232} {arXiv:hep-ph/9510232} \BibitemShut
  {NoStop}%
\bibitem [{\citenamefont {Brodsky}\ and\ \citenamefont
  {Farrar}(1973)}]{Brodsky:1973kr}%
  \BibitemOpen
  \bibfield  {author} {\bibinfo {author} {\bibfnamefont {Stanley~J.}\
  \bibnamefont {Brodsky}}\ and\ \bibinfo {author} {\bibfnamefont {Glennys~R.}\
  \bibnamefont {Farrar}},\ }\bibfield  {title} {\enquote {\bibinfo {title}
  {{Scaling Laws at Large Transverse Momentum}},}\ }\href {\doibase
  10.1103/PhysRevLett.31.1153} {\bibfield  {journal} {\bibinfo  {journal}
  {Phys. Rev. Lett.}\ }\textbf {\bibinfo {volume} {31}},\ \bibinfo {pages}
  {1153--1156} (\bibinfo {year} {1973})}\BibitemShut {NoStop}%
\bibitem [{\citenamefont {Radyushkin}(1977)}]{Radyushkin:1977gp}%
  \BibitemOpen
  \bibfield  {author} {\bibinfo {author} {\bibfnamefont {A.~V.}\ \bibnamefont
  {Radyushkin}},\ }\bibfield  {title} {\enquote {\bibinfo {title} {{Deep
  Elastic Processes of Composite Particles in Field Theory and Asymptotic
  Freedom}},}\ }\href@noop {} {\  (\bibinfo {year} {1977})},\ \Eprint
  {http://arxiv.org/abs/hep-ph/0410276} {arXiv:hep-ph/0410276} \BibitemShut
  {NoStop}%
\bibitem [{\citenamefont {Huber}\ \emph {et~al.}(2008)\citenamefont {Huber}
  \emph {et~al.}}]{JeffersonLab:2008jve}%
  \BibitemOpen
  \bibfield  {author} {\bibinfo {author} {\bibfnamefont {G.~M.}\ \bibnamefont
  {Huber}} \emph {et~al.} (\bibinfo {collaboration} {Jefferson Lab}),\
  }\bibfield  {title} {\enquote {\bibinfo {title} {{Charged pion form-factor
  between Q**2 = 0.60-GeV**2 and 2.45-GeV**2. II. Determination of, and results
  for, the pion form-factor}},}\ }\href {\doibase 10.1103/PhysRevC.78.045203}
  {\bibfield  {journal} {\bibinfo  {journal} {Phys. Rev. C}\ }\textbf {\bibinfo
  {volume} {78}},\ \bibinfo {pages} {045203} (\bibinfo {year} {2008})},\
  \Eprint {http://arxiv.org/abs/0809.3052} {arXiv:0809.3052 [nucl-ex]}
  \BibitemShut {NoStop}%
\bibitem [{\citenamefont {Br\"ommel}\ \emph {et~al.}(2007)\citenamefont
  {Br\"ommel} \emph {et~al.}}]{QCDSFUKQCD:2006gmg}%
  \BibitemOpen
  \bibfield  {author} {\bibinfo {author} {\bibfnamefont {D.}~\bibnamefont
  {Br\"ommel}} \emph {et~al.} (\bibinfo {collaboration} {QCDSF/UKQCD}),\
  }\bibfield  {title} {\enquote {\bibinfo {title} {{The Pion form-factor from
  lattice QCD with two dynamical flavours}},}\ }\href {\doibase
  10.1140/epjc/s10052-007-0295-6} {\bibfield  {journal} {\bibinfo  {journal}
  {Eur. Phys. J. C}\ }\textbf {\bibinfo {volume} {51}},\ \bibinfo {pages}
  {335--345} (\bibinfo {year} {2007})},\ \Eprint
  {http://arxiv.org/abs/hep-lat/0608021} {arXiv:hep-lat/0608021} \BibitemShut
  {NoStop}%
\bibitem [{\citenamefont {Wang}\ \emph {et~al.}(2024)\citenamefont {Wang},
  \citenamefont {He}, \citenamefont {Wang}, \citenamefont {Draper},
  \citenamefont {Liang}, \citenamefont {Liu},\ and\ \citenamefont
  {Yang}}]{Wang:2024lrm}%
  \BibitemOpen
  \bibfield  {author} {\bibinfo {author} {\bibfnamefont {Bigeng}\ \bibnamefont
  {Wang}}, \bibinfo {author} {\bibfnamefont {Fangcheng}\ \bibnamefont {He}},
  \bibinfo {author} {\bibfnamefont {Gen}\ \bibnamefont {Wang}}, \bibinfo
  {author} {\bibfnamefont {Terrence}\ \bibnamefont {Draper}}, \bibinfo {author}
  {\bibfnamefont {Jian}\ \bibnamefont {Liang}}, \bibinfo {author}
  {\bibfnamefont {Keh-Fei}\ \bibnamefont {Liu}}, \ and\ \bibinfo {author}
  {\bibfnamefont {Yi-Bo}\ \bibnamefont {Yang}} (\bibinfo {collaboration}
  {\ensuremath{\chi}QCD}),\ }\bibfield  {title} {\enquote {\bibinfo {title}
  {{Trace anomaly form factors from lattice QCD}},}\ }\href@noop {} {\
  (\bibinfo {year} {2024})},\ \Eprint {http://arxiv.org/abs/2401.05496}
  {arXiv:2401.05496 [hep-lat]} \BibitemShut {NoStop}%
\bibitem [{\citenamefont {Hackett}\ \emph {et~al.}(2023)\citenamefont
  {Hackett}, \citenamefont {Pefkou},\ and\ \citenamefont
  {Shanahan}}]{Hackett:2023rif}%
  \BibitemOpen
  \bibfield  {author} {\bibinfo {author} {\bibfnamefont {Daniel~C.}\
  \bibnamefont {Hackett}}, \bibinfo {author} {\bibfnamefont {Dimitra~A.}\
  \bibnamefont {Pefkou}}, \ and\ \bibinfo {author} {\bibfnamefont {Phiala~E.}\
  \bibnamefont {Shanahan}},\ }\bibfield  {title} {\enquote {\bibinfo {title}
  {{Gravitational form factors of the proton from lattice QCD}},}\ }\href@noop
  {} {\  (\bibinfo {year} {2023})},\ \Eprint {http://arxiv.org/abs/2310.08484}
  {arXiv:2310.08484 [hep-lat]} \BibitemShut {NoStop}%
\bibitem [{\citenamefont {Chen}\ \emph {et~al.}(2006)\citenamefont {Chen} \emph
  {et~al.}}]{Chen:2005mg}%
  \BibitemOpen
  \bibfield  {author} {\bibinfo {author} {\bibfnamefont {Y.}~\bibnamefont
  {Chen}} \emph {et~al.},\ }\bibfield  {title} {\enquote {\bibinfo {title}
  {{Glueball spectrum and matrix elements on anisotropic lattices}},}\ }\href
  {\doibase 10.1103/PhysRevD.73.014516} {\bibfield  {journal} {\bibinfo
  {journal} {Phys. Rev. D}\ }\textbf {\bibinfo {volume} {73}},\ \bibinfo
  {pages} {014516} (\bibinfo {year} {2006})},\ \Eprint
  {http://arxiv.org/abs/hep-lat/0510074} {arXiv:hep-lat/0510074} \BibitemShut
  {NoStop}%
\bibitem [{\citenamefont {Sun}\ \emph {et~al.}(2018)\citenamefont {Sun},
  \citenamefont {Gui}, \citenamefont {Chen}, \citenamefont {Gong},
  \citenamefont {Liu}, \citenamefont {Liu}, \citenamefont {Liu}, \citenamefont
  {Ma},\ and\ \citenamefont {Zhang}}]{Sun:2017ipk}%
  \BibitemOpen
  \bibfield  {author} {\bibinfo {author} {\bibfnamefont {Wei}\ \bibnamefont
  {Sun}}, \bibinfo {author} {\bibfnamefont {Long-Cheng}\ \bibnamefont {Gui}},
  \bibinfo {author} {\bibfnamefont {Ying}\ \bibnamefont {Chen}}, \bibinfo
  {author} {\bibfnamefont {Ming}\ \bibnamefont {Gong}}, \bibinfo {author}
  {\bibfnamefont {Chuan}\ \bibnamefont {Liu}}, \bibinfo {author} {\bibfnamefont
  {Yu-Bin}\ \bibnamefont {Liu}}, \bibinfo {author} {\bibfnamefont {Zhaofeng}\
  \bibnamefont {Liu}}, \bibinfo {author} {\bibfnamefont {Jian-Ping}\
  \bibnamefont {Ma}}, \ and\ \bibinfo {author} {\bibfnamefont {Jian-Bo}\
  \bibnamefont {Zhang}},\ }\bibfield  {title} {\enquote {\bibinfo {title}
  {{Glueball spectrum from $N_f=2$ lattice QCD study on anisotropic
  lattices}},}\ }\href {\doibase 10.1088/1674-1137/42/9/093103} {\bibfield
  {journal} {\bibinfo  {journal} {Chin. Phys. C}\ }\textbf {\bibinfo {volume}
  {42}},\ \bibinfo {pages} {093103} (\bibinfo {year} {2018})},\ \Eprint
  {http://arxiv.org/abs/1702.08174} {arXiv:1702.08174 [hep-lat]} \BibitemShut
  {NoStop}%
\bibitem [{\citenamefont {Yang}\ \emph {et~al.}(2018)\citenamefont {Yang},
  \citenamefont {Liang}, \citenamefont {Bi}, \citenamefont {Chen},
  \citenamefont {Draper}, \citenamefont {Liu},\ and\ \citenamefont
  {Liu}}]{Yang:2018nqn}%
  \BibitemOpen
  \bibfield  {author} {\bibinfo {author} {\bibfnamefont {Yi-Bo}\ \bibnamefont
  {Yang}}, \bibinfo {author} {\bibfnamefont {Jian}\ \bibnamefont {Liang}},
  \bibinfo {author} {\bibfnamefont {Yu-Jiang}\ \bibnamefont {Bi}}, \bibinfo
  {author} {\bibfnamefont {Ying}\ \bibnamefont {Chen}}, \bibinfo {author}
  {\bibfnamefont {Terrence}\ \bibnamefont {Draper}}, \bibinfo {author}
  {\bibfnamefont {Keh-Fei}\ \bibnamefont {Liu}}, \ and\ \bibinfo {author}
  {\bibfnamefont {Zhaofeng}\ \bibnamefont {Liu}},\ }\bibfield  {title}
  {\enquote {\bibinfo {title} {{Proton Mass Decomposition from the QCD Energy
  Momentum Tensor}},}\ }\href {\doibase 10.1103/PhysRevLett.121.212001}
  {\bibfield  {journal} {\bibinfo  {journal} {Phys. Rev. Lett.}\ }\textbf
  {\bibinfo {volume} {121}},\ \bibinfo {pages} {212001} (\bibinfo {year}
  {2018})},\ \Eprint {http://arxiv.org/abs/1808.08677} {arXiv:1808.08677
  [hep-lat]} \BibitemShut {NoStop}%
\bibitem [{\citenamefont {Hoferichter}\ \emph {et~al.}(2016)\citenamefont
  {Hoferichter}, \citenamefont {Ruiz~de Elvira}, \citenamefont {Kubis},\ and\
  \citenamefont {Mei\ss{}ner}}]{Hoferichter:2016ocj}%
  \BibitemOpen
  \bibfield  {author} {\bibinfo {author} {\bibfnamefont {Martin}\ \bibnamefont
  {Hoferichter}}, \bibinfo {author} {\bibfnamefont {Jacobo}\ \bibnamefont
  {Ruiz~de Elvira}}, \bibinfo {author} {\bibfnamefont {Bastian}\ \bibnamefont
  {Kubis}}, \ and\ \bibinfo {author} {\bibfnamefont {Ulf-G.}\ \bibnamefont
  {Mei\ss{}ner}},\ }\bibfield  {title} {\enquote {\bibinfo {title} {{Remarks on
  the pion\textendash{}nucleon \ensuremath{\sigma}-term}},}\ }\href {\doibase
  10.1016/j.physletb.2016.06.038} {\bibfield  {journal} {\bibinfo  {journal}
  {Phys. Lett. B}\ }\textbf {\bibinfo {volume} {760}},\ \bibinfo {pages}
  {74--78} (\bibinfo {year} {2016})},\ \Eprint
  {http://arxiv.org/abs/1602.07688} {arXiv:1602.07688 [hep-lat]} \BibitemShut
  {NoStop}%
\bibitem [{\citenamefont {Alarc\'on}(2021)}]{Alarcon:2021dlz}%
  \BibitemOpen
  \bibfield  {author} {\bibinfo {author} {\bibfnamefont {J.~M.}\ \bibnamefont
  {Alarc\'on}},\ }\bibfield  {title} {\enquote {\bibinfo {title} {{Brief
  history of the pion\textendash{}nucleon sigma term}},}\ }\href {\doibase
  10.1140/epjs/s11734-021-00145-6} {\bibfield  {journal} {\bibinfo  {journal}
  {Eur. Phys. J. ST}\ }\textbf {\bibinfo {volume} {230}},\ \bibinfo {pages}
  {1609--1622} (\bibinfo {year} {2021})},\ \Eprint
  {http://arxiv.org/abs/2205.01108} {arXiv:2205.01108 [hep-ph]} \BibitemShut
  {NoStop}%
\bibitem [{\citenamefont {Mamo}\ and\ \citenamefont
  {Zahed}(2022)}]{Mamo:2022eui}%
  \BibitemOpen
  \bibfield  {author} {\bibinfo {author} {\bibfnamefont {Kiminad~A.}\
  \bibnamefont {Mamo}}\ and\ \bibinfo {author} {\bibfnamefont {Ismail}\
  \bibnamefont {Zahed}},\ }\bibfield  {title} {\enquote {\bibinfo {title}
  {{J/\ensuremath{\psi} near threshold in holographic QCD: A and D
  gravitational form factors}},}\ }\href {\doibase 10.1103/PhysRevD.106.086004}
  {\bibfield  {journal} {\bibinfo  {journal} {Phys. Rev. D}\ }\textbf {\bibinfo
  {volume} {106}},\ \bibinfo {pages} {086004} (\bibinfo {year} {2022})},\
  \Eprint {http://arxiv.org/abs/2204.08857} {arXiv:2204.08857 [hep-ph]}
  \BibitemShut {NoStop}%
\bibitem [{\citenamefont {Wang}\ \emph {et~al.}(2022)\citenamefont {Wang},
  \citenamefont {Yang}, \citenamefont {Liang}, \citenamefont {Draper},\ and\
  \citenamefont {Liu}}]{Wang:2021vqy}%
  \BibitemOpen
  \bibfield  {author} {\bibinfo {author} {\bibfnamefont {Gen}\ \bibnamefont
  {Wang}}, \bibinfo {author} {\bibfnamefont {Yi-Bo}\ \bibnamefont {Yang}},
  \bibinfo {author} {\bibfnamefont {Jian}\ \bibnamefont {Liang}}, \bibinfo
  {author} {\bibfnamefont {Terrence}\ \bibnamefont {Draper}}, \ and\ \bibinfo
  {author} {\bibfnamefont {Keh-Fei}\ \bibnamefont {Liu}} (\bibinfo
  {collaboration} {\ensuremath{\chi}QCD}),\ }\bibfield  {title} {\enquote
  {\bibinfo {title} {{Proton momentum and angular momentum decompositions with
  overlap fermions}},}\ }\href {\doibase 10.1103/PhysRevD.106.014512}
  {\bibfield  {journal} {\bibinfo  {journal} {Phys. Rev. D}\ }\textbf {\bibinfo
  {volume} {106}},\ \bibinfo {pages} {014512} (\bibinfo {year} {2022})},\
  \Eprint {http://arxiv.org/abs/2111.09329} {arXiv:2111.09329 [hep-lat]}
  \BibitemShut {NoStop}%
\bibitem [{\citenamefont {Alexandrou}\ \emph {et~al.}(2020)\citenamefont
  {Alexandrou}, \citenamefont {Bacchio}, \citenamefont {Constantinou},
  \citenamefont {Finkenrath}, \citenamefont {Hadjiyiannakou}, \citenamefont
  {Jansen}, \citenamefont {Koutsou}, \citenamefont {Panagopoulos},\ and\
  \citenamefont {Spanoudes}}]{Alexandrou:2020sml}%
  \BibitemOpen
  \bibfield  {author} {\bibinfo {author} {\bibfnamefont {C.}~\bibnamefont
  {Alexandrou}}, \bibinfo {author} {\bibfnamefont {S.}~\bibnamefont {Bacchio}},
  \bibinfo {author} {\bibfnamefont {M.}~\bibnamefont {Constantinou}}, \bibinfo
  {author} {\bibfnamefont {J.}~\bibnamefont {Finkenrath}}, \bibinfo {author}
  {\bibfnamefont {K.}~\bibnamefont {Hadjiyiannakou}}, \bibinfo {author}
  {\bibfnamefont {K.}~\bibnamefont {Jansen}}, \bibinfo {author} {\bibfnamefont
  {G.}~\bibnamefont {Koutsou}}, \bibinfo {author} {\bibfnamefont
  {H.}~\bibnamefont {Panagopoulos}}, \ and\ \bibinfo {author} {\bibfnamefont
  {G.}~\bibnamefont {Spanoudes}},\ }\bibfield  {title} {\enquote {\bibinfo
  {title} {{Complete flavor decomposition of the spin and momentum fraction of
  the proton using lattice QCD simulations at physical pion mass}},}\ }\href
  {\doibase 10.1103/PhysRevD.101.094513} {\bibfield  {journal} {\bibinfo
  {journal} {Phys. Rev. D}\ }\textbf {\bibinfo {volume} {101}},\ \bibinfo
  {pages} {094513} (\bibinfo {year} {2020})},\ \Eprint
  {http://arxiv.org/abs/2003.08486} {arXiv:2003.08486 [hep-lat]} \BibitemShut
  {NoStop}%
\bibitem [{\citenamefont {Diakonov}(1996)}]{Diakonov:1995ea}%
  \BibitemOpen
  \bibfield  {author} {\bibinfo {author} {\bibfnamefont {Dmitri}\ \bibnamefont
  {Diakonov}},\ }\bibfield  {title} {\enquote {\bibinfo {title} {{Chiral
  symmetry breaking by instantons}},}\ }\href {\doibase
  10.3254/978-1-61499-215-8-397} {\bibfield  {journal} {\bibinfo  {journal}
  {Proc. Int. Sch. Phys. Fermi}\ }\textbf {\bibinfo {volume} {130}},\ \bibinfo
  {pages} {397--432} (\bibinfo {year} {1996})},\ \Eprint
  {http://arxiv.org/abs/hep-ph/9602375} {arXiv:hep-ph/9602375} \BibitemShut
  {NoStop}%
\bibitem [{\citenamefont {Nowak}\ \emph
  {et~al.}(1996{\natexlab{a}})\citenamefont {Nowak}, \citenamefont {Rho},\ and\
  \citenamefont {Zahed}}]{Nowak:1996aj}%
  \BibitemOpen
  \bibfield  {author} {\bibinfo {author} {\bibfnamefont {Maciej~A.}\
  \bibnamefont {Nowak}}, \bibinfo {author} {\bibfnamefont {Mannque}\
  \bibnamefont {Rho}}, \ and\ \bibinfo {author} {\bibfnamefont
  {I.}~\bibnamefont {Zahed}},\ }\href@noop {} {\emph {\bibinfo {title} {{Chiral
  nuclear dynamics}}}}\ (\bibinfo {year} {1996})\BibitemShut {NoStop}%
\bibitem [{\citenamefont {Michael}\ and\ \citenamefont
  {Spencer}(1995{\natexlab{a}})}]{Michael:1994uu}%
  \BibitemOpen
  \bibfield  {author} {\bibinfo {author} {\bibfnamefont {Christopher}\
  \bibnamefont {Michael}}\ and\ \bibinfo {author} {\bibfnamefont {P.~S.}\
  \bibnamefont {Spencer}},\ }\bibfield  {title} {\enquote {\bibinfo {title}
  {{Instanton size distributions from calibrated cooling}},}\ }\href {\doibase
  10.1016/0920-5632(95)00220-4} {\bibfield  {journal} {\bibinfo  {journal}
  {Nucl. Phys. B Proc. Suppl.}\ }\textbf {\bibinfo {volume} {42}},\ \bibinfo
  {pages} {261--263} (\bibinfo {year} {1995}{\natexlab{a}})},\ \Eprint
  {http://arxiv.org/abs/hep-lat/9411015} {arXiv:hep-lat/9411015} \BibitemShut
  {NoStop}%
\bibitem [{\citenamefont {Michael}\ and\ \citenamefont
  {Spencer}(1995{\natexlab{b}})}]{Michael:1995br}%
  \BibitemOpen
  \bibfield  {author} {\bibinfo {author} {\bibfnamefont {Christopher}\
  \bibnamefont {Michael}}\ and\ \bibinfo {author} {\bibfnamefont {P.~S.}\
  \bibnamefont {Spencer}},\ }\bibfield  {title} {\enquote {\bibinfo {title}
  {{Cooling and the SU(2) instanton vacuum}},}\ }\href {\doibase
  10.1103/PhysRevD.52.4691} {\bibfield  {journal} {\bibinfo  {journal} {Phys.
  Rev. D}\ }\textbf {\bibinfo {volume} {52}},\ \bibinfo {pages} {4691--4699}
  (\bibinfo {year} {1995}{\natexlab{b}})},\ \Eprint
  {http://arxiv.org/abs/hep-lat/9503018} {arXiv:hep-lat/9503018} \BibitemShut
  {NoStop}%
\bibitem [{\citenamefont {Leinweber}(1999)}]{Leinweber:1999cw}%
  \BibitemOpen
  \bibfield  {author} {\bibinfo {author} {\bibfnamefont {Derek~B.}\
  \bibnamefont {Leinweber}},\ }\bibfield  {title} {\enquote {\bibinfo {title}
  {{Visualizations of the QCD vacuum}},}\ }in\ \href@noop {} {\emph {\bibinfo
  {booktitle} {{Workshop on Light-Cone QCD and Nonperturbative Hadron
  Physics}}}}\ (\bibinfo {year} {1999})\ pp.\ \bibinfo {pages} {138--143},\
  \Eprint {http://arxiv.org/abs/hep-lat/0004025} {arXiv:hep-lat/0004025}
  \BibitemShut {NoStop}%
\bibitem [{\citenamefont {Chu}\ \emph {et~al.}(1993)\citenamefont {Chu},
  \citenamefont {Grandy}, \citenamefont {Huang},\ and\ \citenamefont
  {Negele}}]{Chu:1993cn}%
  \BibitemOpen
  \bibfield  {author} {\bibinfo {author} {\bibfnamefont {M.~C.}\ \bibnamefont
  {Chu}}, \bibinfo {author} {\bibfnamefont {J.~M.}\ \bibnamefont {Grandy}},
  \bibinfo {author} {\bibfnamefont {S.}~\bibnamefont {Huang}}, \ and\ \bibinfo
  {author} {\bibfnamefont {John~W.}\ \bibnamefont {Negele}},\ }\bibfield
  {title} {\enquote {\bibinfo {title} {{Correlation functions of hadron
  currents in the QCD vacuum calculated in lattice QCD}},}\ }\href {\doibase
  10.1103/PhysRevD.48.3340} {\bibfield  {journal} {\bibinfo  {journal} {Phys.
  Rev. D}\ }\textbf {\bibinfo {volume} {48}},\ \bibinfo {pages} {3340--3353}
  (\bibinfo {year} {1993})},\ \Eprint {http://arxiv.org/abs/hep-lat/9306002}
  {arXiv:hep-lat/9306002} \BibitemShut {NoStop}%
\bibitem [{\citenamefont {Moran}\ and\ \citenamefont
  {Leinweber}(2008)}]{Moran:2008xq}%
  \BibitemOpen
  \bibfield  {author} {\bibinfo {author} {\bibfnamefont {P.~J.}\ \bibnamefont
  {Moran}}\ and\ \bibinfo {author} {\bibfnamefont {D.~B.}\ \bibnamefont
  {Leinweber}},\ }\bibfield  {title} {\enquote {\bibinfo {title} {{Buried
  treasure in the sand of the QCD vacuum}},}\ }in\ \href@noop {} {\emph
  {\bibinfo {booktitle} {{QCD Downunder II}}}}\ (\bibinfo {year} {2008})\
  \Eprint {http://arxiv.org/abs/0805.4246} {arXiv:0805.4246 [hep-lat]}
  \BibitemShut {NoStop}%
\bibitem [{\citenamefont {Zahed}(2021)}]{Zahed:2021fxk}%
  \BibitemOpen
  \bibfield  {author} {\bibinfo {author} {\bibfnamefont {Ismail}\ \bibnamefont
  {Zahed}},\ }\bibfield  {title} {\enquote {\bibinfo {title} {{Mass sum rule of
  hadrons in the QCD instanton vacuum}},}\ }\href {\doibase
  10.1103/PhysRevD.104.054031} {\bibfield  {journal} {\bibinfo  {journal}
  {Phys. Rev. D}\ }\textbf {\bibinfo {volume} {104}},\ \bibinfo {pages}
  {054031} (\bibinfo {year} {2021})},\ \Eprint
  {http://arxiv.org/abs/2102.08191} {arXiv:2102.08191 [hep-ph]} \BibitemShut
  {NoStop}%
\bibitem [{\citenamefont {Novikov}\ \emph {et~al.}(1981)\citenamefont
  {Novikov}, \citenamefont {Shifman}, \citenamefont {Vainshtein},\ and\
  \citenamefont {Zakharov}}]{Novikov:1981xi}%
  \BibitemOpen
  \bibfield  {author} {\bibinfo {author} {\bibfnamefont {V.~A.}\ \bibnamefont
  {Novikov}}, \bibinfo {author} {\bibfnamefont {Mikhail~A.}\ \bibnamefont
  {Shifman}}, \bibinfo {author} {\bibfnamefont {A.~I.}\ \bibnamefont
  {Vainshtein}}, \ and\ \bibinfo {author} {\bibfnamefont {Valentin~I.}\
  \bibnamefont {Zakharov}},\ }\bibfield  {title} {\enquote {\bibinfo {title}
  {{Are All Hadrons Alike?~}},}\ }\href {\doibase 10.1016/0550-3213(81)90303-5}
  {\bibfield  {journal} {\bibinfo  {journal} {Nucl. Phys. B}\ }\textbf
  {\bibinfo {volume} {191}},\ \bibinfo {pages} {301--369} (\bibinfo {year}
  {1981})}\BibitemShut {NoStop}%
\bibitem [{\citenamefont {Witten}(1979)}]{Witten:1979vv}%
  \BibitemOpen
  \bibfield  {author} {\bibinfo {author} {\bibfnamefont {Edward}\ \bibnamefont
  {Witten}},\ }\bibfield  {title} {\enquote {\bibinfo {title} {{Current Algebra
  Theorems for the U(1) Goldstone Boson}},}\ }\href {\doibase
  10.1016/0550-3213(79)90031-2} {\bibfield  {journal} {\bibinfo  {journal}
  {Nucl. Phys. B}\ }\textbf {\bibinfo {volume} {156}},\ \bibinfo {pages}
  {269--283} (\bibinfo {year} {1979})}\BibitemShut {NoStop}%
\bibitem [{\citenamefont {Veneziano}(1979)}]{Veneziano:1979ec}%
  \BibitemOpen
  \bibfield  {author} {\bibinfo {author} {\bibfnamefont {G.}~\bibnamefont
  {Veneziano}},\ }\bibfield  {title} {\enquote {\bibinfo {title} {{U(1) Without
  Instantons}},}\ }\href {\doibase 10.1016/0550-3213(79)90332-8} {\bibfield
  {journal} {\bibinfo  {journal} {Nucl. Phys. B}\ }\textbf {\bibinfo {volume}
  {159}},\ \bibinfo {pages} {213--224} (\bibinfo {year} {1979})}\BibitemShut
  {NoStop}%
\bibitem [{\citenamefont {'t~Hooft}(1976)}]{tHooft:1976snw}%
  \BibitemOpen
  \bibfield  {author} {\bibinfo {author} {\bibfnamefont {Gerard}\ \bibnamefont
  {'t~Hooft}},\ }\bibfield  {title} {\enquote {\bibinfo {title} {{Computation
  of the Quantum Effects Due to a Four-Dimensional Pseudoparticle}},}\ }\href
  {\doibase 10.1103/PhysRevD.14.3432} {\bibfield  {journal} {\bibinfo
  {journal} {Phys. Rev. D}\ }\textbf {\bibinfo {volume} {14}},\ \bibinfo
  {pages} {3432--3450} (\bibinfo {year} {1976})},\ \bibinfo {note} {[Erratum:
  Phys.Rev.D 18, 2199 (1978)]}\BibitemShut {NoStop}%
\bibitem [{\citenamefont {Verbaarschot}\ and\ \citenamefont
  {Zahed}(1993)}]{Verbaarschot:1993pm}%
  \BibitemOpen
  \bibfield  {author} {\bibinfo {author} {\bibfnamefont {J.~J.~M.}\
  \bibnamefont {Verbaarschot}}\ and\ \bibinfo {author} {\bibfnamefont
  {I.}~\bibnamefont {Zahed}},\ }\bibfield  {title} {\enquote {\bibinfo {title}
  {{Spectral density of the QCD Dirac operator near zero virtuality}},}\ }\href
  {\doibase 10.1103/PhysRevLett.70.3852} {\bibfield  {journal} {\bibinfo
  {journal} {Phys. Rev. Lett.}\ }\textbf {\bibinfo {volume} {70}},\ \bibinfo
  {pages} {3852--3855} (\bibinfo {year} {1993})},\ \Eprint
  {http://arxiv.org/abs/hep-th/9303012} {arXiv:hep-th/9303012} \BibitemShut
  {NoStop}%
\bibitem [{\citenamefont {Wittig}(2020)}]{Wittig:2020jtm}%
  \BibitemOpen
  \bibfield  {author} {\bibinfo {author} {\bibfnamefont {Hartmut}\ \bibnamefont
  {Wittig}},\ }\enquote {\bibinfo {title} {{QCD on the Lattice}},}\ in\ \href
  {\doibase 10.1007/978-3-030-38207-0_5} {\emph {\bibinfo {booktitle}
  {{Particle Physics Reference Library}: {Volume 1: Theory and
  Experiments}}}},\ \bibinfo {editor} {edited by\ \bibinfo {editor}
  {\bibfnamefont {Herwig}\ \bibnamefont {Schopper}}}\ (\bibinfo {year} {2020})\
  pp.\ \bibinfo {pages} {137--262}\BibitemShut {NoStop}%
\bibitem [{\citenamefont {Pobylitsa}(1989)}]{Pobylitsa:1989uq}%
  \BibitemOpen
  \bibfield  {author} {\bibinfo {author} {\bibfnamefont {P.~V.}\ \bibnamefont
  {Pobylitsa}},\ }\bibfield  {title} {\enquote {\bibinfo {title} {{The Quark
  Propagator and Correlation Functions in the Instanton Vacuum}},}\ }\href
  {\doibase 10.1016/0370-2693(89)91216-1} {\bibfield  {journal} {\bibinfo
  {journal} {Phys. Lett. B}\ }\textbf {\bibinfo {volume} {226}},\ \bibinfo
  {pages} {387--392} (\bibinfo {year} {1989})}\BibitemShut {NoStop}%
\bibitem [{\citenamefont {Kock}\ \emph {et~al.}(2020)\citenamefont {Kock},
  \citenamefont {Liu},\ and\ \citenamefont {Zahed}}]{Kock:2020frx}%
  \BibitemOpen
  \bibfield  {author} {\bibinfo {author} {\bibfnamefont {Arthur}\ \bibnamefont
  {Kock}}, \bibinfo {author} {\bibfnamefont {Yizhuang}\ \bibnamefont {Liu}}, \
  and\ \bibinfo {author} {\bibfnamefont {Ismail}\ \bibnamefont {Zahed}},\
  }\bibfield  {title} {\enquote {\bibinfo {title} {{Pion and kaon parton
  distributions in the QCD instanton vacuum}},}\ }\href {\doibase
  10.1103/PhysRevD.102.014039} {\bibfield  {journal} {\bibinfo  {journal}
  {Phys. Rev. D}\ }\textbf {\bibinfo {volume} {102}},\ \bibinfo {pages}
  {014039} (\bibinfo {year} {2020})},\ \Eprint
  {http://arxiv.org/abs/2004.01595} {arXiv:2004.01595 [hep-ph]} \BibitemShut
  {NoStop}%
\bibitem [{\citenamefont {Sch\"afer}\ \emph {et~al.}(1994)\citenamefont
  {Sch\"afer}, \citenamefont {Shuryak},\ and\ \citenamefont
  {Verbaarschot}}]{Schafer:1993ra}%
  \BibitemOpen
  \bibfield  {author} {\bibinfo {author} {\bibfnamefont {Thomas}\ \bibnamefont
  {Sch\"afer}}, \bibinfo {author} {\bibfnamefont {Edward~V.}\ \bibnamefont
  {Shuryak}}, \ and\ \bibinfo {author} {\bibfnamefont {J.~J.~M.}\ \bibnamefont
  {Verbaarschot}},\ }\bibfield  {title} {\enquote {\bibinfo {title} {{Baryonic
  correlators in the random instanton vacuum}},}\ }\href {\doibase
  10.1016/0550-3213(94)90497-9} {\bibfield  {journal} {\bibinfo  {journal}
  {Nucl. Phys. B}\ }\textbf {\bibinfo {volume} {412}},\ \bibinfo {pages}
  {143--168} (\bibinfo {year} {1994})},\ \Eprint
  {http://arxiv.org/abs/hep-ph/9306220} {arXiv:hep-ph/9306220} \BibitemShut
  {NoStop}%
\bibitem [{\citenamefont {Bernard}\ \emph {et~al.}(1987)\citenamefont
  {Bernard}, \citenamefont {Meissner},\ and\ \citenamefont
  {Zahed}}]{Bernard:1987ir}%
  \BibitemOpen
  \bibfield  {author} {\bibinfo {author} {\bibfnamefont {V.}~\bibnamefont
  {Bernard}}, \bibinfo {author} {\bibfnamefont {U.~G.}\ \bibnamefont
  {Meissner}}, \ and\ \bibinfo {author} {\bibfnamefont {I.}~\bibnamefont
  {Zahed}},\ }\bibfield  {title} {\enquote {\bibinfo {title} {{Decoupling of
  the Pion at Finite Temperature and Density}},}\ }\href {\doibase
  10.1103/PhysRevD.36.819} {\bibfield  {journal} {\bibinfo  {journal} {Phys.
  Rev. D}\ }\textbf {\bibinfo {volume} {36}},\ \bibinfo {pages} {819} (\bibinfo
  {year} {1987})}\BibitemShut {NoStop}%
\bibitem [{\citenamefont {Klevansky}(1992)}]{Klevansky:1992qe}%
  \BibitemOpen
  \bibfield  {author} {\bibinfo {author} {\bibfnamefont {S.~P.}\ \bibnamefont
  {Klevansky}},\ }\bibfield  {title} {\enquote {\bibinfo {title} {{The
  Nambu-Jona-Lasinio model of quantum chromodynamics}},}\ }\href {\doibase
  10.1103/RevModPhys.64.649} {\bibfield  {journal} {\bibinfo  {journal} {Rev.
  Mod. Phys.}\ }\textbf {\bibinfo {volume} {64}},\ \bibinfo {pages} {649--708}
  (\bibinfo {year} {1992})}\BibitemShut {NoStop}%
\bibitem [{\citenamefont {Roberts}\ \emph {et~al.}(2021)\citenamefont
  {Roberts}, \citenamefont {Richards}, \citenamefont {Horn},\ and\
  \citenamefont {Chang}}]{Roberts:2021nhw}%
  \BibitemOpen
  \bibfield  {author} {\bibinfo {author} {\bibfnamefont {Craig~D.}\
  \bibnamefont {Roberts}}, \bibinfo {author} {\bibfnamefont {David~G.}\
  \bibnamefont {Richards}}, \bibinfo {author} {\bibfnamefont {Tanja}\
  \bibnamefont {Horn}}, \ and\ \bibinfo {author} {\bibfnamefont {Lei}\
  \bibnamefont {Chang}},\ }\bibfield  {title} {\enquote {\bibinfo {title}
  {{Insights into the Emergence of Mass from Studies of Pion and Kaon
  Structure}},}\ }\href@noop {} {\  (\bibinfo {year} {2021})},\ \Eprint
  {http://arxiv.org/abs/2102.01765} {arXiv:2102.01765 [hep-ph]} \BibitemShut
  {NoStop}%
\bibitem [{\citenamefont {Greensite}(2017)}]{Greensite:2016pfc}%
  \BibitemOpen
  \bibfield  {author} {\bibinfo {author} {\bibfnamefont {Jeff}\ \bibnamefont
  {Greensite}},\ }\bibfield  {title} {\enquote {\bibinfo {title} {{Confinement
  from Center Vortices: A review of old and new results}},}\ }\href {\doibase
  10.1051/epjconf/201713701009} {\bibfield  {journal} {\bibinfo  {journal} {EPJ
  Web Conf.}\ }\textbf {\bibinfo {volume} {137}},\ \bibinfo {pages} {01009}
  (\bibinfo {year} {2017})},\ \Eprint {http://arxiv.org/abs/1610.06221}
  {arXiv:1610.06221 [hep-lat]} \BibitemShut {NoStop}%
\bibitem [{\citenamefont {Biddle}\ \emph
  {et~al.}(2020{\natexlab{a}})\citenamefont {Biddle}, \citenamefont {Kamleh},\
  and\ \citenamefont {Leinweber}}]{Biddle:2019gke}%
  \BibitemOpen
  \bibfield  {author} {\bibinfo {author} {\bibfnamefont {James~C.}\
  \bibnamefont {Biddle}}, \bibinfo {author} {\bibfnamefont {Waseem}\
  \bibnamefont {Kamleh}}, \ and\ \bibinfo {author} {\bibfnamefont {Derek~B.}\
  \bibnamefont {Leinweber}},\ }\bibfield  {title} {\enquote {\bibinfo {title}
  {{Visualization of center vortex structure}},}\ }\href {\doibase
  10.1103/PhysRevD.102.034504} {\bibfield  {journal} {\bibinfo  {journal}
  {Phys. Rev. D}\ }\textbf {\bibinfo {volume} {102}},\ \bibinfo {pages}
  {034504} (\bibinfo {year} {2020}{\natexlab{a}})},\ \Eprint
  {http://arxiv.org/abs/1912.09531} {arXiv:1912.09531 [hep-lat]} \BibitemShut
  {NoStop}%
\bibitem [{\citenamefont {Biddle}\ \emph
  {et~al.}(2020{\natexlab{b}})\citenamefont {Biddle}, \citenamefont {Kamleh},\
  and\ \citenamefont {Leinweber}}]{Biddle:2020eec}%
  \BibitemOpen
  \bibfield  {author} {\bibinfo {author} {\bibfnamefont {James~C.}\
  \bibnamefont {Biddle}}, \bibinfo {author} {\bibfnamefont {Waseem}\
  \bibnamefont {Kamleh}}, \ and\ \bibinfo {author} {\bibfnamefont {Derek~B.}\
  \bibnamefont {Leinweber}},\ }\bibfield  {title} {\enquote {\bibinfo {title}
  {{Visualisations of Centre Vortices}},}\ }\href {\doibase
  10.1051/epjconf/202024506010} {\bibfield  {journal} {\bibinfo  {journal} {EPJ
  Web Conf.}\ }\textbf {\bibinfo {volume} {245}},\ \bibinfo {pages} {06010}
  (\bibinfo {year} {2020}{\natexlab{b}})},\ \Eprint
  {http://arxiv.org/abs/2009.12047} {arXiv:2009.12047 [hep-lat]} \BibitemShut
  {NoStop}%
\bibitem [{\citenamefont {Nowak}\ \emph
  {et~al.}(1996{\natexlab{b}})\citenamefont {Nowak}, \citenamefont {Rho},\ and\
  \citenamefont {Zahed}}]{doi:10.1142/1681}%
  \BibitemOpen
  \bibfield  {author} {\bibinfo {author} {\bibfnamefont {Maciej~A}\
  \bibnamefont {Nowak}}, \bibinfo {author} {\bibfnamefont {Mannque}\
  \bibnamefont {Rho}}, \ and\ \bibinfo {author} {\bibfnamefont {Ismail}\
  \bibnamefont {Zahed}},\ }\href {\doibase 10.1142/1681} {\emph {\bibinfo
  {title} {Chiral Nuclear Dynamics}}}\ (\bibinfo  {publisher} {WORLD
  SCIENTIFIC},\ \bibinfo {year} {1996})\ \Eprint
  {http://arxiv.org/abs/https://www.worldscientific.com/doi/pdf/10.1142/1681}
  {https://www.worldscientific.com/doi/pdf/10.1142/1681} \BibitemShut {NoStop}%
\bibitem [{\citenamefont {Vainshtein}\ \emph {et~al.}(1982)\citenamefont
  {Vainshtein}, \citenamefont {Zakharov}, \citenamefont {Novikov},\ and\
  \citenamefont {Shifman}}]{Vainshtein:1981wh}%
  \BibitemOpen
  \bibfield  {author} {\bibinfo {author} {\bibfnamefont {A.~I.}\ \bibnamefont
  {Vainshtein}}, \bibinfo {author} {\bibfnamefont {Valentin~I.}\ \bibnamefont
  {Zakharov}}, \bibinfo {author} {\bibfnamefont {V.~A.}\ \bibnamefont
  {Novikov}}, \ and\ \bibinfo {author} {\bibfnamefont {Mikhail~A.}\
  \bibnamefont {Shifman}},\ }\bibfield  {title} {\enquote {\bibinfo {title}
  {{ABC's of Instantons}},}\ }\href {\doibase 10.1070/PU1982v025n04ABEH004533}
  {\bibfield  {journal} {\bibinfo  {journal} {Sov. Phys. Usp.}\ }\textbf
  {\bibinfo {volume} {25}},\ \bibinfo {pages} {195} (\bibinfo {year}
  {1982})}\BibitemShut {NoStop}%
\bibitem [{\citenamefont {Kochelev}(1998)}]{Kochelev:1996pv}%
  \BibitemOpen
  \bibfield  {author} {\bibinfo {author} {\bibfnamefont {N.~I.}\ \bibnamefont
  {Kochelev}},\ }\bibfield  {title} {\enquote {\bibinfo {title} {{Anomalous
  quark chromomagnetic moment induced by instantons}},}\ }\href {\doibase
  10.1016/S0370-2693(98)00262-7} {\bibfield  {journal} {\bibinfo  {journal}
  {Phys. Lett. B}\ }\textbf {\bibinfo {volume} {426}},\ \bibinfo {pages}
  {149--153} (\bibinfo {year} {1998})},\ \Eprint
  {http://arxiv.org/abs/hep-ph/9610551} {arXiv:hep-ph/9610551} \BibitemShut
  {NoStop}%
\bibitem [{\citenamefont {Diakonov}\ and\ \citenamefont
  {Petrov}(1984)}]{Diakonov:1983hh}%
  \BibitemOpen
  \bibfield  {author} {\bibinfo {author} {\bibfnamefont {Dmitri}\ \bibnamefont
  {Diakonov}}\ and\ \bibinfo {author} {\bibfnamefont {V.~Yu.}\ \bibnamefont
  {Petrov}},\ }\bibfield  {title} {\enquote {\bibinfo {title} {{Instanton Based
  Vacuum from Feynman Variational Principle}},}\ }\href {\doibase
  10.1016/0550-3213(84)90432-2} {\bibfield  {journal} {\bibinfo  {journal}
  {Nucl. Phys. B}\ }\textbf {\bibinfo {volume} {245}},\ \bibinfo {pages}
  {259--292} (\bibinfo {year} {1984})}\BibitemShut {NoStop}%
\bibitem [{\citenamefont {Sch\"afer}\ and\ \citenamefont
  {Shuryak}(1996)}]{Schafer:1995pz}%
  \BibitemOpen
  \bibfield  {author} {\bibinfo {author} {\bibfnamefont {Thomas}\ \bibnamefont
  {Sch\"afer}}\ and\ \bibinfo {author} {\bibfnamefont {Edward~V.}\ \bibnamefont
  {Shuryak}},\ }\bibfield  {title} {\enquote {\bibinfo {title} {{The
  Interacting instanton liquid in QCD at zero and finite temperature}},}\
  }\href {\doibase 10.1103/PhysRevD.53.6522} {\bibfield  {journal} {\bibinfo
  {journal} {Phys. Rev. D}\ }\textbf {\bibinfo {volume} {53}},\ \bibinfo
  {pages} {6522--6542} (\bibinfo {year} {1996})},\ \Eprint
  {http://arxiv.org/abs/hep-ph/9509337} {arXiv:hep-ph/9509337} \BibitemShut
  {NoStop}%
\bibitem [{\citenamefont {Faccioli}\ and\ \citenamefont
  {Shuryak}(2001)}]{Faccioli:2001ug}%
  \BibitemOpen
  \bibfield  {author} {\bibinfo {author} {\bibfnamefont {P.}~\bibnamefont
  {Faccioli}}\ and\ \bibinfo {author} {\bibfnamefont {Edward~V.}\ \bibnamefont
  {Shuryak}},\ }\bibfield  {title} {\enquote {\bibinfo {title} {{Systematic
  study of the single instanton approximation in QCD}},}\ }\href {\doibase
  10.1103/PhysRevD.64.114020} {\bibfield  {journal} {\bibinfo  {journal} {Phys.
  Rev. D}\ }\textbf {\bibinfo {volume} {64}},\ \bibinfo {pages} {114020}
  (\bibinfo {year} {2001})},\ \Eprint {http://arxiv.org/abs/hep-ph/0106019}
  {arXiv:hep-ph/0106019} \BibitemShut {NoStop}%
\bibitem [{\citenamefont {Liu}\ \emph {et~al.}(2023{\natexlab{a}})\citenamefont
  {Liu}, \citenamefont {Shuryak},\ and\ \citenamefont {Zahed}}]{Liu:2023yuj}%
  \BibitemOpen
  \bibfield  {author} {\bibinfo {author} {\bibfnamefont {Wei-Yang}\
  \bibnamefont {Liu}}, \bibinfo {author} {\bibfnamefont {Edward}\ \bibnamefont
  {Shuryak}}, \ and\ \bibinfo {author} {\bibfnamefont {Ismail}\ \bibnamefont
  {Zahed}},\ }\bibfield  {title} {\enquote {\bibinfo {title} {{Hadronic
  structure on the light-front. VII. Pions and kaons and their partonic
  distributions}},}\ }\href {\doibase 10.1103/PhysRevD.107.094024} {\bibfield
  {journal} {\bibinfo  {journal} {Phys. Rev. D}\ }\textbf {\bibinfo {volume}
  {107}},\ \bibinfo {pages} {094024} (\bibinfo {year} {2023}{\natexlab{a}})},\
  \Eprint {http://arxiv.org/abs/2302.03759} {arXiv:2302.03759 [hep-ph]}
  \BibitemShut {NoStop}%
\bibitem [{\citenamefont {Liu}\ \emph {et~al.}(2023{\natexlab{b}})\citenamefont
  {Liu}, \citenamefont {Shuryak},\ and\ \citenamefont {Zahed}}]{Liu:2023fpj}%
  \BibitemOpen
  \bibfield  {author} {\bibinfo {author} {\bibfnamefont {Wei-Yang}\
  \bibnamefont {Liu}}, \bibinfo {author} {\bibfnamefont {Edward}\ \bibnamefont
  {Shuryak}}, \ and\ \bibinfo {author} {\bibfnamefont {Ismail}\ \bibnamefont
  {Zahed}},\ }\bibfield  {title} {\enquote {\bibinfo {title} {{Hadronic
  structure on the light-front VIII. Light scalar and vector mesons}},}\
  }\href@noop {} {\  (\bibinfo {year} {2023}{\natexlab{b}})},\ \Eprint
  {http://arxiv.org/abs/2307.16302} {arXiv:2307.16302 [hep-ph]} \BibitemShut
  {NoStop}%
\bibitem [{\citenamefont {Ioffe}(2003)}]{Ioffe:2002ee}%
  \BibitemOpen
  \bibfield  {author} {\bibinfo {author} {\bibfnamefont {B.~L.}\ \bibnamefont
  {Ioffe}},\ }\bibfield  {title} {\enquote {\bibinfo {title} {{Condensates in
  quantum chromodynamics}},}\ }\href {\doibase 10.1134/1.1540654} {\bibfield
  {journal} {\bibinfo  {journal} {Phys. Atom. Nucl.}\ }\textbf {\bibinfo
  {volume} {66}},\ \bibinfo {pages} {30--43} (\bibinfo {year} {2003})},\
  \Eprint {http://arxiv.org/abs/hep-ph/0207191} {arXiv:hep-ph/0207191}
  \BibitemShut {NoStop}%
\bibitem [{\citenamefont {Aoki}\ \emph {et~al.}(2020)\citenamefont {Aoki} \emph
  {et~al.}}]{FlavourLatticeAveragingGroup:2019iem}%
  \BibitemOpen
  \bibfield  {author} {\bibinfo {author} {\bibfnamefont {S.}~\bibnamefont
  {Aoki}} \emph {et~al.} (\bibinfo {collaboration} {Flavour Lattice Averaging
  Group}),\ }\bibfield  {title} {\enquote {\bibinfo {title} {{FLAG Review 2019:
  Flavour Lattice Averaging Group (FLAG)}},}\ }\href {\doibase
  10.1140/epjc/s10052-019-7354-7} {\bibfield  {journal} {\bibinfo  {journal}
  {Eur. Phys. J. C}\ }\textbf {\bibinfo {volume} {80}},\ \bibinfo {pages} {113}
  (\bibinfo {year} {2020})},\ \Eprint {http://arxiv.org/abs/1902.08191}
  {arXiv:1902.08191 [hep-lat]} \BibitemShut {NoStop}%
\bibitem [{\citenamefont {Qian}\ and\ \citenamefont
  {Zahed}(2016)}]{Qian:2015wyq}%
  \BibitemOpen
  \bibfield  {author} {\bibinfo {author} {\bibfnamefont {Yachao}\ \bibnamefont
  {Qian}}\ and\ \bibinfo {author} {\bibfnamefont {Ismail}\ \bibnamefont
  {Zahed}},\ }\bibfield  {title} {\enquote {\bibinfo {title} {{Spin Physics
  through QCD Instantons}},}\ }\href {\doibase 10.1016/j.aop.2016.09.002}
  {\bibfield  {journal} {\bibinfo  {journal} {Annals Phys.}\ }\textbf {\bibinfo
  {volume} {374}},\ \bibinfo {pages} {314--337} (\bibinfo {year} {2016})},\
  \Eprint {http://arxiv.org/abs/1512.08172} {arXiv:1512.08172 [hep-ph]}
  \BibitemShut {NoStop}%
\bibitem [{\citenamefont {Diakonov}(2003)}]{Diakonov:2002fq}%
  \BibitemOpen
  \bibfield  {author} {\bibinfo {author} {\bibfnamefont {Dmitri}\ \bibnamefont
  {Diakonov}},\ }\bibfield  {title} {\enquote {\bibinfo {title} {{Instantons at
  work}},}\ }\href {\doibase 10.1016/S0146-6410(03)90014-7} {\bibfield
  {journal} {\bibinfo  {journal} {Prog. Part. Nucl. Phys.}\ }\textbf {\bibinfo
  {volume} {51}},\ \bibinfo {pages} {173--222} (\bibinfo {year} {2003})},\
  \Eprint {http://arxiv.org/abs/hep-ph/0212026} {arXiv:hep-ph/0212026}
  \BibitemShut {NoStop}%
\bibitem [{\citenamefont {Shuryak}\ and\ \citenamefont
  {Zahed}(2021)}]{Shuryak:2020ktq}%
  \BibitemOpen
  \bibfield  {author} {\bibinfo {author} {\bibfnamefont {Edward}\ \bibnamefont
  {Shuryak}}\ and\ \bibinfo {author} {\bibfnamefont {Ismail}\ \bibnamefont
  {Zahed}},\ }\bibfield  {title} {\enquote {\bibinfo {title} {{Nonperturbative
  quark-antiquark interactions in mesonic form factors}},}\ }\href {\doibase
  10.1103/PhysRevD.103.054028} {\bibfield  {journal} {\bibinfo  {journal}
  {Phys. Rev. D}\ }\textbf {\bibinfo {volume} {103}},\ \bibinfo {pages}
  {054028} (\bibinfo {year} {2021})},\ \Eprint
  {http://arxiv.org/abs/2008.06169} {arXiv:2008.06169 [hep-ph]} \BibitemShut
  {NoStop}%
\bibitem [{\citenamefont {Weiss}(2021)}]{Weiss:2021kpt}%
  \BibitemOpen
  \bibfield  {author} {\bibinfo {author} {\bibfnamefont {C.}~\bibnamefont
  {Weiss}},\ }\bibfield  {title} {\enquote {\bibinfo {title} {{Nucleon matrix
  element of Weinberg's CP-odd gluon operator from the instanton vacuum}},}\
  }\href {\doibase 10.1016/j.physletb.2021.136447} {\bibfield  {journal}
  {\bibinfo  {journal} {Phys. Lett. B}\ }\textbf {\bibinfo {volume} {819}},\
  \bibinfo {pages} {136447} (\bibinfo {year} {2021})},\ \Eprint
  {http://arxiv.org/abs/2103.13471} {arXiv:2103.13471 [hep-ph]} \BibitemShut
  {NoStop}%
\bibitem [{\citenamefont {Ji}\ \emph {et~al.}(2021)\citenamefont {Ji},
  \citenamefont {Liu},\ and\ \citenamefont {Zahed}}]{Ji:2020bby}%
  \BibitemOpen
  \bibfield  {author} {\bibinfo {author} {\bibfnamefont {Xiangdong}\
  \bibnamefont {Ji}}, \bibinfo {author} {\bibfnamefont {Yizhuang}\ \bibnamefont
  {Liu}}, \ and\ \bibinfo {author} {\bibfnamefont {Ismail}\ \bibnamefont
  {Zahed}},\ }\bibfield  {title} {\enquote {\bibinfo {title} {{Mass structure
  of hadrons and light-front sum rules in the $'t$ Hooft model}},}\ }\href
  {\doibase 10.1103/PhysRevD.103.074002} {\bibfield  {journal} {\bibinfo
  {journal} {Phys. Rev. D}\ }\textbf {\bibinfo {volume} {103}},\ \bibinfo
  {pages} {074002} (\bibinfo {year} {2021})},\ \Eprint
  {http://arxiv.org/abs/2010.06665} {arXiv:2010.06665 [hep-ph]} \BibitemShut
  {NoStop}%
\bibitem [{\citenamefont {Zahed}(2022)}]{Zahed:2022wae}%
  \BibitemOpen
  \bibfield  {author} {\bibinfo {author} {\bibfnamefont {Ismail}\ \bibnamefont
  {Zahed}},\ }\bibfield  {title} {\enquote {\bibinfo {title} {{Spin Sum Rule of
  the Nucleon in the QCD Instanton Vacuum}},}\ }\href {\doibase
  10.3390/sym14050932} {\bibfield  {journal} {\bibinfo  {journal} {Symmetry}\
  }\textbf {\bibinfo {volume} {14}},\ \bibinfo {pages} {932} (\bibinfo {year}
  {2022})}\BibitemShut {NoStop}%
\bibitem [{\citenamefont {Luscher}\ and\ \citenamefont
  {Palombi}(2010)}]{Luscher:2010ik}%
  \BibitemOpen
  \bibfield  {author} {\bibinfo {author} {\bibfnamefont {Martin}\ \bibnamefont
  {Luscher}}\ and\ \bibinfo {author} {\bibfnamefont {Filippo}\ \bibnamefont
  {Palombi}},\ }\bibfield  {title} {\enquote {\bibinfo {title} {{Universality
  of the topological susceptibility in the SU(3) gauge theory}},}\ }\href
  {\doibase 10.1007/JHEP09(2010)110} {\bibfield  {journal} {\bibinfo  {journal}
  {JHEP}\ }\textbf {\bibinfo {volume} {09}},\ \bibinfo {pages} {110} (\bibinfo
  {year} {2010})},\ \Eprint {http://arxiv.org/abs/1008.0732} {arXiv:1008.0732
  [hep-lat]} \BibitemShut {NoStop}%
\bibitem [{\citenamefont {Zahed}(1994)}]{Zahed:1994qh}%
  \BibitemOpen
  \bibfield  {author} {\bibinfo {author} {\bibfnamefont {I.}~\bibnamefont
  {Zahed}},\ }\bibfield  {title} {\enquote {\bibinfo {title} {{QCD instantons
  in vacuum and matter}},}\ }\href {\doibase 10.1016/0550-3213(94)90640-8}
  {\bibfield  {journal} {\bibinfo  {journal} {Nucl. Phys. B}\ }\textbf
  {\bibinfo {volume} {427}},\ \bibinfo {pages} {561--574} (\bibinfo {year}
  {1994})},\ \Eprint {http://arxiv.org/abs/hep-ph/9404326}
  {arXiv:hep-ph/9404326} \BibitemShut {NoStop}%
\bibitem [{\citenamefont {Shuryak}\ and\ \citenamefont
  {Verbaarschot}(1995)}]{Shuryak:1994rr}%
  \BibitemOpen
  \bibfield  {author} {\bibinfo {author} {\bibfnamefont {Edward~V.}\
  \bibnamefont {Shuryak}}\ and\ \bibinfo {author} {\bibfnamefont {J.~J.~M.}\
  \bibnamefont {Verbaarschot}},\ }\bibfield  {title} {\enquote {\bibinfo
  {title} {{Screening of the topological charge in a correlated instanton
  vacuum}},}\ }\href {\doibase 10.1103/PhysRevD.52.295} {\bibfield  {journal}
  {\bibinfo  {journal} {Phys. Rev. D}\ }\textbf {\bibinfo {volume} {52}},\
  \bibinfo {pages} {295--306} (\bibinfo {year} {1995})},\ \Eprint
  {http://arxiv.org/abs/hep-lat/9409020} {arXiv:hep-lat/9409020} \BibitemShut
  {NoStop}%
\bibitem [{\citenamefont {Di~Vecchia}\ and\ \citenamefont
  {Veneziano}(1980)}]{DiVecchia:1980yfw}%
  \BibitemOpen
  \bibfield  {author} {\bibinfo {author} {\bibfnamefont {P.}~\bibnamefont
  {Di~Vecchia}}\ and\ \bibinfo {author} {\bibfnamefont {G.}~\bibnamefont
  {Veneziano}},\ }\bibfield  {title} {\enquote {\bibinfo {title} {{Chiral
  Dynamics in the Large n Limit}},}\ }\href {\doibase
  10.1016/0550-3213(80)90370-3} {\bibfield  {journal} {\bibinfo  {journal}
  {Nucl. Phys. B}\ }\textbf {\bibinfo {volume} {171}},\ \bibinfo {pages}
  {253--272} (\bibinfo {year} {1980})}\BibitemShut {NoStop}%
\bibitem [{\citenamefont {Gasser}\ and\ \citenamefont
  {Leutwyler}(1985)}]{Gasser:1984gg}%
  \BibitemOpen
  \bibfield  {author} {\bibinfo {author} {\bibfnamefont {J.}~\bibnamefont
  {Gasser}}\ and\ \bibinfo {author} {\bibfnamefont {H.}~\bibnamefont
  {Leutwyler}},\ }\bibfield  {title} {\enquote {\bibinfo {title} {{Chiral
  Perturbation Theory: Expansions in the Mass of the Strange Quark}},}\ }\href
  {\doibase 10.1016/0550-3213(85)90492-4} {\bibfield  {journal} {\bibinfo
  {journal} {Nucl. Phys. B}\ }\textbf {\bibinfo {volume} {250}},\ \bibinfo
  {pages} {465--516} (\bibinfo {year} {1985})}\BibitemShut {NoStop}%
\bibitem [{\citenamefont {Leutwyler}\ and\ \citenamefont
  {Smilga}(1992)}]{Leutwyler:1992yt}%
  \BibitemOpen
  \bibfield  {author} {\bibinfo {author} {\bibfnamefont {H.}~\bibnamefont
  {Leutwyler}}\ and\ \bibinfo {author} {\bibfnamefont {Andrei~V.}\ \bibnamefont
  {Smilga}},\ }\bibfield  {title} {\enquote {\bibinfo {title} {{Spectrum of
  Dirac operator and role of winding number in QCD}},}\ }\href {\doibase
  10.1103/PhysRevD.46.5607} {\bibfield  {journal} {\bibinfo  {journal} {Phys.
  Rev. D}\ }\textbf {\bibinfo {volume} {46}},\ \bibinfo {pages} {5607--5632}
  (\bibinfo {year} {1992})}\BibitemShut {NoStop}%
\bibitem [{\citenamefont {Janik}\ \emph {et~al.}(1999)\citenamefont {Janik},
  \citenamefont {Nowak}, \citenamefont {Papp},\ and\ \citenamefont
  {Zahed}}]{Janik:1999ps}%
  \BibitemOpen
  \bibfield  {author} {\bibinfo {author} {\bibfnamefont {Romuald~A.}\
  \bibnamefont {Janik}}, \bibinfo {author} {\bibfnamefont {Maciej~A.}\
  \bibnamefont {Nowak}}, \bibinfo {author} {\bibfnamefont {Gabor}\ \bibnamefont
  {Papp}}, \ and\ \bibinfo {author} {\bibfnamefont {Ismail}\ \bibnamefont
  {Zahed}},\ }\bibfield  {title} {\enquote {\bibinfo {title} {{U(1) problem at
  finite temperature}},}\ }\href {\doibase 10.1063/1.1301690} {\bibfield
  {journal} {\bibinfo  {journal} {AIP Conf. Proc.}\ }\textbf {\bibinfo {volume}
  {494}},\ \bibinfo {pages} {408--422} (\bibinfo {year} {1999})},\ \Eprint
  {http://arxiv.org/abs/hep-lat/9911024} {arXiv:hep-lat/9911024} \BibitemShut
  {NoStop}%
\bibitem [{\citenamefont {Ma}(2003)}]{Ma:2003py}%
  \BibitemOpen
  \bibfield  {author} {\bibinfo {author} {\bibfnamefont {J.~P.}\ \bibnamefont
  {Ma}},\ }\bibfield  {title} {\enquote {\bibinfo {title} {{Diffractive
  photoproduction of eta(c)}},}\ }\href {\doibase
  10.1016/j.nuclphysa.2003.08.016} {\bibfield  {journal} {\bibinfo  {journal}
  {Nucl. Phys. A}\ }\textbf {\bibinfo {volume} {727}},\ \bibinfo {pages}
  {333--352} (\bibinfo {year} {2003})},\ \Eprint
  {http://arxiv.org/abs/hep-ph/0301155} {arXiv:hep-ph/0301155} \BibitemShut
  {NoStop}%
\bibitem [{\citenamefont {Hechenberger}\ \emph {et~al.}(2024)\citenamefont
  {Hechenberger}, \citenamefont {Mamo},\ and\ \citenamefont
  {Zahed}}]{Hechenberger:2024abg}%
  \BibitemOpen
  \bibfield  {author} {\bibinfo {author} {\bibfnamefont {Florian}\ \bibnamefont
  {Hechenberger}}, \bibinfo {author} {\bibfnamefont {Kiminad~A.}\ \bibnamefont
  {Mamo}}, \ and\ \bibinfo {author} {\bibfnamefont {Ismail}\ \bibnamefont
  {Zahed}},\ }\bibfield  {title} {\enquote {\bibinfo {title} {{Threshold
  production of $\eta_{c,b}$ using holographic QCD}},}\ }\href@noop {} {\
  (\bibinfo {year} {2024})},\ \Eprint {http://arxiv.org/abs/2401.12162}
  {arXiv:2401.12162 [hep-ph]} \BibitemShut {NoStop}%
\bibitem [{\citenamefont {Freese}\ \emph {et~al.}(2019)\citenamefont {Freese},
  \citenamefont {Freese}, \citenamefont {Clo\"et},\ and\ \citenamefont
  {Clo\"et}}]{Freese:2019bhb}%
  \BibitemOpen
  \bibfield  {author} {\bibinfo {author} {\bibfnamefont {Adam}\ \bibnamefont
  {Freese}}, \bibinfo {author} {\bibfnamefont {Adam}\ \bibnamefont {Freese}},
  \bibinfo {author} {\bibfnamefont {Ian~C.}\ \bibnamefont {Clo\"et}}, \ and\
  \bibinfo {author} {\bibfnamefont {Ian~C.}\ \bibnamefont {Clo\"et}},\
  }\bibfield  {title} {\enquote {\bibinfo {title} {{Gravitational form factors
  of light mesons}},}\ }\href {\doibase 10.1103/PhysRevC.100.015201} {\bibfield
   {journal} {\bibinfo  {journal} {Phys. Rev. C}\ }\textbf {\bibinfo {volume}
  {100}},\ \bibinfo {pages} {015201} (\bibinfo {year} {2019})},\ \bibinfo
  {note} {[Erratum: Phys.Rev.C 105, 059901 (2022)]},\ \Eprint
  {http://arxiv.org/abs/1903.09222} {arXiv:1903.09222 [nucl-th]} \BibitemShut
  {NoStop}%
\bibitem [{\citenamefont {Ji}(2021)}]{Ji:2021mtz}%
  \BibitemOpen
  \bibfield  {author} {\bibinfo {author} {\bibfnamefont {Xiangdong}\
  \bibnamefont {Ji}},\ }\bibfield  {title} {\enquote {\bibinfo {title} {{Proton
  mass decomposition: naturalness and interpretations}},}\ }\href {\doibase
  10.1007/s11467-021-1065-x} {\bibfield  {journal} {\bibinfo  {journal} {Front.
  Phys. (Beijing)}\ }\textbf {\bibinfo {volume} {16}},\ \bibinfo {pages}
  {64601} (\bibinfo {year} {2021})},\ \Eprint {http://arxiv.org/abs/2102.07830}
  {arXiv:2102.07830 [hep-ph]} \BibitemShut {NoStop}%
\bibitem [{\citenamefont {Ji}(1995{\natexlab{a}})}]{Ji:1995sv}%
  \BibitemOpen
  \bibfield  {author} {\bibinfo {author} {\bibfnamefont {Xiang-Dong}\
  \bibnamefont {Ji}},\ }\bibfield  {title} {\enquote {\bibinfo {title}
  {{Breakup of hadron masses and energy - momentum tensor of QCD}},}\ }\href
  {\doibase 10.1103/PhysRevD.52.271} {\bibfield  {journal} {\bibinfo  {journal}
  {Phys. Rev. D}\ }\textbf {\bibinfo {volume} {52}},\ \bibinfo {pages}
  {271--281} (\bibinfo {year} {1995}{\natexlab{a}})},\ \Eprint
  {http://arxiv.org/abs/hep-ph/9502213} {arXiv:hep-ph/9502213} \BibitemShut
  {NoStop}%
\bibitem [{\citenamefont {Mamo}\ and\ \citenamefont
  {Zahed}(2020)}]{Mamo:2019mka}%
  \BibitemOpen
  \bibfield  {author} {\bibinfo {author} {\bibfnamefont {Kiminad~A.}\
  \bibnamefont {Mamo}}\ and\ \bibinfo {author} {\bibfnamefont {Ismail}\
  \bibnamefont {Zahed}},\ }\bibfield  {title} {\enquote {\bibinfo {title}
  {{Diffractive photoproduction of $J/\psi$ and $\Upsilon$ using holographic
  QCD: gravitational form factors and GPD of gluons in the proton}},}\ }\href
  {\doibase 10.1103/PhysRevD.101.086003} {\bibfield  {journal} {\bibinfo
  {journal} {Phys. Rev. D}\ }\textbf {\bibinfo {volume} {101}},\ \bibinfo
  {pages} {086003} (\bibinfo {year} {2020})},\ \Eprint
  {http://arxiv.org/abs/1910.04707} {arXiv:1910.04707 [hep-ph]} \BibitemShut
  {NoStop}%
\bibitem [{\citenamefont {Hatta}\ \emph {et~al.}(2018)\citenamefont {Hatta},
  \citenamefont {Rajan},\ and\ \citenamefont {Tanaka}}]{Hatta:2018sqd}%
  \BibitemOpen
  \bibfield  {author} {\bibinfo {author} {\bibfnamefont {Yoshitaka}\
  \bibnamefont {Hatta}}, \bibinfo {author} {\bibfnamefont {Abha}\ \bibnamefont
  {Rajan}}, \ and\ \bibinfo {author} {\bibfnamefont {Kazuhiro}\ \bibnamefont
  {Tanaka}},\ }\bibfield  {title} {\enquote {\bibinfo {title} {{Quark and gluon
  contributions to the QCD trace anomaly}},}\ }\href {\doibase
  10.1007/JHEP12(2018)008} {\bibfield  {journal} {\bibinfo  {journal} {JHEP}\
  }\textbf {\bibinfo {volume} {12}},\ \bibinfo {pages} {008} (\bibinfo {year}
  {2018})},\ \Eprint {http://arxiv.org/abs/1810.05116} {arXiv:1810.05116
  [hep-ph]} \BibitemShut {NoStop}%
\bibitem [{\citenamefont {Gross}\ and\ \citenamefont
  {Wilczek}(1974)}]{Gross:1974cs}%
  \BibitemOpen
  \bibfield  {author} {\bibinfo {author} {\bibfnamefont {D.~J.}\ \bibnamefont
  {Gross}}\ and\ \bibinfo {author} {\bibfnamefont {Frank}\ \bibnamefont
  {Wilczek}},\ }\bibfield  {title} {\enquote {\bibinfo {title} {{ASYMPTOTICALLY
  FREE GAUGE THEORIES. 2.}}}\ }\href {\doibase 10.1103/PhysRevD.9.980}
  {\bibfield  {journal} {\bibinfo  {journal} {Phys. Rev. D}\ }\textbf {\bibinfo
  {volume} {9}},\ \bibinfo {pages} {980--993} (\bibinfo {year}
  {1974})}\BibitemShut {NoStop}%
\bibitem [{\citenamefont {Politzer}(1974)}]{Politzer:1974sm}%
  \BibitemOpen
  \bibfield  {author} {\bibinfo {author} {\bibfnamefont {H.~David}\
  \bibnamefont {Politzer}},\ }\bibfield  {title} {\enquote {\bibinfo {title}
  {{Setting the scale for predictions of asymptotic freedom}},}\ }\href
  {\doibase 10.1103/PhysRevD.9.2174} {\bibfield  {journal} {\bibinfo  {journal}
  {Phys. Rev. D}\ }\textbf {\bibinfo {volume} {9}},\ \bibinfo {pages}
  {2174--2175} (\bibinfo {year} {1974})}\BibitemShut {NoStop}%
\bibitem [{\citenamefont {Jung}\ \emph {et~al.}(2014)\citenamefont {Jung},
  \citenamefont {Yakhshiev},\ and\ \citenamefont {Kim}}]{Jung:2013bya}%
  \BibitemOpen
  \bibfield  {author} {\bibinfo {author} {\bibfnamefont {Ju-Hyun}\ \bibnamefont
  {Jung}}, \bibinfo {author} {\bibfnamefont {Ulugbek}\ \bibnamefont
  {Yakhshiev}}, \ and\ \bibinfo {author} {\bibfnamefont {Hyun-Chul}\
  \bibnamefont {Kim}},\ }\bibfield  {title} {\enquote {\bibinfo {title}
  {{Energy\textendash{}momentum tensor form factors of the nucleon within a
  \ensuremath{\pi}\textendash{}\ensuremath{\rho}\textendash{}\ensuremath{\omega}
  soliton model}},}\ }\href {\doibase 10.1088/0954-3899/41/5/055107} {\bibfield
   {journal} {\bibinfo  {journal} {J. Phys. G}\ }\textbf {\bibinfo {volume}
  {41}},\ \bibinfo {pages} {055107} (\bibinfo {year} {2014})},\ \Eprint
  {http://arxiv.org/abs/1310.8064} {arXiv:1310.8064 [hep-ph]} \BibitemShut
  {NoStop}%
\bibitem [{\citenamefont {Shanahan}\ and\ \citenamefont
  {Detmold}(2019)}]{Shanahan:2018pib}%
  \BibitemOpen
  \bibfield  {author} {\bibinfo {author} {\bibfnamefont {P.~E.}\ \bibnamefont
  {Shanahan}}\ and\ \bibinfo {author} {\bibfnamefont {W.}~\bibnamefont
  {Detmold}},\ }\bibfield  {title} {\enquote {\bibinfo {title} {{Gluon
  gravitational form factors of the nucleon and the pion from lattice QCD}},}\
  }\href {\doibase 10.1103/PhysRevD.99.014511} {\bibfield  {journal} {\bibinfo
  {journal} {Phys. Rev. D}\ }\textbf {\bibinfo {volume} {99}},\ \bibinfo
  {pages} {014511} (\bibinfo {year} {2019})},\ \Eprint
  {http://arxiv.org/abs/1810.04626} {arXiv:1810.04626 [hep-lat]} \BibitemShut
  {NoStop}%
\bibitem [{\citenamefont {Pefkou}\ \emph {et~al.}(2022)\citenamefont {Pefkou},
  \citenamefont {Hackett},\ and\ \citenamefont {Shanahan}}]{Pefkou:2021fni}%
  \BibitemOpen
  \bibfield  {author} {\bibinfo {author} {\bibfnamefont {Dimitra~A.}\
  \bibnamefont {Pefkou}}, \bibinfo {author} {\bibfnamefont {Daniel~C.}\
  \bibnamefont {Hackett}}, \ and\ \bibinfo {author} {\bibfnamefont {Phiala~E.}\
  \bibnamefont {Shanahan}},\ }\bibfield  {title} {\enquote {\bibinfo {title}
  {{Gluon gravitational structure of hadrons of different spin}},}\ }\href
  {\doibase 10.1103/PhysRevD.105.054509} {\bibfield  {journal} {\bibinfo
  {journal} {Phys. Rev. D}\ }\textbf {\bibinfo {volume} {105}},\ \bibinfo
  {pages} {054509} (\bibinfo {year} {2022})},\ \Eprint
  {http://arxiv.org/abs/2107.10368} {arXiv:2107.10368 [hep-lat]} \BibitemShut
  {NoStop}%
\bibitem [{\citenamefont {Hou}\ \emph {et~al.}(2021)\citenamefont {Hou} \emph
  {et~al.}}]{Hou:2019efy}%
  \BibitemOpen
  \bibfield  {author} {\bibinfo {author} {\bibfnamefont {Tie-Jiun}\
  \bibnamefont {Hou}} \emph {et~al.},\ }\bibfield  {title} {\enquote {\bibinfo
  {title} {{New CTEQ global analysis of quantum chromodynamics with
  high-precision data from the LHC}},}\ }\href {\doibase
  10.1103/PhysRevD.103.014013} {\bibfield  {journal} {\bibinfo  {journal}
  {Phys. Rev. D}\ }\textbf {\bibinfo {volume} {103}},\ \bibinfo {pages}
  {014013} (\bibinfo {year} {2021})},\ \Eprint
  {http://arxiv.org/abs/1912.10053} {arXiv:1912.10053 [hep-ph]} \BibitemShut
  {NoStop}%
\bibitem [{\citenamefont {Schweitzer}(2004)}]{Schweitzer:2003sb}%
  \BibitemOpen
  \bibfield  {author} {\bibinfo {author} {\bibfnamefont {P.}~\bibnamefont
  {Schweitzer}},\ }\bibfield  {title} {\enquote {\bibinfo {title} {{The Sigma
  term form-factor of the nucleon in the large N(C) limit}},}\ }\href {\doibase
  10.1103/PhysRevD.69.034003} {\bibfield  {journal} {\bibinfo  {journal} {Phys.
  Rev. D}\ }\textbf {\bibinfo {volume} {69}},\ \bibinfo {pages} {034003}
  (\bibinfo {year} {2004})},\ \Eprint {http://arxiv.org/abs/hep-ph/0307336}
  {arXiv:hep-ph/0307336} \BibitemShut {NoStop}%
\bibitem [{\citenamefont {Wakamatsu}(2007)}]{Wakamatsu:2007uc}%
  \BibitemOpen
  \bibfield  {author} {\bibinfo {author} {\bibfnamefont {M.}~\bibnamefont
  {Wakamatsu}},\ }\bibfield  {title} {\enquote {\bibinfo {title} {{On the
  D-term of the nucleon generalized parton distributions}},}\ }\href {\doibase
  10.1016/j.physletb.2007.03.013} {\bibfield  {journal} {\bibinfo  {journal}
  {Phys. Lett. B}\ }\textbf {\bibinfo {volume} {648}},\ \bibinfo {pages}
  {181--185} (\bibinfo {year} {2007})},\ \Eprint
  {http://arxiv.org/abs/hep-ph/0701057} {arXiv:hep-ph/0701057} \BibitemShut
  {NoStop}%
\bibitem [{\citenamefont {Cebulla}\ \emph {et~al.}(2007)\citenamefont
  {Cebulla}, \citenamefont {Goeke}, \citenamefont {Ossmann},\ and\
  \citenamefont {Schweitzer}}]{Cebulla:2007ei}%
  \BibitemOpen
  \bibfield  {author} {\bibinfo {author} {\bibfnamefont {C.}~\bibnamefont
  {Cebulla}}, \bibinfo {author} {\bibfnamefont {K.}~\bibnamefont {Goeke}},
  \bibinfo {author} {\bibfnamefont {J.}~\bibnamefont {Ossmann}}, \ and\
  \bibinfo {author} {\bibfnamefont {P.}~\bibnamefont {Schweitzer}},\ }\bibfield
   {title} {\enquote {\bibinfo {title} {{The Nucleon form-factors of the energy
  momentum tensor in the Skyrme model}},}\ }\href {\doibase
  10.1016/j.nuclphysa.2007.08.004} {\bibfield  {journal} {\bibinfo  {journal}
  {Nucl. Phys. A}\ }\textbf {\bibinfo {volume} {794}},\ \bibinfo {pages}
  {87--114} (\bibinfo {year} {2007})},\ \Eprint
  {http://arxiv.org/abs/hep-ph/0703025} {arXiv:hep-ph/0703025} \BibitemShut
  {NoStop}%
\bibitem [{\citenamefont {Diakonov}(2002)}]{Diakonov:2002mb}%
  \BibitemOpen
  \bibfield  {author} {\bibinfo {author} {\bibfnamefont {Dmitri}\ \bibnamefont
  {Diakonov}},\ }\bibfield  {title} {\enquote {\bibinfo {title} {{Instantons
  and baryon dynamics}},}\ }in\ \href {\doibase 10.1142/9789812704887_0013}
  {\emph {\bibinfo {booktitle} {{9th International Conference on the Structure
  of Baryons}}}}\ (\bibinfo {year} {2002})\ pp.\ \bibinfo {pages} {153--164},\
  \Eprint {http://arxiv.org/abs/hep-ph/0205054} {arXiv:hep-ph/0205054}
  \BibitemShut {NoStop}%
\bibitem [{\citenamefont {Burkert}\ \emph {et~al.}(2018)\citenamefont
  {Burkert}, \citenamefont {Elouadrhiri},\ and\ \citenamefont
  {Girod}}]{Burkert:2018bqq}%
  \BibitemOpen
  \bibfield  {author} {\bibinfo {author} {\bibfnamefont {V.~D.}\ \bibnamefont
  {Burkert}}, \bibinfo {author} {\bibfnamefont {L.}~\bibnamefont
  {Elouadrhiri}}, \ and\ \bibinfo {author} {\bibfnamefont {F.~X.}\ \bibnamefont
  {Girod}},\ }\bibfield  {title} {\enquote {\bibinfo {title} {{The pressure
  distribution inside the proton}},}\ }\href {\doibase
  10.1038/s41586-018-0060-z} {\bibfield  {journal} {\bibinfo  {journal}
  {Nature}\ }\textbf {\bibinfo {volume} {557}},\ \bibinfo {pages} {396--399}
  (\bibinfo {year} {2018})}\BibitemShut {NoStop}%
\bibitem [{\citenamefont {Polyakov}\ and\ \citenamefont
  {Schweitzer}(2018)}]{Polyakov:2018zvc}%
  \BibitemOpen
  \bibfield  {author} {\bibinfo {author} {\bibfnamefont {Maxim~V.}\
  \bibnamefont {Polyakov}}\ and\ \bibinfo {author} {\bibfnamefont {Peter}\
  \bibnamefont {Schweitzer}},\ }\bibfield  {title} {\enquote {\bibinfo {title}
  {{Forces inside hadrons: pressure, surface tension, mechanical radius, and
  all that}},}\ }\href {\doibase 10.1142/S0217751X18300259} {\bibfield
  {journal} {\bibinfo  {journal} {Int. J. Mod. Phys. A}\ }\textbf {\bibinfo
  {volume} {33}},\ \bibinfo {pages} {1830025} (\bibinfo {year} {2018})},\
  \Eprint {http://arxiv.org/abs/1805.06596} {arXiv:1805.06596 [hep-ph]}
  \BibitemShut {NoStop}%
\bibitem [{\citenamefont {Steele}\ \emph {et~al.}(1995)\citenamefont {Steele},
  \citenamefont {Yamagishi},\ and\ \citenamefont {Zahed}}]{Steele:1995yr}%
  \BibitemOpen
  \bibfield  {author} {\bibinfo {author} {\bibfnamefont {James~V.}\
  \bibnamefont {Steele}}, \bibinfo {author} {\bibfnamefont {Hidenaga}\
  \bibnamefont {Yamagishi}}, \ and\ \bibinfo {author} {\bibfnamefont {Ismail}\
  \bibnamefont {Zahed}},\ }\bibfield  {title} {\enquote {\bibinfo {title} {{The
  Pion - nucleon sigma term and the Goldberger-Treiman discrepancy}},}\
  }\href@noop {} {\  (\bibinfo {year} {1995})},\ \Eprint
  {http://arxiv.org/abs/hep-ph/9512233} {arXiv:hep-ph/9512233} \BibitemShut
  {NoStop}%
\bibitem [{\citenamefont {Ji}(1995{\natexlab{b}})}]{Ji:1994av}%
  \BibitemOpen
  \bibfield  {author} {\bibinfo {author} {\bibfnamefont {Xiang-Dong}\
  \bibnamefont {Ji}},\ }\bibfield  {title} {\enquote {\bibinfo {title} {{A QCD
  analysis of the mass structure of the nucleon}},}\ }\href {\doibase
  10.1103/PhysRevLett.74.1071} {\bibfield  {journal} {\bibinfo  {journal}
  {Phys. Rev. Lett.}\ }\textbf {\bibinfo {volume} {74}},\ \bibinfo {pages}
  {1071--1074} (\bibinfo {year} {1995}{\natexlab{b}})},\ \Eprint
  {http://arxiv.org/abs/hep-ph/9410274} {arXiv:hep-ph/9410274} \BibitemShut
  {NoStop}%
\bibitem [{\citenamefont {Lorc\'e}(2018)}]{Lorce:2017xzd}%
  \BibitemOpen
  \bibfield  {author} {\bibinfo {author} {\bibfnamefont {C\'edric}\
  \bibnamefont {Lorc\'e}},\ }\bibfield  {title} {\enquote {\bibinfo {title}
  {{On the hadron mass decomposition}},}\ }\href {\doibase
  10.1140/epjc/s10052-018-5561-2} {\bibfield  {journal} {\bibinfo  {journal}
  {Eur. Phys. J. C}\ }\textbf {\bibinfo {volume} {78}},\ \bibinfo {pages} {120}
  (\bibinfo {year} {2018})},\ \Eprint {http://arxiv.org/abs/1706.05853}
  {arXiv:1706.05853 [hep-ph]} \BibitemShut {NoStop}%
\bibitem [{\citenamefont {Roberts}(2021)}]{Roberts:2021xnz}%
  \BibitemOpen
  \bibfield  {author} {\bibinfo {author} {\bibfnamefont {Craig~D.}\
  \bibnamefont {Roberts}},\ }\bibfield  {title} {\enquote {\bibinfo {title}
  {{On Mass and Matter}},}\ }\href {\doibase 10.1007/s43673-021-00005-4}
  {\bibfield  {journal} {\bibinfo  {journal} {AAPPS Bull.}\ }\textbf {\bibinfo
  {volume} {31}},\ \bibinfo {pages} {6} (\bibinfo {year} {2021})},\ \Eprint
  {http://arxiv.org/abs/2101.08340} {arXiv:2101.08340 [hep-ph]} \BibitemShut
  {NoStop}%
\bibitem [{\citenamefont {Metz}\ \emph {et~al.}(2021)\citenamefont {Metz},
  \citenamefont {Pasquini},\ and\ \citenamefont {Rodini}}]{Metz:2020vxd}%
  \BibitemOpen
  \bibfield  {author} {\bibinfo {author} {\bibfnamefont {Andreas}\ \bibnamefont
  {Metz}}, \bibinfo {author} {\bibfnamefont {Barbara}\ \bibnamefont
  {Pasquini}}, \ and\ \bibinfo {author} {\bibfnamefont {Simone}\ \bibnamefont
  {Rodini}},\ }\bibfield  {title} {\enquote {\bibinfo {title} {{Revisiting the
  proton mass decomposition}},}\ }\href {\doibase 10.1103/PhysRevD.102.114042}
  {\bibfield  {journal} {\bibinfo  {journal} {Phys. Rev. D}\ }\textbf {\bibinfo
  {volume} {102}},\ \bibinfo {pages} {114042} (\bibinfo {year} {2021})},\
  \Eprint {http://arxiv.org/abs/2006.11171} {arXiv:2006.11171 [hep-ph]}
  \BibitemShut {NoStop}%
\bibitem [{\citenamefont {Deur}\ \emph {et~al.}(2018)\citenamefont {Deur},
  \citenamefont {Brodsky},\ and\ \citenamefont {De~T\'eramond}}]{Deur:2018roz}%
  \BibitemOpen
  \bibfield  {author} {\bibinfo {author} {\bibfnamefont {Alexandre}\
  \bibnamefont {Deur}}, \bibinfo {author} {\bibfnamefont {Stanley~J.}\
  \bibnamefont {Brodsky}}, \ and\ \bibinfo {author} {\bibfnamefont {Guy~F.}\
  \bibnamefont {De~T\'eramond}},\ }\bibfield  {title} {\enquote {\bibinfo
  {title} {{The Spin Structure of the Nucleon}},}\ }\href {\doibase
  10.1088/1361-6633/ab0b8f} {\  (\bibinfo {year} {2018}),\
  10.1088/1361-6633/ab0b8f},\ \Eprint {http://arxiv.org/abs/1807.05250}
  {arXiv:1807.05250 [hep-ph]} \BibitemShut {NoStop}%
\bibitem [{\citenamefont {Ji}(1997)}]{Ji:1996ek}%
  \BibitemOpen
  \bibfield  {author} {\bibinfo {author} {\bibfnamefont {Xiang-Dong}\
  \bibnamefont {Ji}},\ }\bibfield  {title} {\enquote {\bibinfo {title}
  {{Gauge-Invariant Decomposition of Nucleon Spin}},}\ }\href {\doibase
  10.1103/PhysRevLett.78.610} {\bibfield  {journal} {\bibinfo  {journal} {Phys.
  Rev. Lett.}\ }\textbf {\bibinfo {volume} {78}},\ \bibinfo {pages} {610--613}
  (\bibinfo {year} {1997})},\ \Eprint {http://arxiv.org/abs/hep-ph/9603249}
  {arXiv:hep-ph/9603249} \BibitemShut {NoStop}%
\bibitem [{\citenamefont {Adler}(1969)}]{Adler:1969gk}%
  \BibitemOpen
  \bibfield  {author} {\bibinfo {author} {\bibfnamefont {Stephen~L.}\
  \bibnamefont {Adler}},\ }\bibfield  {title} {\enquote {\bibinfo {title}
  {{Axial vector vertex in spinor electrodynamics}},}\ }\href {\doibase
  10.1103/PhysRev.177.2426} {\bibfield  {journal} {\bibinfo  {journal} {Phys.
  Rev.}\ }\textbf {\bibinfo {volume} {177}},\ \bibinfo {pages} {2426--2438}
  (\bibinfo {year} {1969})}\BibitemShut {NoStop}%
\bibitem [{\citenamefont {Ashman}\ \emph {et~al.}(1989)\citenamefont {Ashman}
  \emph {et~al.}}]{EuropeanMuon:1989yki}%
  \BibitemOpen
  \bibfield  {author} {\bibinfo {author} {\bibfnamefont {J.}~\bibnamefont
  {Ashman}} \emph {et~al.} (\bibinfo {collaboration} {European Muon}),\
  }\bibfield  {title} {\enquote {\bibinfo {title} {{An Investigation of the
  Spin Structure of the Proton in Deep Inelastic Scattering of Polarized Muons
  on Polarized Protons}},}\ }\href {\doibase 10.1016/0550-3213(89)90089-8}
  {\bibfield  {journal} {\bibinfo  {journal} {Nucl. Phys. B}\ }\textbf
  {\bibinfo {volume} {328}},\ \bibinfo {pages} {1} (\bibinfo {year}
  {1989})}\BibitemShut {NoStop}%
\bibitem [{\citenamefont {Brodsky}\ \emph {et~al.}(1988)\citenamefont
  {Brodsky}, \citenamefont {Ellis},\ and\ \citenamefont
  {Karliner}}]{Brodsky:1988ip}%
  \BibitemOpen
  \bibfield  {author} {\bibinfo {author} {\bibfnamefont {Stanley~J.}\
  \bibnamefont {Brodsky}}, \bibinfo {author} {\bibfnamefont {John~R.}\
  \bibnamefont {Ellis}}, \ and\ \bibinfo {author} {\bibfnamefont {Marek}\
  \bibnamefont {Karliner}},\ }\bibfield  {title} {\enquote {\bibinfo {title}
  {{Chiral Symmetry and the Spin of the Proton}},}\ }\href {\doibase
  10.1016/0370-2693(88)91511-0} {\bibfield  {journal} {\bibinfo  {journal}
  {Phys. Lett. B}\ }\textbf {\bibinfo {volume} {206}},\ \bibinfo {pages}
  {309--315} (\bibinfo {year} {1988})}\BibitemShut {NoStop}%
\bibitem [{\citenamefont {Ellis}\ and\ \citenamefont
  {Karliner}(1995)}]{Ellis:1995jx}%
  \BibitemOpen
  \bibfield  {author} {\bibinfo {author} {\bibfnamefont {John~R.}\ \bibnamefont
  {Ellis}}\ and\ \bibinfo {author} {\bibfnamefont {Marek}\ \bibnamefont
  {Karliner}},\ }\bibfield  {title} {\enquote {\bibinfo {title} {{Nucleon
  spin}},}\ }in\ \href@noop {} {\emph {\bibinfo {booktitle} {{Workshop on the
  Prospects of Spin Physics at HERA}}}}\ (\bibinfo {year} {1995})\ pp.\
  \bibinfo {pages} {141--152},\ \Eprint {http://arxiv.org/abs/hep-ph/9510402}
  {arXiv:hep-ph/9510402} \BibitemShut {NoStop}%
\bibitem [{\citenamefont {Alexakhin}\ \emph {et~al.}(2007)\citenamefont
  {Alexakhin} \emph {et~al.}}]{COMPASS:2006mhr}%
  \BibitemOpen
  \bibfield  {author} {\bibinfo {author} {\bibfnamefont {V.~Yu.}\ \bibnamefont
  {Alexakhin}} \emph {et~al.} (\bibinfo {collaboration} {COMPASS}),\ }\bibfield
   {title} {\enquote {\bibinfo {title} {{The Deuteron Spin-dependent Structure
  Function g1(d) and its First Moment}},}\ }\href {\doibase
  10.1016/j.physletb.2006.12.076} {\bibfield  {journal} {\bibinfo  {journal}
  {Phys. Lett. B}\ }\textbf {\bibinfo {volume} {647}},\ \bibinfo {pages}
  {8--17} (\bibinfo {year} {2007})},\ \Eprint
  {http://arxiv.org/abs/hep-ex/0609038} {arXiv:hep-ex/0609038} \BibitemShut
  {NoStop}%
\bibitem [{\citenamefont {Airapetian}\ \emph {et~al.}(2007)\citenamefont
  {Airapetian} \emph {et~al.}}]{HERMES:2006jyl}%
  \BibitemOpen
  \bibfield  {author} {\bibinfo {author} {\bibfnamefont {A.}~\bibnamefont
  {Airapetian}} \emph {et~al.} (\bibinfo {collaboration} {HERMES}),\ }\bibfield
   {title} {\enquote {\bibinfo {title} {{Precise determination of the spin
  structure function g(1) of the proton, deuteron and neutron}},}\ }\href
  {\doibase 10.1103/PhysRevD.75.012007} {\bibfield  {journal} {\bibinfo
  {journal} {Phys. Rev. D}\ }\textbf {\bibinfo {volume} {75}},\ \bibinfo
  {pages} {012007} (\bibinfo {year} {2007})},\ \Eprint
  {http://arxiv.org/abs/hep-ex/0609039} {arXiv:hep-ex/0609039} \BibitemShut
  {NoStop}%
\bibitem [{\citenamefont {Alexandrou}\ \emph
  {et~al.}(2021{\natexlab{a}})\citenamefont {Alexandrou}, \citenamefont
  {Constantinou}, \citenamefont {Hadjiyiannakou}, \citenamefont {Jansen},\ and\
  \citenamefont {Manigrasso}}]{Alexandrou:2021oih}%
  \BibitemOpen
  \bibfield  {author} {\bibinfo {author} {\bibfnamefont {Constantia}\
  \bibnamefont {Alexandrou}}, \bibinfo {author} {\bibfnamefont {Martha}\
  \bibnamefont {Constantinou}}, \bibinfo {author} {\bibfnamefont {Kyriakos}\
  \bibnamefont {Hadjiyiannakou}}, \bibinfo {author} {\bibfnamefont {Karl}\
  \bibnamefont {Jansen}}, \ and\ \bibinfo {author} {\bibfnamefont {Floriano}\
  \bibnamefont {Manigrasso}},\ }\bibfield  {title} {\enquote {\bibinfo {title}
  {{Flavor decomposition of the nucleon unpolarized, helicity, and transversity
  parton distribution functions from lattice QCD simulations}},}\ }\href
  {\doibase 10.1103/PhysRevD.104.054503} {\bibfield  {journal} {\bibinfo
  {journal} {Phys. Rev. D}\ }\textbf {\bibinfo {volume} {104}},\ \bibinfo
  {pages} {054503} (\bibinfo {year} {2021}{\natexlab{a}})},\ \Eprint
  {http://arxiv.org/abs/2106.16065} {arXiv:2106.16065 [hep-lat]} \BibitemShut
  {NoStop}%
\bibitem [{\citenamefont {Alexandrou}\ \emph
  {et~al.}(2021{\natexlab{b}})\citenamefont {Alexandrou}, \citenamefont
  {Bacchio}, \citenamefont {Constantinou}, \citenamefont {Hadjiyiannakou},
  \citenamefont {Jansen},\ and\ \citenamefont {Koutsou}}]{Alexandrou:2021wzv}%
  \BibitemOpen
  \bibfield  {author} {\bibinfo {author} {\bibfnamefont {C.}~\bibnamefont
  {Alexandrou}}, \bibinfo {author} {\bibfnamefont {S.}~\bibnamefont {Bacchio}},
  \bibinfo {author} {\bibfnamefont {M.}~\bibnamefont {Constantinou}}, \bibinfo
  {author} {\bibfnamefont {K.}~\bibnamefont {Hadjiyiannakou}}, \bibinfo
  {author} {\bibfnamefont {K.}~\bibnamefont {Jansen}}, \ and\ \bibinfo {author}
  {\bibfnamefont {G.}~\bibnamefont {Koutsou}},\ }\bibfield  {title} {\enquote
  {\bibinfo {title} {{Quark flavor decomposition of the nucleon axial form
  factors}},}\ }\href {\doibase 10.1103/PhysRevD.104.074503} {\bibfield
  {journal} {\bibinfo  {journal} {Phys. Rev. D}\ }\textbf {\bibinfo {volume}
  {104}},\ \bibinfo {pages} {074503} (\bibinfo {year} {2021}{\natexlab{b}})},\
  \Eprint {http://arxiv.org/abs/2106.13468} {arXiv:2106.13468 [hep-lat]}
  \BibitemShut {NoStop}%
\bibitem [{\citenamefont {Liu}(2016)}]{Liu:2015nva}%
  \BibitemOpen
  \bibfield  {author} {\bibinfo {author} {\bibfnamefont {Keh-Fei}\ \bibnamefont
  {Liu}},\ }\bibfield  {title} {\enquote {\bibinfo {title} {{Quark and Glue
  Components of the Proton Spin from Lattice Calculation}},}\ }\href {\doibase
  10.1142/S2010194516600053} {\bibfield  {journal} {\bibinfo  {journal} {Int.
  J. Mod. Phys. Conf. Ser.}\ }\textbf {\bibinfo {volume} {40}},\ \bibinfo
  {pages} {1660005} (\bibinfo {year} {2016})},\ \Eprint
  {http://arxiv.org/abs/1504.06601} {arXiv:1504.06601 [hep-ph]} \BibitemShut
  {NoStop}%
\bibitem [{\citenamefont {Anselmino}\ \emph {et~al.}(1993)\citenamefont
  {Anselmino}, \citenamefont {Predazzi}, \citenamefont {Ekelin}, \citenamefont
  {Fredriksson},\ and\ \citenamefont {Lichtenberg}}]{Anselmino:1992vg}%
  \BibitemOpen
  \bibfield  {author} {\bibinfo {author} {\bibfnamefont {Mauro}\ \bibnamefont
  {Anselmino}}, \bibinfo {author} {\bibfnamefont {Enrico}\ \bibnamefont
  {Predazzi}}, \bibinfo {author} {\bibfnamefont {Svante}\ \bibnamefont
  {Ekelin}}, \bibinfo {author} {\bibfnamefont {Sverker}\ \bibnamefont
  {Fredriksson}}, \ and\ \bibinfo {author} {\bibfnamefont {D.~B.}\ \bibnamefont
  {Lichtenberg}},\ }\bibfield  {title} {\enquote {\bibinfo {title}
  {{Diquarks}},}\ }\href {\doibase 10.1103/RevModPhys.65.1199} {\bibfield
  {journal} {\bibinfo  {journal} {Rev. Mod. Phys.}\ }\textbf {\bibinfo {volume}
  {65}},\ \bibinfo {pages} {1199--1234} (\bibinfo {year} {1993})}\BibitemShut
  {NoStop}%
\bibitem [{\citenamefont {Alkofer}\ \emph {et~al.}(1989)\citenamefont
  {Alkofer}, \citenamefont {Nowak}, \citenamefont {Verbaarschot},\ and\
  \citenamefont {Zahed}}]{Alkofer:1989uj}%
  \BibitemOpen
  \bibfield  {author} {\bibinfo {author} {\bibfnamefont {Reinhard}\
  \bibnamefont {Alkofer}}, \bibinfo {author} {\bibfnamefont {Maciej~A.}\
  \bibnamefont {Nowak}}, \bibinfo {author} {\bibfnamefont {J.~J.~M.}\
  \bibnamefont {Verbaarschot}}, \ and\ \bibinfo {author} {\bibfnamefont
  {I.}~\bibnamefont {Zahed}},\ }\bibfield  {title} {\enquote {\bibinfo {title}
  {{Pseudoscalars in the Instanton Liquid Model}},}\ }\href {\doibase
  10.1016/0370-2693(89)90643-6} {\bibfield  {journal} {\bibinfo  {journal}
  {Phys. Lett. B}\ }\textbf {\bibinfo {volume} {233}},\ \bibinfo {pages}
  {205--209} (\bibinfo {year} {1989})}\BibitemShut {NoStop}%
\bibitem [{\citenamefont {Luscher}(2010)}]{Luscher:2009eq}%
  \BibitemOpen
  \bibfield  {author} {\bibinfo {author} {\bibfnamefont {Martin}\ \bibnamefont
  {Luscher}},\ }\bibfield  {title} {\enquote {\bibinfo {title} {{Trivializing
  maps, the Wilson flow and the HMC algorithm}},}\ }\href {\doibase
  10.1007/s00220-009-0953-7} {\bibfield  {journal} {\bibinfo  {journal}
  {Commun. Math. Phys.}\ }\textbf {\bibinfo {volume} {293}},\ \bibinfo {pages}
  {899--919} (\bibinfo {year} {2010})},\ \Eprint
  {http://arxiv.org/abs/0907.5491} {arXiv:0907.5491 [hep-lat]} \BibitemShut
  {NoStop}%
\bibitem [{\citenamefont {Luscher}\ and\ \citenamefont
  {Weisz}(2011)}]{Luscher:2011bx}%
  \BibitemOpen
  \bibfield  {author} {\bibinfo {author} {\bibfnamefont {Martin}\ \bibnamefont
  {Luscher}}\ and\ \bibinfo {author} {\bibfnamefont {Peter}\ \bibnamefont
  {Weisz}},\ }\bibfield  {title} {\enquote {\bibinfo {title} {{Perturbative
  analysis of the gradient flow in non-abelian gauge theories}},}\ }\href
  {\doibase 10.1007/JHEP02(2011)051} {\bibfield  {journal} {\bibinfo  {journal}
  {JHEP}\ }\textbf {\bibinfo {volume} {02}},\ \bibinfo {pages} {051} (\bibinfo
  {year} {2011})},\ \Eprint {http://arxiv.org/abs/1101.0963} {arXiv:1101.0963
  [hep-th]} \BibitemShut {NoStop}%
\bibitem [{\citenamefont {Dimopoulos}\ \emph {et~al.}(2019)\citenamefont
  {Dimopoulos} \emph {et~al.}}]{Dimopoulos:2018xkm}%
  \BibitemOpen
  \bibfield  {author} {\bibinfo {author} {\bibfnamefont {Petros}\ \bibnamefont
  {Dimopoulos}} \emph {et~al.},\ }\bibfield  {title} {\enquote {\bibinfo
  {title} {{Topological susceptibility and $\eta'$ meson mass from $N_f=2$
  lattice QCD at the physical point}},}\ }\href {\doibase
  10.1103/PhysRevD.99.034511} {\bibfield  {journal} {\bibinfo  {journal} {Phys.
  Rev. D}\ }\textbf {\bibinfo {volume} {99}},\ \bibinfo {pages} {034511}
  (\bibinfo {year} {2019})},\ \Eprint {http://arxiv.org/abs/1812.08787}
  {arXiv:1812.08787 [hep-lat]} \BibitemShut {NoStop}%
\bibitem [{\citenamefont {Creutz}(1978)}]{creutz1978invariant}%
  \BibitemOpen
  \bibfield  {author} {\bibinfo {author} {\bibfnamefont {Michael}\ \bibnamefont
  {Creutz}},\ }\bibfield  {title} {\enquote {\bibinfo {title} {On invariant
  integration over su (n)},}\ }\href@noop {} {\bibfield  {journal} {\bibinfo
  {journal} {Journal of Mathematical Physics}\ }\textbf {\bibinfo {volume}
  {19}},\ \bibinfo {pages} {2043--2046} (\bibinfo {year} {1978})}\BibitemShut
  {NoStop}%
\bibitem [{\citenamefont {Chernyshev}\ \emph {et~al.}(1996)\citenamefont
  {Chernyshev}, \citenamefont {Nowak},\ and\ \citenamefont
  {Zahed}}]{Chernyshev:1995gj}%
  \BibitemOpen
  \bibfield  {author} {\bibinfo {author} {\bibfnamefont {S.}~\bibnamefont
  {Chernyshev}}, \bibinfo {author} {\bibfnamefont {Maciej~A.}\ \bibnamefont
  {Nowak}}, \ and\ \bibinfo {author} {\bibfnamefont {I.}~\bibnamefont
  {Zahed}},\ }\bibfield  {title} {\enquote {\bibinfo {title} {{Heavy hadrons
  and QCD instantons}},}\ }\href {\doibase 10.1103/PhysRevD.53.5176} {\bibfield
   {journal} {\bibinfo  {journal} {Phys. Rev. D}\ }\textbf {\bibinfo {volume}
  {53}},\ \bibinfo {pages} {5176--5184} (\bibinfo {year} {1996})},\ \Eprint
  {http://arxiv.org/abs/hep-ph/9510326} {arXiv:hep-ph/9510326} \BibitemShut
  {NoStop}%
\bibitem [{\citenamefont {Nowak}\ \emph {et~al.}(1989)\citenamefont {Nowak},
  \citenamefont {Verbaarschot},\ and\ \citenamefont {Zahed}}]{nowak1989flavor}%
  \BibitemOpen
  \bibfield  {author} {\bibinfo {author} {\bibfnamefont {Maciej~A}\
  \bibnamefont {Nowak}}, \bibinfo {author} {\bibfnamefont {JJM}\ \bibnamefont
  {Verbaarschot}}, \ and\ \bibinfo {author} {\bibfnamefont {I}~\bibnamefont
  {Zahed}},\ }\bibfield  {title} {\enquote {\bibinfo {title} {Flavor mixing in
  the instanton vacuum},}\ }\href@noop {} {\bibfield  {journal} {\bibinfo
  {journal} {Nuclear Physics B}\ }\textbf {\bibinfo {volume} {324}},\ \bibinfo
  {pages} {1--33} (\bibinfo {year} {1989})}\BibitemShut {NoStop}%
\bibitem [{\citenamefont {Miesch}\ \emph {et~al.}(2023)\citenamefont {Miesch},
  \citenamefont {Shuryak},\ and\ \citenamefont {Zahed}}]{Miesch:2023hjt}%
  \BibitemOpen
  \bibfield  {author} {\bibinfo {author} {\bibfnamefont {Nicholas}\
  \bibnamefont {Miesch}}, \bibinfo {author} {\bibfnamefont {Edward}\
  \bibnamefont {Shuryak}}, \ and\ \bibinfo {author} {\bibfnamefont {Ismail}\
  \bibnamefont {Zahed}},\ }\bibfield  {title} {\enquote {\bibinfo {title}
  {{Baryons and tetraquarks using instanton-induced interactions}},}\
  }\href@noop {} {\  (\bibinfo {year} {2023})},\ \Eprint
  {http://arxiv.org/abs/2308.05638} {arXiv:2308.05638 [hep-ph]} \BibitemShut
  {NoStop}%
\end{thebibliography}%
\end{document}